# Pseudorandomness in Central Force Optimization

**Richard A. Formato[1]**

***Abstract:*** Central Force Optimization is a deterministic metaheuristic for an evolutionary algorithm that searches a decision space by flying probes whose trajectories are computed using a gravitational metaphor. CFO benefits substantially from the inclusion of a pseudorandom component (a numerical sequence that is precisely known by specification or calculation but otherwise arbitrary). The essential requirement is that the sequence is uncorrelated with the decision space topology, so that its effect is to pseudorandomly distribute probes throughout the landscape. While this process may appear to be similar to the randomness in an inherently stochastic algorithm, it is in fact fundamentally different because CFO remains deterministic at every step. Three pseudorandom methods are discussed (initial probe distribution, repositioning factor, and decision space adaptation). A sample problem is presented in detail and summary data included for a 23-function benchmark suite. CFO's performance is quite good compared to other highly developed, state-of-the-art algorithms.

***Ver. 2*** *(typographical errors in Step (c) of Fig. 1 and discussion of Fig. 8 corrected; clarification of definition of $\vec{R}_{best}$ ; error in Table 2 test function $f_{12}$ corrected; $f_{12}$ source code in Appendix 3, page 60 corrected).*

***3 February 2010***
***Saint Augustine, Florida***

# Pseudorandomness in Central Force Optimization

## Richard A. Formato[1]

## 1. Introduction

This note examines the role of pseudorandomness in Central Force Optimization. CFO is a deterministic Nature-inspired search and optimization metaheuristic for an evolutionary algorithm (EA) based on gravitational kinematics [1-10]. CFO analogizes Newton's mathematically precise laws of motion and gravity, so that its underlying equations are equally precise. The algorithm locates the global maxima of an *objective function* defined on a *decision space* $\Omega$ with unknown topology ("landscape"). CFO searches $\Omega$ by "flying" a group of "probes" whose trajectories are computed from two deterministic *equations of motion* at a series of discrete "time" steps (iterations). Details of the CFO metaheuristic are in Appendix 1.

CFO is fundamentally different from Nature-inspired EAs whose underlying equations are inherently stochastic. Particle Swarm Optimization (PSO) and Ant Colony Optimization (ACO) are examples. Their equations are formulated in terms of random variables, and removing randomness causes these algorithms to fail completely. By contrast, CFO's equations are inherently deterministic. Every CFO run with the same setup returns precisely the same values step-by-step throughout the entire run. Nevertheless, effective implementations substantially benefit from a "pseudorandom" component that enters the algorithm indirectly, not through its basic equations. Although pseudorandomness is not required in CFO, numerical experiments show that it is an important feature in effective implementations.

A *pseudorandom* variable is defined here as one whose value is precisely known but arbitrarily assigned. The value can be specified in advance (for example, an arbitrary sequence of numbers) or it can be calculated in a prescribed manner. The variable's "randomness" derives from the fact that its value is arbitrary and uncorrelated with $\Omega$'s topology, not that it is uncertain in the sense of a true random variable. A *random* variable's value is calculated from a probability distribution with successive calculations yielding different values that cannot be known in advance. This type of randomness is fundamentally different from CFO's pseudorandomness. Even when CFO includes pseudorandomness, it remains deterministic, always yielding the same result for runs with the same setup. Once a pseudorandom variable is specified, either explicitly or by calculation, its value is known with absolute precision, so that CFO's trajectory calculations are deterministic even in the presence of pseudorandomness. There are many ways pseudorandomness can be added to CFO; three simple methods are described here.

## 2. CFO Implementation

This note discusses a CFO implementation with pseudorandomness injected in the following ways: (1) the initial probe distribution; (2) the "repositioning factor;" and (3) changing the decision space boundaries. The algorithm is referred to as "CFO-PR." Every CFO run starts with a user-specified initial probe distribution (total number and locations in $\Omega$ at the beginning of the run, step 0). An arbitrary, variable initial probe distribution is a convenient way to inject pseudorandomness, the effect of which is to provide better sampling of $\Omega$'s topology than a static distribution. Each initial probe distribution in a set of distributions provides different information about $\Omega$'s landscape. As the results below show, certain ones perform much better than others.

The second method of injecting pseudorandomness is the use of a step-by-step variable repositioning factor $\Delta F_{rep} \leq F_{rep} \leq 1$ where $\Delta F_{rep}$ is the step increment. "Repositioning" refers to the process of retrieving a probe that has flown outside the decision space (discussed in detail in Appendix 1). A variable $F_{rep}$ has the effect of pseudorandomly distributing probes throughout $\Omega$, which provides better sampling of the decision space landscape as a run progresses.

---

**For** $N_p / N_d = 2$ to $8$ by $2$:

**For** $\gamma = \gamma_{start}$ to $\gamma_{stop}$ by $\Delta\gamma$:

(a.1)   Compute initial probe distribution.

(a.2)   Compute initial fitness matrix.

(a.3)   Assign initial probe accelerations.

(a.4)   Set initial $F_{rep}$.

**For** $j = 0$ to $N_t$ (or earlier termination criterion):

(b)   Compute probe position vectors,

   $\vec{R}_j^p, 1 \leq p \leq N_p$ [eq.(2)].

(c)   Retrieve errant probes ($1 \leq p \leq N_p$):

   If $\vec{R}_j^p \cdot \hat{e}_i < x_i^{min}$ ∴

   $\vec{R}_j^p \cdot \hat{e}_i = x_i^{min} + F_{rep}(\vec{R}_{j-1}^p \cdot \hat{e}_i - x_i^{min})$

   If $\vec{R}_j^p \cdot \hat{e}_i > x_i^{max}$ ∴

   $\vec{R}_j^p \cdot \hat{e}_i = x_i^{max} - F_{rep}(x_i^{max} - \vec{R}_{j-1}^p \cdot \hat{e}_i)$

(d)   Compute fitness matrix for current probe distribution.

(e)   Compute accelerations using current probe distribution and fitnesses [eq. (1)].

(f)   Increment $F_{rep}$ by $\Delta F_{rep}$:

   If $F_{rep} > 1 \therefore F_{rep} = \Delta F_{rep}$.

(g)   If $j \, MOD \, 20 = 0$ ∴

   Shrink $\Omega$ around $\vec{R}_{best}$.

**Next** $j$

**Next** $\gamma$

**Next** $N_p / N_d$

---

Fig. 1.  CFO-PR Pseudocode with Variable Initial
Probes and $F_{rep}$ and DS Adaptation

The third way pseudorandomness is injected is by shrinking the decision space around the best probe's location. This process coupled with variable $F_{rep}$ redistributes probes in the smaller $\Omega$ in an arbitrary but precise, hence pseudorandom, manner. The effect is to speed CFO's convergence, but at the risk of premature convergence [on an empirical basis, this issue does not appear to be significant using the procedures described below].

Pseudocode for CFO-PR appears in Fig. 1 and source code in Appendix 3. The inner time step loop ("$j$" loop) is common to all CFO implementations, but the two outer loops inject initial probe pseudorandomness. The $\gamma$ loop controls where initial probes are deployed, and the $N_p/N_d$ loop determines their number. These two parameters define the initial probe distribution, and the two loops together create a variable, pseudorandom distribution (see Appendix 1 for notation and equations). The variable $F_{rep}$ procedure appears in step (f) of the pseudocode. And the pseudorandom decision space adaptation is in step (g). $\Omega$'s boundaries shrink around the then best probe position vector every 20$^{th}$ step as discussed in detail below. A two-dimensional example is used to illustrate these techniques because it provides a concrete visualization of the different methods. In the actual CFO-PR implementation, of course, these techniques are generalized to the $N_d$-dimensional case as shown in the pseudocode and source code.

***Initial Probes:*** The manner in which initial probes are deployed using $\gamma$ is shown schematically in Fig. 2, which provides a 2-dimensional (2D) schematic representation of a variable initial probe distribution comprising an orthogonal array of $N_p/N_d$ probes per axis deployed uniformly on "probe lines" parallel to the coordinate axes that intersect at a point along $\Omega$'s principal diagonal. $N_d$ is $\Omega$'s dimensionality (here $N_d = 2$), and $x_i^{min} \leq x_i \leq x_i^{max}$, $1 \leq i \leq N_d$ define $\Omega$'s domain.

For illustrative purposes, Fig. 2 shows nine probes on each probe line. The lines, which are parallel to the $x_1$ and $x_2$ axes, intersect at a point on $\Omega$'s principal diagonal marked by position vector $\vec{D} = \vec{X}_{min} + \gamma(\vec{X}_{max} - \vec{X}_{min})$, where $\vec{X}_{min} = \sum_{i=1}^{N_d} x_i^{min}\hat{e}_i$ and $\vec{X}_{max} = \sum_{i=1}^{N_d} x_i^{max}\hat{e}_i$ are the diagonal's endpoint vectors. Parameter $0 \leq \gamma \leq 1$ specifies where along the diagonal the orthogonal probe array is placed by locating the probe lines' intersection point.

While Fig. 2 shows an equal number of probes on each line, a different number of probes per axis can be used instead. For example, if equal probe spacing were desired in a decision space with unequal boundaries, or if overlapping probes were to be excluded in a symmetrical space, then unequal numbers would be used. Unequal numbers also might be appropriate if *a priori* knowledge of $\Omega$'s landscape, however obtained, suggests denser sampling in one area. The initial probe distribution in Fig. 2 with variable $N_p/N_d$ was used for the CFO-PR runs reported here, but any number of other variable initial probe distributions could be used instead. The key idea is that the initial probe distribution must be pseudorandom, that is, arbitrary and therefore uncorrelated with the decision space landscape.

**Repositioning Factor:** A variable value for $F_{rep}$ also adds pseudorandomness. $F_{rep}$ starts at some arbitrary initial value that is incremented at each iteration by an arbitrary amount $\Delta F_{rep}$ so that $\Delta F_{rep} \le F_{rep} \le 1$. This errant probe retrieval scheme is an example of an arbitrary sequence of calculated numbers deterministically assigned to a CFO parameter. CFO's ability to search the decision space depends on where errant probes are reinserted in $\Omega$, and this process is pseudorandom in nature because $F_{rep}$ is deterministic but arbitrary and uncorrelated with $\Omega$'s topology. By placing errant probes pseudorandomly throughout the decision space, more information is developed about its topology as the run progresses. The scheme used here was empirically determined, and appears to work well across a wide range of objective functions. But, of course, there are many other procedures for setting $F_{rep}$'s value, some no doubt better than others.

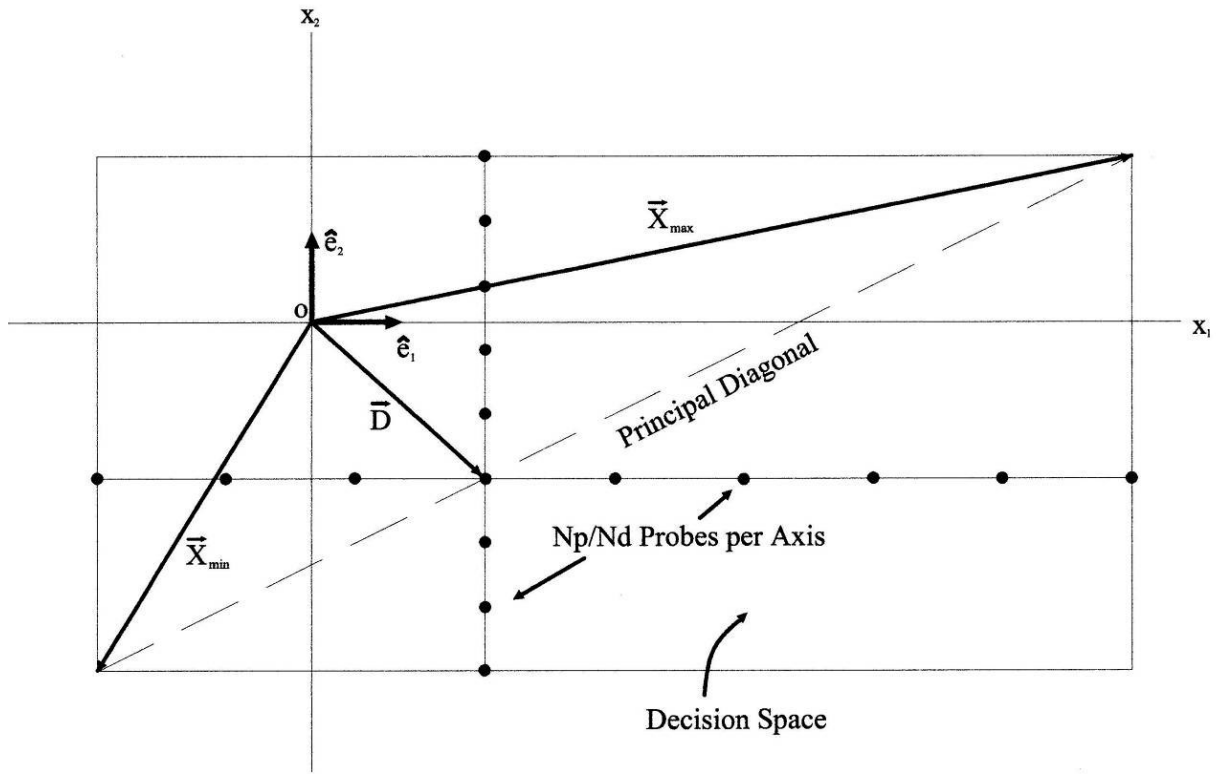

Fig. 2. Variable 2-D Initial Probe Distribution used for CFO Runs Reported Here

**Decision Space Adaptation:** CFO-PR also includes adaptive reconfiguration of the decision space in order to improve convergence speed. This feature also is pseudorandom in nature because the way $\Omega$'s boundaries are changed is arbitrary and uncorrelated with the landscape. Fig. 3 illustrates in 2D how $\Omega$'s size is adaptively reduced in this case every $20^{th}$ step around the probe's location with the then best fitness throughout the run up to the current iteration, $\vec{R}_{best}$. $\Omega$'s boundary coordinates are reduced by one-half the distance from the best probe's position to the each boundary on a coordinate-by-coordinate basis, that is,

$x_i'^{\min} = x_i^{\min} + \dfrac{\vec{R}_{best} \cdot \hat{e}_i - x_i^{\min}}{2}$ and $x_i'^{\max} = x_i^{\max} - \dfrac{x_i^{\max} - \vec{R}_{best} \cdot \hat{e}_i}{2}$, where the primed coordinate is the new decision space boundary, and the dot denotes vector inner product. For clarity, Fig. 3 shows $\vec{R}_{best}$ as fixed, whereas generally it varies throughout a run. Changing $\Omega$'s boundary every twenty steps instead of some other interval was chosen arbitrarily (another, probably better approach, might be a reactive adaptation based on performance measures such as convergence speed or fitness saturation).

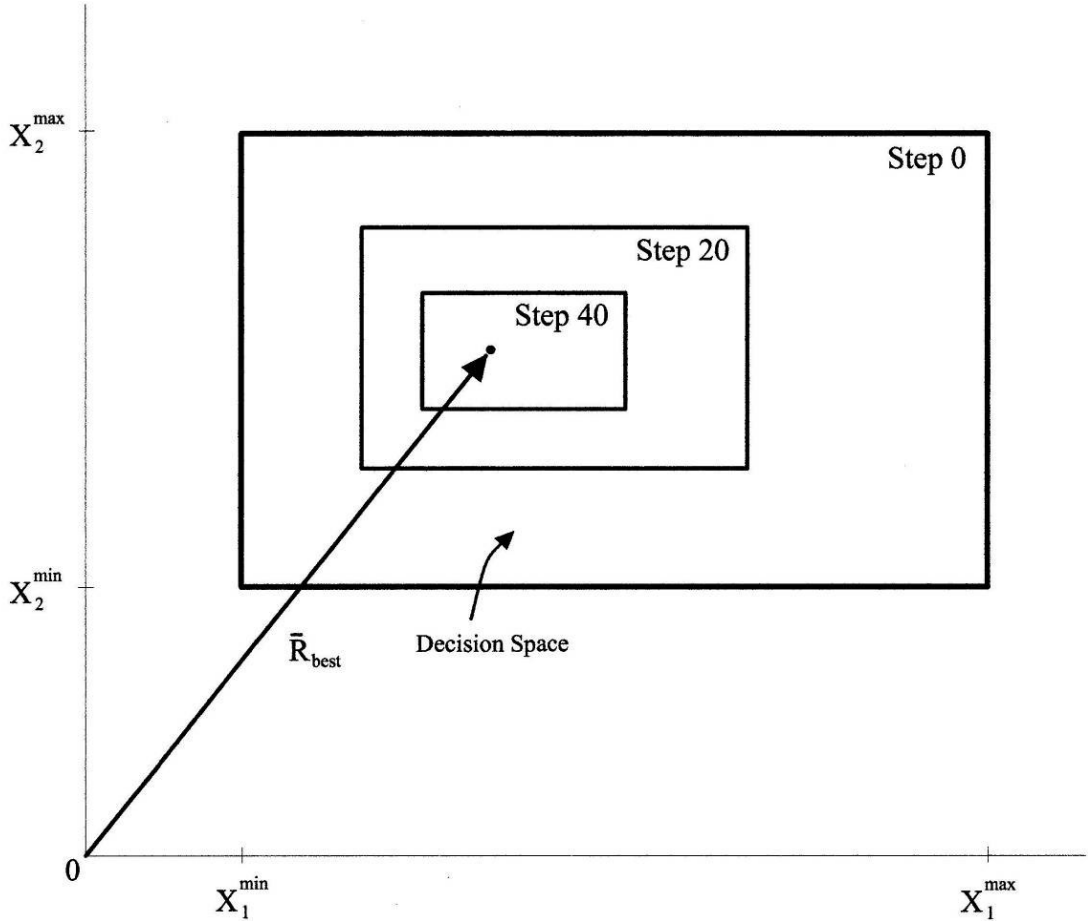

Fig. 3. Schematic 2-D Decision Space Adaptation (with constant $\vec{R}_{best}$)

## 3. A Sample Problem

The effectiveness of injecting pseudorandomness into CFO-PR will be illustrated with the 2D Goldstein-Price function ("GP") plotted in Fig. 4. GP is defined as

$$f(x_1, x_2) = -[1 + (x_1 + x_2 + 1)^2 \cdot (19 - 14x_1 + 3x_1^2 - 14x_2 + 6x_1x_2 + 3x_2^2)]$$
$$\times [30 + (2x_1 - 3x_2)^2 \cdot (18 - 32x_1 + 12x_1^2 + 48x_2 - 36x_1x_2 + 27x_2^2)]$$
,

where $\Omega$: $-100 \le x_1, x_2 \le 100$ [note that in most published reports $\Omega$ is much smaller, *viz.*, $-2 \le x_1, x_2 \le 2$]. GP's global maximum is $-3$ at $(0,-1)$. This function is multimodal with few local maxima, and it varies over nearly nineteen orders of magnitude as shown in Fig. 4.

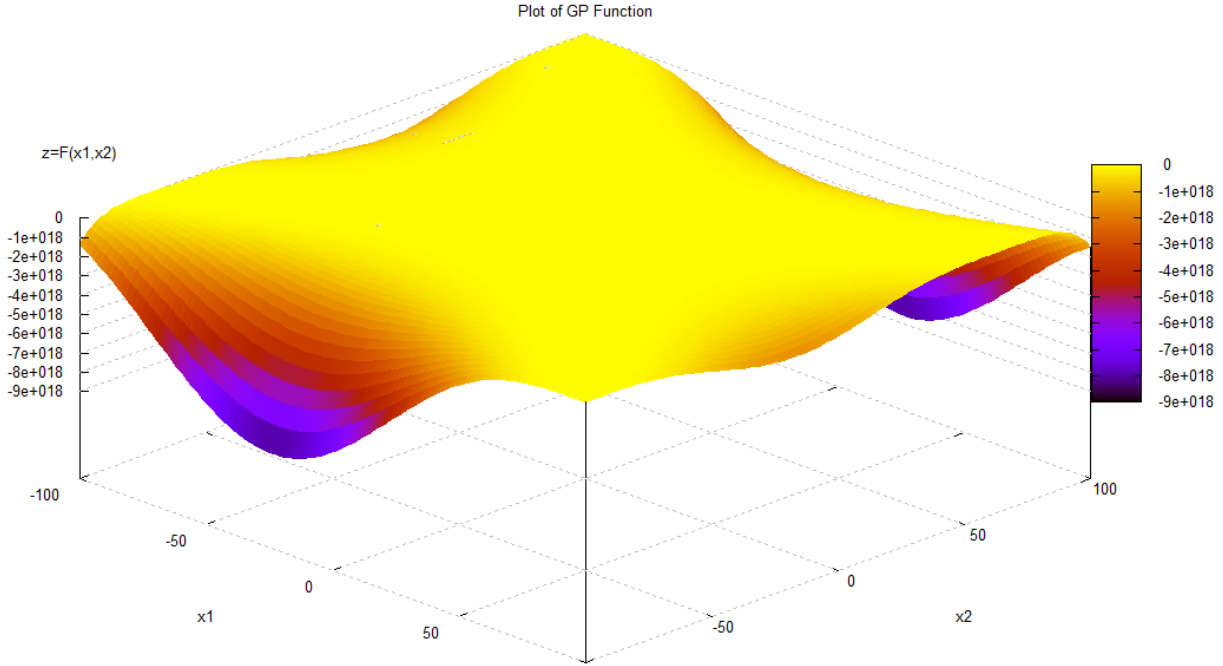

Fig. 4. Goldstein-Price Function

The following parameter values were used for all runs reported in this note: $\alpha = 2$, $\beta = 2$, $G = 2$, $\Delta t = 1$, initial acceleration of zero, initial $F_{rep} = 0.5$, $\Delta F_{rep} = 0.05$, $\gamma_{start} = 0$, $\gamma_{stop} = 1$ with $\Delta \gamma = 0.1$ (eleven runs). For GP, $N_p / N_d = 4$ to 14 by 2, with different ranges of this parameter for the other test functions as described below. In all cases, a run was terminated early if the average best fitness over 50 steps (including the current step) and the current best fitness differed by less than $10^{-6}$. The pseudorandom initial probes distributions computed using the procedure illustrated in Fig. 2 are plotted in Fig. 5.

Table 1 shows a summary of the results for the GP function. A total of sixty-six optimization runs were made in six groups of eleven runs each [data for "Run #0" are starting values]. The column headings are for the most part self-explanatory. Each run began with $N_t = 500$, but, as the *#Steps* column shows, in no case were 500 iterations used because every run terminated early. $N_{eval}$ is the number of function evaluations performed for the shortened run, and the total number of evaluations over all runs appears at the bottom of this column. The $F_{rep}$ column lists $F_{rep}$'s value at the end of the run, and the

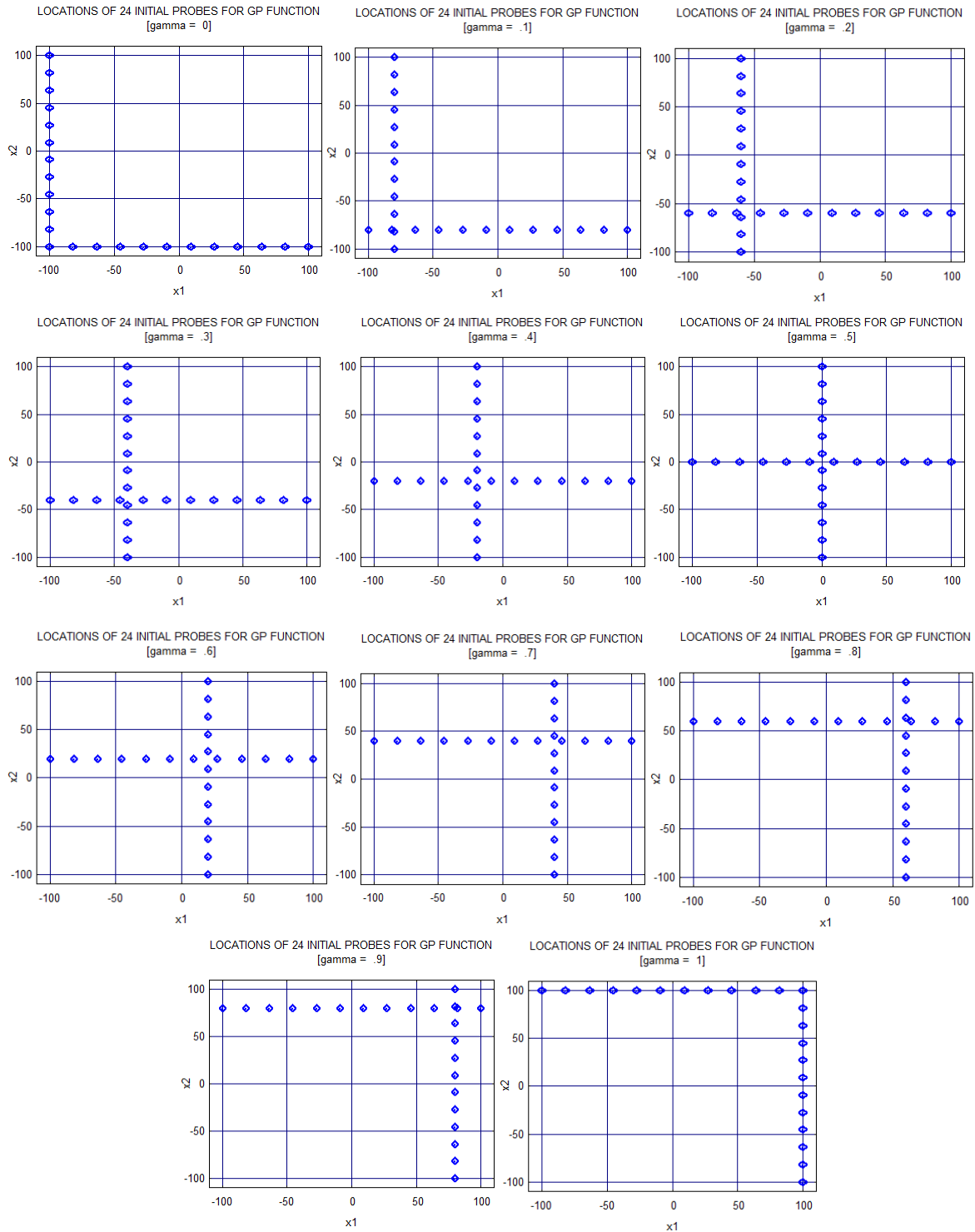

Fig. 5.  GP Best Run Initial Probe Distributions

```
Run ID: 12-26-2009, 14:02:08

FUNCTION: GP
```

| Run # | Gamma | Nt | Nd | Np | G | DelT | Alpha | Beta | #Steps | Neval | Frep | Fitness | Initial Probes |
|---|---|---|---|---|---|---|---|---|---|---|---|---|---|
| 0 | 0.000 | 500 | 2 | 2 | 1.0 | 1.0 | 2.00 | | 0 | 0 | 0.50000V | -9999.00000000 | UNIFORM I-AXIS |
| 1 | 0.000 | 500 | 2 | 8 | 2.0 | 1.0 | 2.00 | 2.00 | 78 | 632 | 0.60000V | -5.48169471 | UNIFORM I-AXIS |
| 2 | 0.100 | 500 | 2 | 8 | 2.0 | 1.0 | 2.00 | 2.00 | 134 | 1080 | 0.55000V | -84.78003234 | UNIFORM I-AXIS |
| 3 | 0.200 | 500 | 2 | 8 | 2.0 | 1.0 | 2.00 | 2.00 | 191 | 1536 | 0.55000V | -1.19822531 | UNIFORM I-AXIS |
| 4 | 0.300 | 500 | 2 | 8 | 2.0 | 1.0 | 2.00 | 2.00 | 79 | 640 | 0.65000V | -8.62285433 | UNIFORM I-AXIS |
| 5 | 0.400 | 500 | 2 | 8 | 2.0 | 1.0 | 2.00 | 2.00 | 403 | 3232 | 0.70000V | -84.47139703 | UNIFORM I-AXIS |
| 6 | 0.500 | 500 | 2 | 8 | 2.0 | 1.0 | 2.00 | 2.00 | 261 | 2096 | 0.25000V | -10.56054743 | UNIFORM I-AXIS |
| 7 | 0.600 | 500 | 2 | 8 | 2.0 | 1.0 | 2.00 | 2.00 | 135 | 1088 | 0.60000V | -4.98190912 | UNIFORM I-AXIS |
| 8 | 0.700 | 500 | 2 | 8 | 2.0 | 1.0 | 2.00 | 2.00 | 229 | 1840 | 0.55000V | -3.00130536 | UNIFORM I-AXIS |
| 9 | 0.800 | 500 | 2 | 8 | 2.0 | 1.0 | 2.00 | 2.00 | 133 | 1072 | 0.50000V | -89.42718008 | UNIFORM I-AXIS |
| 10 | 0.900 | 500 | 2 | 8 | 2.0 | 1.0 | 2.00 | 2.00 | 116 | 936 | 0.60000V | -8.57642421 | UNIFORM I-AXIS |
| 11 | 1.000 | 500 | 2 | 8 | 2.0 | 1.0 | 2.00 | 2.00 | 193 | 1552 | 0.65000V | -3.08573605 | UNIFORM I-AXIS |
| 12 | 0.000 | 500 | 2 | 12 | 2.0 | 1.0 | 2.00 | 2.00 | 118 | 948 | 0.60000V | -236.66437724 | UNIFORM I-AXIS |
| 13 | 0.100 | 500 | 2 | 12 | 2.0 | 1.0 | 2.00 | 2.00 | 263 | 3168 | 0.35000V | -3.00574349 | UNIFORM I-AXIS |
| 14 | 0.200 | 500 | 2 | 12 | 2.0 | 1.0 | 2.00 | 2.00 | 78 | 948 | 0.65000V | -57.61728445 | UNIFORM I-AXIS |
| 15 | 0.300 | 500 | 2 | 12 | 2.0 | 1.0 | 2.00 | 2.00 | 304 | 3660 | 0.55000V | -3.00017384 | UNIFORM I-AXIS |
| 16 | 0.400 | 500 | 2 | 12 | 2.0 | 1.0 | 2.00 | 2.00 | 154 | 1860 | 0.60000V | -3.55376552 | UNIFORM I-AXIS |
| 17 | 0.500 | 500 | 2 | 12 | 2.0 | 1.0 | 2.00 | 2.00 | 196 | 2364 | 0.80000V | -3.00014046 | UNIFORM I-AXIS |
| 18 | 0.600 | 500 | 2 | 12 | 2.0 | 1.0 | 2.00 | 2.00 | 154 | 1860 | 0.60000V | -6.35080630 | UNIFORM I-AXIS |
| 19 | 0.700 | 500 | 2 | 12 | 2.0 | 1.0 | 2.00 | 2.00 | 212 | 2556 | 0.65000V | -3.03478155 | UNIFORM I-AXIS |
| 20 | 0.800 | 500 | 2 | 12 | 2.0 | 1.0 | 2.00 | 2.00 | 78 | 948 | 0.60000V | -16.66444362 | UNIFORM I-AXIS |
| 21 | 0.900 | 500 | 2 | 12 | 2.0 | 1.0 | 2.00 | 2.00 | 99 | 1200 | 0.70000V | -6.52063755 | UNIFORM I-AXIS |
| 22 | 1.000 | 500 | 2 | 12 | 2.0 | 1.0 | 2.00 | 2.00 | 250 | 3012 | 0.65000V | -3.01375068 | UNIFORM I-AXIS |
| 23 | 0.000 | 500 | 2 | 16 | 2.0 | 1.0 | 2.00 | 2.00 | 78 | 1264 | 0.60000V | -22.18779159 | UNIFORM I-AXIS |
| 24 | 0.100 | 500 | 2 | 16 | 2.0 | 1.0 | 2.00 | 2.00 | 60 | 976 | 0.65000V | -16.89097168 | UNIFORM I-AXIS |
| 25 | 0.200 | 500 | 2 | 16 | 2.0 | 1.0 | 2.00 | 2.00 | 173 | 2784 | 0.60000V | -3.01613943 | UNIFORM I-AXIS |
| 26 | 0.300 | 500 | 2 | 16 | 2.0 | 1.0 | 2.00 | 2.00 | 79 | 1280 | 0.65000V | -76.69765275 | UNIFORM I-AXIS |
| 27 | 0.400 | 500 | 2 | 16 | 2.0 | 1.0 | 2.00 | 2.00 | 189 | 3040 | 0.45000V | -3.00972155 | UNIFORM I-AXIS |
| 28 | 0.500 | 500 | 2 | 16 | 2.0 | 1.0 | 2.00 | 2.00 | 191 | 3072 | 0.55000V | -6.65915188 | UNIFORM I-AXIS |
| 29 | 0.600 | 500 | 2 | 16 | 2.0 | 1.0 | 2.00 | 2.00 | 122 | 1968 | 0.90000V | -15.75030525 | UNIFORM I-AXIS |
| 30 | 0.700 | 500 | 2 | 16 | 2.0 | 1.0 | 2.00 | 2.00 | 60 | 976 | 0.65000V | -31.95935579 | UNIFORM I-AXIS |
| 31 | 0.800 | 500 | 2 | 16 | 2.0 | 1.0 | 2.00 | 2.00 | 95 | 1536 | 0.55000V | -49.88324658 | UNIFORM I-AXIS |
| 32 | 0.900 | 500 | 2 | 16 | 2.0 | 1.0 | 2.00 | 2.00 | 304 | 4880 | 0.50000V | -3.00045306 | UNIFORM I-AXIS |
| 33 | 1.000 | 500 | 2 | 16 | 2.0 | 1.0 | 2.00 | 2.00 | 318 | 5104 | 0.25000V | -3.00010350 | UNIFORM I-AXIS |
| 34 | 0.000 | 500 | 2 | 20 | 2.0 | 1.0 | 2.00 | 2.00 | 78 | 1580 | 0.60000V | -76.94725945 | UNIFORM I-AXIS |
| 35 | 0.100 | 500 | 2 | 20 | 2.0 | 1.0 | 2.00 | 2.00 | 116 | 2340 | 0.60000V | -3.12194540 | UNIFORM I-AXIS |
| 36 | 0.200 | 500 | 2 | 20 | 2.0 | 1.0 | 2.00 | 2.00 | 193 | 3840 | 0.55000V | -3.03175136 | UNIFORM I-AXIS |
| 37 | 0.300 | 500 | 2 | 20 | 2.0 | 1.0 | 2.00 | 2.00 | 78 | 1580 | 0.60000V | -182.65850920 | UNIFORM I-AXIS |
| 38 | 0.400 | 500 | 2 | 20 | 2.0 | 1.0 | 2.00 | 2.00 | 193 | 3880 | 0.65000V | -3.14860260 | UNIFORM I-AXIS |
| 39 | 0.500 | 500 | 2 | 20 | 2.0 | 1.0 | 2.00 | 2.00 | 98 | 1980 | 0.65000V | -3.62879802 | UNIFORM I-AXIS |
| 40 | 0.600 | 500 | 2 | 20 | 2.0 | 1.0 | 2.00 | 2.00 | 192 | 3860 | 0.60000V | -3.04067143 | UNIFORM I-AXIS |
| 41 | 0.700 | 500 | 2 | 20 | 2.0 | 1.0 | 2.00 | 2.00 | 139 | 2800 | 0.80000V | -3.54277590 | UNIFORM I-AXIS |
| 42 | 0.800 | 500 | 2 | 20 | 2.0 | 1.0 | 2.00 | 2.00 | 171 | 3440 | 0.55000V | -3.00072472 | UNIFORM I-AXIS |
| 43 | 0.900 | 500 | 2 | 20 | 2.0 | 1.0 | 2.00 | 2.00 | 101 | 2040 | 0.80000V | -4.37891524 | UNIFORM I-AXIS |
| 44 | 1.000 | 500 | 2 | 20 | 2.0 | 1.0 | 2.00 | 2.00 | 79 | 1600 | 0.65000V | -35.88632429 | UNIFORM I-AXIS |
| 45 | 0.000 | 500 | 2 | 24 | 2.0 | 1.0 | 2.00 | 2.00 | 226 | 5448 | 0.40000V | -3.00031923 | UNIFORM I-AXIS |
| 46 | 0.100 | 500 | 2 | 24 | 2.0 | 1.0 | 2.00 | 2.00 | 78 | 1896 | 0.60000V | -3.06529259 | UNIFORM I-AXIS |
| 47 | 0.200 | 500 | 2 | 24 | 2.0 | 1.0 | 2.00 | 2.00 | 209 | 5040 | 0.50000V | -3.03088087 | UNIFORM I-AXIS |
| 48 | 0.300 | 500 | 2 | 24 | 2.0 | 1.0 | 2.00 | 2.00 | 79 | 1920 | 0.65000V | -30.56252660 | UNIFORM I-AXIS |
| 49 | 0.400 | 500 | 2 | 24 | 2.0 | 1.0 | 2.00 | 2.00 | 306 | 7368 | 0.60000V | -3.00075380 | UNIFORM I-AXIS |
| 50 | 0.500 | 500 | 2 | 24 | 2.0 | 1.0 | 2.00 | 2.00 | 248 | 5976 | 0.55000V | -3.00968878 | UNIFORM I-AXIS |
| 51 | 0.600 | 500 | 2 | 24 | 2.0 | 1.0 | 2.00 | 2.00 | 97 | 2352 | 0.60000V | -3.34890591 | UNIFORM I-AXIS |
| 52 | 0.700 | 500 | 2 | 24 | 2.0 | 1.0 | 2.00 | 2.00 | 246 | 5928 | 0.45000V | -3.01530923 | UNIFORM I-AXIS |
| 53 | 0.800 | 500 | 2 | 24 | 2.0 | 1.0 | 2.00 | 2.00 | 78 | 1896 | 0.60000V | -31.34561763 | UNIFORM I-AXIS |
| **54** | **0.900** | **500** | **2** | **24** | **2.0** | **1.0** | **2.00** | **2.00** | **60** | **1464** | **0.65000V** | **-3.00000000** | **UNIFORM I-AXIS** |
| 55 | 1.000 | 500 | 2 | 24 | 2.0 | 1.0 | 2.00 | 2.00 | 101 | 2448 | 0.80000V | -15.41042957 | UNIFORM I-AXIS |
| 56 | 0.000 | 500 | 2 | 28 | 2.0 | 1.0 | 2.00 | 2.00 | 266 | 7476 | 0.55000V | -3.00178783 | UNIFORM I-AXIS |
| 57 | 0.100 | 500 | 2 | 28 | 2.0 | 1.0 | 2.00 | 2.00 | 257 | 7224 | 0.55000V | -3.02198109 | UNIFORM I-AXIS |
| 58 | 0.200 | 500 | 2 | 28 | 2.0 | 1.0 | 2.00 | 2.00 | 60 | 1708 | 0.65000V | -102.95525789 | UNIFORM I-AXIS |
| 59 | 0.300 | 500 | 2 | 28 | 2.0 | 1.0 | 2.00 | 2.00 | 154 | 4340 | 0.60000V | -3.40567728 | UNIFORM I-AXIS |
| 60 | 0.400 | 500 | 2 | 28 | 2.0 | 1.0 | 2.00 | 2.00 | 79 | 2240 | 0.65000V | -40.28712313 | UNIFORM I-AXIS |
| 61 | 0.500 | 500 | 2 | 28 | 2.0 | 1.0 | 2.00 | 2.00 | 210 | 5908 | 0.55000V | -3.00334670 | UNIFORM I-AXIS |
| 62 | 0.600 | 500 | 2 | 28 | 2.0 | 1.0 | 2.00 | 2.00 | 79 | 2240 | 0.65000V | -32.25789818 | UNIFORM I-AXIS |
| 63 | 0.700 | 500 | 2 | 28 | 2.0 | 1.0 | 2.00 | 2.00 | 60 | 1708 | 0.65000V | -36.68532659 | UNIFORM I-AXIS |
| 64 | 0.800 | 500 | 2 | 28 | 2.0 | 1.0 | 2.00 | 2.00 | 60 | 1708 | 0.65000V | -14.65270155 | UNIFORM I-AXIS |
| 65 | 0.900 | 500 | 2 | 28 | 2.0 | 1.0 | 2.00 | 2.00 | 78 | 2212 | 0.60000V | -31.47896038 | UNIFORM I-AXIS |
| 66 | 1.000 | 500 | 2 | 28 | 2.0 | 1.0 | 2.00 | 2.00 | 282 | 7924 | 0.35000V | -3.00151179 | UNIFORM I-AXIS |

```
                              Total Function Evaluations:    180472
```

| **54** | **0.900** | **500** | **2** | **24** | **2.0** | **1.0** | **2.00** | **2.00** | **60** | **1464** | **0.65000V** | **-3.00000000** | **UNIFORM I-AXIS** |

"V" denotes that $F_{rep}$ was variable as discussed above. *Fitness* tabulates the best fitness returned during the run. The *Initial Probes* column shows the type of initial probe distribution, in this case probes uniformly spaced along probe lines parallel to $\Omega$'s axes (notated "I-AXIS") as described above and shown in Fig. 5.

The best fitness ranged from a low of $-236.6643...$ in run #12 to the global maximum of $-3$ at $(0,-1)$ returned in run #54 (best results highlighted in blue). Parameters for run #54 were $N_p/N_d = 12$, $N_p = 24$, and $\gamma = 0.9$. The total number of function evaluations over all runs was 180,472 while $N_{eval}$ for the best run was 1,464 (60 iterations). Fig. 6 plots the evolution of GP's best fitness, which in only two steps increases from $-2.992268247672 \times 10^{-12}$ to GP's actual global maximum of $-3$. This seems to be quite remarkable in view of the initial probe distribution for $\gamma = 0.9$ (Fig. 5) in which all probes are far removed from the maximum's location at $(0, -1)$. As the data in Table 1 clearly show, some sets of parameters are much better than others. Without pseudorandom initial probes, one run would be made with, in this example, only 13.6% probability of locating the global maximum with a fractional accuracy of 0.03% (9 of 66 runs). This statistic highlights the importance of pseudorandomness in CFO.

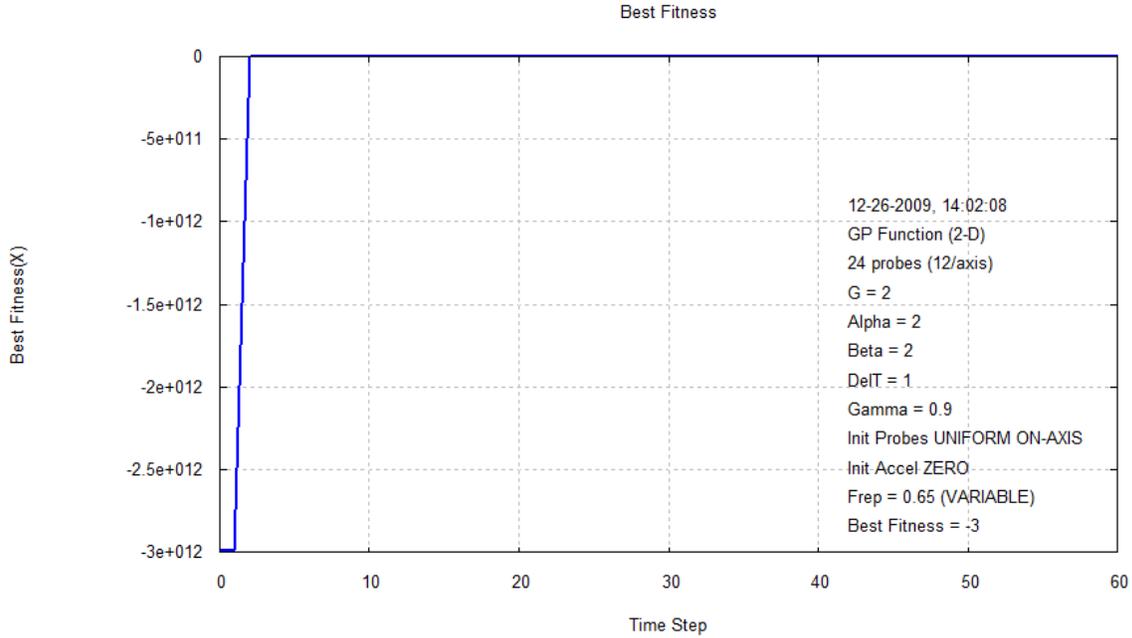

Fig. 6. Evolution of GP Function Best Fitness

Fig. 7 plots CFO's "$D_{avg}$ curve" for GP. $D_{avg}$ is the normalized average distance between the probe with the best fitness and all other probes at each time step, *viz.*,

$$D_{avg} = \frac{1}{L \cdot (N_p - 1)} \sum_{p=1}^{N_p} \sqrt{\sum_{i=1}^{N_d} (x_i^{p,j} - x_i^{p*,j})^2}$$ where $p*$ is the number of the probe with the best

fitness, and $L = \sqrt{\sum_{i=1}^{N_d} (x_i^{max} - x_i^{min})^2}$ is the length of $\Omega$'s principal diagonal (see Appendix 1 for definitions). $D_{avg}$ decreases monotonically through step 10 to 0.0496977, then increases very quickly to a peak of 0.4767301 at step 11, followed by another quasi-monotonic decrease through step 29 to 0.0274355. This cycle repeats through step 48 where $D_{avg}$ is 0.0169657, followed by a jump to 0.1194779 at step 49. After a slight dip through step 53, $D_{avg}$ flattens out around a value of 0.11… The quasi-oscillatory behavior in $D_{avg}$ usually correlates with local trapping, which in this case happens to be at the global maximum. Oscillation in $D_{avg}$ may be a similar phenomenon to oscillation seen in "$\Delta V$" curves for gravitationally trapped Near Earth Objects (NEOs), strongly suggesting that NEO theory may hold the key to analytical mitigation or elimination of local trapping at local maxima and possibly a proof of convergence for CFO [7-9].

Because CFO-PR converges so quickly on GP's global maximum, the number of the probe with the best fitness is constant after step #1 as seen in the best probe plot in Fig. 8. The best probe number (#14) is the same for steps 0 and 1, but it switches to probe #2 at step 2. Neither the number of the best probe nor the fitness change after step 2. Of course, for most functions the best probe number varies throughout a run, often quite erratically.

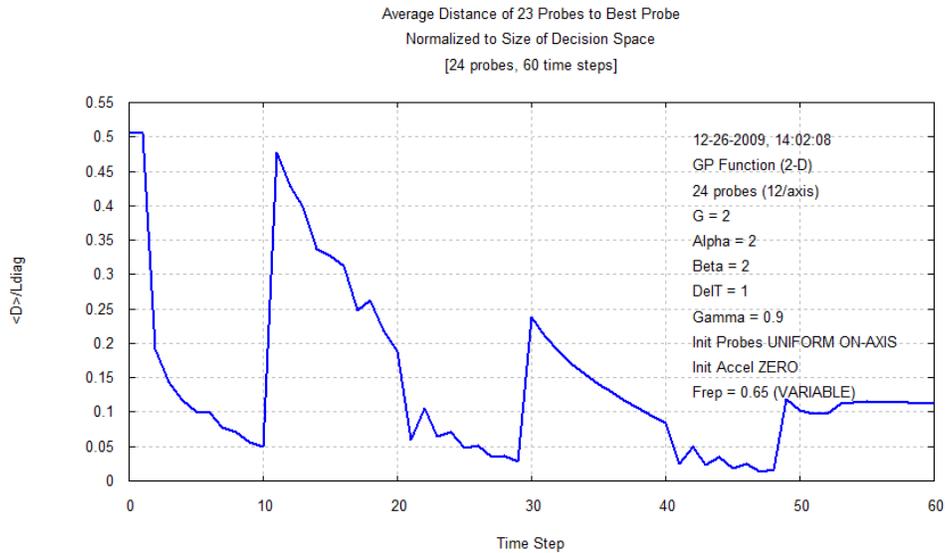

Fig. 7. Evolution of GP Function $D_{avg}$

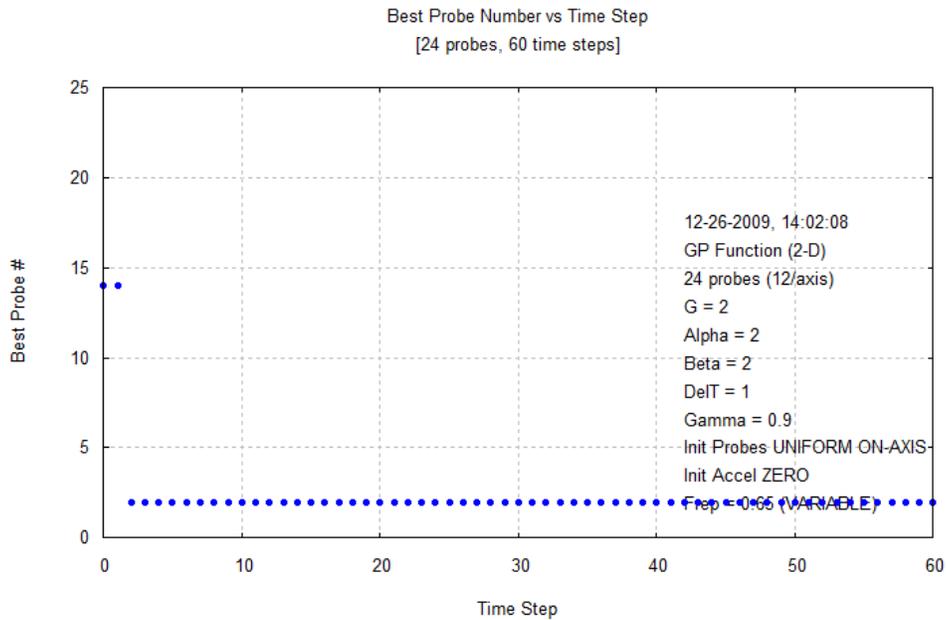

Fig. 8. GP Function Best Probe Number

Figs. 9 and 10, respectively, plot the probe trajectories for the probes with the best ten fitnesses ordered by fitness and for the first sixteen individual probes ordered by probe number [number of trajectories plotted chosen as a matter of convenience]. Both plots are very chaotic with no obvious sign of regularity in how probes gravitate to the global maximum. Nevertheless, there is some measure of regularity as reflected in the $D_{avg}$ curve because its appearance is not nearly as chaotic as the trajectory plots. In fact, in many cases $D_{avg}$ exhibits a mathematically precise oscillation even when the probe trajectories look like Figs. 9 and 10 (see in particular [7,9]).

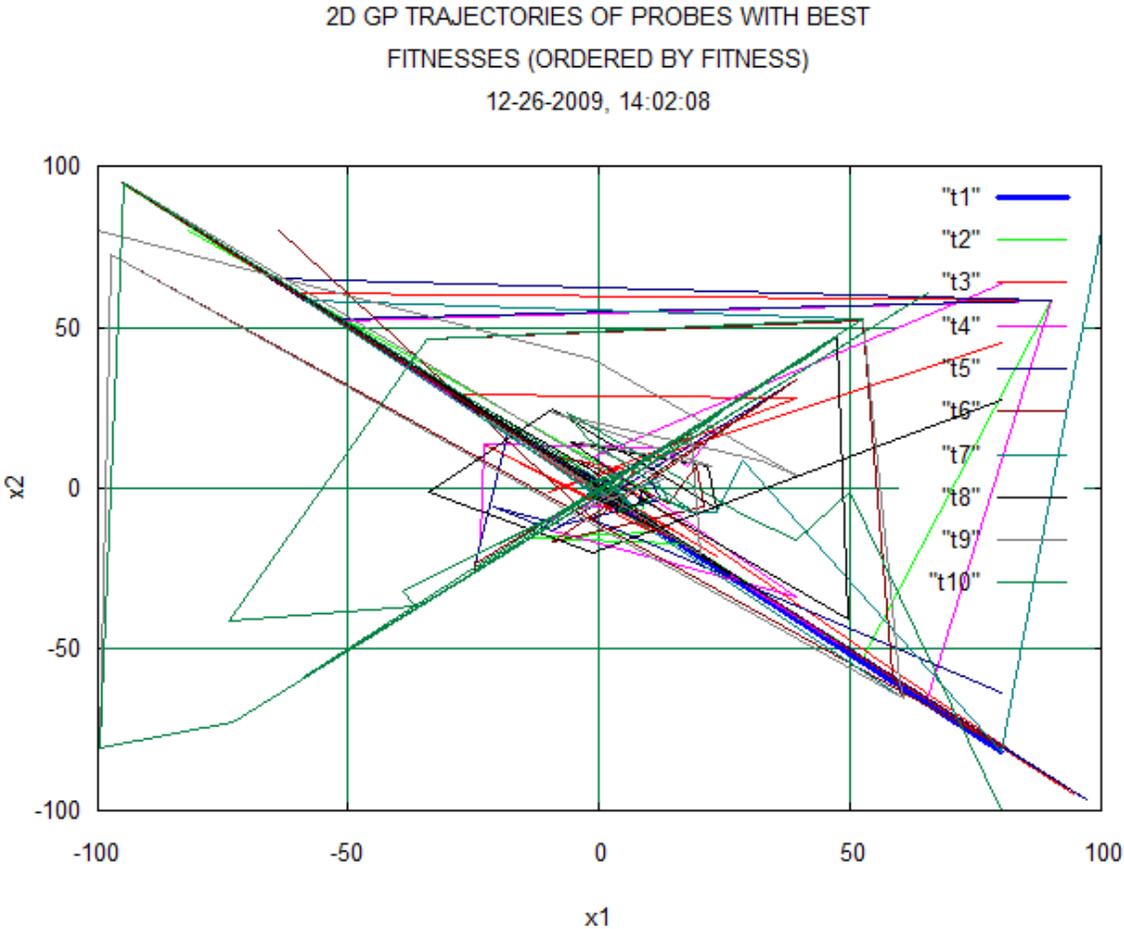

Fig. 9. GP Function Trajectories of Probes with Best Fitness

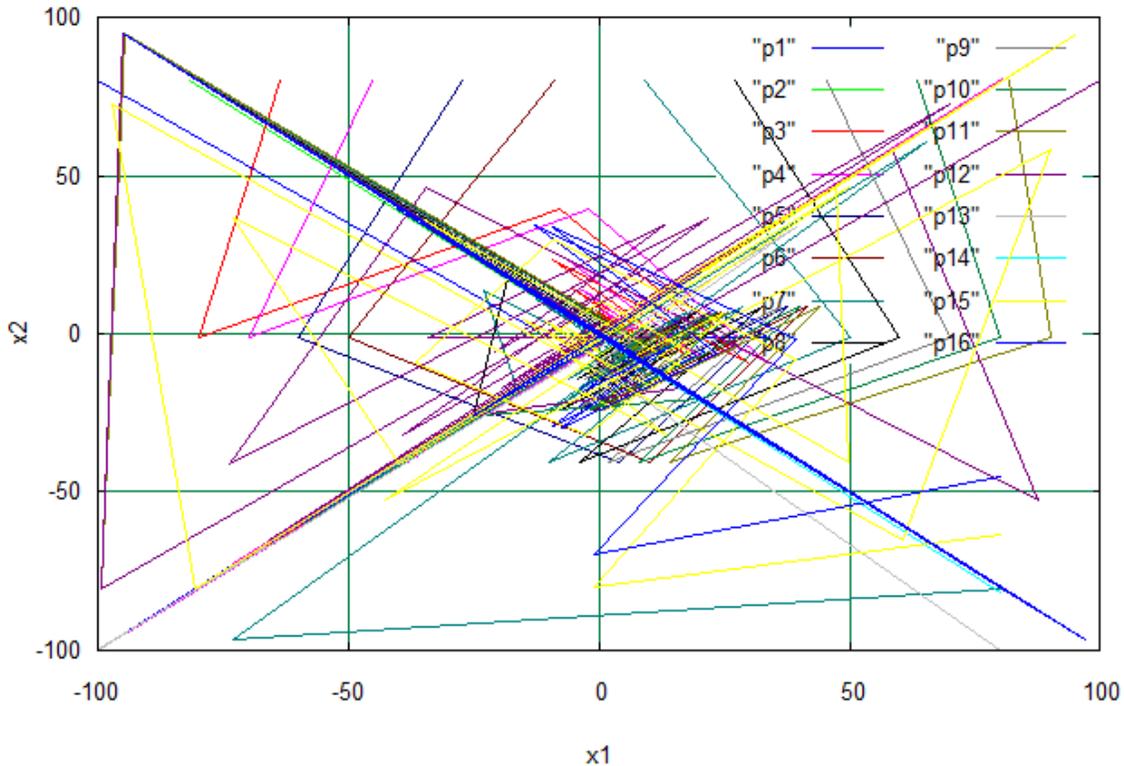

Fig. 10. GP Function Probe Trajectories by Probe Number

## 4. A Benchmark Suite

CFO-PR was tested against the twenty-three function benchmark suite in [11], and the results compared to the other algorithms' reported in that paper. The benchmarks, their decision spaces, and the three other algorithms ("GSO," "PSO," "GA") are discussed in detail in [11]. GSO (Group Search Optimizer) is a novel, highly refined Nature-inspired metaheuristic that mimics animal searching behavior based on a "producer-scrounger" model. PSO was implemented using "PSOt," a MATLAB-based toolbox that includes standard and variant PSO algorithms. The genetic algorithm in [11] was implemented using the GAOT toolbox (genetic algorithm optimization toolbox). Recommended default parameter values were used for the PSOt and GA algorithms as described in [11].

Table 2 summarizes CFO-PR's results using the same function numbering as [11] (more detailed data appear in Appendix 2). $f_{max}$ is the known global *maximum* (note that the negative of each benchmark in [11] is used because, unlike the other algorithms, CFO locates maxima, not minima). $< \cdot >$ denotes average value. Because GSO, PSO and GA are inherently stochastic, evaluating their performance requires a statistical assessment. Statistical data for those algorithms in Table 2 are reproduced from [11]. CFO's results, of course, are repeatable over runs with the same parameters because it is deterministic; no statistical description is necessary.

The CFO-PR data correspond to the best fitness returned by the single best run in the set of runs with variable $N_p/N_d$ and variable $\gamma$. For functions $f_{14}$ - $f_{23}$ eleven runs were made with $0 \leq \gamma \leq 1$ in increments of 0.1 and $4 \leq N_p/N_d \leq 14$ by 2 (66 runs total). For $f_1$ - $f_{13}$ the same procedure was used, but with $2 \leq N_p/N_d \leq 6$ by 2 (33 runs total) in order to avoid excessive runtimes. Table 2 shows the $\gamma$ value corresponding to the best fitness, $\gamma_{best}$, and the corresponding best value of $N_p/N_d$. $N_{eval}$ is the number of function evaluations, and it is tabulated for the single best run and for the group of runs used to determine $\gamma_{best}$ and the best number of probes per axis.

In the first group of high dimensionality unimodal functions, $f_1$ - $f_7$, CFO-PR returned the best fitness on five of the seven functions ($f_3$ - $f_7$). PSO performed best on the first two. In the second set of high dimensionality multimodal functions with many local maxima, $f_8$ – $f_{13}$, CFO-PR performed best on two ($f_9$, $f_{10}$) and essentially the same as the best other algorithm (GSO) on $f_8$. In last group of ten multimodal functions with few local maxima, $f_{14}$ - $f_{23}$, CFO-PR returned the best fitness on four ($f_{20}$ - $f_{23}$), equal fitnesses on three ($f_{14,}$ $f_{17,}$ $f_{18}$), and very slightly lower fitnesses on the remaining three.

Even though it is in its infancy, CFO-PR performed very well against three other highly sophisticated algorithms. It returned the best, equal, essentially equal or very slightly lower fitnesses on eighteen of the twenty three test functions. It is reasonable to conclude that, overall, CFO-PR performed as well or better than GSO, which in turn performed better than PSO or GA.

## 5. Conclusion

This note suggests that pseudorandomness is an important, indeed perhaps essential, aspect of effective CFO implementations. A pseudorandom variable has an arbitrary but precisely known value that may be assigned or calculated. Its essential characteristic is that the value is uncorrelated with the decision space's topology, so that it has the effect of distributing probes pseudorandomly throughout the landscape.

While in a general sense this process may appear to be similar to the randomness in an inherently stochastic algorithm, it is in fact fundamentally different. The equations underlying stochastic algorithms are formulated in terms of true random variables whose values are computed from probability distributions and consequently unknowable until the calculation is made. Therefore successive calculations yield different values, and every optimization run has a different outcome.

By contrast, a pseudorandom variable in the context of CFO is known with absolute precision because of how its value is determined (assignment or deterministic calculation). This property allows CFO to compute probe trajectories precisely because it is inherently deterministic. Every CFO with the same setup, with or without a pseudorandom component, yields exactly the same results step-by-step throughout an entire run. Importantly, CFO's reproducibility lends itself well to "reactive" implementations in which run parameters are "tuned" in response to performance metrics such as rate of convergence or fitness saturation, for example.  Reactive stochastic algorithms, on the other hand, are not easily implemented.

Table 2.  CFP-PR Comparative Results for 23 Benchmark Functions

( $N_d$ = Function Dimension,  $f_{max}$ = Known Global Maximum)

| Test Function* | $N_d$ | $f_{max}$* | <Best Fitness>/ Other Algorithm | ------------- CFO ------------- | | | | |
|---|---|---|---|---|---|---|---|---|
| | | | | Best Fitness | $\gamma_{best}$ | Best $N_p/N_d$ | $N_{eval}$ Best Run | $N_{eval}$ Total |
| Unimodal Functions (other algorithms: average of 1000 runs) | | | | | | | | |
| $f_1$ | 30 | 0 | **-3.6927x10⁻³⁷** / PSO | -4.8438x10⁻⁴ | 0.1 | 4 | 20,640 | 507,060 |
| $f_2$ | 30 | 0 | **-2.9168x10⁻²⁴** / PSO | -4x10⁻⁸ | 0.5 | 2 | 5,040 | 716,400 |
| $f_3$ | 30 | 0 | -1.1979x10⁻³ / PSO | **-6x10⁻⁸** | 0.5 | 2 | 10,260 | 1,534,260 |
| $f_4$ | 30 | 0 | -0.1078 / GSO | **-4.2x10⁻⁷** | 0.5 | 2 | 5,160 | 332,340 |
| $f_5$ | 30 | 0 | -37.3582 / PSO | **-1.09289x10⁻³** | 0.9 | 6 | 34,560 | 845,640 |
| $f_6$ | 30 | 0 | -1.6000x10⁻² / PSO | **0** | 1.0 | 6 | 10,980 | 350,280 |
| $f_7$ | 30 | 0 | -9.9024x10⁻³ / PSO | **-4.249x10⁻⁵** | 0.1 | 4 | 60,120 | 1,983,960 |
| Multimodal Functions, Many Local Maxima (other algorithms: avg 1000 runs) | | | | | | | | |
| $f_8$ | 30 | 12,569.5 | **12,569.4882** / PSO | 12,569.4866 | 0.5 | 4 | 12,720 | 448,800 |
| $f_9$ | 30 | 0 | -0.6509 / GA | **-2.05x10⁻⁶** | 0.7 | 4 | 16,440 | 680,640 |
| $f_{10}$ | 30 | 0 | -2.6548x10⁻⁵ / GSO | **-1.5x10⁻⁷** | 0.5 | 2 | 5,100 | 904,980 |
| $f_{11}$ | 30 | 0 | **-3.0792x10⁻²** / PSO | -9.97293x10⁻² | 0.1 | 6 | 42,660 | 489,060 |
| $f_{12}$ | 30 | 0 | **-2.7648x10⁻¹¹** / GSO | -2.067x10⁻⁵ | 0.5 | 2 | 3,660 | 341,400 |
| $f_{13}$ | 30 | 0 | **-4.6948x10⁻⁵** / GSO | -3.2853x10⁻³ | 0.6 | 6 | 16,920 | 679,620 |
| Multimodal Functions, Few Local Maxima (other algorithms: avg 50 runs) | | | | | | | | |
| $f_{14}$ | 2 | -1 | **-0.9980** / GSO | **-0.9980** | 0.2 | 12 | 5,952 | 141,076 |
| $f_{15}$ | 4 | -0.0003075 | **-3.7713x10⁻⁴** / GSO | -4.889x10⁻⁴ | 0 | 12 | 3,360 | 304,664 |
| $f_{16}$ | 2 | 1.0316285 | **1.031628** / GSO | 1.031626 | 0.4 | 12 | 6,288 | 124,340 |
| $f_{17}$ | 2 | -0.398 | **-0.3979** / GSO | **-0.3979** | 0 | 8 | 1,872 | 108,340 |
| $f_{18}$ | 2 | -3 | **-3** / GSO | **-3** | 0.9 | 12 | 1,464 | 180,472 |
| $f_{19}$ | 3 | 3.86 | **3.8628** / GSO | 3.8627 | 0.2 | 14 | 3,150 | 200,268 |
| $f_{20}$ | 6 | 3.32 | 3.2697 / GSO | **3.32173** | 0.3 | 12 | 18,072 | 730,212 |
| $f_{21}$ | 4 | 10 | 7.5439 / PSO | **10.1532** | 0.4 | 6 | 1,896 | 336,712 |
| $f_{22}$ | 4 | 10 | 8.3553 / PSO | **10.4029** | 0.8 | 6 | 2,208 | 386,176 |
| $f_{23}$ | 4 | 10 | 8.9439 / PSO | **10.5363** | 0.8 | 6 | 2,256 | 394,320 |

*Note: negative of the functions in [11] are computed by CFO because CFO searches for maxima instead of minima.

This note provides examples of how pseudorandomness can improve CFO's performance. Three different approaches are used (initial probe distribution, repositioning factor, and decision space adaptation), and each was discussed in detail. A sample CFO problem was presented in detail, and summary data included for two versions of the algorithm tested against a 23-function benchmark suite. CFO's performance is quite good compared to other highly developed, state-of-the-art algorithms.

Hopefully these results will encourage further work on improved methodologies for injecting pseudorandomness into CFO. Of course, any or all of CFO's run parameters can be pseudorandomized, not only the three considered here. But even with respect to those parameters, different approaches to how they are pseudorandomized may yield better results or faster runtimes. There are many fruitful areas of research on CFO, and it is the author's hope that this and the other CFO papers will provide the foundation and catalyst for that work.

*2 January 2010*
*Brewster, Massachusetts*

**Ver. 2** *(typographical errors in Step (c) of Fig. 1 and discussion of Fig. 8 corrected; clarification of definition of $\vec{R}_{best}$; error in Table 2 test function $f_{12}$ corrected; $f_{12}$ source code in Appendix 3, page 60 corrected).*

*3 February 2010*
*Saint Augustine, Florida*

[1]Richard A. Formato, JD, PhD
Registered Patent Attorney & Consulting Engineer
P.O. Box 1714, Harwich, MA 02645 USA
rf2@ieee.org

**Appendix 1. CFO Metaheuristic**

CFO searches an $N_d$-dimensional decision space $\Omega$ for the global *maxima* of an *objective function* $f(x_1, x_2, ..., x_{N_d})$ defined on $\Omega$: $x_i^{min} \leq x_i \leq x_i^{max}$, $1 \leq i \leq N_d$. The $x_i$ are the *decision variables*, and $i$ the coordinate number. The term *fitness* refers to the value of $f(\vec{x})$ at point $\vec{x}$ in $\Omega$. There is no *a priori* information about the objective function's maxima, that is, $f(\vec{x})$'s topology or "landscape" is unknown.

CFO searches $\Omega$ by flying "probes" through the space at discrete "time" steps (iterations). Each probe's location is specified by its position vector computed from two *equations of motion* that analogize their real-world counterparts for material objects moving through physical space under the influence of gravity without energy dissipation.

Probe $p$'s position vector at step $j$ is $\vec{R}_j^p = \sum_{k=1}^{N_d} x_k^{p,j} \hat{e}_k$, where the $x_k^{p,j}$ are its coordinates and $\hat{e}_k$ the unit vector along the $x_k$-axis. The indices $p$, $1 \leq p \leq N_p$, and $j$, $0 \leq j \leq N_t$, respectively, are the probe number and iteration number, where $N_p$ and $N_t$ are the corresponding *total* number of probes and *total* number of time steps.

***Equations of Motion.*** In metaphorical "CFO space" each of the $N_p$ probes experiences an acceleration created by the "gravitational pull" of "masses" in $\Omega$. Probe $p$'s acceleration at step $j-1$ is given by

$$\vec{a}_{j-1}^p = G \sum_{\substack{k=1 \\ k \neq p}}^{N_p} U(M_{j-1}^k - M_{j-1}^p) \cdot (M_{j-1}^k - M_{j-1}^p)^\alpha \times \frac{(\vec{R}_{j-1}^k - \vec{R}_{j-1}^p)}{\left\| \vec{R}_{j-1}^k - \vec{R}_{j-1}^p \right\|^\beta}, \quad (1)$$

which is the first of CFO's two equations of motion. In equation (1), $M_{j-1}^p = f(x_1^{p,j-1}, x_2^{p,j-1}, ..., x_{N_d}^{p,j-1})$ is the objective function's fitness at probe $p$'s location at time step $j-1$. Each of the other probes at that step (iteration) has associated with it fitness $M_{j-1}^k, k = 1, ..., p-1, p+1, ..., N_p$. $G$ is CFO's *gravitational constant*, and $U(\cdot)$ is the Unit Step function, $U(z) = \begin{cases} 1, & z \geq 0 \\ 0, & otherwise \end{cases}$.

The acceleration $\vec{a}_{j-1}^p$ causes probe $p$ to move from position $\vec{R}_{j-1}^p$ at step $j-1$ to $\vec{R}_j^p$ at step $j$ according to the trajectory equation

$$\vec{R}_j^p = \vec{R}_{j-1}^p + \frac{1}{2}\vec{a}_{j-1}^p \Delta t^2, \quad j \geq 1, \quad (2)$$

which is CFO's second equation of motion.

The CFO equations of motion, (1) and (2), combine to compute a new probe distribution at each time step using "masses" discovered by the probe distribution at the previous step. $\Delta t$ is the "time" interval between steps during which the acceleration is constant. Note that CFO's terminology has no significance beyond being a reflection of CFO's kinematic roots, as is the factor ½ in eq. (2). The gravitational constant, $G$, and time increment, $\Delta t$, have direct analogues in Newton's equations of motion for real masses moving under real gravity through three-dimensional physical space. The CFO exponents $\alpha$ and $\beta$, by contrast, have no analogues in Nature. They provide added flexibility to the algorithm designer who, in metaphorical "CFO space," is free to change how "gravity" varies with distance, or mass, or both, if doing so creates a more effective algorithm.

**"Mass."** The concept of "mass" in CFO space is very important and quite different than it is in real space. Mass in the physical Universe is an inherent, immutable property of matter, whereas in CFO space it is a positive-definite *user-defined function* of the objective function's fitness [not (necessarily) the fitness itself]. For example, in equation (1) mass is defined as $MASS_{CFO} = U(M^k_{j-1} - M^p_{j-1}) \cdot (M^k_{j-1} - M^p_{j-1})^\alpha$ [difference in fitness values raised to the $\alpha$ power multiplied by the Unit Step]. A different function can be used if it results in a better performing CFO algorithm. In this specific implementation the Unit Step is a critical element because it prevents negative mass. Without the Unit Step CFO mass could be negative depending on which fitness is greater. But mass in the real Universe always is positive, and as a consequence the force of gravity always attractive. By contrast, mass can be positive or negative in metaphorical CFO space, depending on how it is defined, and undesirable effects may result from the wrong definition. Negative mass creates a *repulsive* gravitational force that flies probes away from maxima instead of toward them, thus defeating the very purpose of the algorithm.

**Errant Probes.** At any iteration in a CFO run, it is possible that a probe's acceleration computed from eq. (2) may be too great to keep it inside $\Omega$. If any coordinate $x_i < x_i^{min}$ or $x_i > x_i^{max}$, the probe enters a region of *unfeasible* solutions that are not valid for the problem at hand. The question is what to do with an errant probe, and it arises in many algorithms. There are many approaches s. While many schemes are possible, a simple, empirically determined one is used here. On a coordinate-by-coordinate basis, probes flying out of the decision space are placed a fraction $\Delta F_{rep} \leq F_{rep} \leq 1$ of the distance between the probe's starting coordinate and the corresponding boundary coordinate. $F_{rep}$ is the variable "repositioning factor" introduced in [2]. Its value, as well as those of all the CFO parameters, were determined empirically.

# Appendix 2.  CFO Summary Data for the GSO Benchmark Suite [11]

Note:  In the tables below, "UNIFORM P-AXIS" means "I-AXIS" as discussed above.

## F1

Run ID: 12-27-2009, 13:28:55

FUNCTION: F1

| Run # | Gamma | Nt | Nd | Np | G | DelT | Alpha | Beta | #Steps | Neval | Frep | Fitness | Initial Probes |
|-------|-------|-----|-----|-----|-----|------|-------|------|--------|--------|---------|--------------|----------------|
| 0 | 0.000 | 500 | 30 | 60 | 2.0 | 1.0 | 2.00 | 2.00 | 0 | 0 | 0.50000V | -9999.00000000 | UNIFORM P-AXIS |
| 1 | 0.000 | 500 | 30 | 60 | 2.0 | 1.0 | 2.00 | 2.00 | 85 | 5160 | 0.95000V | -0.06315972 | UNIFORM P-AXIS |
| 2 | 0.100 | 500 | 30 | 60 | 2.0 | 1.0 | 2.00 | 2.00 | 232 | 13980 | 0.70000V | -135.67004943 | UNIFORM P-AXIS |
| 3 | 0.200 | 500 | 30 | 60 | 2.0 | 1.0 | 2.00 | 2.00 | 238 | 14340 | 0.05000V | -345.75281239 | UNIFORM P-AXIS |
| 4 | 0.300 | 500 | 30 | 60 | 2.0 | 1.0 | 2.00 | 2.00 | 162 | 9780 | 0.05000V | -85.84498630 | UNIFORM P-AXIS |
| 5 | 0.400 | 500 | 30 | 60 | 2.0 | 1.0 | 2.00 | 2.00 | 97 | 5880 | 0.60000V | -163.14259732 | UNIFORM P-AXIS |
| 6 | 0.500 | 500 | 30 | 60 | 2.0 | 1.0 | 2.00 | 2.00 | 115 | 6960 | 0.55000V | -0.02659219 | UNIFORM P-AXIS |
| 7 | 0.600 | 500 | 30 | 60 | 2.0 | 1.0 | 2.00 | 2.00 | 72 | 4380 | 0.30000V | -212.60250000 | UNIFORM P-AXIS |
| 8 | 0.700 | 500 | 30 | 60 | 2.0 | 1.0 | 2.00 | 2.00 | 222 | 13380 | 0.20000V | -559.30076908 | UNIFORM P-AXIS |
| 9 | 0.800 | 500 | 30 | 60 | 2.0 | 1.0 | 2.00 | 2.00 | 160 | 9660 | 0.90000V | -20.19453374 | UNIFORM P-AXIS |
| 10 | 0.900 | 500 | 30 | 60 | 2.0 | 1.0 | 2.00 | 2.00 | 197 | 11880 | 0.85000V | -2.43693893 | UNIFORM P-AXIS |
| 11 | 1.000 | 500 | 30 | 60 | 2.0 | 1.0 | 2.00 | 2.00 | 220 | 13260 | 0.10000V | -2.41967196 | UNIFORM P-AXIS |
| 12 | 0.000 | 500 | 30 | 120 | 2.0 | 1.0 | 2.00 | 2.00 | 110 | 13320 | 0.30000V | -1.23728667 | UNIFORM P-AXIS |
| 13 | 0.100 | 500 | 30 | 120 | 2.0 | 1.0 | 2.00 | 2.00 | 171 | 20640 | 0.50000V | -0.00048438 | UNIFORM P-AXIS |
| 14 | 0.200 | 500 | 30 | 120 | 2.0 | 1.0 | 2.00 | 2.00 | 128 | 15480 | 0.25000V | -0.00954054 | UNIFORM P-AXIS |
| 15 | 0.300 | 500 | 30 | 120 | 2.0 | 1.0 | 2.00 | 2.00 | 136 | 16440 | 0.65000V | -0.00386787 | UNIFORM P-AXIS |
| 16 | 0.400 | 500 | 30 | 120 | 2.0 | 1.0 | 2.00 | 2.00 | 98 | 11880 | 0.65000V | -0.03513774 | UNIFORM P-AXIS |
| 17 | 0.500 | 500 | 30 | 120 | 2.0 | 1.0 | 2.00 | 2.00 | 117 | 14160 | 0.65000V | -2.48228919 | UNIFORM P-AXIS |
| 18 | 0.600 | 500 | 30 | 120 | 2.0 | 1.0 | 2.00 | 2.00 | 98 | 11880 | 0.65000V | -0.03513774 | UNIFORM P-AXIS |
| 19 | 0.700 | 500 | 30 | 120 | 2.0 | 1.0 | 2.00 | 2.00 | 134 | 16200 | 0.55000V | -0.31826953 | UNIFORM P-AXIS |
| 20 | 0.800 | 500 | 30 | 120 | 2.0 | 1.0 | 2.00 | 2.00 | 126 | 15240 | 0.15000V | -0.00992779 | UNIFORM P-AXIS |
| 21 | 0.900 | 500 | 30 | 120 | 2.0 | 1.0 | 2.00 | 2.00 | 98 | 11880 | 0.65000V | -0.00915096 | UNIFORM P-AXIS |
| 22 | 1.000 | 500 | 30 | 120 | 2.0 | 1.0 | 2.00 | 2.00 | 110 | 13320 | 0.30000V | -1.23728667 | UNIFORM P-AXIS |
| 23 | 0.000 | 500 | 30 | 180 | 2.0 | 1.0 | 2.00 | 2.00 | 130 | 23580 | 0.35000V | -0.00413390 | UNIFORM P-AXIS |
| 24 | 0.100 | 500 | 30 | 180 | 2.0 | 1.0 | 2.00 | 2.00 | 150 | 27180 | 0.40000V | -2.11477561 | UNIFORM P-AXIS |
| 25 | 0.200 | 500 | 30 | 180 | 2.0 | 1.0 | 2.00 | 2.00 | 78 | 14220 | 0.60000V | -0.33188771 | UNIFORM P-AXIS |
| 26 | 0.300 | 500 | 30 | 180 | 2.0 | 1.0 | 2.00 | 2.00 | 78 | 14220 | 0.60000V | -2.12513165 | UNIFORM P-AXIS |
| 27 | 0.400 | 500 | 30 | 180 | 2.0 | 1.0 | 2.00 | 2.00 | 150 | 27180 | 0.40000V | -0.00219997 | UNIFORM P-AXIS |
| 28 | 0.500 | 500 | 30 | 180 | 2.0 | 1.0 | 2.00 | 2.00 | 136 | 24660 | 0.65000V | -0.06292008 | UNIFORM P-AXIS |
| 29 | 0.600 | 500 | 30 | 180 | 2.0 | 1.0 | 2.00 | 2.00 | 153 | 27720 | 0.55000V | -0.06039108 | UNIFORM P-AXIS |
| 30 | 0.700 | 500 | 30 | 180 | 2.0 | 1.0 | 2.00 | 2.00 | 78 | 14220 | 0.60000V | -2.12513165 | UNIFORM P-AXIS |
| 31 | 0.800 | 500 | 30 | 180 | 2.0 | 1.0 | 2.00 | 2.00 | 78 | 14220 | 0.60000V | -0.33188771 | UNIFORM P-AXIS |
| 32 | 0.900 | 500 | 30 | 180 | 2.0 | 1.0 | 2.00 | 2.00 | 150 | 27180 | 0.40000V | -2.11477561 | UNIFORM P-AXIS |
| 33 | 1.000 | 500 | 30 | 180 | 2.0 | 1.0 | 2.00 | 2.00 | 130 | 23580 | 0.35000V | -0.00413390 | UNIFORM P-AXIS |

Total Function Evaluations: 507060

| 13 | 0.100 | 500 | 30 | 120 | 2.0 | 1.0 | 2.00 | 2.00 | 171 | 20640 | 0.50000V | -0.00048438 | UNIFORM P-AXIS |

## F2

Run ID: 12-27-2009, 13:29:23

FUNCTION: F2

| Run # | Gamma | Nt | Nd | Np | G | DelT | Alpha | Beta | #Steps | Neval | Frep | Fitness | Initial Probes |
|-------|-------|-----|-----|-----|-----|------|-------|------|--------|--------|---------|--------------|----------------|
| 0 | 0.000 | 500 | 30 | 60 | 2.0 | 1.0 | 2.00 | 2.00 | 0 | 0 | 0.50000V | -9999.00000000 | UNIFORM P-AXIS |
| 1 | 0.000 | 500 | 30 | 60 | 2.0 | 1.0 | 2.00 | 2.00 | 175 | 10560 | 0.70000V | -6.60617435 | UNIFORM P-AXIS |
| 2 | 0.100 | 500 | 30 | 60 | 2.0 | 1.0 | 2.00 | 2.00 | 370 | 22260 | 0.95000V | -3.81990738 | UNIFORM P-AXIS |
| 3 | 0.200 | 500 | 30 | 60 | 2.0 | 1.0 | 2.00 | 2.00 | 324 | 19500 | 0.55000V | -2.06811449 | UNIFORM P-AXIS |
| 4 | 0.300 | 500 | 30 | 60 | 2.0 | 1.0 | 2.00 | 2.00 | 190 | 11460 | 0.50000V | -2.09728545 | UNIFORM P-AXIS |
| 5 | 0.400 | 500 | 30 | 60 | 2.0 | 1.0 | 2.00 | 2.00 | 126 | 7620 | 0.15000V | -2.73598457 | UNIFORM P-AXIS |
| 6 | 0.500 | 500 | 30 | 60 | 2.0 | 1.0 | 2.00 | 2.00 | 83 | 5040 | 0.85000V | -0.00000004 | UNIFORM P-AXIS |
| 7 | 0.600 | 500 | 30 | 60 | 2.0 | 1.0 | 2.00 | 2.00 | 174 | 10500 | 0.65000V | -1.24666633 | UNIFORM P-AXIS |
| 8 | 0.700 | 500 | 30 | 60 | 2.0 | 1.0 | 2.00 | 2.00 | 157 | 9480 | 0.75000V | -0.70580201 | UNIFORM P-AXIS |
| 9 | 0.800 | 500 | 30 | 60 | 2.0 | 1.0 | 2.00 | 2.00 | 123 | 7440 | 0.95000V | -3.00749022 | UNIFORM P-AXIS |
| 10 | 0.900 | 500 | 30 | 60 | 2.0 | 1.0 | 2.00 | 2.00 | 461 | 27720 | 0.75000V | -3.47243586 | UNIFORM P-AXIS |
| 11 | 1.000 | 500 | 30 | 60 | 2.0 | 1.0 | 2.00 | 2.00 | 500 | 30060 | 0.80000V | -0.19552788 | UNIFORM P-AXIS |
| 12 | 0.000 | 500 | 30 | 120 | 2.0 | 1.0 | 2.00 | 2.00 | 173 | 20880 | 0.60000V | -0.50882324 | UNIFORM P-AXIS |
| 13 | 0.100 | 500 | 30 | 120 | 2.0 | 1.0 | 2.00 | 2.00 | 97 | 11760 | 0.60000V | -1.38178169 | UNIFORM P-AXIS |
| 14 | 0.200 | 500 | 30 | 120 | 2.0 | 1.0 | 2.00 | 2.00 | 80 | 9720 | 0.70000V | -0.42701634 | UNIFORM P-AXIS |
| 15 | 0.300 | 500 | 30 | 120 | 2.0 | 1.0 | 2.00 | 2.00 | 79 | 9600 | 0.65000V | -0.31587227 | UNIFORM P-AXIS |
| 16 | 0.400 | 500 | 30 | 120 | 2.0 | 1.0 | 2.00 | 2.00 | 118 | 14280 | 0.70000V | -0.18736453 | UNIFORM P-AXIS |
| 17 | 0.500 | 500 | 30 | 120 | 2.0 | 1.0 | 2.00 | 2.00 | 60 | 7320 | 0.65000V | -0.06929315 | UNIFORM P-AXIS |
| 18 | 0.600 | 500 | 30 | 120 | 2.0 | 1.0 | 2.00 | 2.00 | 353 | 42480 | 0.10000V | -0.06374008 | UNIFORM P-AXIS |
| 19 | 0.700 | 500 | 30 | 120 | 2.0 | 1.0 | 2.00 | 2.00 | 134 | 16200 | 0.55000V | -0.06659569 | UNIFORM P-AXIS |
| 20 | 0.800 | 500 | 30 | 120 | 2.0 | 1.0 | 2.00 | 2.00 | 380 | 45720 | 0.50000V | -0.11070981 | UNIFORM P-AXIS |
| 21 | 0.900 | 500 | 30 | 120 | 2.0 | 1.0 | 2.00 | 2.00 | 184 | 22200 | 0.20000V | -0.13620894 | UNIFORM P-AXIS |
| 22 | 1.000 | 500 | 30 | 120 | 2.0 | 1.0 | 2.00 | 2.00 | 98 | 11880 | 0.65000V | -0.09823314 | UNIFORM P-AXIS |
| 23 | 0.000 | 500 | 30 | 180 | 2.0 | 1.0 | 2.00 | 2.00 | 96 | 17460 | 0.55000V | -0.01628443 | UNIFORM P-AXIS |
| 24 | 0.100 | 500 | 30 | 180 | 2.0 | 1.0 | 2.00 | 2.00 | 499 | 90000 | 0.75000V | -0.00051367 | UNIFORM P-AXIS |
| 25 | 0.200 | 500 | 30 | 180 | 2.0 | 1.0 | 2.00 | 2.00 | 216 | 39060 | 0.85000V | -0.02264350 | UNIFORM P-AXIS |
| 26 | 0.300 | 500 | 30 | 180 | 2.0 | 1.0 | 2.00 | 2.00 | 81 | 14760 | 0.75000V | -0.08109305 | UNIFORM P-AXIS |
| 27 | 0.400 | 500 | 30 | 180 | 2.0 | 1.0 | 2.00 | 2.00 | 78 | 14220 | 0.60000V | -0.16869854 | UNIFORM P-AXIS |
| 28 | 0.500 | 500 | 30 | 180 | 2.0 | 1.0 | 2.00 | 2.00 | 60 | 10980 | 0.65000V | -0.03716355 | UNIFORM P-AXIS |
| 29 | 0.600 | 500 | 30 | 180 | 2.0 | 1.0 | 2.00 | 2.00 | 78 | 14220 | 0.60000V | -0.16869854 | UNIFORM P-AXIS |
| 30 | 0.700 | 500 | 30 | 180 | 2.0 | 1.0 | 2.00 | 2.00 | 81 | 14760 | 0.75000V | -0.08109305 | UNIFORM P-AXIS |
| 31 | 0.800 | 500 | 30 | 180 | 2.0 | 1.0 | 2.00 | 2.00 | 210 | 37980 | 0.55000V | -0.03264013 | UNIFORM P-AXIS |
| 32 | 0.900 | 500 | 30 | 180 | 2.0 | 1.0 | 2.00 | 2.00 | 398 | 71820 | 0.45000V | -0.00642562 | UNIFORM P-AXIS |
| 33 | 1.000 | 500 | 30 | 180 | 2.0 | 1.0 | 2.00 | 2.00 | 96 | 17460 | 0.55000V | -0.22930127 | UNIFORM P-AXIS |

Total Function Evaluations: 716400

| 6 | 0.500 | 500 | 30 | 60 | 2.0 | 1.0 | 2.00 | 2.00 | 83 | 5040 | 0.85000V | -0.00000004 | UNIFORM P-AXIS |

# F3

Run ID: 12-27-2009, 14:55:32

FUNCTION: F3

| Run # | Gamma | Nt | Nd | Np | G | DelT | Alpha | Beta | #Steps | Neval | Frep | Fitness | Initial Probes |
|-------|-------|----|----|----|----|------|-------|------|--------|-------|------|---------|----------------|
| 0 | 0.000 | 500 | 30 | 60 | 2.0 | 1.0 | 2.00 | 2.00 | 0 | 0 | 0.50000V | -9999.00000000 | UNIFORM P-AXIS |
| 1 | 0.000 | 500 | 30 | 60 | 2.0 | 1.0 | 2.00 | 2.00 | 362 | 21780 | 0.55000V | -981.62841988 | UNIFORM P-AXIS |
| 2 | 0.100 | 500 | 30 | 60 | 2.0 | 1.0 | 2.00 | 2.00 | 383 | 23040 | 0.65000V | -91.16096360 | UNIFORM P-AXIS |
| 3 | 0.200 | 500 | 30 | 60 | 2.0 | 1.0 | 2.00 | 2.00 | 354 | 21300 | 0.15000V | -43.94966422 | UNIFORM P-AXIS |
| 4 | 0.300 | 500 | 30 | 60 | 2.0 | 1.0 | 2.00 | 2.00 | 358 | 21540 | 0.35000V | -88.81550993 | UNIFORM P-AXIS |
| 5 | 0.400 | 500 | 30 | 60 | 2.0 | 1.0 | 2.00 | 2.00 | 348 | 20940 | 0.80000V | -66.55740578 | UNIFORM P-AXIS |
| 6 | 0.500 | 500 | 30 | 60 | 2.0 | 1.0 | 2.00 | 2.00 | 170 | 10260 | 0.45000V | -0.00000006 | UNIFORM P-AXIS |
| 7 | 0.600 | 500 | 30 | 60 | 2.0 | 1.0 | 2.00 | 2.00 | 456 | 27420 | 0.50000V | -66.39589627 | UNIFORM P-AXIS |
| 8 | 0.700 | 500 | 30 | 60 | 2.0 | 1.0 | 2.00 | 2.00 | 331 | 19920 | 0.90000V | -86.26322188 | UNIFORM P-AXIS |
| 9 | 0.800 | 500 | 30 | 60 | 2.0 | 1.0 | 2.00 | 2.00 | 383 | 23040 | 0.65000V | -42.03391601 | UNIFORM P-AXIS |
| 10 | 0.900 | 500 | 30 | 60 | 2.0 | 1.0 | 2.00 | 2.00 | 333 | 20040 | 0.05000V | -75.23554855 | UNIFORM P-AXIS |
| 11 | 1.000 | 500 | 30 | 60 | 2.0 | 1.0 | 2.00 | 2.00 | 500 | 30060 | 0.80000V | -979.00439553 | UNIFORM P-AXIS |
| 12 | 0.000 | 500 | 30 | 120 | 2.0 | 1.0 | 2.00 | 2.00 | 388 | 46680 | 0.90000V | -74.16162186 | UNIFORM P-AXIS |
| 13 | 0.100 | 500 | 30 | 120 | 2.0 | 1.0 | 2.00 | 2.00 | 420 | 50520 | 0.60000V | -31.07754473 | UNIFORM P-AXIS |
| 14 | 0.200 | 500 | 30 | 120 | 2.0 | 1.0 | 2.00 | 2.00 | 363 | 43680 | 0.60000V | -44.64307545 | UNIFORM P-AXIS |
| 15 | 0.300 | 500 | 30 | 120 | 2.0 | 1.0 | 2.00 | 2.00 | 390 | 46920 | 0.05000V | -134.76969539 | UNIFORM P-AXIS |
| 16 | 0.400 | 500 | 30 | 120 | 2.0 | 1.0 | 2.00 | 2.00 | 407 | 48960 | 0.90000V | -47.09603099 | UNIFORM P-AXIS |
| 17 | 0.500 | 500 | 30 | 120 | 2.0 | 1.0 | 2.00 | 2.00 | 370 | 44520 | 0.95000V | -1.54893769 | UNIFORM P-AXIS |
| 18 | 0.600 | 500 | 30 | 120 | 2.0 | 1.0 | 2.00 | 2.00 | 446 | 53640 | 0.95000V | -36.81437550 | UNIFORM P-AXIS |
| 19 | 0.700 | 500 | 30 | 120 | 2.0 | 1.0 | 2.00 | 2.00 | 389 | 46800 | 0.95000V | -136.56793953 | UNIFORM P-AXIS |
| 20 | 0.800 | 500 | 30 | 120 | 2.0 | 1.0 | 2.00 | 2.00 | 380 | 45720 | 0.50000V | -43.06679509 | UNIFORM P-AXIS |
| 21 | 0.900 | 500 | 30 | 120 | 2.0 | 1.0 | 2.00 | 2.00 | 345 | 41520 | 0.65000V | -25.74778799 | UNIFORM P-AXIS |
| 22 | 1.000 | 500 | 30 | 120 | 2.0 | 1.0 | 2.00 | 2.00 | 417 | 50160 | 0.45000V | -76.55528086 | UNIFORM P-AXIS |
| 23 | 0.000 | 500 | 30 | 180 | 2.0 | 1.0 | 2.00 | 2.00 | 364 | 65700 | 0.65000V | -39.37989070 | UNIFORM P-AXIS |
| 24 | 0.100 | 500 | 30 | 180 | 2.0 | 1.0 | 2.00 | 2.00 | 419 | 75600 | 0.55000V | -10.05898847 | UNIFORM P-AXIS |
| 25 | 0.200 | 500 | 30 | 180 | 2.0 | 1.0 | 2.00 | 2.00 | 425 | 76680 | 0.85000V | -0.83905075 | UNIFORM P-AXIS |
| 26 | 0.300 | 500 | 30 | 180 | 2.0 | 1.0 | 2.00 | 2.00 | 364 | 65700 | 0.65000V | -92.48498686 | UNIFORM P-AXIS |
| 27 | 0.400 | 500 | 30 | 180 | 2.0 | 1.0 | 2.00 | 2.00 | 383 | 69120 | 0.65000V | -55.05339898 | UNIFORM P-AXIS |
| 28 | 0.500 | 500 | 30 | 180 | 2.0 | 1.0 | 2.00 | 2.00 | 403 | 72720 | 0.70000V | -0.22231755 | UNIFORM P-AXIS |
| 29 | 0.600 | 500 | 30 | 180 | 2.0 | 1.0 | 2.00 | 2.00 | 366 | 66060 | 0.75000V | -55.86579942 | UNIFORM P-AXIS |
| 30 | 0.700 | 500 | 30 | 180 | 2.0 | 1.0 | 2.00 | 2.00 | 402 | 72540 | 0.65000V | -101.88858589 | UNIFORM P-AXIS |
| 31 | 0.800 | 500 | 30 | 180 | 2.0 | 1.0 | 2.00 | 2.00 | 389 | 70200 | 0.95000V | -7.12140022 | UNIFORM P-AXIS |
| 32 | 0.900 | 500 | 30 | 180 | 2.0 | 1.0 | 2.00 | 2.00 | 397 | 71640 | 0.40000V | -9.52128172 | UNIFORM P-AXIS |
| 33 | 1.000 | 500 | 30 | 180 | 2.0 | 1.0 | 2.00 | 2.00 | 387 | 69840 | 0.85000V | -54.11204028 | UNIFORM P-AXIS |

Total Function Evaluations: 1534260

| 6 | 0.500 | 500 | 30 | 60 | 2.0 | 1.0 | 2.00 | 2.00 | 170 | 10260 | 0.45000V | -0.00000006 | UNIFORM P-AXIS |

# F4

Run ID: 12-27-2009, 15:30:00

FUNCTION: F4

| Run # | Gamma | Nt | Nd | Np | G | DelT | Alpha | Beta | #Steps | Neval | Frep | Fitness | Initial Probes |
|-------|-------|----|----|----|----|------|-------|------|--------|-------|------|---------|----------------|
| 0 | 0.000 | 500 | 30 | 60 | 2.0 | 1.0 | 2.00 | 2.00 | 0 | 0 | 0.50000V | -9999.00000000 | UNIFORM P-AXIS |
| 1 | 0.000 | 500 | 30 | 60 | 2.0 | 1.0 | 2.00 | 2.00 | 60 | 3660 | 0.65000V | -100.00000000 | UNIFORM P-AXIS |
| 2 | 0.100 | 500 | 30 | 60 | 2.0 | 1.0 | 2.00 | 2.00 | 71 | 4320 | 0.25000V | -78.34901257 | UNIFORM P-AXIS |
| 3 | 0.200 | 500 | 30 | 60 | 2.0 | 1.0 | 2.00 | 2.00 | 71 | 4320 | 0.25000V | -54.46679317 | UNIFORM P-AXIS |
| 4 | 0.300 | 500 | 30 | 60 | 2.0 | 1.0 | 2.00 | 2.00 | 100 | 6060 | 0.75000V | -37.42713572 | UNIFORM P-AXIS |
| 5 | 0.400 | 500 | 30 | 60 | 2.0 | 1.0 | 2.00 | 2.00 | 121 | 7320 | 0.85000V | -12.05143023 | UNIFORM P-AXIS |
| 6 | 0.500 | 500 | 30 | 60 | 2.0 | 1.0 | 2.00 | 2.00 | 85 | 5160 | 0.95000V | -0.00000042 | UNIFORM P-AXIS |
| 7 | 0.600 | 500 | 30 | 60 | 2.0 | 1.0 | 2.00 | 2.00 | 110 | 6660 | 0.30000V | -12.81211639 | UNIFORM P-AXIS |
| 8 | 0.700 | 500 | 30 | 60 | 2.0 | 1.0 | 2.00 | 2.00 | 155 | 9360 | 0.65000V | -29.63311265 | UNIFORM P-AXIS |
| 9 | 0.800 | 500 | 30 | 60 | 2.0 | 1.0 | 2.00 | 2.00 | 71 | 4320 | 0.25000V | -54.46679317 | UNIFORM P-AXIS |
| 10 | 0.900 | 500 | 30 | 60 | 2.0 | 1.0 | 2.00 | 2.00 | 71 | 4320 | 0.25000V | -77.68060588 | UNIFORM P-AXIS |
| 11 | 1.000 | 500 | 30 | 60 | 2.0 | 1.0 | 2.00 | 2.00 | 60 | 3660 | 0.65000V | -100.00000000 | UNIFORM P-AXIS |
| 12 | 0.000 | 500 | 30 | 120 | 2.0 | 1.0 | 2.00 | 2.00 | 60 | 7320 | 0.65000V | -100.00000000 | UNIFORM P-AXIS |
| 13 | 0.100 | 500 | 30 | 120 | 2.0 | 1.0 | 2.00 | 2.00 | 60 | 7320 | 0.65000V | -69.33580190 | UNIFORM P-AXIS |
| 14 | 0.200 | 500 | 30 | 120 | 2.0 | 1.0 | 2.00 | 2.00 | 60 | 7320 | 0.65000V | -27.05570292 | UNIFORM P-AXIS |
| 15 | 0.300 | 500 | 30 | 120 | 2.0 | 1.0 | 2.00 | 2.00 | 75 | 9120 | 0.45000V | -5.57949024 | UNIFORM P-AXIS |
| 16 | 0.400 | 500 | 30 | 120 | 2.0 | 1.0 | 2.00 | 2.00 | 60 | 7320 | 0.65000V | -7.97580490 | UNIFORM P-AXIS |
| 17 | 0.500 | 500 | 30 | 120 | 2.0 | 1.0 | 2.00 | 2.00 | 101 | 12240 | 0.80000V | -0.05938789 | UNIFORM P-AXIS |
| 18 | 0.600 | 500 | 30 | 120 | 2.0 | 1.0 | 2.00 | 2.00 | 60 | 7320 | 0.65000V | -7.97580490 | UNIFORM P-AXIS |
| 19 | 0.700 | 500 | 30 | 120 | 2.0 | 1.0 | 2.00 | 2.00 | 75 | 9120 | 0.45000V | -5.57949024 | UNIFORM P-AXIS |
| 20 | 0.800 | 500 | 30 | 120 | 2.0 | 1.0 | 2.00 | 2.00 | 60 | 7320 | 0.65000V | -27.05570292 | UNIFORM P-AXIS |
| 21 | 0.900 | 500 | 30 | 120 | 2.0 | 1.0 | 2.00 | 2.00 | 60 | 7320 | 0.65000V | -69.33580190 | UNIFORM P-AXIS |
| 22 | 1.000 | 500 | 30 | 120 | 2.0 | 1.0 | 2.00 | 2.00 | 60 | 7320 | 0.65000V | -4.10445402 | UNIFORM P-AXIS |
| 23 | 0.000 | 500 | 30 | 180 | 2.0 | 1.0 | 2.00 | 2.00 | 60 | 10980 | 0.65000V | -100.00000000 | UNIFORM P-AXIS |
| 24 | 0.100 | 500 | 30 | 180 | 2.0 | 1.0 | 2.00 | 2.00 | 60 | 10980 | 0.65000V | -53.09339766 | UNIFORM P-AXIS |
| 25 | 0.200 | 500 | 30 | 180 | 2.0 | 1.0 | 2.00 | 2.00 | 75 | 13680 | 0.45000V | -2.16245242 | UNIFORM P-AXIS |
| 26 | 0.300 | 500 | 30 | 180 | 2.0 | 1.0 | 2.00 | 2.00 | 98 | 17820 | 0.65000V | -0.02337879 | UNIFORM P-AXIS |
| 27 | 0.400 | 500 | 30 | 180 | 2.0 | 1.0 | 2.00 | 2.00 | 108 | 19620 | 0.20000V | -0.70650669 | UNIFORM P-AXIS |
| 28 | 0.500 | 500 | 30 | 180 | 2.0 | 1.0 | 2.00 | 2.00 | 77 | 14040 | 0.55000V | -0.12574949 | UNIFORM P-AXIS |
| 29 | 0.600 | 500 | 30 | 180 | 2.0 | 1.0 | 2.00 | 2.00 | 190 | 34380 | 0.50000V | -0.01553213 | UNIFORM P-AXIS |
| 30 | 0.700 | 500 | 30 | 180 | 2.0 | 1.0 | 2.00 | 2.00 | 98 | 17820 | 0.65000V | -0.02337879 | UNIFORM P-AXIS |
| 31 | 0.800 | 500 | 30 | 180 | 2.0 | 1.0 | 2.00 | 2.00 | 114 | 20700 | 0.65000V | -0.00451056 | UNIFORM P-AXIS |
| 32 | 0.900 | 500 | 30 | 180 | 2.0 | 1.0 | 2.00 | 2.00 | 60 | 10980 | 0.65000V | -53.09339766 | UNIFORM P-AXIS |
| 33 | 1.000 | 500 | 30 | 180 | 2.0 | 1.0 | 2.00 | 2.00 | 72 | 13140 | 0.30000V | -1.23280952 | UNIFORM P-AXIS |

Total Function Evaluations: 332340

| 6 | 0.500 | 500 | 30 | 60 | 2.0 | 1.0 | 2.00 | 2.00 | 85 | 5160 | 0.95000V | -0.00000042 | UNIFORM P-AXIS |

## F5

FUNCTION: F5

| Run # | Gamma | Nt | Nd | Np | G | DelT | Alpha | Beta | #Steps | Neval | Frep | Fitness | Initial Probes |
|-------|-------|-----|-----|-----|-----|------|-------|------|--------|-------|----------|------------------|----------------|
| 0 | 0.000 | 500 | 30 | 60 | 2.0 | 1.0 | 2.00 | 2.00 | 0 | 0 | 0.50000V | -9999.00000000 | UNIFORM P-AXIS |
| 1 | 0.000 | 500 | 30 | 60 | 2.0 | 1.0 | 2.00 | 2.00 | 98 | 5940 | 0.65000V | -3.70063828 | UNIFORM P-AXIS |
| 2 | 0.100 | 500 | 30 | 60 | 2.0 | 1.0 | 2.00 | 2.00 | 191 | 11520 | 0.55000V | -13.15019089 | UNIFORM P-AXIS |
| 3 | 0.200 | 500 | 30 | 60 | 2.0 | 1.0 | 2.00 | 2.00 | 359 | 21600 | 0.40000V | -0.02171873 | UNIFORM P-AXIS |
| 4 | 0.300 | 500 | 30 | 60 | 2.0 | 1.0 | 2.00 | 2.00 | 320 | 19260 | 0.35000V | -0.06060966 | UNIFORM P-AXIS |
| 5 | 0.400 | 500 | 30 | 60 | 2.0 | 1.0 | 2.00 | 2.00 | 347 | 20880 | 0.75000V | -0.99902885 | UNIFORM P-AXIS |
| 6 | 0.500 | 500 | 30 | 60 | 2.0 | 1.0 | 2.00 | 2.00 | 334 | 20100 | 0.10000V | -1.16157214 | UNIFORM P-AXIS |
| 7 | 0.600 | 500 | 30 | 60 | 2.0 | 1.0 | 2.00 | 2.00 | 97 | 5880 | 0.60000V | -19026.84403442 | UNIFORM P-AXIS |
| 8 | 0.700 | 500 | 30 | 60 | 2.0 | 1.0 | 2.00 | 2.00 | 79 | 4800 | 0.65000V | -495.91491702 | UNIFORM P-AXIS |
| 9 | 0.800 | 500 | 30 | 60 | 2.0 | 1.0 | 2.00 | 2.00 | 441 | 26520 | 0.70000V | -2.54248267 | UNIFORM P-AXIS |
| 10 | 0.900 | 500 | 30 | 60 | 2.0 | 1.0 | 2.00 | 2.00 | 399 | 24000 | 0.50000V | -0.58852057 | UNIFORM P-AXIS |
| 11 | 1.000 | 500 | 30 | 60 | 2.0 | 1.0 | 2.00 | 2.00 | 60 | 3660 | 0.65000V | -44159.15850962 | UNIFORM P-AXIS |
| 12 | 0.000 | 500 | 30 | 120 | 2.0 | 1.0 | 2.00 | 2.00 | 264 | 31800 | 0.40000V | -48.73428954 | UNIFORM P-AXIS |
| 13 | 0.100 | 500 | 30 | 120 | 2.0 | 1.0 | 2.00 | 2.00 | 96 | 11640 | 0.55000V | -4.72035934 | UNIFORM P-AXIS |
| 14 | 0.200 | 500 | 30 | 120 | 2.0 | 1.0 | 2.00 | 2.00 | 208 | 25080 | 0.45000V | -0.00163122 | UNIFORM P-AXIS |
| 15 | 0.300 | 500 | 30 | 120 | 2.0 | 1.0 | 2.00 | 2.00 | 99 | 12000 | 0.70000V | -1.16925621 | UNIFORM P-AXIS |
| 16 | 0.400 | 500 | 30 | 120 | 2.0 | 1.0 | 2.00 | 2.00 | 365 | 43920 | 0.70000V | -1.06926420 | UNIFORM P-AXIS |
| 17 | 0.500 | 500 | 30 | 120 | 2.0 | 1.0 | 2.00 | 2.00 | 79 | 9600 | 0.65000V | -4.80953659 | UNIFORM P-AXIS |
| 18 | 0.600 | 500 | 30 | 120 | 2.0 | 1.0 | 2.00 | 2.00 | 392 | 47160 | 0.15000V | -2.19512324 | UNIFORM P-AXIS |
| 19 | 0.700 | 500 | 30 | 120 | 2.0 | 1.0 | 2.00 | 2.00 | 343 | 41280 | 0.55000V | -0.02220271 | UNIFORM P-AXIS |
| 20 | 0.800 | 500 | 30 | 120 | 2.0 | 1.0 | 2.00 | 2.00 | 400 | 48120 | 0.55000V | -0.48415875 | UNIFORM P-AXIS |
| 21 | 0.900 | 500 | 30 | 120 | 2.0 | 1.0 | 2.00 | 2.00 | 192 | 23160 | 0.60000V | -0.33830043 | UNIFORM P-AXIS |
| 22 | 1.000 | 500 | 30 | 120 | 2.0 | 1.0 | 2.00 | 2.00 | 173 | 20880 | 0.60000V | -2586.52615560 | UNIFORM P-AXIS |
| 23 | 0.000 | 500 | 30 | 180 | 2.0 | 1.0 | 2.00 | 2.00 | 338 | 61020 | 0.30000V | -0.11427428 | UNIFORM P-AXIS |
| 24 | 0.100 | 500 | 30 | 180 | 2.0 | 1.0 | 2.00 | 2.00 | 78 | 14220 | 0.60000V | -0.86881815 | UNIFORM P-AXIS |
| 25 | 0.200 | 500 | 30 | 180 | 2.0 | 1.0 | 2.00 | 2.00 | 191 | 34560 | 0.55000V | -0.06891125 | UNIFORM P-AXIS |
| 26 | 0.300 | 500 | 30 | 180 | 2.0 | 1.0 | 2.00 | 2.00 | 79 | 14400 | 0.65000V | -1.47748671 | UNIFORM P-AXIS |
| 27 | 0.400 | 500 | 30 | 180 | 2.0 | 1.0 | 2.00 | 2.00 | 80 | 14580 | 0.70000V | -13.38325143 | UNIFORM P-AXIS |
| 28 | 0.500 | 500 | 30 | 180 | 2.0 | 1.0 | 2.00 | 2.00 | 98 | 17820 | 0.65000V | -2.14030613 | UNIFORM P-AXIS |
| 29 | 0.600 | 500 | 30 | 180 | 2.0 | 1.0 | 2.00 | 2.00 | 192 | 34740 | 0.60000V | -294.28499320 | UNIFORM P-AXIS |
| 30 | 0.700 | 500 | 30 | 180 | 2.0 | 1.0 | 2.00 | 2.00 | 349 | 63000 | 0.80000V | -0.53537670 | UNIFORM P-AXIS |
| 31 | 0.800 | 500 | 30 | 180 | 2.0 | 1.0 | 2.00 | 2.00 | 77 | 14040 | 0.55000V | -2.50318155 | UNIFORM P-AXIS |
| 32 | 0.900 | 500 | 30 | 180 | 2.0 | 1.0 | 2.00 | 2.00 | 191 | 34560 | 0.55000V | -0.00109289 | UNIFORM P-AXIS |
| 33 | 1.000 | 500 | 30 | 180 | 2.0 | 1.0 | 2.00 | 2.00 | 354 | 63900 | 0.15000V | -5.02357451 | UNIFORM P-AXIS |

Total Function Evaluations: 845640

| 32 | 0.900 | 500 | 30 | 180 | 2.0 | 1.0 | 2.00 | 2.00 | 191 | 34560 | 0.55000V | -0.00109289 | UNIFORM P-AXIS |

## F6

FUNCTION: F6

| Run # | Gamma | Nt | Nd | Np | G | DelT | Alpha | Beta | #Steps | Neval | Frep | Fitness | Initial Probes |
|-------|-------|-----|-----|-----|-----|------|-------|------|--------|-------|----------|----------------|----------------|
| 0 | 0.000 | 500 | 30 | 60 | 2.0 | 1.0 | 2.00 | 2.00 | 0 | 0 | 0.50000V | -9999.00000000 | UNIFORM P-AXIS |
| 1 | 0.000 | 500 | 30 | 60 | 2.0 | 1.0 | 2.00 | 2.00 | 82 | 4980 | 0.80000V | 0.00000000 | UNIFORM P-AXIS |
| 2 | 0.100 | 500 | 30 | 60 | 2.0 | 1.0 | 2.00 | 2.00 | 117 | 7080 | 0.65000V | -150.00000000 | UNIFORM P-AXIS |
| 3 | 0.200 | 500 | 30 | 60 | 2.0 | 1.0 | 2.00 | 2.00 | 119 | 7200 | 0.75000V | -324.00000000 | UNIFORM P-AXIS |
| 4 | 0.300 | 500 | 30 | 60 | 2.0 | 1.0 | 2.00 | 2.00 | 123 | 7440 | 0.95000V | -81.00000000 | UNIFORM P-AXIS |
| 5 | 0.400 | 500 | 30 | 60 | 2.0 | 1.0 | 2.00 | 2.00 | 109 | 6600 | 0.25000V | -150.00000000 | UNIFORM P-AXIS |
| 6 | 0.500 | 500 | 30 | 60 | 2.0 | 1.0 | 2.00 | 2.00 | 111 | 6720 | 0.35000V | -256.00000000 | UNIFORM P-AXIS |
| 7 | 0.600 | 500 | 30 | 60 | 2.0 | 1.0 | 2.00 | 2.00 | 137 | 8280 | 0.70000V | -169.00000000 | UNIFORM P-AXIS |
| 8 | 0.700 | 500 | 30 | 60 | 2.0 | 1.0 | 2.00 | 2.00 | 92 | 5580 | 0.35000V | -870.00000000 | UNIFORM P-AXIS |
| 9 | 0.800 | 500 | 30 | 60 | 2.0 | 1.0 | 2.00 | 2.00 | 92 | 5580 | 0.35000V | -390.00000000 | UNIFORM P-AXIS |
| 10 | 0.900 | 500 | 30 | 60 | 2.0 | 1.0 | 2.00 | 2.00 | 117 | 7080 | 0.65000V | -150.00000000 | UNIFORM P-AXIS |
| 11 | 1.000 | 500 | 30 | 60 | 2.0 | 1.0 | 2.00 | 2.00 | 117 | 7080 | 0.65000V | -480.00000000 | UNIFORM P-AXIS |
| 12 | 0.000 | 500 | 30 | 120 | 2.0 | 1.0 | 2.00 | 2.00 | 110 | 13320 | 0.30000V | -1.00000000 | UNIFORM P-AXIS |
| 13 | 0.100 | 500 | 30 | 120 | 2.0 | 1.0 | 2.00 | 2.00 | 130 | 15720 | 0.35000V | 0.00000000 | UNIFORM P-AXIS |
| 14 | 0.200 | 500 | 30 | 120 | 2.0 | 1.0 | 2.00 | 2.00 | 116 | 14040 | 0.60000V | 0.00000000 | UNIFORM P-AXIS |
| 15 | 0.300 | 500 | 30 | 120 | 2.0 | 1.0 | 2.00 | 2.00 | 78 | 9480 | 0.60000V | -1.00000000 | UNIFORM P-AXIS |
| 16 | 0.400 | 500 | 30 | 120 | 2.0 | 1.0 | 2.00 | 2.00 | 79 | 9600 | 0.65000V | 0.00000000 | UNIFORM P-AXIS |
| 17 | 0.500 | 500 | 30 | 120 | 2.0 | 1.0 | 2.00 | 2.00 | 78 | 9480 | 0.60000V | 0.00000000 | UNIFORM P-AXIS |
| 18 | 0.600 | 500 | 30 | 120 | 2.0 | 1.0 | 2.00 | 2.00 | 79 | 9600 | 0.65000V | 0.00000000 | UNIFORM P-AXIS |
| 19 | 0.700 | 500 | 30 | 120 | 2.0 | 1.0 | 2.00 | 2.00 | 78 | 9480 | 0.60000V | 0.00000000 | UNIFORM P-AXIS |
| 20 | 0.800 | 500 | 30 | 120 | 2.0 | 1.0 | 2.00 | 2.00 | 78 | 9480 | 0.60000V | -1.00000000 | UNIFORM P-AXIS |
| 21 | 0.900 | 500 | 30 | 120 | 2.0 | 1.0 | 2.00 | 2.00 | 79 | 9600 | 0.65000V | -1.00000000 | UNIFORM P-AXIS |
| 22 | 1.000 | 500 | 30 | 120 | 2.0 | 1.0 | 2.00 | 2.00 | 110 | 13320 | 0.60000V | -1.00000000 | UNIFORM P-AXIS |
| 23 | 0.000 | 500 | 30 | 180 | 2.0 | 1.0 | 2.00 | 2.00 | 60 | 10980 | 0.65000V | 0.00000000 | UNIFORM P-AXIS |
| 24 | 0.100 | 500 | 30 | 180 | 2.0 | 1.0 | 2.00 | 2.00 | 78 | 14220 | 0.60000V | -4.00000000 | UNIFORM P-AXIS |
| 25 | 0.200 | 500 | 30 | 180 | 2.0 | 1.0 | 2.00 | 2.00 | 78 | 14220 | 0.60000V | 0.00000000 | UNIFORM P-AXIS |
| 26 | 0.300 | 500 | 30 | 180 | 2.0 | 1.0 | 2.00 | 2.00 | 97 | 17640 | 0.60000V | -1.00000000 | UNIFORM P-AXIS |
| 27 | 0.400 | 500 | 30 | 180 | 2.0 | 1.0 | 2.00 | 2.00 | 78 | 14220 | 0.60000V | 0.00000000 | UNIFORM P-AXIS |
| 28 | 0.500 | 500 | 30 | 180 | 2.0 | 1.0 | 2.00 | 2.00 | 78 | 14220 | 0.60000V | -1.00000000 | UNIFORM P-AXIS |
| 29 | 0.600 | 500 | 30 | 180 | 2.0 | 1.0 | 2.00 | 2.00 | 60 | 10980 | 0.65000V | 0.00000000 | UNIFORM P-AXIS |
| 30 | 0.700 | 500 | 30 | 180 | 2.0 | 1.0 | 2.00 | 2.00 | 97 | 17640 | 0.60000V | -1.00000000 | UNIFORM P-AXIS |
| 31 | 0.800 | 500 | 30 | 180 | 2.0 | 1.0 | 2.00 | 2.00 | 78 | 14220 | 0.60000V | -4.00000000 | UNIFORM P-AXIS |
| 32 | 0.900 | 500 | 30 | 180 | 2.0 | 1.0 | 2.00 | 2.00 | 78 | 14220 | 0.60000V | 0.00000000 | UNIFORM P-AXIS |
| 33 | 1.000 | 500 | 30 | 180 | 2.0 | 1.0 | 2.00 | 2.00 | 60 | 10980 | 0.65000V | 0.00000000 | UNIFORM P-AXIS |

Total Function Evaluations: 350280

| 33 | 1.000 | 500 | 30 | 180 | 2.0 | 1.0 | 2.00 | 2.00 | 60 | 10980 | 0.65000V | 0.00000000 | UNIFORM P-AXIS |

# F7

Run ID: 12-31-2009, 11:31:18

FUNCTION: F7

| Run # | Gamma | Nt | Nd | Np | G | DelT | Alpha | Beta | #Steps | Neval | Frep | Fitness | Initial Probes |
|-------|-------|-----|-----|-----|-----|------|-------|------|--------|-------|-----------|------------------|
| 0 | 0.000 | 500 | 30 | 60 | 2.0 | 1.0 | 2.00 | 2.00 | 0 | 0 | 0.50000V | -9999.00000000 | UNIFORM P-AXIS |
| 1 | 0.000 | 500 | 30 | 60 | 2.0 | 1.0 | 2.00 | 2.00 | 500 | 30060 | 0.80000V | -0.01097246 | UNIFORM P-AXIS |
| 2 | 0.100 | 500 | 30 | 60 | 2.0 | 1.0 | 2.00 | 2.00 | 500 | 30060 | 0.80000V | -0.00474254 | UNIFORM P-AXIS |
| 3 | 0.200 | 500 | 30 | 60 | 2.0 | 1.0 | 2.00 | 2.00 | 500 | 30060 | 0.80000V | -0.00105431 | UNIFORM P-AXIS |
| 4 | 0.300 | 500 | 30 | 60 | 2.0 | 1.0 | 2.00 | 2.00 | 500 | 30060 | 0.80000V | -0.00191198 | UNIFORM P-AXIS |
| 5 | 0.400 | 500 | 30 | 60 | 2.0 | 1.0 | 2.00 | 2.00 | 500 | 30060 | 0.80000V | -0.00207141 | UNIFORM P-AXIS |
| 6 | 0.500 | 500 | 30 | 60 | 2.0 | 1.0 | 2.00 | 2.00 | 500 | 30060 | 0.80000V | -0.00075131 | UNIFORM P-AXIS |
| 7 | 0.600 | 500 | 30 | 60 | 2.0 | 1.0 | 2.00 | 2.00 | 500 | 30060 | 0.80000V | -0.00308822 | UNIFORM P-AXIS |
| 8 | 0.700 | 500 | 30 | 60 | 2.0 | 1.0 | 2.00 | 2.00 | 500 | 30060 | 0.80000V | -0.00122170 | UNIFORM P-AXIS |
| 9 | 0.800 | 500 | 30 | 60 | 2.0 | 1.0 | 2.00 | 2.00 | 500 | 30060 | 0.80000V | -0.00161902 | UNIFORM P-AXIS |
| 10 | 0.900 | 500 | 30 | 60 | 2.0 | 1.0 | 2.00 | 2.00 | 500 | 30060 | 0.80000V | -0.00311318 | UNIFORM P-AXIS |
| 11 | 1.000 | 500 | 30 | 60 | 2.0 | 1.0 | 2.00 | 2.00 | 500 | 30060 | 0.80000V | -0.00373317 | UNIFORM P-AXIS |
| 12 | 0.000 | 500 | 30 | 120 | 2.0 | 1.0 | 2.00 | 2.00 | 500 | 60120 | 0.80000V | -0.00332989 | UNIFORM P-AXIS |
| 13 | 0.100 | 500 | 30 | 120 | 2.0 | 1.0 | 2.00 | 2.00 | 500 | 60120 | 0.80000V | -0.00004249 | UNIFORM P-AXIS |
| 14 | 0.200 | 500 | 30 | 120 | 2.0 | 1.0 | 2.00 | 2.00 | 500 | 60120 | 0.80000V | -0.00135119 | UNIFORM P-AXIS |
| 15 | 0.300 | 500 | 30 | 120 | 2.0 | 1.0 | 2.00 | 2.00 | 500 | 60120 | 0.80000V | -0.00196345 | UNIFORM P-AXIS |
| 16 | 0.400 | 500 | 30 | 120 | 2.0 | 1.0 | 2.00 | 2.00 | 500 | 60120 | 0.80000V | -0.00169604 | UNIFORM P-AXIS |
| 17 | 0.500 | 500 | 30 | 120 | 2.0 | 1.0 | 2.00 | 2.00 | 500 | 60120 | 0.80000V | -0.00094857 | UNIFORM P-AXIS |
| 18 | 0.600 | 500 | 30 | 120 | 2.0 | 1.0 | 2.00 | 2.00 | 500 | 60120 | 0.80000V | -0.00065703 | UNIFORM P-AXIS |
| 19 | 0.700 | 500 | 30 | 120 | 2.0 | 1.0 | 2.00 | 2.00 | 500 | 60120 | 0.80000V | -0.00084987 | UNIFORM P-AXIS |
| 20 | 0.800 | 500 | 30 | 120 | 2.0 | 1.0 | 2.00 | 2.00 | 500 | 60120 | 0.80000V | -0.00099223 | UNIFORM P-AXIS |
| 21 | 0.900 | 500 | 30 | 120 | 2.0 | 1.0 | 2.00 | 2.00 | 500 | 60120 | 0.80000V | -0.00033298 | UNIFORM P-AXIS |
| 22 | 1.000 | 500 | 30 | 120 | 2.0 | 1.0 | 2.00 | 2.00 | 500 | 60120 | 0.80000V | -0.00078422 | UNIFORM P-AXIS |
| 23 | 0.000 | 500 | 30 | 180 | 2.0 | 1.0 | 2.00 | 2.00 | 500 | 90180 | 0.80000V | -0.00150250 | UNIFORM P-AXIS |
| 24 | 0.100 | 500 | 30 | 180 | 2.0 | 1.0 | 2.00 | 2.00 | 500 | 90180 | 0.80000V | -0.00021166 | UNIFORM P-AXIS |
| 25 | 0.200 | 500 | 30 | 180 | 2.0 | 1.0 | 2.00 | 2.00 | 500 | 90180 | 0.80000V | -0.00057275 | UNIFORM P-AXIS |
| 26 | 0.300 | 500 | 30 | 180 | 2.0 | 1.0 | 2.00 | 2.00 | 500 | 90180 | 0.80000V | -0.00044470 | UNIFORM P-AXIS |
| 27 | 0.400 | 500 | 30 | 180 | 2.0 | 1.0 | 2.00 | 2.00 | 500 | 90180 | 0.80000V | -0.00202352 | UNIFORM P-AXIS |
| 28 | 0.500 | 500 | 30 | 180 | 2.0 | 1.0 | 2.00 | 2.00 | 500 | 90180 | 0.80000V | -0.00025176 | UNIFORM P-AXIS |
| 29 | 0.600 | 500 | 30 | 180 | 2.0 | 1.0 | 2.00 | 2.00 | 500 | 90180 | 0.80000V | -0.00179181 | UNIFORM P-AXIS |
| 30 | 0.700 | 500 | 30 | 180 | 2.0 | 1.0 | 2.00 | 2.00 | 500 | 90180 | 0.80000V | -0.00033893 | UNIFORM P-AXIS |
| 31 | 0.800 | 500 | 30 | 180 | 2.0 | 1.0 | 2.00 | 2.00 | 500 | 90180 | 0.80000V | -0.00100772 | UNIFORM P-AXIS |
| 32 | 0.900 | 500 | 30 | 180 | 2.0 | 1.0 | 2.00 | 2.00 | 500 | 90180 | 0.80000V | -0.00060779 | UNIFORM P-AXIS |
| 33 | 1.000 | 500 | 30 | 180 | 2.0 | 1.0 | 2.00 | 2.00 | 500 | 90180 | 0.80000V | -0.00118826 | UNIFORM P-AXIS |

Total Function Evaluations:    1983960

| 13 | 0.100 | 500 | 30 | 120 | 2.0 | 1.0 | 2.00 | 2.00 | 500 | 60120 | 0.80000V | -0.00004249 | UNIFORM P-AXIS |

NOTE: THIS FILE HAS BEEN EDITED TO SHOW THE CORRECT BEST RUN ON THE BOTTOM LINE BECAUSE FUNCTION F7 IS RANDOM.

# F8

Run ID: 12-31-2009, 16:01:36

FUNCTION: F8

| Run # | Gamma | Nt | Nd | Np | G | DelT | Alpha | Beta | #Steps | Neval | Frep | Fitness | Initial Probes |
|-------|-------|-----|-----|-----|-----|------|-------|------|--------|-------|-----------|------------------|
| 0 | 0.000 | 500 | 30 | 60 | 2.0 | 1.0 | 2.00 | 2.00 | 0 | 0 | 0.50000V | -9999.00000000 | UNIFORM P-AXIS |
| 1 | 0.000 | 500 | 30 | 60 | 2.0 | 1.0 | 2.00 | 2.00 | 110 | 6660 | 0.30000V | 5537.61608593 | UNIFORM P-AXIS |
| 2 | 0.100 | 500 | 30 | 60 | 2.0 | 1.0 | 2.00 | 2.00 | 85 | 5160 | 0.95000V | 12568.82823317 | UNIFORM P-AXIS |
| 3 | 0.200 | 500 | 30 | 60 | 2.0 | 1.0 | 2.00 | 2.00 | 86 | 5220 | 0.05000V | 12569.28760395 | UNIFORM P-AXIS |
| 4 | 0.300 | 500 | 30 | 60 | 2.0 | 1.0 | 2.00 | 2.00 | 88 | 5340 | 0.15000V | 12569.48093048 | UNIFORM P-AXIS |
| 5 | 0.400 | 500 | 30 | 60 | 2.0 | 1.0 | 2.00 | 2.00 | 86 | 5220 | 0.05000V | 12569.43736670 | UNIFORM P-AXIS |
| 6 | 0.500 | 500 | 30 | 60 | 2.0 | 1.0 | 2.00 | 2.00 | 95 | 5760 | 0.50000V | 12561.34875424 | UNIFORM P-AXIS |
| 7 | 0.600 | 500 | 30 | 60 | 2.0 | 1.0 | 2.00 | 2.00 | 137 | 8280 | 0.70000V | 12559.51991759 | UNIFORM P-AXIS |
| 8 | 0.700 | 500 | 30 | 60 | 2.0 | 1.0 | 2.00 | 2.00 | 97 | 5880 | 0.60000V | 9016.10961344 | UNIFORM P-AXIS |
| 9 | 0.800 | 500 | 30 | 60 | 2.0 | 1.0 | 2.00 | 2.00 | 132 | 7980 | 0.45000V | 12569.48515193 | UNIFORM P-AXIS |
| 10 | 0.900 | 500 | 30 | 60 | 2.0 | 1.0 | 2.00 | 2.00 | 105 | 6360 | 0.05000V | 12569.46303573 | UNIFORM P-AXIS |
| 11 | 1.000 | 500 | 30 | 60 | 2.0 | 1.0 | 2.00 | 2.00 | 130 | 7860 | 0.35000V | 8652.91904178 | UNIFORM P-AXIS |
| 12 | 0.000 | 500 | 30 | 120 | 2.0 | 1.0 | 2.00 | 2.00 | 110 | 13320 | 0.30000V | 5536.79333863 | UNIFORM P-AXIS |
| 13 | 0.100 | 500 | 30 | 120 | 2.0 | 1.0 | 2.00 | 2.00 | 124 | 15000 | 0.05000V | 12569.48174634 | UNIFORM P-AXIS |
| 14 | 0.200 | 500 | 30 | 120 | 2.0 | 1.0 | 2.00 | 2.00 | 87 | 10560 | 0.10000V | 12569.47740007 | UNIFORM P-AXIS |
| 15 | 0.300 | 500 | 30 | 120 | 2.0 | 1.0 | 2.00 | 2.00 | 106 | 12840 | 0.10000V | 12569.48333599 | UNIFORM P-AXIS |
| 16 | 0.400 | 500 | 30 | 120 | 2.0 | 1.0 | 2.00 | 2.00 | 86 | 10440 | 0.05000V | 12569.40587582 | UNIFORM P-AXIS |
| 17 | 0.500 | 500 | 30 | 120 | 2.0 | 1.0 | 2.00 | 2.00 | 105 | 12720 | 0.05000V | 12569.48661484 | UNIFORM P-AXIS |
| 18 | 0.600 | 500 | 30 | 120 | 2.0 | 1.0 | 2.00 | 2.00 | 124 | 15000 | 0.05000V | 12569.43670675 | UNIFORM P-AXIS |
| 19 | 0.700 | 500 | 30 | 120 | 2.0 | 1.0 | 2.00 | 2.00 | 87 | 10560 | 0.10000V | 12569.47329569 | UNIFORM P-AXIS |
| 20 | 0.800 | 500 | 30 | 120 | 2.0 | 1.0 | 2.00 | 2.00 | 73 | 8880 | 0.35000V | 8995.25108485 | UNIFORM P-AXIS |
| 21 | 0.900 | 500 | 30 | 120 | 2.0 | 1.0 | 2.00 | 2.00 | 105 | 12720 | 0.05000V | 12569.48652992 | UNIFORM P-AXIS |
| 22 | 1.000 | 500 | 30 | 120 | 2.0 | 1.0 | 2.00 | 2.00 | 194 | 23400 | 0.70000V | 12568.87307882 | UNIFORM P-AXIS |
| 23 | 0.000 | 500 | 30 | 180 | 2.0 | 1.0 | 2.00 | 2.00 | 95 | 17280 | 0.50000V | 12569.35060751 | UNIFORM P-AXIS |
| 24 | 0.100 | 500 | 30 | 180 | 2.0 | 1.0 | 2.00 | 2.00 | 144 | 26100 | 0.10000V | 12569.43649774 | UNIFORM P-AXIS |
| 25 | 0.200 | 500 | 30 | 180 | 2.0 | 1.0 | 2.00 | 2.00 | 162 | 29340 | 0.05000V | 12351.27895503 | UNIFORM P-AXIS |
| 26 | 0.300 | 500 | 30 | 180 | 2.0 | 1.0 | 2.00 | 2.00 | 87 | 15840 | 0.10000V | 12569.45150324 | UNIFORM P-AXIS |
| 27 | 0.400 | 500 | 30 | 180 | 2.0 | 1.0 | 2.00 | 2.00 | 78 | 14220 | 0.60000V | 12174.51472557 | UNIFORM P-AXIS |
| 28 | 0.500 | 500 | 30 | 180 | 2.0 | 1.0 | 2.00 | 2.00 | 125 | 22680 | 0.10000V | 12569.47757504 | UNIFORM P-AXIS |
| 29 | 0.600 | 500 | 30 | 180 | 2.0 | 1.0 | 2.00 | 2.00 | 134 | 24300 | 0.55000V | 12569.20312794 | UNIFORM P-AXIS |
| 30 | 0.700 | 500 | 30 | 180 | 2.0 | 1.0 | 2.00 | 2.00 | 148 | 26820 | 0.30000V | 12569.39892325 | UNIFORM P-AXIS |
| 31 | 0.800 | 500 | 30 | 180 | 2.0 | 1.0 | 2.00 | 2.00 | 110 | 19980 | 0.30000V | 12569.45496162 | UNIFORM P-AXIS |
| 32 | 0.900 | 500 | 30 | 180 | 2.0 | 1.0 | 2.00 | 2.00 | 86 | 15660 | 0.05000V | 12569.48420468 | UNIFORM P-AXIS |
| 33 | 1.000 | 500 | 30 | 180 | 2.0 | 1.0 | 2.00 | 2.00 | 118 | 21420 | 0.70000V | 12569.46181648 | UNIFORM P-AXIS |

Total Function Evaluations:    448800

| 17 | 0.500 | 500 | 30 | 120 | 2.0 | 1.0 | 2.00 | 2.00 | 105 | 12720 | 0.05000V | 12569.48661484 | UNIFORM P-AXIS |

# F9

FUNCTION: F9

| Run # | Gamma | Nt | Nd | Np | G | DelT | Alpha | Beta | #Steps | Neval | Frep | Fitness | Initial Probes |
|-------|-------|-----|-----|-----|-----|------|-------|------|--------|-------|----------|------------------|----------------|
| 0 | 0.000 | 500 | 30 | 60 | 2.0 | 1.0 | 2.00 | 2.00 | 0 | 0 | 0.50000V | -9999.00000000 | UNIFORM P-AXIS |
| 1 | 0.000 | 500 | 30 | 60 | 2.0 | 1.0 | 2.00 | 2.00 | 97 | 5880 | 0.60000V | -31.70098539 | UNIFORM P-AXIS |
| 2 | 0.100 | 500 | 30 | 60 | 2.0 | 1.0 | 2.00 | 2.00 | 99 | 6000 | 0.70000V | -34.19310095 | UNIFORM P-AXIS |
| 3 | 0.200 | 500 | 30 | 60 | 2.0 | 1.0 | 2.00 | 2.00 | 311 | 18720 | 0.85000V | -123.74637495 | UNIFORM P-AXIS |
| 4 | 0.300 | 500 | 30 | 60 | 2.0 | 1.0 | 2.00 | 2.00 | 239 | 14400 | 0.10000V | -108.91030539 | UNIFORM P-AXIS |
| 5 | 0.400 | 500 | 30 | 60 | 2.0 | 1.0 | 2.00 | 2.00 | 103 | 6240 | 0.90000V | -31.69341889 | UNIFORM P-AXIS |
| 6 | 0.500 | 500 | 30 | 60 | 2.0 | 1.0 | 2.00 | 2.00 | 113 | 6840 | 0.45000V | -0.99112119 | UNIFORM P-AXIS |
| 7 | 0.600 | 500 | 30 | 60 | 2.0 | 1.0 | 2.00 | 2.00 | 500 | 30060 | 0.80000V | -41.36476630 | UNIFORM P-AXIS |
| 8 | 0.700 | 500 | 30 | 60 | 2.0 | 1.0 | 2.00 | 2.00 | 172 | 10380 | 0.55000V | -44.82618586 | UNIFORM P-AXIS |
| 9 | 0.800 | 500 | 30 | 60 | 2.0 | 1.0 | 2.00 | 2.00 | 116 | 7020 | 0.60000V | -112.60669341 | UNIFORM P-AXIS |
| 10 | 0.900 | 500 | 30 | 60 | 2.0 | 1.0 | 2.00 | 2.00 | 78 | 4740 | 0.60000V | -483.01739630 | UNIFORM P-AXIS |
| 11 | 1.000 | 500 | 30 | 60 | 2.0 | 1.0 | 2.00 | 2.00 | 113 | 6840 | 0.45000V | -0.00000352 | UNIFORM P-AXIS |
| 12 | 0.000 | 500 | 30 | 120 | 2.0 | 1.0 | 2.00 | 2.00 | 60 | 7320 | 0.65000V | -40.51140239 | UNIFORM P-AXIS |
| 13 | 0.100 | 500 | 30 | 120 | 2.0 | 1.0 | 2.00 | 2.00 | 247 | 29760 | 0.50000V | -29.69982488 | UNIFORM P-AXIS |
| 14 | 0.200 | 500 | 30 | 120 | 2.0 | 1.0 | 2.00 | 2.00 | 116 | 14040 | 0.60000V | -0.00383322 | UNIFORM P-AXIS |
| 15 | 0.300 | 500 | 30 | 120 | 2.0 | 1.0 | 2.00 | 2.00 | 151 | 18240 | 0.45000V | -0.00133000 | UNIFORM P-AXIS |
| 16 | 0.400 | 500 | 30 | 120 | 2.0 | 1.0 | 2.00 | 2.00 | 96 | 11640 | 0.55000V | -0.00003333 | UNIFORM P-AXIS |
| 17 | 0.500 | 500 | 30 | 120 | 2.0 | 1.0 | 2.00 | 2.00 | 165 | 19920 | 0.20000V | -0.99011355 | UNIFORM P-AXIS |
| 18 | 0.600 | 500 | 30 | 120 | 2.0 | 1.0 | 2.00 | 2.00 | 96 | 11640 | 0.55000V | -0.00003333 | UNIFORM P-AXIS |
| 19 | 0.700 | 500 | 30 | 120 | 2.0 | 1.0 | 2.00 | 2.00 | 136 | 16440 | 0.65000V | -0.00000205 | UNIFORM P-AXIS |
| 20 | 0.800 | 500 | 30 | 120 | 2.0 | 1.0 | 2.00 | 2.00 | 116 | 14040 | 0.60000V | -0.00383322 | UNIFORM P-AXIS |
| 21 | 0.900 | 500 | 30 | 120 | 2.0 | 1.0 | 2.00 | 2.00 | 155 | 18720 | 0.65000V | -30.01506187 | UNIFORM P-AXIS |
| 22 | 1.000 | 500 | 30 | 120 | 2.0 | 1.0 | 2.00 | 2.00 | 500 | 60120 | 0.80000V | -0.22468142 | UNIFORM P-AXIS |
| 23 | 0.000 | 500 | 30 | 180 | 2.0 | 1.0 | 2.00 | 2.00 | 116 | 21060 | 0.60000V | -0.00005257 | UNIFORM P-AXIS |
| 24 | 0.100 | 500 | 30 | 180 | 2.0 | 1.0 | 2.00 | 2.00 | 116 | 21060 | 0.60000V | -0.00035429 | UNIFORM P-AXIS |
| 25 | 0.200 | 500 | 30 | 180 | 2.0 | 1.0 | 2.00 | 2.00 | 182 | 32940 | 0.10000V | -0.99080917 | UNIFORM P-AXIS |
| 26 | 0.300 | 500 | 30 | 180 | 2.0 | 1.0 | 2.00 | 2.00 | 176 | 31860 | 0.75000V | -0.99013929 | UNIFORM P-AXIS |
| 27 | 0.400 | 500 | 30 | 180 | 2.0 | 1.0 | 2.00 | 2.00 | 155 | 28080 | 0.65000V | -0.00054147 | UNIFORM P-AXIS |
| 28 | 0.500 | 500 | 30 | 180 | 2.0 | 1.0 | 2.00 | 2.00 | 96 | 17460 | 0.55000V | -0.00014569 | UNIFORM P-AXIS |
| 29 | 0.600 | 500 | 30 | 180 | 2.0 | 1.0 | 2.00 | 2.00 | 500 | 90180 | 0.80000V | -0.28712215 | UNIFORM P-AXIS |
| 30 | 0.700 | 500 | 30 | 180 | 2.0 | 1.0 | 2.00 | 2.00 | 116 | 21060 | 0.60000V | -0.00015616 | UNIFORM P-AXIS |
| 31 | 0.800 | 500 | 30 | 180 | 2.0 | 1.0 | 2.00 | 2.00 | 143 | 25920 | 0.05000V | -15.84039530 | UNIFORM P-AXIS |
| 32 | 0.900 | 500 | 30 | 180 | 2.0 | 1.0 | 2.00 | 2.00 | 171 | 30960 | 0.50000V | -0.00051308 | UNIFORM P-AXIS |
| 33 | 1.000 | 500 | 30 | 180 | 2.0 | 1.0 | 2.00 | 2.00 | 116 | 21060 | 0.60000V | -0.00005257 | UNIFORM P-AXIS |

Total Function Evaluations: 680640

| 19 | 0.700 | 500 | 30 | 120 | 2.0 | 1.0 | 2.00 | 2.00 | 136 | 16440 | 0.65000V | -0.00000205 | UNIFORM P-AXIS |

# F10

FUNCTION: F10

| Run # | Gamma | Nt | Nd | Np | G | DelT | Alpha | Beta | #Steps | Neval | Frep | Fitness | Initial Probes |
|-------|-------|-----|-----|-----|-----|------|-------|------|--------|-------|----------|------------------|----------------|
| 0 | 0.000 | 500 | 30 | 60 | 2.0 | 1.0 | 2.00 | 2.00 | 0 | 0 | 0.50000V | -9999.00000000 | UNIFORM P-AXIS |
| 1 | 0.000 | 500 | 30 | 60 | 2.0 | 1.0 | 2.00 | 2.00 | 171 | 10320 | 0.50000V | -19.95554498 | UNIFORM P-AXIS |
| 2 | 0.100 | 500 | 30 | 60 | 2.0 | 1.0 | 2.00 | 2.00 | 196 | 11820 | 0.80000V | -21.94028909 | UNIFORM P-AXIS |
| 3 | 0.200 | 500 | 30 | 60 | 2.0 | 1.0 | 2.00 | 2.00 | 179 | 10800 | 0.90000V | -18.77780599 | UNIFORM P-AXIS |
| 4 | 0.300 | 500 | 30 | 60 | 2.0 | 1.0 | 2.00 | 2.00 | 273 | 16440 | 0.85000V | -14.99021318 | UNIFORM P-AXIS |
| 5 | 0.400 | 500 | 30 | 60 | 2.0 | 1.0 | 2.00 | 2.00 | 164 | 9900 | 0.15000V | -12.68499222 | UNIFORM P-AXIS |
| 6 | 0.500 | 500 | 30 | 60 | 2.0 | 1.0 | 2.00 | 2.00 | 84 | 5100 | 0.90000V | -0.00000015 | UNIFORM P-AXIS |
| 7 | 0.600 | 500 | 30 | 60 | 2.0 | 1.0 | 2.00 | 2.00 | 120 | 7260 | 0.80000V | -16.10493237 | UNIFORM P-AXIS |
| 8 | 0.700 | 500 | 30 | 60 | 2.0 | 1.0 | 2.00 | 2.00 | 156 | 9420 | 0.70000V | -17.47681504 | UNIFORM P-AXIS |
| 9 | 0.800 | 500 | 30 | 60 | 2.0 | 1.0 | 2.00 | 2.00 | 171 | 10320 | 0.50000V | -19.67837086 | UNIFORM P-AXIS |
| 10 | 0.900 | 500 | 30 | 60 | 2.0 | 1.0 | 2.00 | 2.00 | 196 | 11820 | 0.80000V | -21.97779101 | UNIFORM P-AXIS |
| 11 | 1.000 | 500 | 30 | 60 | 2.0 | 1.0 | 2.00 | 2.00 | 140 | 8460 | 0.85000V | -0.65731364 | UNIFORM P-AXIS |
| 12 | 0.000 | 500 | 30 | 120 | 2.0 | 1.0 | 2.00 | 2.00 | 163 | 19680 | 0.10000V | -19.95577490 | UNIFORM P-AXIS |
| 13 | 0.100 | 500 | 30 | 120 | 2.0 | 1.0 | 2.00 | 2.00 | 225 | 27120 | 0.35000V | -21.88619837 | UNIFORM P-AXIS |
| 14 | 0.200 | 500 | 30 | 120 | 2.0 | 1.0 | 2.00 | 2.00 | 229 | 27600 | 0.55000V | -18.91894002 | UNIFORM P-AXIS |
| 15 | 0.300 | 500 | 30 | 120 | 2.0 | 1.0 | 2.00 | 2.00 | 207 | 24960 | 0.40000V | -4.27574675 | UNIFORM P-AXIS |
| 16 | 0.400 | 500 | 30 | 120 | 2.0 | 1.0 | 2.00 | 2.00 | 155 | 18720 | 0.65000V | -12.54005560 | UNIFORM P-AXIS |
| 17 | 0.500 | 500 | 30 | 120 | 2.0 | 1.0 | 2.00 | 2.00 | 155 | 18720 | 0.65000V | -0.01693374 | UNIFORM P-AXIS |
| 18 | 0.600 | 500 | 30 | 120 | 2.0 | 1.0 | 2.00 | 2.00 | 332 | 39960 | 0.95000V | -6.87736931 | UNIFORM P-AXIS |
| 19 | 0.700 | 500 | 30 | 120 | 2.0 | 1.0 | 2.00 | 2.00 | 290 | 34920 | 0.75000V | -15.81273556 | UNIFORM P-AXIS |
| 20 | 0.800 | 500 | 30 | 120 | 2.0 | 1.0 | 2.00 | 2.00 | 283 | 34080 | 0.40000V | -19.29625344 | UNIFORM P-AXIS |
| 21 | 0.900 | 500 | 30 | 120 | 2.0 | 1.0 | 2.00 | 2.00 | 226 | 27240 | 0.40000V | -21.87441323 | UNIFORM P-AXIS |
| 22 | 1.000 | 500 | 30 | 120 | 2.0 | 1.0 | 2.00 | 2.00 | 426 | 51240 | 0.90000V | -0.00004949 | UNIFORM P-AXIS |
| 23 | 0.000 | 500 | 30 | 180 | 2.0 | 1.0 | 2.00 | 2.00 | 138 | 25020 | 0.75000V | -19.95643855 | UNIFORM P-AXIS |
| 24 | 0.100 | 500 | 30 | 180 | 2.0 | 1.0 | 2.00 | 2.00 | 206 | 37260 | 0.35000V | -21.74579561 | UNIFORM P-AXIS |
| 25 | 0.200 | 500 | 30 | 180 | 2.0 | 1.0 | 2.00 | 2.00 | 203 | 36720 | 0.20000V | -19.62244329 | UNIFORM P-AXIS |
| 26 | 0.300 | 500 | 30 | 180 | 2.0 | 1.0 | 2.00 | 2.00 | 500 | 90180 | 0.40000V | -17.80088045 | UNIFORM P-AXIS |
| 27 | 0.400 | 500 | 30 | 180 | 2.0 | 1.0 | 2.00 | 2.00 | 232 | 41940 | 0.70000V | -10.88650930 | UNIFORM P-AXIS |
| 28 | 0.500 | 500 | 30 | 180 | 2.0 | 1.0 | 2.00 | 2.00 | 78 | 14220 | 0.60000V | -0.39610444 | UNIFORM P-AXIS |
| 29 | 0.600 | 500 | 30 | 180 | 2.0 | 1.0 | 2.00 | 2.00 | 125 | 22680 | 0.10000V | -0.01473085 | UNIFORM P-AXIS |
| 30 | 0.700 | 500 | 30 | 180 | 2.0 | 1.0 | 2.00 | 2.00 | 194 | 35100 | 0.70000V | -6.47287042 | UNIFORM P-AXIS |
| 31 | 0.800 | 500 | 30 | 180 | 2.0 | 1.0 | 2.00 | 2.00 | 393 | 70920 | 0.20000V | -0.01545910 | UNIFORM P-AXIS |
| 32 | 0.900 | 500 | 30 | 180 | 2.0 | 1.0 | 2.00 | 2.00 | 386 | 69660 | 0.80000V | -19.96912371 | UNIFORM P-AXIS |
| 33 | 1.000 | 500 | 30 | 180 | 2.0 | 1.0 | 2.00 | 2.00 | 140 | 25380 | 0.85000V | -0.09944870 | UNIFORM P-AXIS |

Total Function Evaluations: 904980

| 6 | 0.500 | 500 | 30 | 60 | 2.0 | 1.0 | 2.00 | 2.00 | 84 | 5100 | 0.90000V | -0.00000015 | UNIFORM P-AXIS |

# F11

Run ID: 12-31-2009, 19:27:23

FUNCTION: F11

| Run # | Gamma | Nt | Nd | Np | G | DelT | Alpha | Beta | #Steps | Neval | Frep | Fitness | Initial Probes |
|-------|-------|-----|-----|-----|-----|------|-------|------|--------|-------|----------|-----------------|----------------|
| 0 | 0.000 | 500 | 30 | 60 | 2.0 | 1.0 | 2.00 | 2.00 | 0 | 0 | 0.50000V | -9999.00000000 | UNIFORM P-AXIS |
| 1 | 0.000 | 500 | 30 | 60 | 2.0 | 1.0 | 2.00 | 2.00 | 161 | 9720 | 0.95000V | -3349.89947839 | UNIFORM P-AXIS |
| 2 | 0.100 | 500 | 30 | 60 | 2.0 | 1.0 | 2.00 | 2.00 | 109 | 6600 | 0.25000V | -682.73550403 | UNIFORM P-AXIS |
| 3 | 0.200 | 500 | 30 | 60 | 2.0 | 1.0 | 2.00 | 2.00 | 121 | 7320 | 0.85000V | -787.80355520 | UNIFORM P-AXIS |
| 4 | 0.300 | 500 | 30 | 60 | 2.0 | 1.0 | 2.00 | 2.00 | 242 | 14580 | 0.25000V | -779.13514096 | UNIFORM P-AXIS |
| 5 | 0.400 | 500 | 30 | 60 | 2.0 | 1.0 | 2.00 | 2.00 | 500 | 30060 | 0.80000V | -328.25766449 | UNIFORM P-AXIS |
| 6 | 0.500 | 500 | 30 | 60 | 2.0 | 1.0 | 2.00 | 2.00 | 500 | 30060 | 0.80000V | -64.46413940 | UNIFORM P-AXIS |
| 7 | 0.600 | 500 | 30 | 60 | 2.0 | 1.0 | 2.00 | 2.00 | 473 | 28440 | 0.40000V | -4.00165096 | UNIFORM P-AXIS |
| 8 | 0.700 | 500 | 30 | 60 | 2.0 | 1.0 | 2.00 | 2.00 | 122 | 7380 | 0.90000V | -142.36479694 | UNIFORM P-AXIS |
| 9 | 0.800 | 500 | 30 | 60 | 2.0 | 1.0 | 2.00 | 2.00 | 103 | 6240 | 0.90000V | -487.25760030 | UNIFORM P-AXIS |
| 10 | 0.900 | 500 | 30 | 60 | 2.0 | 1.0 | 2.00 | 2.00 | 104 | 6300 | 0.95000V | -1024.52777584 | UNIFORM P-AXIS |
| 11 | 1.000 | 500 | 30 | 60 | 2.0 | 1.0 | 2.00 | 2.00 | 232 | 13980 | 0.70000V | -2.00124049 | UNIFORM P-AXIS |
| 12 | 0.000 | 500 | 30 | 120 | 2.0 | 1.0 | 2.00 | 2.00 | 174 | 21000 | 0.65000V | -1042.16470301 | UNIFORM P-AXIS |
| 13 | 0.100 | 500 | 30 | 120 | 2.0 | 1.0 | 2.00 | 2.00 | 60 | 7320 | 0.65000V | -3.03975464 | UNIFORM P-AXIS |
| 14 | 0.200 | 500 | 30 | 120 | 2.0 | 1.0 | 2.00 | 2.00 | 60 | 7320 | 0.65000V | -2.57331594 | UNIFORM P-AXIS |
| 15 | 0.300 | 500 | 30 | 120 | 2.0 | 1.0 | 2.00 | 2.00 | 60 | 7320 | 0.65000V | -1.39086441 | UNIFORM P-AXIS |
| 16 | 0.400 | 500 | 30 | 120 | 2.0 | 1.0 | 2.00 | 2.00 | 60 | 7320 | 0.65000V | -5.67090092 | UNIFORM P-AXIS |
| 17 | 0.500 | 500 | 30 | 120 | 2.0 | 1.0 | 2.00 | 2.00 | 140 | 16920 | 0.85000V | -64.20929684 | UNIFORM P-AXIS |
| 18 | 0.600 | 500 | 30 | 120 | 2.0 | 1.0 | 2.00 | 2.00 | 60 | 7320 | 0.65000V | -6.40000000 | UNIFORM P-AXIS |
| 19 | 0.700 | 500 | 30 | 120 | 2.0 | 1.0 | 2.00 | 2.00 | 178 | 21480 | 0.85000V | -97.02717321 | UNIFORM P-AXIS |
| 20 | 0.800 | 500 | 30 | 120 | 2.0 | 1.0 | 2.00 | 2.00 | 60 | 7320 | 0.65000V | -445.43143407 | UNIFORM P-AXIS |
| 21 | 0.900 | 500 | 30 | 120 | 2.0 | 1.0 | 2.00 | 2.00 | 60 | 7320 | 0.65000V | -40.34010306 | UNIFORM P-AXIS |
| 22 | 1.000 | 500 | 30 | 120 | 2.0 | 1.0 | 2.00 | 2.00 | 60 | 7320 | 0.65000V | -4.05519280 | UNIFORM P-AXIS |
| 23 | 0.000 | 500 | 30 | 180 | 2.0 | 1.0 | 2.00 | 2.00 | 111 | 20160 | 0.35000V | -1127.36887599 | UNIFORM P-AXIS |
| 24 | 0.100 | 500 | 30 | 180 | 2.0 | 1.0 | 2.00 | 2.00 | 236 | 42660 | 0.90000V | -0.09972929 | UNIFORM P-AXIS |
| 25 | 0.200 | 500 | 30 | 180 | 2.0 | 1.0 | 2.00 | 2.00 | 251 | 45360 | 0.70000V | -1.89829119 | UNIFORM P-AXIS |
| 26 | 0.300 | 500 | 30 | 180 | 2.0 | 1.0 | 2.00 | 2.00 | 60 | 10980 | 0.65000V | -4.42021222 | UNIFORM P-AXIS |
| 27 | 0.400 | 500 | 30 | 180 | 2.0 | 1.0 | 2.00 | 2.00 | 60 | 10980 | 0.65000V | -8.01901276 | UNIFORM P-AXIS |
| 28 | 0.500 | 500 | 30 | 180 | 2.0 | 1.0 | 2.00 | 2.00 | 60 | 10980 | 0.65000V | -68.05499564 | UNIFORM P-AXIS |
| 29 | 0.600 | 500 | 30 | 180 | 2.0 | 1.0 | 2.00 | 2.00 | 60 | 10980 | 0.65000V | -3.84611240 | UNIFORM P-AXIS |
| 30 | 0.700 | 500 | 30 | 180 | 2.0 | 1.0 | 2.00 | 2.00 | 140 | 25380 | 0.85000V | -121.65998851 | UNIFORM P-AXIS |
| 31 | 0.800 | 500 | 30 | 180 | 2.0 | 1.0 | 2.00 | 2.00 | 60 | 10980 | 0.65000V | -12.45943847 | UNIFORM P-AXIS |
| 32 | 0.900 | 500 | 30 | 180 | 2.0 | 1.0 | 2.00 | 2.00 | 60 | 10980 | 0.65000V | -3.38969420 | UNIFORM P-AXIS |
| 33 | 1.000 | 500 | 30 | 180 | 2.0 | 1.0 | 2.00 | 2.00 | 60 | 10980 | 0.65000V | -3.56479100 | UNIFORM P-AXIS |

Total Function Evaluations:  489060

| 24 | 0.100 | 500 | 30 | 180 | 2.0 | 1.0 | 2.00 | 2.00 | 236 | 42660 | 0.90000V | -0.09972929 | UNIFORM P-AXIS |

# F12

Run ID: 02-03-2010, 14:55:15 [CORRECTED 02-03-2010; SEE F12 SOURCE LISTING PAGE 60]

FUNCTION: F12

| Run # | Gamma | Nt | Nd | Np | G | DelT | Alpha | Beta | #Steps | Neval | Frep | Fitness | Initial Probes |
|-------|-------|-----|-----|-----|-----|------|-------|------|--------|-------|----------|-------------|----------------|
| 0 | 0.000 | 500 | 30 | 60 | 2.0 | 1.0 | 2.00 | 2.00 | 0 | 0 | 0.50000V | -9999.00000000 | UNIFORM P-AXIS |
| 1 | 0.000 | 500 | 30 | 60 | 2.0 | 1.0 | 2.00 | 2.00 | 78 | 4740 | 0.60000V | -0.79539561 | UNIFORM P-AXIS |
| 2 | 0.100 | 500 | 30 | 60 | 2.0 | 1.0 | 2.00 | 2.00 | 99 | 6000 | 0.70000V | -0.74543651 | UNIFORM P-AXIS |
| 3 | 0.200 | 500 | 30 | 60 | 2.0 | 1.0 | 2.00 | 2.00 | 92 | 5580 | 0.35000V | -0.42992625 | UNIFORM P-AXIS |
| 4 | 0.300 | 500 | 30 | 60 | 2.0 | 1.0 | 2.00 | 2.00 | 96 | 5820 | 0.55000V | -0.43773493 | UNIFORM P-AXIS |
| 5 | 0.400 | 500 | 30 | 60 | 2.0 | 1.0 | 2.00 | 2.00 | 156 | 9420 | 0.70000V | -0.25675666 | UNIFORM P-AXIS |
| 6 | 0.500 | 500 | 30 | 60 | 2.0 | 1.0 | 2.00 | 2.00 | 60 | 3660 | 0.65000V | -0.00002067 | UNIFORM P-AXIS |
| 7 | 0.600 | 500 | 30 | 60 | 2.0 | 1.0 | 2.00 | 2.00 | 60 | 3660 | 0.65000V | -0.24925810 | UNIFORM P-AXIS |
| 8 | 0.700 | 500 | 30 | 60 | 2.0 | 1.0 | 2.00 | 2.00 | 60 | 3660 | 0.65000V | -0.72371775 | UNIFORM P-AXIS |
| 9 | 0.800 | 500 | 30 | 60 | 2.0 | 1.0 | 2.00 | 2.00 | 60 | 3660 | 0.65000V | -2.19256987 | UNIFORM P-AXIS |
| 10 | 0.900 | 500 | 30 | 60 | 2.0 | 1.0 | 2.00 | 2.00 | 115 | 6960 | 0.55000V | -0.13285456 | UNIFORM P-AXIS |
| 11 | 1.000 | 500 | 30 | 60 | 2.0 | 1.0 | 2.00 | 2.00 | 82 | 4980 | 0.80000V | -1.60665754 | UNIFORM P-AXIS |
| 12 | 0.000 | 500 | 30 | 120 | 2.0 | 1.0 | 2.00 | 2.00 | 135 | 16320 | 0.60000V | -0.39097490 | UNIFORM P-AXIS |
| 13 | 0.100 | 500 | 30 | 120 | 2.0 | 1.0 | 2.00 | 2.00 | 60 | 7320 | 0.65000V | -0.00368253 | UNIFORM P-AXIS |
| 14 | 0.200 | 500 | 30 | 120 | 2.0 | 1.0 | 2.00 | 2.00 | 62 | 7560 | 0.75000V | -0.25600092 | UNIFORM P-AXIS |
| 15 | 0.300 | 500 | 30 | 120 | 2.0 | 1.0 | 2.00 | 2.00 | 133 | 16080 | 0.50000V | -0.02016903 | UNIFORM P-AXIS |
| 16 | 0.400 | 500 | 30 | 120 | 2.0 | 1.0 | 2.00 | 2.00 | 78 | 9480 | 0.65000V | -0.25920627 | UNIFORM P-AXIS |
| 17 | 0.500 | 500 | 30 | 120 | 2.0 | 1.0 | 2.00 | 2.00 | 96 | 11640 | 0.55000V | -0.01298711 | UNIFORM P-AXIS |
| 18 | 0.600 | 500 | 30 | 120 | 2.0 | 1.0 | 2.00 | 2.00 | 135 | 16320 | 0.60000V | -0.26677253 | UNIFORM P-AXIS |
| 19 | 0.700 | 500 | 30 | 120 | 2.0 | 1.0 | 2.00 | 2.00 | 115 | 13920 | 0.55000V | -0.36764888 | UNIFORM P-AXIS |
| 20 | 0.800 | 500 | 30 | 120 | 2.0 | 1.0 | 2.00 | 2.00 | 60 | 7320 | 0.65000V | -0.02887759 | UNIFORM P-AXIS |
| 21 | 0.900 | 500 | 30 | 120 | 2.0 | 1.0 | 2.00 | 2.00 | 97 | 11760 | 0.60000V | -0.03222183 | UNIFORM P-AXIS |
| 22 | 1.000 | 500 | 30 | 120 | 2.0 | 1.0 | 2.00 | 2.00 | 78 | 9480 | 0.65000V | -0.06522233 | UNIFORM P-AXIS |
| 23 | 0.000 | 500 | 30 | 180 | 2.0 | 1.0 | 2.00 | 2.00 | 117 | 21240 | 0.65000V | -0.14434782 | UNIFORM P-AXIS |
| 24 | 0.100 | 500 | 30 | 180 | 2.0 | 1.0 | 2.00 | 2.00 | 60 | 10980 | 0.65000V | -0.06696541 | UNIFORM P-AXIS |
| 25 | 0.200 | 500 | 30 | 180 | 2.0 | 1.0 | 2.00 | 2.00 | 60 | 10980 | 0.65000V | -0.07738490 | UNIFORM P-AXIS |
| 26 | 0.300 | 500 | 30 | 180 | 2.0 | 1.0 | 2.00 | 2.00 | 116 | 21060 | 0.60000V | -0.16057613 | UNIFORM P-AXIS |
| 27 | 0.400 | 500 | 30 | 180 | 2.0 | 1.0 | 2.00 | 2.00 | 60 | 10980 | 0.65000V | -0.03082085 | UNIFORM P-AXIS |
| 28 | 0.500 | 500 | 30 | 180 | 2.0 | 1.0 | 2.00 | 2.00 | 97 | 17640 | 0.60000V | -0.00446426 | UNIFORM P-AXIS |
| 29 | 0.600 | 500 | 30 | 180 | 2.0 | 1.0 | 2.00 | 2.00 | 96 | 17460 | 0.55000V | -0.11565986 | UNIFORM P-AXIS |
| 30 | 0.700 | 500 | 30 | 180 | 2.0 | 1.0 | 2.00 | 2.00 | 60 | 10980 | 0.65000V | -0.00606971 | UNIFORM P-AXIS |
| 31 | 0.800 | 500 | 30 | 180 | 2.0 | 1.0 | 2.00 | 2.00 | 60 | 10980 | 0.65000V | -0.01177648 | UNIFORM P-AXIS |
| 32 | 0.900 | 500 | 30 | 180 | 2.0 | 1.0 | 2.00 | 2.00 | 60 | 10980 | 0.65000V | -0.01952312 | UNIFORM P-AXIS |
| 33 | 1.000 | 500 | 30 | 180 | 2.0 | 1.0 | 2.00 | 2.00 | 70 | 12780 | 0.20000V | -0.26244867 | UNIFORM P-AXIS |

Total Function Evaluations:  341400

| 6 | 0.500 | 500 | 30 | 60 | 2.0 | 1.0 | 2.00 | 2.00 | 60 | 3660 | 0.65000V | -0.00002067 | UNIFORM P-AXIS |

# F13

FUNCTION: F13

| Run # | Gamma | Nt | Nd | Np | G | DelT | Alpha | Beta | #Steps | Neval | Frep | Fitness | Initial Probes |
|-------|-------|----|----|----|----|------|-------|------|--------|-------|------|---------|----------------|
| 0 | 0.000 | 500 | 30 | 60 | 2.0 | 1.0 | 2.00 | 2.00 | 0 | 0 | 0.50000V | -9999.00000000 | UNIFORM P-AXIS |
| 1 | 0.000 | 500 | 30 | 60 | 2.0 | 1.0 | 2.00 | 2.00 | 91 | 5520 | 0.30000V | -1.98285152 | UNIFORM P-AXIS |
| 2 | 0.100 | 500 | 30 | 60 | 2.0 | 1.0 | 2.00 | 2.00 | 312 | 18780 | 0.90000V | -0.23909156 | UNIFORM P-AXIS |
| 3 | 0.200 | 500 | 30 | 60 | 2.0 | 1.0 | 2.00 | 2.00 | 90 | 5460 | 0.25000V | -0.15479827 | UNIFORM P-AXIS |
| 4 | 0.300 | 500 | 30 | 60 | 2.0 | 1.0 | 2.00 | 2.00 | 81 | 4920 | 0.75000V | -0.28720755 | UNIFORM P-AXIS |
| 5 | 0.400 | 500 | 30 | 60 | 2.0 | 1.0 | 2.00 | 2.00 | 233 | 14040 | 0.75000V | -0.14883921 | UNIFORM P-AXIS |
| 6 | 0.500 | 500 | 30 | 60 | 2.0 | 1.0 | 2.00 | 2.00 | 184 | 11100 | 0.20000V | -0.06596601 | UNIFORM P-AXIS |
| 7 | 0.600 | 500 | 30 | 60 | 2.0 | 1.0 | 2.00 | 2.00 | 226 | 13620 | 0.40000V | -0.28878418 | UNIFORM P-AXIS |
| 8 | 0.700 | 500 | 30 | 60 | 2.0 | 1.0 | 2.00 | 2.00 | 122 | 7380 | 0.90000V | -0.76742123 | UNIFORM P-AXIS |
| 9 | 0.800 | 500 | 30 | 60 | 2.0 | 1.0 | 2.00 | 2.00 | 90 | 5460 | 0.25000V | -0.60350146 | UNIFORM P-AXIS |
| 10 | 0.900 | 500 | 30 | 60 | 2.0 | 1.0 | 2.00 | 2.00 | 134 | 8100 | 0.55000V | -0.50142794 | UNIFORM P-AXIS |
| 11 | 1.000 | 500 | 30 | 60 | 2.0 | 1.0 | 2.00 | 2.00 | 148 | 8940 | 0.30000V | -0.26195645 | UNIFORM P-AXIS |
| 12 | 0.000 | 500 | 30 | 120 | 2.0 | 1.0 | 2.00 | 2.00 | 331 | 39840 | 0.90000V | -0.17132369 | UNIFORM P-AXIS |
| 13 | 0.100 | 500 | 30 | 120 | 2.0 | 1.0 | 2.00 | 2.00 | 130 | 15720 | 0.35000V | -0.63280216 | UNIFORM P-AXIS |
| 14 | 0.200 | 500 | 30 | 120 | 2.0 | 1.0 | 2.00 | 2.00 | 312 | 37560 | 0.90000V | -0.33885975 | UNIFORM P-AXIS |
| 15 | 0.300 | 500 | 30 | 120 | 2.0 | 1.0 | 2.00 | 2.00 | 204 | 24600 | 0.25000V | -0.18849309 | UNIFORM P-AXIS |
| 16 | 0.400 | 500 | 30 | 120 | 2.0 | 1.0 | 2.00 | 2.00 | 60 | 7320 | 0.65000V | -0.41319188 | UNIFORM P-AXIS |
| 17 | 0.500 | 500 | 30 | 120 | 2.0 | 1.0 | 2.00 | 2.00 | 98 | 11880 | 0.65000V | -0.03399272 | UNIFORM P-AXIS |
| 18 | 0.600 | 500 | 30 | 120 | 2.0 | 1.0 | 2.00 | 2.00 | 165 | 19920 | 0.20000V | -0.25527275 | UNIFORM P-AXIS |
| 19 | 0.700 | 500 | 30 | 120 | 2.0 | 1.0 | 2.00 | 2.00 | 383 | 46080 | 0.65000V | -0.27622507 | UNIFORM P-AXIS |
| 20 | 0.800 | 500 | 30 | 120 | 2.0 | 1.0 | 2.00 | 2.00 | 172 | 20760 | 0.55000V | -0.17442617 | UNIFORM P-AXIS |
| 21 | 0.900 | 500 | 30 | 120 | 2.0 | 1.0 | 2.00 | 2.00 | 212 | 25560 | 0.65000V | -0.02638101 | UNIFORM P-AXIS |
| 22 | 1.000 | 500 | 30 | 120 | 2.0 | 1.0 | 2.00 | 2.00 | 215 | 25920 | 0.80000V | -0.27534519 | UNIFORM P-AXIS |
| 23 | 0.000 | 500 | 30 | 180 | 2.0 | 1.0 | 2.00 | 2.00 | 97 | 17640 | 0.60000V | -0.07220597 | UNIFORM P-AXIS |
| 24 | 0.100 | 500 | 30 | 180 | 2.0 | 1.0 | 2.00 | 2.00 | 60 | 10980 | 0.65000V | -0.07174276 | UNIFORM P-AXIS |
| 25 | 0.200 | 500 | 30 | 180 | 2.0 | 1.0 | 2.00 | 2.00 | 370 | 66780 | 0.95000V | -0.15772989 | UNIFORM P-AXIS |
| 26 | 0.300 | 500 | 30 | 180 | 2.0 | 1.0 | 2.00 | 2.00 | 151 | 27360 | 0.45000V | -0.08917428 | UNIFORM P-AXIS |
| 27 | 0.400 | 500 | 30 | 180 | 2.0 | 1.0 | 2.00 | 2.00 | 195 | 35280 | 0.75000V | -0.06522780 | UNIFORM P-AXIS |
| 28 | 0.500 | 500 | 30 | 180 | 2.0 | 1.0 | 2.00 | 2.00 | 134 | 24300 | 0.55000V | -0.00713353 | UNIFORM P-AXIS |
| 29 | 0.600 | 500 | 30 | 180 | 2.0 | 1.0 | 2.00 | 2.00 | 93 | 16920 | 0.40000V | -0.00328529 | UNIFORM P-AXIS |
| 30 | 0.700 | 500 | 30 | 180 | 2.0 | 1.0 | 2.00 | 2.00 | 186 | 33660 | 0.30000V | -0.19308067 | UNIFORM P-AXIS |
| 31 | 0.800 | 500 | 30 | 180 | 2.0 | 1.0 | 2.00 | 2.00 | 170 | 30780 | 0.45000V | -0.07304213 | UNIFORM P-AXIS |
| 32 | 0.900 | 500 | 30 | 180 | 2.0 | 1.0 | 2.00 | 2.00 | 60 | 10980 | 0.65000V | -0.24282441 | UNIFORM P-AXIS |
| 33 | 1.000 | 500 | 30 | 180 | 2.0 | 1.0 | 2.00 | 2.00 | 146 | 26460 | 0.20000V | -0.02567155 | UNIFORM P-AXIS |

Total Function Evaluations: 679620

| 29 | 0.600 | 500 | 30 | 180 | 2.0 | 1.0 | 2.00 | 2.00 | 93 | 16920 | 0.40000V | -0.00328529 | UNIFORM P-AXIS |

# F14

FUNCTION: F14

| Run # | Gamma | Nt | Nd | Np | G | DelT | Alpha | Beta | #Steps | Neval | Frep | Fitness | Initial Probes |
|-------|-------|----|----|----|----|------|-------|------|--------|-------|------|---------|----------------|
| 0 | 0.000 | 500 | 2 | 8 | 2.0 | 1.0 | 2.00 | 2.00 | 0 | 0 | 0.50000V | -9999.00000000 | UNIFORM P-AXIS |
| 1 | 0.000 | 500 | 2 | 8 | 2.0 | 1.0 | 2.00 | 2.00 | 213 | 1712 | 0.70000V | -5.93292124 | UNIFORM P-AXIS |
| 2 | 0.100 | 500 | 2 | 8 | 2.0 | 1.0 | 2.00 | 2.00 | 500 | 4008 | 0.80000V | -2.08127846 | UNIFORM P-AXIS |
| 3 | 0.200 | 500 | 2 | 8 | 2.0 | 1.0 | 2.00 | 2.00 | 97 | 784 | 0.60000V | -2.98238078 | UNIFORM P-AXIS |
| 4 | 0.300 | 500 | 2 | 8 | 2.0 | 1.0 | 2.00 | 2.00 | 114 | 920 | 0.50000V | -1.00528367 | UNIFORM P-AXIS |
| 5 | 0.400 | 500 | 2 | 8 | 2.0 | 1.0 | 2.00 | 2.00 | 78 | 632 | 0.60000V | -6.90336530 | UNIFORM P-AXIS |
| 6 | 0.500 | 500 | 2 | 8 | 2.0 | 1.0 | 2.00 | 2.00 | 123 | 992 | 0.95000V | -1.99203466 | UNIFORM P-AXIS |
| 7 | 0.600 | 500 | 2 | 8 | 2.0 | 1.0 | 2.00 | 2.00 | 70 | 568 | 0.20000V | -18.31709957 | UNIFORM P-AXIS |
| 8 | 0.700 | 500 | 2 | 8 | 2.0 | 1.0 | 2.00 | 2.00 | 118 | 952 | 0.70000V | -2.98345327 | UNIFORM P-AXIS |
| 9 | 0.800 | 500 | 2 | 8 | 2.0 | 1.0 | 2.00 | 2.00 | 108 | 872 | 0.20000V | -7.87400969 | UNIFORM P-AXIS |
| 10 | 0.900 | 500 | 2 | 8 | 2.0 | 1.0 | 2.00 | 2.00 | 95 | 768 | 0.50000V | -9.80389837 | UNIFORM P-AXIS |
| 11 | 1.000 | 500 | 2 | 8 | 2.0 | 1.0 | 2.00 | 2.00 | 76 | 616 | 0.50000V | -10.76397606 | UNIFORM P-AXIS |
| 12 | 0.000 | 500 | 2 | 12 | 2.0 | 1.0 | 2.00 | 2.00 | 213 | 2568 | 0.70000V | -1.99245403 | UNIFORM P-AXIS |
| 13 | 0.100 | 500 | 2 | 12 | 2.0 | 1.0 | 2.00 | 2.00 | 139 | 1680 | 0.80000V | -1.00776304 | UNIFORM P-AXIS |
| 14 | 0.200 | 500 | 2 | 12 | 2.0 | 1.0 | 2.00 | 2.00 | 114 | 1380 | 0.50000V | -10.76325056 | UNIFORM P-AXIS |
| 15 | 0.300 | 500 | 2 | 12 | 2.0 | 1.0 | 2.00 | 2.00 | 92 | 1116 | 0.35000V | -1.00708019 | UNIFORM P-AXIS |
| 16 | 0.400 | 500 | 2 | 12 | 2.0 | 1.0 | 2.00 | 2.00 | 98 | 1188 | 0.65000V | -5.92895444 | UNIFORM P-AXIS |
| 17 | 0.500 | 500 | 2 | 12 | 2.0 | 1.0 | 2.00 | 2.00 | 72 | 876 | 0.30000V | -3.03091277 | UNIFORM P-AXIS |
| 18 | 0.600 | 500 | 2 | 12 | 2.0 | 1.0 | 2.00 | 2.00 | 82 | 996 | 0.80000V | -6.90589693 | UNIFORM P-AXIS |
| 19 | 0.700 | 500 | 2 | 12 | 2.0 | 1.0 | 2.00 | 2.00 | 110 | 1332 | 0.30000V | -8.84107275 | UNIFORM P-AXIS |
| 20 | 0.800 | 500 | 2 | 12 | 2.0 | 1.0 | 2.00 | 2.00 | 136 | 1644 | 0.65000V | -1.08554446 | UNIFORM P-AXIS |
| 21 | 0.900 | 500 | 2 | 12 | 2.0 | 1.0 | 2.00 | 2.00 | 87 | 1056 | 0.10000V | -2.98244726 | UNIFORM P-AXIS |
| 22 | 1.000 | 500 | 2 | 12 | 2.0 | 1.0 | 2.00 | 2.00 | 76 | 924 | 0.50000V | -10.76397665 | UNIFORM P-AXIS |
| 23 | 0.000 | 500 | 2 | 16 | 2.0 | 1.0 | 2.00 | 2.00 | 130 | 2096 | 0.35000V | -1.99739732 | UNIFORM P-AXIS |
| 24 | 0.100 | 500 | 2 | 16 | 2.0 | 1.0 | 2.00 | 2.00 | 115 | 1856 | 0.55000V | -1.99209759 | UNIFORM P-AXIS |
| 25 | 0.200 | 500 | 2 | 16 | 2.0 | 1.0 | 2.00 | 2.00 | 98 | 1584 | 0.65000V | -0.99806031 | UNIFORM P-AXIS |
| 26 | 0.300 | 500 | 2 | 16 | 2.0 | 1.0 | 2.00 | 2.00 | 79 | 1280 | 0.65000V | -3.96825029 | UNIFORM P-AXIS |
| 27 | 0.400 | 500 | 2 | 16 | 2.0 | 1.0 | 2.00 | 2.00 | 60 | 976 | 0.65000V | -13.61860920 | UNIFORM P-AXIS |
| 28 | 0.500 | 500 | 2 | 16 | 2.0 | 1.0 | 2.00 | 2.00 | 93 | 1504 | 0.40000V | -1.00064288 | UNIFORM P-AXIS |
| 29 | 0.600 | 500 | 2 | 16 | 2.0 | 1.0 | 2.00 | 2.00 | 81 | 1312 | 0.75000V | -4.95050169 | UNIFORM P-AXIS |
| 30 | 0.700 | 500 | 2 | 16 | 2.0 | 1.0 | 2.00 | 2.00 | 96 | 1552 | 0.55000V | -4.95057264 | UNIFORM P-AXIS |
| 31 | 0.800 | 500 | 2 | 16 | 2.0 | 1.0 | 2.00 | 2.00 | 95 | 1536 | 0.50000V | -13.61905412 | UNIFORM P-AXIS |
| 32 | 0.900 | 500 | 2 | 16 | 2.0 | 1.0 | 2.00 | 2.00 | 82 | 1328 | 0.80000V | -2.98333672 | UNIFORM P-AXIS |
| 33 | 1.000 | 500 | 2 | 16 | 2.0 | 1.0 | 2.00 | 2.00 | 153 | 2464 | 0.55000V | -0.99800512 | UNIFORM P-AXIS |
| 34 | 0.000 | 500 | 2 | 20 | 2.0 | 1.0 | 2.00 | 2.00 | 128 | 2580 | 0.25000V | -6.90577771 | UNIFORM P-AXIS |
| 35 | 0.100 | 500 | 2 | 20 | 2.0 | 1.0 | 2.00 | 2.00 | 155 | 3120 | 0.65000V | -11.71875390 | UNIFORM P-AXIS |
| 36 | 0.200 | 500 | 2 | 20 | 2.0 | 1.0 | 2.00 | 2.00 | 124 | 2500 | 0.05000V | -5.92888566 | UNIFORM P-AXIS |
| 37 | 0.300 | 500 | 2 | 20 | 2.0 | 1.0 | 2.00 | 2.00 | 60 | 1220 | 0.65000V | -2.00372774 | UNIFORM P-AXIS |
| 38 | 0.400 | 500 | 2 | 20 | 2.0 | 1.0 | 2.00 | 2.00 | 110 | 2220 | 0.30000V | -6.90334175 | UNIFORM P-AXIS |
| 39 | 0.500 | 500 | 2 | 20 | 2.0 | 1.0 | 2.00 | 2.00 | 77 | 1560 | 0.55000V | -1.99362004 | UNIFORM P-AXIS |
| 40 | 0.600 | 500 | 2 | 20 | 2.0 | 1.0 | 2.00 | 2.00 | 94 | 1900 | 0.45000V | -4.95052579 | UNIFORM P-AXIS |
| 41 | 0.700 | 500 | 2 | 20 | 2.0 | 1.0 | 2.00 | 2.00 | 122 | 2460 | 0.90000V | -1.00012042 | UNIFORM P-AXIS |
| 42 | 0.800 | 500 | 2 | 20 | 2.0 | 1.0 | 2.00 | 2.00 | 97 | 1960 | 0.60000V | -12.67050646 | UNIFORM P-AXIS |
| 43 | 0.900 | 500 | 2 | 20 | 2.0 | 1.0 | 2.00 | 2.00 | 95 | 1920 | 0.50000V | -2.98355029 | UNIFORM P-AXIS |
| 44 | 1.000 | 500 | 2 | 20 | 2.0 | 1.0 | 2.00 | 2.00 | 116 | 2340 | 0.60000V | -6.90750161 | UNIFORM P-AXIS |
| 45 | 0.000 | 500 | 2 | 24 | 2.0 | 1.0 | 2.00 | 2.00 | 103 | 2496 | 0.90000V | -13.62018424 | UNIFORM P-AXIS |
| 46 | 0.100 | 500 | 2 | 24 | 2.0 | 1.0 | 2.00 | 2.00 | 469 | 11280 | 0.70000V | -1.01402257 | UNIFORM P-AXIS |
| 47 | 0.200 | 500 | 2 | 24 | 2.0 | 1.0 | 2.00 | 2.00 | 247 | 5952 | 0.50000V | -0.99800389 | UNIFORM P-AXIS |
| 48 | 0.300 | 500 | 2 | 24 | 2.0 | 1.0 | 2.00 | 2.00 | 60 | 1464 | 0.65000V | -12.67050583 | UNIFORM P-AXIS |

| Run # | Gamma | Nt | Nd | Np | G | DelT | Alpha | Beta | #Steps | Neval | Frep | Fitness | Initial Probes |
|---|---|---|---|---|---|---|---|---|---|---|---|---|---|
| 49 | 0.400 | 500 | 2 | 24 | 2.0 | 1.0 | 2.00 | 2.00 | 114 | 2760 | 0.50000V | -1.99259236 | UNIFORM P-AXIS |
| 50 | 0.500 | 500 | 2 | 24 | 2.0 | 1.0 | 2.00 | 2.00 | 135 | 3264 | 0.60000V | -2.98494679 | UNIFORM P-AXIS |
| 51 | 0.600 | 500 | 2 | 24 | 2.0 | 1.0 | 2.00 | 2.00 | 112 | 2712 | 0.40000V | -2.98305135 | UNIFORM P-AXIS |
| 52 | 0.700 | 500 | 2 | 24 | 2.0 | 1.0 | 2.00 | 2.00 | 60 | 1464 | 0.65000V | -12.67050581 | UNIFORM P-AXIS |
| 53 | 0.800 | 500 | 2 | 24 | 2.0 | 1.0 | 2.00 | 2.00 | 78 | 1896 | 0.60000V | -8.84121987 | UNIFORM P-AXIS |
| 54 | 0.900 | 500 | 2 | 24 | 2.0 | 1.0 | 2.00 | 2.00 | 129 | 3120 | 0.30000V | -9.80393440 | UNIFORM P-AXIS |
| 55 | 1.000 | 500 | 2 | 24 | 2.0 | 1.0 | 2.00 | 2.00 | 75 | 1824 | 0.45000V | -6.90339260 | UNIFORM P-AXIS |
| 56 | 0.000 | 500 | 2 | 28 | 2.0 | 1.0 | 2.00 | 2.00 | 105 | 2968 | 0.05000V | -12.72309266 | UNIFORM P-AXIS |
| 57 | 0.100 | 500 | 2 | 28 | 2.0 | 1.0 | 2.00 | 2.00 | 291 | 8176 | 0.80000V | -2.05112505 | UNIFORM P-AXIS |
| 58 | 0.200 | 500 | 2 | 28 | 2.0 | 1.0 | 2.00 | 2.00 | 77 | 2184 | 0.35000V | -0.99800500 | UNIFORM P-AXIS |
| 59 | 0.300 | 500 | 2 | 28 | 2.0 | 1.0 | 2.00 | 2.00 | 78 | 2212 | 0.60000V | -12.68239387 | UNIFORM P-AXIS |
| 60 | 0.400 | 500 | 2 | 28 | 2.0 | 1.0 | 2.00 | 2.00 | 77 | 2184 | 0.55000V | -5.92884534 | UNIFORM P-AXIS |
| 61 | 0.500 | 500 | 2 | 28 | 2.0 | 1.0 | 2.00 | 2.00 | 60 | 1708 | 0.65000V | -7.87399305 | UNIFORM P-AXIS |
| 62 | 0.600 | 500 | 2 | 28 | 2.0 | 1.0 | 2.00 | 2.00 | 79 | 2240 | 0.65000V | -7.87399303 | UNIFORM P-AXIS |
| 63 | 0.700 | 500 | 2 | 28 | 2.0 | 1.0 | 2.00 | 2.00 | 123 | 3472 | 0.95000V | -7.89231597 | UNIFORM P-AXIS |
| 64 | 0.800 | 500 | 2 | 28 | 2.0 | 1.0 | 2.00 | 2.00 | 92 | 2604 | 0.35000V | -1.00752170 | UNIFORM P-AXIS |
| 65 | 0.900 | 500 | 2 | 28 | 2.0 | 1.0 | 2.00 | 2.00 | 110 | 3108 | 0.30000V | -8.84100679 | UNIFORM P-AXIS |
| 66 | 1.000 | 500 | 2 | 28 | 2.0 | 1.0 | 2.00 | 2.00 | 111 | 3136 | 0.35000V | -0.99828689 | UNIFORM P-AXIS |

Total Function Evaluations: 141076

---

| 47 | 0.200 | 500 | 2 | 24 | 2.0 | 1.0 | 2.00 | 2.00 | 247 | 5952 | 0.50000V | -0.99800389 | UNIFORM P-AXIS |

## **F15**

Run ID: 12-26-2009, 23:44:46

FUNCTION: F15

| Run # | Gamma | Nt | Nd | Np | G | DelT | Alpha | Beta | #Steps | Neval | Frep | Fitness | Initial Probes |
|---|---|---|---|---|---|---|---|---|---|---|---|---|---|
| 0 | 0.000 | 500 | 4 | 16 | 2.0 | 1.0 | 2.00 | 2.00 | 0 | 0 | 0.50000V | -9999.00000000 | UNIFORM P-AXIS |

---

| 1 | 0.000 | 500 | 4 | 16 | 2.0 | 1.0 | 2.00 | 2.00 | 60 | 976 | 0.65000V | -0.00969087 | UNIFORM P-AXIS |
| 2 | 0.100 | 500 | 4 | 16 | 2.0 | 1.0 | 2.00 | 2.00 | 60 | 976 | 0.65000V | -0.04147696 | UNIFORM P-AXIS |
| 3 | 0.200 | 500 | 4 | 16 | 2.0 | 1.0 | 2.00 | 2.00 | 293 | 4704 | 0.90000V | -0.01107829 | UNIFORM P-AXIS |
| 4 | 0.300 | 500 | 4 | 16 | 2.0 | 1.0 | 2.00 | 2.00 | 174 | 2800 | 0.65000V | -0.01340605 | UNIFORM P-AXIS |
| 5 | 0.400 | 500 | 4 | 16 | 2.0 | 1.0 | 2.00 | 2.00 | 60 | 976 | 0.65000V | -0.04063055 | UNIFORM P-AXIS |
| 6 | 0.500 | 500 | 4 | 16 | 2.0 | 1.0 | 2.00 | 2.00 | 98 | 1584 | 0.65000V | -0.00241631 | UNIFORM P-AXIS |
| 7 | 0.600 | 500 | 4 | 16 | 2.0 | 1.0 | 2.00 | 2.00 | 60 | 976 | 0.65000V | -0.12870474 | UNIFORM P-AXIS |
| 8 | 0.700 | 500 | 4 | 16 | 2.0 | 1.0 | 2.00 | 2.00 | 60 | 976 | 0.65000V | -0.11830207 | UNIFORM P-AXIS |
| 9 | 0.800 | 500 | 4 | 16 | 2.0 | 1.0 | 2.00 | 2.00 | 60 | 976 | 0.65000V | -0.06996965 | UNIFORM P-AXIS |
| 10 | 0.900 | 500 | 4 | 16 | 2.0 | 1.0 | 2.00 | 2.00 | 135 | 2176 | 0.65000V | -0.00326219 | UNIFORM P-AXIS |
| 11 | 1.000 | 500 | 4 | 16 | 2.0 | 1.0 | 2.00 | 2.00 | 174 | 2800 | 0.65000V | -0.01224564 | UNIFORM P-AXIS |
| 12 | 0.000 | 500 | 4 | 24 | 2.0 | 1.0 | 2.00 | 2.00 | 60 | 1464 | 0.65000V | -0.01602030 | UNIFORM P-AXIS |
| 13 | 0.100 | 500 | 4 | 24 | 2.0 | 1.0 | 2.00 | 2.00 | 60 | 1464 | 0.65000V | -0.00973842 | UNIFORM P-AXIS |
| 14 | 0.200 | 500 | 4 | 24 | 2.0 | 1.0 | 2.00 | 2.00 | 60 | 1464 | 0.65000V | -0.01306855 | UNIFORM P-AXIS |
| 15 | 0.300 | 500 | 4 | 24 | 2.0 | 1.0 | 2.00 | 2.00 | 60 | 1464 | 0.65000V | -0.00676815 | UNIFORM P-AXIS |
| 16 | 0.400 | 500 | 4 | 24 | 2.0 | 1.0 | 2.00 | 2.00 | 60 | 1464 | 0.65000V | -0.04798806 | UNIFORM P-AXIS |
| 17 | 0.500 | 500 | 4 | 24 | 2.0 | 1.0 | 2.00 | 2.00 | 60 | 1464 | 0.65000V | -0.02226235 | UNIFORM P-AXIS |
| 18 | 0.600 | 500 | 4 | 24 | 2.0 | 1.0 | 2.00 | 2.00 | 60 | 1464 | 0.65000V | -0.09216877 | UNIFORM P-AXIS |
| 19 | 0.700 | 500 | 4 | 24 | 2.0 | 1.0 | 2.00 | 2.00 | 174 | 4200 | 0.65000V | -0.00277113 | UNIFORM P-AXIS |
| 20 | 0.800 | 500 | 4 | 24 | 2.0 | 1.0 | 2.00 | 2.00 | 256 | 6168 | 0.95000V | -0.04058785 | UNIFORM P-AXIS |
| 21 | 0.900 | 500 | 4 | 24 | 2.0 | 1.0 | 2.00 | 2.00 | 128 | 3096 | 0.25000V | -0.00937831 | UNIFORM P-AXIS |
| 22 | 1.000 | 500 | 4 | 24 | 2.0 | 1.0 | 2.00 | 2.00 | 193 | 4656 | 0.65000V | -0.01337984 | UNIFORM P-AXIS |
| 23 | 0.000 | 500 | 4 | 32 | 2.0 | 1.0 | 2.00 | 2.00 | 60 | 1952 | 0.65000V | -0.02197297 | UNIFORM P-AXIS |
| 24 | 0.100 | 500 | 4 | 32 | 2.0 | 1.0 | 2.00 | 2.00 | 60 | 1952 | 0.65000V | -0.00559270 | UNIFORM P-AXIS |
| 25 | 0.200 | 500 | 4 | 32 | 2.0 | 1.0 | 2.00 | 2.00 | 311 | 9984 | 0.85000V | -0.00272084 | UNIFORM P-AXIS |
| 26 | 0.300 | 500 | 4 | 32 | 2.0 | 1.0 | 2.00 | 2.00 | 350 | 11232 | 0.90000V | -0.00190484 | UNIFORM P-AXIS |
| 27 | 0.400 | 500 | 4 | 32 | 2.0 | 1.0 | 2.00 | 2.00 | 60 | 1952 | 0.65000V | -0.02345614 | UNIFORM P-AXIS |
| 28 | 0.500 | 500 | 4 | 32 | 2.0 | 1.0 | 2.00 | 2.00 | 92 | 2976 | 0.35000V | -0.00265986 | UNIFORM P-AXIS |
| 29 | 0.600 | 500 | 4 | 32 | 2.0 | 1.0 | 2.00 | 2.00 | 98 | 3168 | 0.65000V | -0.01874733 | UNIFORM P-AXIS |
| 30 | 0.700 | 500 | 4 | 32 | 2.0 | 1.0 | 2.00 | 2.00 | 231 | 7424 | 0.65000V | -0.01974054 | UNIFORM P-AXIS |
| 31 | 0.800 | 500 | 4 | 32 | 2.0 | 1.0 | 2.00 | 2.00 | 249 | 8000 | 0.60000V | -0.03410112 | UNIFORM P-AXIS |
| 32 | 0.900 | 500 | 4 | 32 | 2.0 | 1.0 | 2.00 | 2.00 | 60 | 1952 | 0.65000V | -0.02925196 | UNIFORM P-AXIS |
| 33 | 1.000 | 500 | 4 | 32 | 2.0 | 1.0 | 2.00 | 2.00 | 271 | 8704 | 0.75000V | -0.01358199 | UNIFORM P-AXIS |
| 34 | 0.000 | 500 | 4 | 40 | 2.0 | 1.0 | 2.00 | 2.00 | 60 | 2440 | 0.65000V | -0.00760705 | UNIFORM P-AXIS |
| 35 | 0.100 | 500 | 4 | 40 | 2.0 | 1.0 | 2.00 | 2.00 | 285 | 11440 | 0.50000V | -0.00496634 | UNIFORM P-AXIS |
| 36 | 0.200 | 500 | 4 | 40 | 2.0 | 1.0 | 2.00 | 2.00 | 60 | 2440 | 0.65000V | -0.00822103 | UNIFORM P-AXIS |
| 37 | 0.300 | 500 | 4 | 40 | 2.0 | 1.0 | 2.00 | 2.00 | 231 | 9280 | 0.65000V | -0.00222432 | UNIFORM P-AXIS |
| 38 | 0.400 | 500 | 4 | 40 | 2.0 | 1.0 | 2.00 | 2.00 | 60 | 2440 | 0.65000V | -0.05848026 | UNIFORM P-AXIS |
| 39 | 0.500 | 500 | 4 | 40 | 2.0 | 1.0 | 2.00 | 2.00 | 117 | 4720 | 0.65000V | -0.00290803 | UNIFORM P-AXIS |
| 40 | 0.600 | 500 | 4 | 40 | 2.0 | 1.0 | 2.00 | 2.00 | 60 | 2440 | 0.65000V | -0.04223306 | UNIFORM P-AXIS |
| 41 | 0.700 | 500 | 4 | 40 | 2.0 | 1.0 | 2.00 | 2.00 | 60 | 2440 | 0.65000V | -0.03074483 | UNIFORM P-AXIS |
| 42 | 0.800 | 500 | 4 | 40 | 2.0 | 1.0 | 2.00 | 2.00 | 175 | 7040 | 0.70000V | -0.01799522 | UNIFORM P-AXIS |
| 43 | 0.900 | 500 | 4 | 40 | 2.0 | 1.0 | 2.00 | 2.00 | 270 | 10840 | 0.70000V | -0.00111803 | UNIFORM P-AXIS |
| 44 | 1.000 | 500 | 4 | 40 | 2.0 | 1.0 | 2.00 | 2.00 | 210 | 8440 | 0.55000V | -0.00191557 | UNIFORM P-AXIS |
| 45 | 0.000 | 500 | 4 | 48 | 2.0 | 1.0 | 2.00 | 2.00 | 69 | 3360 | 0.15000V | -0.00048890 | UNIFORM P-AXIS |
| 46 | 0.100 | 500 | 4 | 48 | 2.0 | 1.0 | 2.00 | 2.00 | 174 | 8400 | 0.65000V | -0.00194328 | UNIFORM P-AXIS |
| 47 | 0.200 | 500 | 4 | 48 | 2.0 | 1.0 | 2.00 | 2.00 | 70 | 3408 | 0.20000V | -0.00907792 | UNIFORM P-AXIS |
| 48 | 0.300 | 500 | 4 | 48 | 2.0 | 1.0 | 2.00 | 2.00 | 121 | 5856 | 0.85000V | -0.00352392 | UNIFORM P-AXIS |
| 49 | 0.400 | 500 | 4 | 48 | 2.0 | 1.0 | 2.00 | 2.00 | 60 | 2928 | 0.65000V | -0.05265765 | UNIFORM P-AXIS |
| 50 | 0.500 | 500 | 4 | 48 | 2.0 | 1.0 | 2.00 | 2.00 | 74 | 3600 | 0.40000V | -0.00340201 | UNIFORM P-AXIS |
| 51 | 0.600 | 500 | 4 | 48 | 2.0 | 1.0 | 2.00 | 2.00 | 60 | 2928 | 0.65000V | -0.03430604 | UNIFORM P-AXIS |
| 52 | 0.700 | 500 | 4 | 48 | 2.0 | 1.0 | 2.00 | 2.00 | 99 | 4800 | 0.70000V | -0.01095233 | UNIFORM P-AXIS |
| 53 | 0.800 | 500 | 4 | 48 | 2.0 | 1.0 | 2.00 | 2.00 | 117 | 5664 | 0.65000V | -0.02417579 | UNIFORM P-AXIS |
| 54 | 0.900 | 500 | 4 | 48 | 2.0 | 1.0 | 2.00 | 2.00 | 110 | 5328 | 0.30000V | -0.00077444 | UNIFORM P-AXIS |
| 55 | 1.000 | 500 | 4 | 48 | 2.0 | 1.0 | 2.00 | 2.00 | 231 | 11136 | 0.65000V | -0.01333495 | UNIFORM P-AXIS |
| 56 | 0.000 | 500 | 4 | 56 | 2.0 | 1.0 | 2.00 | 2.00 | 82 | 4648 | 0.80000V | -0.00177343 | UNIFORM P-AXIS |
| 57 | 0.100 | 500 | 4 | 56 | 2.0 | 1.0 | 2.00 | 2.00 | 136 | 7672 | 0.65000V | -0.00246125 | UNIFORM P-AXIS |
| 58 | 0.200 | 500 | 4 | 56 | 2.0 | 1.0 | 2.00 | 2.00 | 60 | 3416 | 0.65000V | -0.01356890 | UNIFORM P-AXIS |
| 59 | 0.300 | 500 | 4 | 56 | 2.0 | 1.0 | 2.00 | 2.00 | 60 | 3416 | 0.65000V | -0.03136321 | UNIFORM P-AXIS |
| 60 | 0.400 | 500 | 4 | 56 | 2.0 | 1.0 | 2.00 | 2.00 | 175 | 9856 | 0.70000V | -0.00734608 | UNIFORM P-AXIS |
| 61 | 0.500 | 500 | 4 | 56 | 2.0 | 1.0 | 2.00 | 2.00 | 78 | 4424 | 0.60000V | -0.00702747 | UNIFORM P-AXIS |
| 62 | 0.600 | 500 | 4 | 56 | 2.0 | 1.0 | 2.00 | 2.00 | 279 | 15680 | 0.20000V | -0.00472297 | UNIFORM P-AXIS |
| 63 | 0.700 | 500 | 4 | 56 | 2.0 | 1.0 | 2.00 | 2.00 | 156 | 8792 | 0.70000V | -0.01750298 | UNIFORM P-AXIS |
| 64 | 0.800 | 500 | 4 | 56 | 2.0 | 1.0 | 2.00 | 2.00 | 99 | 5600 | 0.70000V | -0.01215245 | UNIFORM P-AXIS |
| 65 | 0.900 | 500 | 4 | 56 | 2.0 | 1.0 | 2.00 | 2.00 | 98 | 5544 | 0.65000V | -0.00050140 | UNIFORM P-AXIS |
| 66 | 1.000 | 500 | 4 | 56 | 2.0 | 1.0 | 2.00 | 2.00 | 118 | 6664 | 0.70000V | -0.00970616 | UNIFORM P-AXIS |

Total Function Evaluations: 304664

---

| 45 | 0.000 | 500 | 4 | 48 | 2.0 | 1.0 | 2.00 | 2.00 | 69 | 3360 | 0.15000V | -0.00048890 | UNIFORM P-AXIS |

# F16

Run ID: 12-26-2009, 23:48:35

FUNCTION: F16

| Run # | Gamma | Nt | Nd | Np | G | DelT | Alpha | Beta | #Steps | Neval | Frep | Fitness | Initial Probes |
|-------|-------|----|----|----|----|------|-------|------|--------|-------|------|---------|----------------|
| 0 | 0.000 | 500 | 2 | 8 | 2.0 | 1.0 | 2.00 | 2.00 | 0 | 0 | 0.50000V | -9999.00000000 | UNIFORM P-AXIS |
| 1 | 0.000 | 500 | 2 | 8 | 2.0 | 1.0 | 2.00 | 2.00 | 78 | 632 | 0.60000V | 1.02587776 | UNIFORM P-AXIS |
| 2 | 0.100 | 500 | 2 | 8 | 2.0 | 1.0 | 2.00 | 2.00 | 111 | 896 | 0.35000V | 1.03160668 | UNIFORM P-AXIS |
| 3 | 0.200 | 500 | 2 | 8 | 2.0 | 1.0 | 2.00 | 2.00 | 60 | 488 | 0.65000V | 0.99305889 | UNIFORM P-AXIS |
| 4 | 0.300 | 500 | 2 | 8 | 2.0 | 1.0 | 2.00 | 2.00 | 137 | 1104 | 0.70000V | 1.03148086 | UNIFORM P-AXIS |
| 5 | 0.400 | 500 | 2 | 8 | 2.0 | 1.0 | 2.00 | 2.00 | 115 | 928 | 0.35000V | 1.03125790 | UNIFORM P-AXIS |
| 6 | 0.500 | 500 | 2 | 8 | 2.0 | 1.0 | 2.00 | 2.00 | 144 | 1160 | 0.10000V | 1.03070958 | UNIFORM P-AXIS |
| 7 | 0.600 | 500 | 2 | 8 | 2.0 | 1.0 | 2.00 | 2.00 | 115 | 928 | 0.55000V | 1.03125790 | UNIFORM P-AXIS |
| 8 | 0.700 | 500 | 2 | 8 | 2.0 | 1.0 | 2.00 | 2.00 | 137 | 1104 | 0.70000V | 1.03148086 | UNIFORM P-AXIS |
| 9 | 0.800 | 500 | 2 | 8 | 2.0 | 1.0 | 2.00 | 2.00 | 60 | 488 | 0.65000V | 0.99305889 | UNIFORM P-AXIS |
| 10 | 0.900 | 500 | 2 | 8 | 2.0 | 1.0 | 2.00 | 2.00 | 111 | 896 | 0.35000V | 1.03160668 | UNIFORM P-AXIS |
| 11 | 1.000 | 500 | 2 | 8 | 2.0 | 1.0 | 2.00 | 2.00 | 78 | 632 | 0.60000V | 1.02587776 | UNIFORM P-AXIS |
| 12 | 0.000 | 500 | 2 | 12 | 2.0 | 1.0 | 2.00 | 2.00 | 96 | 1164 | 0.55000V | 1.03013810 | UNIFORM P-AXIS |
| 13 | 0.100 | 500 | 2 | 12 | 2.0 | 1.0 | 2.00 | 2.00 | 73 | 888 | 0.35000V | 1.02492533 | UNIFORM P-AXIS |
| 14 | 0.200 | 500 | 2 | 12 | 2.0 | 1.0 | 2.00 | 2.00 | 77 | 936 | 0.55000V | 1.03161663 | UNIFORM P-AXIS |
| 15 | 0.300 | 500 | 2 | 12 | 2.0 | 1.0 | 2.00 | 2.00 | 111 | 1344 | 0.35000V | 1.02941460 | UNIFORM P-AXIS |
| 16 | 0.400 | 500 | 2 | 12 | 2.0 | 1.0 | 2.00 | 2.00 | 114 | 1380 | 0.45000V | 1.03130657 | UNIFORM P-AXIS |
| 17 | 0.500 | 500 | 2 | 12 | 2.0 | 1.0 | 2.00 | 2.00 | 60 | 732 | 0.65000V | 0.99998652 | UNIFORM P-AXIS |
| 18 | 0.600 | 500 | 2 | 12 | 2.0 | 1.0 | 2.00 | 2.00 | 114 | 1380 | 0.50000V | 1.03130657 | UNIFORM P-AXIS |
| 19 | 0.700 | 500 | 2 | 12 | 2.0 | 1.0 | 2.00 | 2.00 | 111 | 1344 | 0.35000V | 1.02941460 | UNIFORM P-AXIS |
| 20 | 0.800 | 500 | 2 | 12 | 2.0 | 1.0 | 2.00 | 2.00 | 60 | 732 | 0.65000V | 1.03031188 | UNIFORM P-AXIS |
| 21 | 0.900 | 500 | 2 | 12 | 2.0 | 1.0 | 2.00 | 2.00 | 73 | 888 | 0.35000V | 1.02492533 | UNIFORM P-AXIS |
| 22 | 1.000 | 500 | 2 | 12 | 2.0 | 1.0 | 2.00 | 2.00 | 96 | 1164 | 0.55000V | 1.03013810 | UNIFORM P-AXIS |
| 23 | 0.000 | 500 | 2 | 16 | 2.0 | 1.0 | 2.00 | 2.00 | 148 | 2384 | 0.30000V | 1.03018603 | UNIFORM P-AXIS |
| 24 | 0.100 | 500 | 2 | 16 | 2.0 | 1.0 | 2.00 | 2.00 | 131 | 2112 | 0.40000V | 1.02973102 | UNIFORM P-AXIS |
| 25 | 0.200 | 500 | 2 | 16 | 2.0 | 1.0 | 2.00 | 2.00 | 130 | 2096 | 0.35000V | 1.02998274 | UNIFORM P-AXIS |
| 26 | 0.300 | 500 | 2 | 16 | 2.0 | 1.0 | 2.00 | 2.00 | 122 | 1968 | 0.90000V | 1.02837012 | UNIFORM P-AXIS |
| 27 | 0.400 | 500 | 2 | 16 | 2.0 | 1.0 | 2.00 | 2.00 | 97 | 1568 | 0.60000V | 1.01423368 | UNIFORM P-AXIS |
| 28 | 0.500 | 500 | 2 | 16 | 2.0 | 1.0 | 2.00 | 2.00 | 132 | 2128 | 0.45000V | 1.02874055 | UNIFORM P-AXIS |
| 29 | 0.600 | 500 | 2 | 16 | 2.0 | 1.0 | 2.00 | 2.00 | 97 | 1568 | 0.60000V | 1.01423368 | UNIFORM P-AXIS |
| 30 | 0.700 | 500 | 2 | 16 | 2.0 | 1.0 | 2.00 | 2.00 | 122 | 1968 | 0.90000V | 1.02837016 | UNIFORM P-AXIS |
| 31 | 0.800 | 500 | 2 | 16 | 2.0 | 1.0 | 2.00 | 2.00 | 130 | 2096 | 0.35000V | 1.02998274 | UNIFORM P-AXIS |
| 32 | 0.900 | 500 | 2 | 16 | 2.0 | 1.0 | 2.00 | 2.00 | 131 | 2112 | 0.40000V | 1.02973102 | UNIFORM P-AXIS |
| 33 | 1.000 | 500 | 2 | 16 | 2.0 | 1.0 | 2.00 | 2.00 | 96 | 1552 | 0.55000V | 1.02313447 | UNIFORM P-AXIS |
| 34 | 0.000 | 500 | 2 | 20 | 2.0 | 1.0 | 2.00 | 2.00 | 131 | 2240 | 0.35000V | 1.01250976 | UNIFORM P-AXIS |
| 35 | 0.100 | 500 | 2 | 20 | 2.0 | 1.0 | 2.00 | 2.00 | 60 | 1220 | 0.65000V | 1.02425714 | UNIFORM P-AXIS |
| 36 | 0.200 | 500 | 2 | 20 | 2.0 | 1.0 | 2.00 | 2.00 | 151 | 3040 | 0.45000V | 1.03113113 | UNIFORM P-AXIS |
| 37 | 0.300 | 500 | 2 | 20 | 2.0 | 1.0 | 2.00 | 2.00 | 77 | 1560 | 0.55000V | 1.02996462 | UNIFORM P-AXIS |
| 38 | 0.400 | 500 | 2 | 20 | 2.0 | 1.0 | 2.00 | 2.00 | 78 | 1580 | 0.60000V | 1.03120923 | UNIFORM P-AXIS |
| 39 | 0.500 | 500 | 2 | 20 | 2.0 | 1.0 | 2.00 | 2.00 | 155 | 3120 | 0.65000V | 1.02700912 | UNIFORM P-AXIS |
| 40 | 0.600 | 500 | 2 | 20 | 2.0 | 1.0 | 2.00 | 2.00 | 78 | 1580 | 0.60000V | 1.03120923 | UNIFORM P-AXIS |
| 41 | 0.700 | 500 | 2 | 20 | 2.0 | 1.0 | 2.00 | 2.00 | 77 | 1560 | 0.55000V | 1.02996462 | UNIFORM P-AXIS |
| 42 | 0.800 | 500 | 2 | 20 | 2.0 | 1.0 | 2.00 | 2.00 | 151 | 3040 | 0.45000V | 1.03114606 | UNIFORM P-AXIS |
| 43 | 0.900 | 500 | 2 | 20 | 2.0 | 1.0 | 2.00 | 2.00 | 60 | 1220 | 0.65000V | 1.02425714 | UNIFORM P-AXIS |
| 44 | 1.000 | 500 | 2 | 20 | 2.0 | 1.0 | 2.00 | 2.00 | 60 | 1220 | 0.65000V | 1.02062108 | UNIFORM P-AXIS |
| 45 | 0.000 | 500 | 2 | 24 | 2.0 | 1.0 | 2.00 | 2.00 | 60 | 1464 | 0.65000V | 1.02305918 | UNIFORM P-AXIS |
| 46 | 0.100 | 500 | 2 | 24 | 2.0 | 1.0 | 2.00 | 2.00 | 113 | 2736 | 0.45000V | 1.02572173 | UNIFORM P-AXIS |
| 47 | 0.200 | 500 | 2 | 24 | 2.0 | 1.0 | 2.00 | 2.00 | 130 | 3144 | 0.35000V | 1.02370473 | UNIFORM P-AXIS |
| 48 | 0.300 | 500 | 2 | 24 | 2.0 | 1.0 | 2.00 | 2.00 | 96 | 2328 | 0.55000V | 1.02984598 | UNIFORM P-AXIS |
| 49 | 0.400 | 500 | 2 | 24 | 2.0 | 1.0 | 2.00 | 2.00 | 261 | 6288 | 0.25000V | 1.03162595 | UNIFORM P-AXIS |
| 50 | 0.500 | 500 | 2 | 24 | 2.0 | 1.0 | 2.00 | 2.00 | 97 | 2352 | 0.60000V | 1.03047576 | UNIFORM P-AXIS |
| 51 | 0.600 | 500 | 2 | 24 | 2.0 | 1.0 | 2.00 | 2.00 | 95 | 2304 | 0.50000V | 1.03079194 | UNIFORM P-AXIS |
| 52 | 0.700 | 500 | 2 | 24 | 2.0 | 1.0 | 2.00 | 2.00 | 115 | 2784 | 0.55000V | 1.02989343 | UNIFORM P-AXIS |
| 53 | 0.800 | 500 | 2 | 24 | 2.0 | 1.0 | 2.00 | 2.00 | 129 | 3120 | 0.30000V | 1.02930539 | UNIFORM P-AXIS |
| 54 | 0.900 | 500 | 2 | 24 | 2.0 | 1.0 | 2.00 | 2.00 | 303 | 7296 | 0.45000V | 1.03149759 | UNIFORM P-AXIS |
| 55 | 1.000 | 500 | 2 | 24 | 2.0 | 1.0 | 2.00 | 2.00 | 60 | 1464 | 0.65000V | 1.01912590 | UNIFORM P-AXIS |
| 56 | 0.000 | 500 | 2 | 28 | 2.0 | 1.0 | 2.00 | 2.00 | 60 | 1708 | 0.65000V | 1.02614496 | UNIFORM P-AXIS |
| 57 | 0.100 | 500 | 2 | 28 | 2.0 | 1.0 | 2.00 | 2.00 | 74 | 2100 | 0.40000V | 1.03161154 | UNIFORM P-AXIS |
| 58 | 0.200 | 500 | 2 | 28 | 2.0 | 1.0 | 2.00 | 2.00 | 80 | 2268 | 0.70000V | 1.03032871 | UNIFORM P-AXIS |
| 59 | 0.300 | 500 | 2 | 28 | 2.0 | 1.0 | 2.00 | 2.00 | 126 | 3556 | 0.15000V | 1.03155024 | UNIFORM P-AXIS |
| 60 | 0.400 | 500 | 2 | 28 | 2.0 | 1.0 | 2.00 | 2.00 | 60 | 1708 | 0.65000V | 1.02949203 | UNIFORM P-AXIS |
| 61 | 0.500 | 500 | 2 | 28 | 2.0 | 1.0 | 2.00 | 2.00 | 60 | 1708 | 0.65000V | 0.99911920 | UNIFORM P-AXIS |
| 62 | 0.600 | 500 | 2 | 28 | 2.0 | 1.0 | 2.00 | 2.00 | 60 | 1708 | 0.65000V | 1.02949203 | UNIFORM P-AXIS |
| 63 | 0.700 | 500 | 2 | 28 | 2.0 | 1.0 | 2.00 | 2.00 | 126 | 3556 | 0.15000V | 1.03155024 | UNIFORM P-AXIS |
| 64 | 0.800 | 500 | 2 | 28 | 2.0 | 1.0 | 2.00 | 2.00 | 80 | 2268 | 0.70000V | 1.03032871 | UNIFORM P-AXIS |
| 65 | 0.900 | 500 | 2 | 28 | 2.0 | 1.0 | 2.00 | 2.00 | 74 | 2100 | 0.40000V | 1.03161154 | UNIFORM P-AXIS |
| 66 | 1.000 | 500 | 2 | 28 | 2.0 | 1.0 | 2.00 | 2.00 | 79 | 2240 | 0.65000V | 1.02808103 | UNIFORM P-AXIS |

Total Function Evaluations: 124340

| 49 | 0.400 | 500 | 2 | 24 | 2.0 | 1.0 | 2.00 | 2.00 | 261 | 6288 | 0.25000V | 1.03162595 | UNIFORM P-AXIS |

# F17

Run ID: 12-26-2009, 23:52:43

FUNCTION: F17

| Run # | Gamma | Nt | Nd | Np | G | DelT | Alpha | Beta | #Steps | Neval | Frep | Fitness | Initial Probes |
|-------|-------|----|----|----|----|------|-------|------|--------|-------|------|---------|----------------|
| 0 | 0.000 | 500 | 2 | 8 | 2.0 | 1.0 | 2.00 | 2.00 | 0 | 0 | 0.50000V | -9999.00000000 | UNIFORM P-AXIS |
| 1 | 0.000 | 500 | 2 | 8 | 2.0 | 1.0 | 2.00 | 2.00 | 100 | 808 | 0.75000V | -0.46439597 | UNIFORM P-AXIS |
| 2 | 0.100 | 500 | 2 | 8 | 2.0 | 1.0 | 2.00 | 2.00 | 82 | 664 | 0.80000V | -0.39953210 | UNIFORM P-AXIS |
| 3 | 0.200 | 500 | 2 | 8 | 2.0 | 1.0 | 2.00 | 2.00 | 96 | 776 | 0.55000V | -0.42672486 | UNIFORM P-AXIS |
| 4 | 0.300 | 500 | 2 | 8 | 2.0 | 1.0 | 2.00 | 2.00 | 82 | 664 | 0.80000V | -0.46602832 | UNIFORM P-AXIS |
| 5 | 0.400 | 500 | 2 | 8 | 2.0 | 1.0 | 2.00 | 2.00 | 270 | 2168 | 0.70000V | -0.42249455 | UNIFORM P-AXIS |
| 6 | 0.500 | 500 | 2 | 8 | 2.0 | 1.0 | 2.00 | 2.00 | 72 | 584 | 0.30000V | -0.41429073 | UNIFORM P-AXIS |
| 7 | 0.600 | 500 | 2 | 8 | 2.0 | 1.0 | 2.00 | 2.00 | 76 | 616 | 0.50000V | -0.39795354 | UNIFORM P-AXIS |
| 8 | 0.700 | 500 | 2 | 8 | 2.0 | 1.0 | 2.00 | 2.00 | 60 | 488 | 0.65000V | -0.44406252 | UNIFORM P-AXIS |
| 9 | 0.800 | 500 | 2 | 8 | 2.0 | 1.0 | 2.00 | 2.00 | 89 | 720 | 0.20000V | -0.41551430 | UNIFORM P-AXIS |
| 10 | 0.900 | 500 | 2 | 8 | 2.0 | 1.0 | 2.00 | 2.00 | 95 | 768 | 0.50000V | -0.44055628 | UNIFORM P-AXIS |
| 11 | 1.000 | 500 | 2 | 8 | 2.0 | 1.0 | 2.00 | 2.00 | 70 | 568 | 0.20000V | -0.42345094 | UNIFORM P-AXIS |
| 12 | 0.000 | 500 | 2 | 12 | 2.0 | 1.0 | 2.00 | 2.00 | 92 | 1116 | 0.35000V | -0.41691841 | UNIFORM P-AXIS |
| 13 | 0.100 | 500 | 2 | 12 | 2.0 | 1.0 | 2.00 | 2.00 | 99 | 1200 | 0.70000V | -0.50131512 | UNIFORM P-AXIS |
| 14 | 0.200 | 500 | 2 | 12 | 2.0 | 1.0 | 2.00 | 2.00 | 103 | 1248 | 0.90000V | -0.40440758 | UNIFORM P-AXIS |
| 15 | 0.300 | 500 | 2 | 12 | 2.0 | 1.0 | 2.00 | 2.00 | 60 | 732 | 0.65000V | -0.40024733 | UNIFORM P-AXIS |
| 16 | 0.400 | 500 | 2 | 12 | 2.0 | 1.0 | 2.00 | 2.00 | 60 | 732 | 0.65000V | -0.44377180 | UNIFORM P-AXIS |
| 17 | 0.500 | 500 | 2 | 12 | 2.0 | 1.0 | 2.00 | 2.00 | 114 | 1380 | 0.50000V | -0.45485690 | UNIFORM P-AXIS |
| 18 | 0.600 | 500 | 2 | 12 | 2.0 | 1.0 | 2.00 | 2.00 | 60 | 732 | 0.65000V | -0.42645784 | UNIFORM P-AXIS |
| 19 | 0.700 | 500 | 2 | 12 | 2.0 | 1.0 | 2.00 | 2.00 | 72 | 876 | 0.30000V | -0.39934165 | UNIFORM P-AXIS |
| 20 | 0.800 | 500 | 2 | 12 | 2.0 | 1.0 | 2.00 | 2.00 | 60 | 732 | 0.65000V | -0.46538381 | UNIFORM P-AXIS |
| 21 | 0.900 | 500 | 2 | 12 | 2.0 | 1.0 | 2.00 | 2.00 | 82 | 996 | 0.80000V | -0.41912809 | UNIFORM P-AXIS |
| 22 | 1.000 | 500 | 2 | 12 | 2.0 | 1.0 | 2.00 | 2.00 | 95 | 1152 | 0.50000V | -0.40267002 | UNIFORM P-AXIS |
| 23 | 0.000 | 500 | 2 | 16 | 2.0 | 1.0 | 2.00 | 2.00 | 116 | 1872 | 0.60000V | -0.39790238 | UNIFORM P-AXIS |
| 24 | 0.100 | 500 | 2 | 16 | 2.0 | 1.0 | 2.00 | 2.00 | 137 | 2208 | 0.70000V | -0.39963599 | UNIFORM P-AXIS |
| 25 | 0.200 | 500 | 2 | 16 | 2.0 | 1.0 | 2.00 | 2.00 | 77 | 1248 | 0.55000V | -0.40617340 | UNIFORM P-AXIS |
| 26 | 0.300 | 500 | 2 | 16 | 2.0 | 1.0 | 2.00 | 2.00 | 85 | 1376 | 0.95000V | -0.39852121 | UNIFORM P-AXIS |
| 27 | 0.400 | 500 | 2 | 16 | 2.0 | 1.0 | 2.00 | 2.00 | 112 | 1808 | 0.40000V | -0.39840760 | UNIFORM P-AXIS |
| 28 | 0.500 | 500 | 2 | 16 | 2.0 | 1.0 | 2.00 | 2.00 | 79 | 1280 | 0.65000V | -0.40838689 | UNIFORM P-AXIS |
| 29 | 0.600 | 500 | 2 | 16 | 2.0 | 1.0 | 2.00 | 2.00 | 133 | 2144 | 0.50000V | -0.41414181 | UNIFORM P-AXIS |
| 30 | 0.700 | 500 | 2 | 16 | 2.0 | 1.0 | 2.00 | 2.00 | 76 | 1232 | 0.50000V | -0.39932652 | UNIFORM P-AXIS |
| 31 | 0.800 | 500 | 2 | 16 | 2.0 | 1.0 | 2.00 | 2.00 | 60 | 976 | 0.65000V | -0.42335713 | UNIFORM P-AXIS |
| 32 | 0.900 | 500 | 2 | 16 | 2.0 | 1.0 | 2.00 | 2.00 | 60 | 976 | 0.65000V | -0.40134318 | UNIFORM P-AXIS |
| 33 | 1.000 | 500 | 2 | 16 | 2.0 | 1.0 | 2.00 | 2.00 | 309 | 4960 | 0.75000V | -0.39989906 | UNIFORM P-AXIS |
| 34 | 0.000 | 500 | 2 | 20 | 2.0 | 1.0 | 2.00 | 2.00 | 78 | 1580 | 0.60000V | -0.40090871 | UNIFORM P-AXIS |
| 35 | 0.100 | 500 | 2 | 20 | 2.0 | 1.0 | 2.00 | 2.00 | 60 | 1220 | 0.65000V | -0.40240145 | UNIFORM P-AXIS |
| 36 | 0.200 | 500 | 2 | 20 | 2.0 | 1.0 | 2.00 | 2.00 | 76 | 1540 | 0.50000V | -0.40601199 | UNIFORM P-AXIS |
| 37 | 0.300 | 500 | 2 | 20 | 2.0 | 1.0 | 2.00 | 2.00 | 95 | 1920 | 0.50000V | -0.40997353 | UNIFORM P-AXIS |
| 38 | 0.400 | 500 | 2 | 20 | 2.0 | 1.0 | 2.00 | 2.00 | 76 | 1540 | 0.50000V | -0.40543885 | UNIFORM P-AXIS |
| 39 | 0.500 | 500 | 2 | 20 | 2.0 | 1.0 | 2.00 | 2.00 | 76 | 1540 | 0.50000V | -0.40212945 | UNIFORM P-AXIS |
| 40 | 0.600 | 500 | 2 | 20 | 2.0 | 1.0 | 2.00 | 2.00 | 80 | 1620 | 0.70000V | -0.39803955 | UNIFORM P-AXIS |
| 41 | 0.700 | 500 | 2 | 20 | 2.0 | 1.0 | 2.00 | 2.00 | 85 | 1720 | 0.95000V | -0.47106855 | UNIFORM P-AXIS |
| 42 | 0.800 | 500 | 2 | 20 | 2.0 | 1.0 | 2.00 | 2.00 | 60 | 1220 | 0.65000V | -0.40920220 | UNIFORM P-AXIS |
| 43 | 0.900 | 500 | 2 | 20 | 2.0 | 1.0 | 2.00 | 2.00 | 74 | 1500 | 0.40000V | -0.40471949 | UNIFORM P-AXIS |
| 44 | 1.000 | 500 | 2 | 20 | 2.0 | 1.0 | 2.00 | 2.00 | 74 | 1500 | 0.40000V | -0.40068500 | UNIFORM P-AXIS |
| 45 | 0.000 | 500 | 2 | 24 | 2.0 | 1.0 | 2.00 | 2.00 | 93 | 2256 | 0.40000V | -0.39872669 | UNIFORM P-AXIS |
| 46 | 0.100 | 500 | 2 | 24 | 2.0 | 1.0 | 2.00 | 2.00 | 75 | 1824 | 0.45000V | -0.39898701 | UNIFORM P-AXIS |
| 47 | 0.200 | 500 | 2 | 24 | 2.0 | 1.0 | 2.00 | 2.00 | 78 | 1896 | 0.60000V | -0.40251366 | UNIFORM P-AXIS |
| 48 | 0.300 | 500 | 2 | 24 | 2.0 | 1.0 | 2.00 | 2.00 | 114 | 2760 | 0.50000V | -0.41446304 | UNIFORM P-AXIS |
| 49 | 0.400 | 500 | 2 | 24 | 2.0 | 1.0 | 2.00 | 2.00 | 97 | 2352 | 0.60000V | -0.40190883 | UNIFORM P-AXIS |
| 50 | 0.500 | 500 | 2 | 24 | 2.0 | 1.0 | 2.00 | 2.00 | 79 | 1920 | 0.65000V | -0.40045620 | UNIFORM P-AXIS |
| 51 | 0.600 | 500 | 2 | 24 | 2.0 | 1.0 | 2.00 | 2.00 | 135 | 3264 | 0.50000V | -0.42255103 | UNIFORM P-AXIS |
| 52 | 0.700 | 500 | 2 | 24 | 2.0 | 1.0 | 2.00 | 2.00 | 74 | 1800 | 0.40000V | -0.41756267 | UNIFORM P-AXIS |
| 53 | 0.800 | 500 | 2 | 24 | 2.0 | 1.0 | 2.00 | 2.00 | 92 | 2232 | 0.35000V | -0.40648412 | UNIFORM P-AXIS |
| 54 | 0.900 | 500 | 2 | 24 | 2.0 | 1.0 | 2.00 | 2.00 | 77 | 1872 | 0.55000V | -0.40305464 | UNIFORM P-AXIS |
| 55 | 1.000 | 500 | 2 | 24 | 2.0 | 1.0 | 2.00 | 2.00 | 74 | 1800 | 0.40000V | -0.39811469 | UNIFORM P-AXIS |
| 56 | 0.000 | 500 | 2 | 28 | 2.0 | 1.0 | 2.00 | 2.00 | 111 | 3136 | 0.35000V | -0.40172094 | UNIFORM P-AXIS |
| 57 | 0.100 | 500 | 2 | 28 | 2.0 | 1.0 | 2.00 | 2.00 | 90 | 2548 | 0.25000V | -0.40193575 | UNIFORM P-AXIS |
| 58 | 0.200 | 500 | 2 | 28 | 2.0 | 1.0 | 2.00 | 2.00 | 101 | 2856 | 0.80000V | -0.39848716 | UNIFORM P-AXIS |
| 59 | 0.300 | 500 | 2 | 28 | 2.0 | 1.0 | 2.00 | 2.00 | 73 | 2072 | 0.35000V | -0.41375115 | UNIFORM P-AXIS |
| 60 | 0.400 | 500 | 2 | 28 | 2.0 | 1.0 | 2.00 | 2.00 | 60 | 1708 | 0.65000V | -0.40399916 | UNIFORM P-AXIS |
| 61 | 0.500 | 500 | 2 | 28 | 2.0 | 1.0 | 2.00 | 2.00 | 96 | 2716 | 0.55000V | -0.40467652 | UNIFORM P-AXIS |
| 62 | 0.600 | 500 | 2 | 28 | 2.0 | 1.0 | 2.00 | 2.00 | 60 | 1708 | 0.65000V | -0.40740570 | UNIFORM P-AXIS |
| 63 | 0.700 | 500 | 2 | 28 | 2.0 | 1.0 | 2.00 | 2.00 | 110 | 3108 | 0.30000V | -0.39803164 | UNIFORM P-AXIS |
| 64 | 0.800 | 500 | 2 | 28 | 2.0 | 1.0 | 2.00 | 2.00 | 106 | 2996 | 0.10000V | -0.39793899 | UNIFORM P-AXIS |
| 65 | 0.900 | 500 | 2 | 28 | 2.0 | 1.0 | 2.00 | 2.00 | 77 | 2184 | 0.55000V | -0.39935405 | UNIFORM P-AXIS |
| 66 | 1.000 | 500 | 2 | 28 | 2.0 | 1.0 | 2.00 | 2.00 | 93 | 2632 | 0.40000V | -0.39935800 | UNIFORM P-AXIS |

Total Function Evaluations:    108340

---

| | | | | | | | | | | | | | |
|---|---|---|---|---|---|---|---|---|---|---|---|---|---|
| 23 | 0.000 | 500 | 2 | 16 | 2.0 | 1.0 | 2.00 | 2.00 | 116 | 1872 | 0.60000V | -0.39790238 | UNIFORM P-AXIS |

---

# F18

Run ID: 12-26-2009, 22:19:28

FUNCTION: F18

| Run # | Gamma | Nt | Nd | Np | G | DelT | Alpha | Beta | #Steps | Neval | Frep | Fitness | Initial Probes |
|---|---|---|---|---|---|---|---|---|---|---|---|---|---|
| 0 | 0.000 | 500 | 2 | 8 | 2.0 | 1.0 | 2.00 | 2.00 | 0 | 0 | 0.50000V | -9999.00000000 | UNIFORM P-AXIS |
| 1 | 0.000 | 500 | 2 | 8 | 2.0 | 1.0 | 2.00 | 2.00 | 78 | 632 | 0.60000V | -5.48169471 | UNIFORM P-AXIS |
| 2 | 0.100 | 500 | 2 | 8 | 2.0 | 1.0 | 2.00 | 2.00 | 134 | 1080 | 0.55000V | -84.78003234 | UNIFORM P-AXIS |
| 3 | 0.200 | 500 | 2 | 8 | 2.0 | 1.0 | 2.00 | 2.00 | 191 | 1536 | 0.55000V | -3.19822531 | UNIFORM P-AXIS |
| 4 | 0.300 | 500 | 2 | 8 | 2.0 | 1.0 | 2.00 | 2.00 | 79 | 640 | 0.65000V | -8.62285433 | UNIFORM P-AXIS |
| 5 | 0.400 | 500 | 2 | 8 | 2.0 | 1.0 | 2.00 | 2.00 | 403 | 3232 | 0.70000V | -84.47139703 | UNIFORM P-AXIS |
| 6 | 0.500 | 500 | 2 | 8 | 2.0 | 1.0 | 2.00 | 2.00 | 261 | 2096 | 0.25000V | -10.56054743 | UNIFORM P-AXIS |
| 7 | 0.600 | 500 | 2 | 8 | 2.0 | 1.0 | 2.00 | 2.00 | 135 | 1088 | 0.60000V | -4.98190912 | UNIFORM P-AXIS |
| 8 | 0.700 | 500 | 2 | 8 | 2.0 | 1.0 | 2.00 | 2.00 | 229 | 1840 | 0.55000V | -3.00130536 | UNIFORM P-AXIS |
| 9 | 0.800 | 500 | 2 | 8 | 2.0 | 1.0 | 2.00 | 2.00 | 133 | 1072 | 0.50000V | -89.42718008 | UNIFORM P-AXIS |
| 10 | 0.900 | 500 | 2 | 8 | 2.0 | 1.0 | 2.00 | 2.00 | 116 | 936 | 0.60000V | -8.57642421 | UNIFORM P-AXIS |
| 11 | 1.000 | 500 | 2 | 8 | 2.0 | 1.0 | 2.00 | 2.00 | 193 | 1552 | 0.65000V | -3.08573605 | UNIFORM P-AXIS |
| 12 | 0.000 | 500 | 2 | 12 | 2.0 | 1.0 | 2.00 | 2.00 | 78 | 948 | 0.60000V | -236.66437724 | UNIFORM P-AXIS |
| 13 | 0.100 | 500 | 2 | 12 | 2.0 | 1.0 | 2.00 | 2.00 | 263 | 3168 | 0.35000V | -3.00574349 | UNIFORM P-AXIS |
| 14 | 0.200 | 500 | 2 | 12 | 2.0 | 1.0 | 2.00 | 2.00 | 78 | 948 | 0.60000V | -57.61728445 | UNIFORM P-AXIS |
| 15 | 0.300 | 500 | 2 | 12 | 2.0 | 1.0 | 2.00 | 2.00 | 304 | 3660 | 0.50000V | -3.00017384 | UNIFORM P-AXIS |
| 16 | 0.400 | 500 | 2 | 12 | 2.0 | 1.0 | 2.00 | 2.00 | 154 | 1860 | 0.60000V | -3.55376552 | UNIFORM P-AXIS |
| 17 | 0.500 | 500 | 2 | 12 | 2.0 | 1.0 | 2.00 | 2.00 | 196 | 2364 | 0.80000V | -3.00014046 | UNIFORM P-AXIS |
| 18 | 0.600 | 500 | 2 | 12 | 2.0 | 1.0 | 2.00 | 2.00 | 154 | 1860 | 0.60000V | -3.35080630 | UNIFORM P-AXIS |
| 19 | 0.700 | 500 | 2 | 12 | 2.0 | 1.0 | 2.00 | 2.00 | 212 | 2556 | 0.65000V | -3.03478155 | UNIFORM P-AXIS |
| 20 | 0.800 | 500 | 2 | 12 | 2.0 | 1.0 | 2.00 | 2.00 | 78 | 948 | 0.60000V | -16.66444362 | UNIFORM P-AXIS |
| 21 | 0.900 | 500 | 2 | 12 | 2.0 | 1.0 | 2.00 | 2.00 | 99 | 1200 | 0.70000V | -6.52063755 | UNIFORM P-AXIS |
| 22 | 1.000 | 500 | 2 | 12 | 2.0 | 1.0 | 2.00 | 2.00 | 250 | 3012 | 0.65000V | -3.01375068 | UNIFORM P-AXIS |
| 23 | 0.000 | 500 | 2 | 16 | 2.0 | 1.0 | 2.00 | 2.00 | 78 | 1264 | 0.60000V | -221.18779159 | UNIFORM P-AXIS |
| 24 | 0.100 | 500 | 2 | 16 | 2.0 | 1.0 | 2.00 | 2.00 | 60 | 976 | 0.60000V | -16.89097168 | UNIFORM P-AXIS |
| 25 | 0.200 | 500 | 2 | 16 | 2.0 | 1.0 | 2.00 | 2.00 | 173 | 2784 | 0.60000V | -3.01613943 | UNIFORM P-AXIS |
| 26 | 0.300 | 500 | 2 | 16 | 2.0 | 1.0 | 2.00 | 2.00 | 79 | 1280 | 0.65000V | -76.69765275 | UNIFORM P-AXIS |
| 27 | 0.400 | 500 | 2 | 16 | 2.0 | 1.0 | 2.00 | 2.00 | 189 | 3040 | 0.45000V | -3.00972155 | UNIFORM P-AXIS |
| 28 | 0.500 | 500 | 2 | 16 | 2.0 | 1.0 | 2.00 | 2.00 | 191 | 3072 | 0.55000V | -6.65915188 | UNIFORM P-AXIS |
| 29 | 0.600 | 500 | 2 | 16 | 2.0 | 1.0 | 2.00 | 2.00 | 122 | 1968 | 0.30000V | -15.75050525 | UNIFORM P-AXIS |
| 30 | 0.700 | 500 | 2 | 16 | 2.0 | 1.0 | 2.00 | 2.00 | 60 | 976 | 0.60000V | -31.95935579 | UNIFORM P-AXIS |
| 31 | 0.800 | 500 | 2 | 16 | 2.0 | 1.0 | 2.00 | 2.00 | 95 | 1536 | 0.50000V | -49.89324658 | UNIFORM P-AXIS |
| 32 | 0.900 | 500 | 2 | 16 | 2.0 | 1.0 | 2.00 | 2.00 | 304 | 4880 | 0.50000V | -3.00045306 | UNIFORM P-AXIS |
| 33 | 1.000 | 500 | 2 | 16 | 2.0 | 1.0 | 2.00 | 2.00 | 318 | 5104 | 0.25000V | -3.00010350 | UNIFORM P-AXIS |
| 34 | 0.000 | 500 | 2 | 20 | 2.0 | 1.0 | 2.00 | 2.00 | 78 | 1580 | 0.60000V | -76.94725945 | UNIFORM P-AXIS |
| 35 | 0.100 | 500 | 2 | 20 | 2.0 | 1.0 | 2.00 | 2.00 | 116 | 2340 | 0.60000V | -3.12194540 | UNIFORM P-AXIS |
| 36 | 0.200 | 500 | 2 | 20 | 2.0 | 1.0 | 2.00 | 2.00 | 191 | 3840 | 0.55000V | -3.03175136 | UNIFORM P-AXIS |
| 37 | 0.300 | 500 | 2 | 20 | 2.0 | 1.0 | 2.00 | 2.00 | 78 | 1580 | 0.60000V | -182.65850920 | UNIFORM P-AXIS |
| 38 | 0.400 | 500 | 2 | 20 | 2.0 | 1.0 | 2.00 | 2.00 | 193 | 3880 | 0.65000V | -3.14860263 | UNIFORM P-AXIS |
| 39 | 0.500 | 500 | 2 | 20 | 2.0 | 1.0 | 2.00 | 2.00 | 98 | 1980 | 0.65000V | -3.62879802 | UNIFORM P-AXIS |
| 40 | 0.600 | 500 | 2 | 20 | 2.0 | 1.0 | 2.00 | 2.00 | 192 | 3860 | 0.60000V | -3.04067143 | UNIFORM P-AXIS |
| 41 | 0.700 | 500 | 2 | 20 | 2.0 | 1.0 | 2.00 | 2.00 | 139 | 2800 | 0.80000V | -3.54277590 | UNIFORM P-AXIS |
| 42 | 0.800 | 500 | 2 | 20 | 2.0 | 1.0 | 2.00 | 2.00 | 171 | 3440 | 0.50000V | -3.00072472 | UNIFORM P-AXIS |
| 43 | 0.900 | 500 | 2 | 20 | 2.0 | 1.0 | 2.00 | 2.00 | 101 | 2040 | 0.40000V | -4.37891524 | UNIFORM P-AXIS |
| 44 | 1.000 | 500 | 2 | 20 | 2.0 | 1.0 | 2.00 | 2.00 | 79 | 1600 | 0.65000V | -35.88632429 | UNIFORM P-AXIS |

```
45   0.000  500   2   24   2.0   1.0   2.00   2.00    226    5448   0.40000V     -3.00031923   UNIFORM P-AXIS
46   0.100  500   2   24   2.0   1.0   2.00   2.00     78    1896   0.60000V     -9.06529259   UNIFORM P-AXIS
47   0.200  500   2   24   2.0   1.0   2.00   2.00    209    5040   0.50000V     -3.03088087   UNIFORM P-AXIS
48   0.300  500   2   24   2.0   1.0   2.00   2.00     79    1920   0.65000V    -30.56252660   UNIFORM P-AXIS
49   0.400  500   2   24   2.0   1.0   2.00   2.00    306    7368   0.60000V     -3.00075380   UNIFORM P-AXIS
50   0.500  500   2   24   2.0   1.0   2.00   2.00    248    5976   0.55000V     -3.00968878   UNIFORM P-AXIS
51   0.600  500   2   24   2.0   1.0   2.00   2.00     97    2352   0.60000V     -3.34890591   UNIFORM P-AXIS
52   0.700  500   2   24   2.0   1.0   2.00   2.00    246    5928   0.45000V     -3.01530923   UNIFORM P-AXIS
53   0.800  500   2   24   2.0   1.0   2.00   2.00     78    1896   0.60000V    -31.34561763   UNIFORM P-AXIS
54   0.900  500   2   24   2.0   1.0   2.00   2.00     60    1464   0.65000V     -3.00000000   UNIFORM P-AXIS
55   1.000  500   2   24   2.0   1.0   2.00   2.00    101    2448   0.80000V    -15.41042957   UNIFORM P-AXIS
56   0.000  500   2   28   2.0   1.0   2.00   2.00    266    7476   0.50000V     -3.00178783   UNIFORM P-AXIS
57   0.100  500   2   28   2.0   1.0   2.00   2.00    257    7224   0.05000V     -3.02198109   UNIFORM P-AXIS
58   0.200  500   2   28   2.0   1.0   2.00   2.00     60    1708   0.65000V   -102.95525789   UNIFORM P-AXIS
59   0.300  500   2   28   2.0   1.0   2.00   2.00    154    4340   0.60000V     -3.40567728   UNIFORM P-AXIS
60   0.400  500   2   28   2.0   1.0   2.00   2.00     79    2240   0.65000V    -40.28712313   UNIFORM P-AXIS
61   0.500  500   2   28   2.0   1.0   2.00   2.00    210    5908   0.55000V     -3.00034670   UNIFORM P-AXIS
62   0.600  500   2   28   2.0   1.0   2.00   2.00     79    2240   0.65000V     -3.22578918   UNIFORM P-AXIS
63   0.700  500   2   28   2.0   1.0   2.00   2.00     60    1708   0.65000V    -36.68532659   UNIFORM P-AXIS
64   0.800  500   2   28   2.0   1.0   2.00   2.00     60    1708   0.65000V    -14.65270155   UNIFORM P-AXIS
65   0.900  500   2   28   2.0   1.0   2.00   2.00     78    2212   0.60000V    -31.47896038   UNIFORM P-AXIS
66   1.000  500   2   28   2.0   1.0   2.00   2.00    282    7924   0.35000V     -3.00151179   UNIFORM P-AXIS

                                         Total Function Evaluations:   180472

-----------------------------------------------------------------------------------------------------------
54   0.900  500   2   24   2.0   1.0   2.00   2.00     60    1464   0.65000V     -3.00000000   UNIFORM P-AXIS
```

## F19

Run ID: 12-26-2009, 23:57:27

FUNCTION: F19

```
Run #   Gamma   Nt    Nd   Np    G    DelT   Alpha   Beta   #Steps   Neval    Prep       Fitness     Initial Probes
-----   -----   --    --   --   ---   ----   -----   ----   ------   -----   -------   -----------   --------------
0       0.000   500   3    12   2.0   1.0   2.00    2.00      0        0     0.50000V  -9999.00000000  UNIFORM P-AXIS
-----------------------------------------------------------------------------------------------------------
1       0.000   500   3    12   2.0   1.0   2.00    2.00     92      1116    0.35000V     3.81087000   UNIFORM P-AXIS
2       0.100   500   3    12   2.0   1.0   2.00    2.00    156      1884    0.70000V     3.80252096   UNIFORM P-AXIS
3       0.200   500   3    12   2.0   1.0   2.00    2.00     92      1116    0.35000V     3.85625317   UNIFORM P-AXIS
4       0.300   500   3    12   2.0   1.0   2.00    2.00     90      1092    0.25000V     3.82032439   UNIFORM P-AXIS
5       0.400   500   3    12   2.0   1.0   2.00    2.00     76       924    0.50000V     3.86157246   UNIFORM P-AXIS
6       0.500   500   3    12   2.0   1.0   2.00    2.00     77       936    0.55000V     3.84618140   UNIFORM P-AXIS
7       0.600   500   3    12   2.0   1.0   2.00    2.00     74       900    0.40000V     3.83838507   UNIFORM P-AXIS
8       0.700   500   3    12   2.0   1.0   2.00    2.00    112      1356    0.40000V     3.83554332   UNIFORM P-AXIS
9       0.800   500   3    12   2.0   1.0   2.00    2.00     60       732    0.65000V     3.84711642   UNIFORM P-AXIS
10      0.900   500   3    12   2.0   1.0   2.00    2.00    137      1656    0.70000V     3.85599159   UNIFORM P-AXIS
11      1.000   500   3    12   2.0   1.0   2.00    2.00     74       900    0.40000V     3.70557750   UNIFORM P-AXIS
12      0.000   500   3    18   2.0   1.0   2.00    2.00    149      2700    0.35000V     3.78695126   UNIFORM P-AXIS
13      0.100   500   3    18   2.0   1.0   2.00    2.00     91      1656    0.30000V     3.85424146   UNIFORM P-AXIS
14      0.200   500   3    18   2.0   1.0   2.00    2.00    249      4500    0.60000V     3.85002422   UNIFORM P-AXIS
15      0.300   500   3    18   2.0   1.0   2.00    2.00     93      1692    0.40000V     3.85668197   UNIFORM P-AXIS
16      0.400   500   3    18   2.0   1.0   2.00    2.00     90      1638    0.25000V     3.84336439   UNIFORM P-AXIS
17      0.500   500   3    18   2.0   1.0   2.00    2.00     77      1404    0.35000V     3.85075229   UNIFORM P-AXIS
18      0.600   500   3    18   2.0   1.0   2.00    2.00    356      6426    0.25000V     3.84476719   UNIFORM P-AXIS
19      0.700   500   3    18   2.0   1.0   2.00    2.00    114      2070    0.50000V     3.84820906   UNIFORM P-AXIS
20      0.800   500   3    18   2.0   1.0   2.00    2.00    114      2070    0.50000V     3.83793620   UNIFORM P-AXIS
21      0.900   500   3    18   2.0   1.0   2.00    2.00     93      1692    0.40000V     3.83173651   UNIFORM P-AXIS
22      1.000   500   3    18   2.0   1.0   2.00    2.00    116      2106    0.60000V     3.79007044   UNIFORM P-AXIS
23      0.000   500   3    24   2.0   1.0   2.00    2.00    103      2496    0.90000V     3.77571774   UNIFORM P-AXIS
24      0.100   500   3    24   2.0   1.0   2.00    2.00    193      4656    0.65000V     3.85998185   UNIFORM P-AXIS
25      0.200   500   3    24   2.0   1.0   2.00    2.00     60      1464    0.65000V     3.85827634   UNIFORM P-AXIS
26      0.300   500   3    24   2.0   1.0   2.00    2.00    112      2712    0.40000V     3.85863108   UNIFORM P-AXIS
27      0.400   500   3    24   2.0   1.0   2.00    2.00     61      1488    0.70000V     3.80982656   UNIFORM P-AXIS
28      0.500   500   3    24   2.0   1.0   2.00    2.00    275      6624    0.95000V     3.86090728   UNIFORM P-AXIS
29      0.600   500   3    24   2.0   1.0   2.00    2.00     94      2280    0.45000V     3.81276278   UNIFORM P-AXIS
30      0.700   500   3    24   2.0   1.0   2.00    2.00     75      1824    0.45000V     3.84748287   UNIFORM P-AXIS
31      0.800   500   3    24   2.0   1.0   2.00    2.00     77      1872    0.55000V     3.84487382   UNIFORM P-AXIS
32      0.900   500   3    24   2.0   1.0   2.00    2.00    117      2832    0.65000V     3.85440945   UNIFORM P-AXIS
33      1.000   500   3    24   2.0   1.0   2.00    2.00    111      2688    0.35000V     3.85064729   UNIFORM P-AXIS
34      0.000   500   3    30   2.0   1.0   2.00    2.00     74      2250    0.40000V     3.83941990   UNIFORM P-AXIS
35      0.100   500   3    30   2.0   1.0   2.00    2.00    113      3420    0.45000V     3.84742951   UNIFORM P-AXIS
36      0.200   500   3    30   2.0   1.0   2.00    2.00    231      6960    0.45000V     3.82489621   UNIFORM P-AXIS
37      0.300   500   3    30   2.0   1.0   2.00    2.00     77      2340    0.55000V     3.85474673   UNIFORM P-AXIS
38      0.400   500   3    30   2.0   1.0   2.00    2.00     60      1830    0.65000V     3.86095502   UNIFORM P-AXIS
39      0.500   500   3    30   2.0   1.0   2.00    2.00    174      5250    0.65000V     3.84955030   UNIFORM P-AXIS
40      0.600   500   3    30   2.0   1.0   2.00    2.00     96      2910    0.35000V     3.83954812   UNIFORM P-AXIS
41      0.700   500   3    30   2.0   1.0   2.00    2.00     96      2910    0.55000V     3.83746687   UNIFORM P-AXIS
42      0.800   500   3    30   2.0   1.0   2.00    2.00    272      8190    0.80000V     3.83739135   UNIFORM P-AXIS
43      0.900   500   3    30   2.0   1.0   2.00    2.00     60      1830    0.65000V     3.85079537   UNIFORM P-AXIS
44      1.000   500   3    30   2.0   1.0   2.00    2.00    231      6960    0.65000V     3.79786366   UNIFORM P-AXIS
45      0.000   500   3    36   2.0   1.0   2.00    2.00     60      2196    0.65000V     3.84685849   UNIFORM P-AXIS
46      0.100   500   3    36   2.0   1.0   2.00    2.00     60      2196    0.65000V     3.85060158   UNIFORM P-AXIS
47      0.200   500   3    36   2.0   1.0   2.00    2.00    108      3924    0.20000V     3.84686078   UNIFORM P-AXIS
48      0.300   500   3    36   2.0   1.0   2.00    2.00     76      2772    0.50000V     3.84532644   UNIFORM P-AXIS
49      0.400   500   3    36   2.0   1.0   2.00    2.00     79      2880    0.65000V     3.85460096   UNIFORM P-AXIS
50      0.500   500   3    36   2.0   1.0   2.00    2.00    113      4104    0.45000V     3.84320651   UNIFORM P-AXIS
51      0.600   500   3    36   2.0   1.0   2.00    2.00     74      2700    0.40000V     3.84383985   UNIFORM P-AXIS
52      0.700   500   3    36   2.0   1.0   2.00    2.00    286     10332    0.35000V     3.84332469   UNIFORM P-AXIS
53      0.800   500   3    36   2.0   1.0   2.00    2.00     96      3492    0.55000V     3.85185414   UNIFORM P-AXIS
54      0.900   500   3    36   2.0   1.0   2.00    2.00     94      3420    0.45000V     3.84805588   UNIFORM P-AXIS
55      1.000   500   3    36   2.0   1.0   2.00    2.00    126      4572    0.15000V     3.80228240   UNIFORM P-AXIS
56      0.000   500   3    42   2.0   1.0   2.00    2.00     60      2562    0.65000V     3.83661732   UNIFORM P-AXIS
57      0.100   500   3    42   2.0   1.0   2.00    2.00     79      3360    0.65000V     3.81761193   UNIFORM P-AXIS
58      0.200   500   3    42   2.0   1.0   2.00    2.00     74      3150    0.40000V     3.86268376   UNIFORM P-AXIS
59      0.300   500   3    42   2.0   1.0   2.00    2.00    101      4284    0.80000V     3.85655395   UNIFORM P-AXIS
60      0.400   500   3    42   2.0   1.0   2.00    2.00    111      4704    0.35000V     3.84157656   UNIFORM P-AXIS
61      0.500   500   3    42   2.0   1.0   2.00    2.00     60      2562    0.65000V     3.85432960   UNIFORM P-AXIS
62      0.600   500   3    42   2.0   1.0   2.00    2.00     93      3948    0.40000V     3.83483193   UNIFORM P-AXIS
63      0.700   500   3    42   2.0   1.0   2.00    2.00     60      2562    0.65000V     3.85431079   UNIFORM P-AXIS
64      0.800   500   3    42   2.0   1.0   2.00    2.00    112      4746    0.40000V     3.84765623   UNIFORM P-AXIS
65      0.900   500   3    42   2.0   1.0   2.00    2.00     77      3276    0.55000V     3.85815731   UNIFORM P-AXIS
66      1.000   500   3    42   2.0   1.0   2.00    2.00    146      6174    0.20000V     3.84384847   UNIFORM P-AXIS

                                         Total Function Evaluations:   200268

-----------------------------------------------------------------------------------------------------------
58      0.200   500   3    42   2.0   1.0   2.00    2.00     74      3150    0.40000V     3.86268376   UNIFORM P-AXIS
```

# F20

Run ID: 12-27-2009, 00:00:52

FUNCTION: F20

| Run # | Gamma | Nt | Nd | Np | G | DelT | Alpha | Beta | #Steps | Neval | Frep | Fitness | Initial Probes |
|-------|-------|-----|----|----|-----|------|-------|------|--------|-------|----------|------------------|-----------------|
| 0 | 0.000 | 500 | 6 | 24 | 2.0 | 1.0 | 2.00 | 2.00 | 0 | 0 | 0.50000V | -9999.00000000 | UNIFORM P-AXIS |
| 1 | 0.000 | 500 | 6 | 24 | 2.0 | 1.0 | 2.00 | 2.00 | 154 | 3720 | 0.60000V | 1.79481098 | UNIFORM P-AXIS |
| 2 | 0.100 | 500 | 6 | 24 | 2.0 | 1.0 | 2.00 | 2.00 | 168 | 4056 | 0.35000V | 3.29593022 | UNIFORM P-AXIS |
| 3 | 0.200 | 500 | 6 | 24 | 2.0 | 1.0 | 2.00 | 2.00 | 124 | 3000 | 0.05000V | 3.31976377 | UNIFORM P-AXIS |
| 4 | 0.300 | 500 | 6 | 24 | 2.0 | 1.0 | 2.00 | 2.00 | 250 | 6024 | 0.65000V | 3.29485054 | UNIFORM P-AXIS |
| 5 | 0.400 | 500 | 6 | 24 | 2.0 | 1.0 | 2.00 | 2.00 | 284 | 6840 | 0.45000V | 3.31728832 | UNIFORM P-AXIS |
| 6 | 0.500 | 500 | 6 | 24 | 2.0 | 1.0 | 2.00 | 2.00 | 129 | 3120 | 0.30000V | 3.31388945 | UNIFORM P-AXIS |
| 7 | 0.600 | 500 | 6 | 24 | 2.0 | 1.0 | 2.00 | 2.00 | 250 | 6024 | 0.65000V | 3.28047153 | UNIFORM P-AXIS |
| 8 | 0.700 | 500 | 6 | 24 | 2.0 | 1.0 | 2.00 | 2.00 | 284 | 6840 | 0.45000V | 3.15797322 | UNIFORM P-AXIS |
| 9 | 0.800 | 500 | 6 | 24 | 2.0 | 1.0 | 2.00 | 2.00 | 162 | 3912 | 0.05000V | 3.25303766 | UNIFORM P-AXIS |
| 10 | 0.900 | 500 | 6 | 24 | 2.0 | 1.0 | 2.00 | 2.00 | 142 | 3432 | 0.95000V | 1.44877557 | UNIFORM P-AXIS |
| 11 | 1.000 | 500 | 6 | 24 | 2.0 | 1.0 | 2.00 | 2.00 | 214 | 5160 | 0.75000V | 3.21996273 | UNIFORM P-AXIS |
| 12 | 0.000 | 500 | 6 | 36 | 2.0 | 1.0 | 2.00 | 2.00 | 250 | 9036 | 0.65000V | 3.30877277 | UNIFORM P-AXIS |
| 13 | 0.100 | 500 | 6 | 36 | 2.0 | 1.0 | 2.00 | 2.00 | 194 | 7020 | 0.70000V | 3.31923851 | UNIFORM P-AXIS |
| 14 | 0.200 | 500 | 6 | 36 | 2.0 | 1.0 | 2.00 | 2.00 | 175 | 6336 | 0.70000V | 3.31659640 | UNIFORM P-AXIS |
| 15 | 0.300 | 500 | 6 | 36 | 2.0 | 1.0 | 2.00 | 2.00 | 191 | 6912 | 0.55000V | 3.30506311 | UNIFORM P-AXIS |
| 16 | 0.400 | 500 | 6 | 36 | 2.0 | 1.0 | 2.00 | 2.00 | 175 | 6336 | 0.70000V | 3.28854629 | UNIFORM P-AXIS |
| 17 | 0.500 | 500 | 6 | 36 | 2.0 | 1.0 | 2.00 | 2.00 | 231 | 8352 | 0.65000V | 3.31583530 | UNIFORM P-AXIS |
| 18 | 0.600 | 500 | 6 | 36 | 2.0 | 1.0 | 2.00 | 2.00 | 155 | 5616 | 0.65000V | 3.11598206 | UNIFORM P-AXIS |
| 19 | 0.700 | 500 | 6 | 36 | 2.0 | 1.0 | 2.00 | 2.00 | 232 | 8388 | 0.70000V | 3.13277320 | UNIFORM P-AXIS |
| 20 | 0.800 | 500 | 6 | 36 | 2.0 | 1.0 | 2.00 | 2.00 | 193 | 6984 | 0.65000V | 3.31407408 | UNIFORM P-AXIS |
| 21 | 0.900 | 500 | 6 | 36 | 2.0 | 1.0 | 2.00 | 2.00 | 109 | 3960 | 0.25000V | 3.63932000 | UNIFORM P-AXIS |
| 22 | 1.000 | 500 | 6 | 36 | 2.0 | 1.0 | 2.00 | 2.00 | 179 | 6480 | 0.90000V | 3.26629451 | UNIFORM P-AXIS |
| 23 | 0.000 | 500 | 6 | 48 | 2.0 | 1.0 | 2.00 | 2.00 | 250 | 12048 | 0.65000V | 3.32048934 | UNIFORM P-AXIS |
| 24 | 0.100 | 500 | 6 | 48 | 2.0 | 1.0 | 2.00 | 2.00 | 213 | 10272 | 0.70000V | 3.31842289 | UNIFORM P-AXIS |
| 25 | 0.200 | 500 | 6 | 48 | 2.0 | 1.0 | 2.00 | 2.00 | 193 | 9312 | 0.65000V | 3.31716803 | UNIFORM P-AXIS |
| 26 | 0.300 | 500 | 6 | 48 | 2.0 | 1.0 | 2.00 | 2.00 | 277 | 13344 | 0.10000V | 3.31881259 | UNIFORM P-AXIS |
| 27 | 0.400 | 500 | 6 | 48 | 2.0 | 1.0 | 2.00 | 2.00 | 231 | 11136 | 0.65000V | 3.31690037 | UNIFORM P-AXIS |
| 28 | 0.500 | 500 | 6 | 48 | 2.0 | 1.0 | 2.00 | 2.00 | 286 | 13776 | 0.55000V | 3.31917080 | UNIFORM P-AXIS |
| 29 | 0.600 | 500 | 6 | 48 | 2.0 | 1.0 | 2.00 | 2.00 | 151 | 7296 | 0.45000V | 3.12577992 | UNIFORM P-AXIS |
| 30 | 0.700 | 500 | 6 | 48 | 2.0 | 1.0 | 2.00 | 2.00 | 168 | 8112 | 0.35000V | 3.13969728 | UNIFORM P-AXIS |
| 31 | 0.800 | 500 | 6 | 48 | 2.0 | 1.0 | 2.00 | 2.00 | 144 | 6960 | 0.10000V | 3.31609910 | UNIFORM P-AXIS |
| 32 | 0.900 | 500 | 6 | 48 | 2.0 | 1.0 | 2.00 | 2.00 | 180 | 8688 | 0.95000V | 3.21817480 | UNIFORM P-AXIS |
| 33 | 1.000 | 500 | 6 | 48 | 2.0 | 1.0 | 2.00 | 2.00 | 250 | 12048 | 0.65000V | 3.16479410 | UNIFORM P-AXIS |
| 34 | 0.000 | 500 | 6 | 60 | 2.0 | 1.0 | 2.00 | 2.00 | 134 | 8100 | 0.55000V | 3.31659991 | UNIFORM P-AXIS |
| 35 | 0.100 | 500 | 6 | 60 | 2.0 | 1.0 | 2.00 | 2.00 | 193 | 11640 | 0.65000V | 3.31372407 | UNIFORM P-AXIS |
| 36 | 0.200 | 500 | 6 | 60 | 2.0 | 1.0 | 2.00 | 2.00 | 126 | 7620 | 0.15000V | 3.31638228 | UNIFORM P-AXIS |
| 37 | 0.300 | 500 | 6 | 60 | 2.0 | 1.0 | 2.00 | 2.00 | 139 | 8400 | 0.80000V | 3.31640953 | UNIFORM P-AXIS |
| 38 | 0.400 | 500 | 6 | 60 | 2.0 | 1.0 | 2.00 | 2.00 | 145 | 8760 | 0.15000V | 3.31306608 | UNIFORM P-AXIS |
| 39 | 0.500 | 500 | 6 | 60 | 2.0 | 1.0 | 2.00 | 2.00 | 251 | 15120 | 0.70000V | 3.31365808 | UNIFORM P-AXIS |
| 40 | 0.600 | 500 | 6 | 60 | 2.0 | 1.0 | 2.00 | 2.00 | 249 | 15000 | 0.60000V | 3.31607817 | UNIFORM P-AXIS |
| 41 | 0.700 | 500 | 6 | 60 | 2.0 | 1.0 | 2.00 | 2.00 | 158 | 9540 | 0.40000V | 3.17449059 | UNIFORM P-AXIS |
| 42 | 0.800 | 500 | 6 | 60 | 2.0 | 1.0 | 2.00 | 2.00 | 193 | 11640 | 0.65000V | 3.32110070 | UNIFORM P-AXIS |
| 43 | 0.900 | 500 | 6 | 60 | 2.0 | 1.0 | 2.00 | 2.00 | 167 | 10080 | 0.30000V | 3.23623635 | UNIFORM P-AXIS |
| 44 | 1.000 | 500 | 6 | 60 | 2.0 | 1.0 | 2.00 | 2.00 | 282 | 16980 | 0.35000V | 3.26645604 | UNIFORM P-AXIS |
| 45 | 0.000 | 500 | 6 | 72 | 2.0 | 1.0 | 2.00 | 2.00 | 193 | 13968 | 0.65000V | 3.31934262 | UNIFORM P-AXIS |
| 46 | 0.100 | 500 | 6 | 72 | 2.0 | 1.0 | 2.00 | 2.00 | 157 | 11376 | 0.75000V | 3.31520000 | UNIFORM P-AXIS |
| 47 | 0.200 | 500 | 6 | 72 | 2.0 | 1.0 | 2.00 | 2.00 | 250 | 18072 | 0.65000V | 3.31085989 | UNIFORM P-AXIS |
| 48 | 0.300 | 500 | 6 | 72 | 2.0 | 1.0 | 2.00 | 2.00 | 250 | 18072 | 0.65000V | 3.32173273 | UNIFORM P-AXIS |
| 49 | 0.400 | 500 | 6 | 72 | 2.0 | 1.0 | 2.00 | 2.00 | 250 | 18072 | 0.65000V | 3.30166220 | UNIFORM P-AXIS |
| 50 | 0.500 | 500 | 6 | 72 | 2.0 | 1.0 | 2.00 | 2.00 | 230 | 16632 | 0.60000V | 3.31879682 | UNIFORM P-AXIS |
| 51 | 0.600 | 500 | 6 | 72 | 2.0 | 1.0 | 2.00 | 2.00 | 267 | 19296 | 0.55000V | 3.31773563 | UNIFORM P-AXIS |
| 52 | 0.700 | 500 | 6 | 72 | 2.0 | 1.0 | 2.00 | 2.00 | 160 | 11592 | 0.90000V | 3.17735937 | UNIFORM P-AXIS |
| 53 | 0.800 | 500 | 6 | 72 | 2.0 | 1.0 | 2.00 | 2.00 | 227 | 16416 | 0.45000V | 3.31701910 | UNIFORM P-AXIS |
| 54 | 0.900 | 500 | 6 | 72 | 2.0 | 1.0 | 2.00 | 2.00 | 174 | 12600 | 0.65000V | 3.25654442 | UNIFORM P-AXIS |
| 55 | 1.000 | 500 | 6 | 72 | 2.0 | 1.0 | 2.00 | 2.00 | 270 | 19512 | 0.70000V | 3.31105409 | UNIFORM P-AXIS |
| 56 | 0.000 | 500 | 6 | 84 | 2.0 | 1.0 | 2.00 | 2.00 | 283 | 23856 | 0.40000V | 3.31541041 | UNIFORM P-AXIS |
| 57 | 0.100 | 500 | 6 | 84 | 2.0 | 1.0 | 2.00 | 2.00 | 160 | 13524 | 0.90000V | 3.31043298 | UNIFORM P-AXIS |
| 58 | 0.200 | 500 | 6 | 84 | 2.0 | 1.0 | 2.00 | 2.00 | 206 | 17388 | 0.35000V | 3.31145519 | UNIFORM P-AXIS |
| 59 | 0.300 | 500 | 6 | 84 | 2.0 | 1.0 | 2.00 | 2.00 | 103 | 8736 | 0.90000V | 3.32074934 | UNIFORM P-AXIS |
| 60 | 0.400 | 500 | 6 | 84 | 2.0 | 1.0 | 2.00 | 2.00 | 231 | 19488 | 0.65000V | 3.32093772 | UNIFORM P-AXIS |
| 61 | 0.500 | 500 | 6 | 84 | 2.0 | 1.0 | 2.00 | 2.00 | 249 | 21000 | 0.60000V | 3.29768846 | UNIFORM P-AXIS |
| 62 | 0.600 | 500 | 6 | 84 | 2.0 | 1.0 | 2.00 | 2.00 | 250 | 21084 | 0.65000V | 3.31956844 | UNIFORM P-AXIS |
| 63 | 0.700 | 500 | 6 | 84 | 2.0 | 1.0 | 2.00 | 2.00 | 99 | 8400 | 0.70000V | 3.18238200 | UNIFORM P-AXIS |
| 64 | 0.800 | 500 | 6 | 84 | 2.0 | 1.0 | 2.00 | 2.00 | 165 | 13944 | 0.20000V | 3.31971372 | UNIFORM P-AXIS |
| 65 | 0.900 | 500 | 6 | 84 | 2.0 | 1.0 | 2.00 | 2.00 | 268 | 22596 | 0.60000V | 3.30768144 | UNIFORM P-AXIS |
| 66 | 1.000 | 500 | 6 | 84 | 2.0 | 1.0 | 2.00 | 2.00 | 251 | 21168 | 0.70000V | 3.30198521 | UNIFORM P-AXIS |

Total Function Evaluations: 730212

| 48 | 0.300 | 500 | 6 | 72 | 2.0 | 1.0 | 2.00 | 2.00 | 250 | 18072 | 0.65000V | 3.32173273 | UNIFORM P-AXIS |

# F21

Run ID: 12-27-2009, 00:08:36

FUNCTION: F21

| Run # | Gamma | Nt | Nd | Np | G | DelT | Alpha | Beta | #Steps | Neval | Frep | Fitness | Initial Probes |
|-------|-------|-----|----|----|-----|------|-------|------|--------|-------|----------|------------------|-----------------|
| 0 | 0.000 | 500 | 4 | 16 | 2.0 | 1.0 | 2.00 | 2.00 | 0 | 0 | 0.65000V | -9999.00000000 | UNIFORM P-AXIS |
| 1 | 0.000 | 500 | 4 | 16 | 2.0 | 1.0 | 2.00 | 2.00 | 60 | 976 | 0.65000V | 0.27311534 | UNIFORM P-AXIS |
| 2 | 0.100 | 500 | 4 | 16 | 2.0 | 1.0 | 2.00 | 2.00 | 96 | 1552 | 0.55000V | 5.04133435 | UNIFORM P-AXIS |
| 3 | 0.200 | 500 | 4 | 16 | 2.0 | 1.0 | 2.00 | 2.00 | 138 | 2224 | 0.75000V | 0.45919710 | UNIFORM P-AXIS |
| 4 | 0.300 | 500 | 4 | 16 | 2.0 | 1.0 | 2.00 | 2.00 | 138 | 2224 | 0.75000V | 0.67224136 | UNIFORM P-AXIS |
| 5 | 0.400 | 500 | 4 | 16 | 2.0 | 1.0 | 2.00 | 2.00 | 172 | 2768 | 0.55000V | 10.14144490 | UNIFORM P-AXIS |
| 6 | 0.500 | 500 | 4 | 16 | 2.0 | 1.0 | 2.00 | 2.00 | 156 | 2512 | 0.70000V | 0.61614905 | UNIFORM P-AXIS |
| 7 | 0.600 | 500 | 4 | 16 | 2.0 | 1.0 | 2.00 | 2.00 | 272 | 4368 | 0.80000V | 10.14663737 | UNIFORM P-AXIS |
| 8 | 0.700 | 500 | 4 | 16 | 2.0 | 1.0 | 2.00 | 2.00 | 60 | 976 | 0.65000V | 0.53087283 | UNIFORM P-AXIS |
| 9 | 0.800 | 500 | 4 | 16 | 2.0 | 1.0 | 2.00 | 2.00 | 96 | 1552 | 0.55000V | 4.85279463 | UNIFORM P-AXIS |
| 10 | 0.900 | 500 | 4 | 16 | 2.0 | 1.0 | 2.00 | 2.00 | 196 | 3152 | 0.80000V | 1.25937046 | UNIFORM P-AXIS |
| 11 | 1.000 | 500 | 4 | 16 | 2.0 | 1.0 | 2.00 | 2.00 | 191 | 3072 | 0.55000V | 10.14038367 | UNIFORM P-AXIS |
| 12 | 0.000 | 500 | 4 | 24 | 2.0 | 1.0 | 2.00 | 2.00 | 171 | 4128 | 0.50000V | 5.05394549 | UNIFORM P-AXIS |
| 13 | 0.100 | 500 | 4 | 24 | 2.0 | 1.0 | 2.00 | 2.00 | 127 | 3072 | 0.20000V | 10.15293741 | UNIFORM P-AXIS |
| 14 | 0.200 | 500 | 4 | 24 | 2.0 | 1.0 | 2.00 | 2.00 | 173 | 4176 | 0.60000V | 10.15309558 | UNIFORM P-AXIS |
| 15 | 0.300 | 500 | 4 | 24 | 2.0 | 1.0 | 2.00 | 2.00 | 153 | 3696 | 0.55000V | 1.97500752 | UNIFORM P-AXIS |
| 16 | 0.400 | 500 | 4 | 24 | 2.0 | 1.0 | 2.00 | 2.00 | 78 | 1896 | 0.60000V | 10.15319585 | UNIFORM P-AXIS |

| Run # | Gamma | Nt | Nd | Np | G | DelT | Alpha | Beta | #Steps | Neval | Frep | Fitness | Initial Probes |
|---|---|---|---|---|---|---|---|---|---|---|---|---|---|
| 17 | 0.500 | 500 | 4 | 24 | 2.0 | 1.0 | 2.00 | 2.00 | 60 | 1464 | 0.65000V | 0.57297331 | UNIFORM P-AXIS |
| 18 | 0.600 | 500 | 4 | 24 | 2.0 | 1.0 | 2.00 | 2.00 | 500 | 12024 | 0.80000V | 8.82793293 | UNIFORM P-AXIS |
| 19 | 0.700 | 500 | 4 | 24 | 2.0 | 1.0 | 2.00 | 2.00 | 174 | 4200 | 0.65000V | 0.53158719 | UNIFORM P-AXIS |
| 20 | 0.800 | 500 | 4 | 24 | 2.0 | 1.0 | 2.00 | 2.00 | 117 | 2832 | 0.65000V | 10.14658660 | UNIFORM P-AXIS |
| 21 | 0.900 | 500 | 4 | 24 | 2.0 | 1.0 | 2.00 | 2.00 | 208 | 5016 | 0.45000V | 2.05090467 | UNIFORM P-AXIS |
| 22 | 1.000 | 500 | 4 | 24 | 2.0 | 1.0 | 2.00 | 2.00 | 204 | 4920 | 0.25000V | 10.15142802 | UNIFORM P-AXIS |
| 23 | 0.000 | 500 | 4 | 32 | 2.0 | 1.0 | 2.00 | 2.00 | 134 | 4320 | 0.55000V | 5.04863711 | UNIFORM P-AXIS |
| 24 | 0.100 | 500 | 4 | 32 | 2.0 | 1.0 | 2.00 | 2.00 | 96 | 3104 | 0.55000V | 5.05472330 | UNIFORM P-AXIS |
| 25 | 0.200 | 500 | 4 | 32 | 2.0 | 1.0 | 2.00 | 2.00 | 138 | 4448 | 0.75000V | 0.42839721 | UNIFORM P-AXIS |
| 26 | 0.300 | 500 | 4 | 32 | 2.0 | 1.0 | 2.00 | 2.00 | 212 | 6816 | 0.65000V | 0.85458904 | UNIFORM P-AXIS |
| 27 | 0.400 | 500 | 4 | 32 | 2.0 | 1.0 | 2.00 | 2.00 | 114 | 3680 | 0.50000V | 10.15296968 | UNIFORM P-AXIS |
| 28 | 0.500 | 500 | 4 | 32 | 2.0 | 1.0 | 2.00 | 2.00 | 60 | 1952 | 0.65000V | 0.58456509 | UNIFORM P-AXIS |
| 29 | 0.600 | 500 | 4 | 32 | 2.0 | 1.0 | 2.00 | 2.00 | 98 | 3168 | 0.65000V | 5.02590739 | UNIFORM P-AXIS |
| 30 | 0.700 | 500 | 4 | 32 | 2.0 | 1.0 | 2.00 | 2.00 | 60 | 1952 | 0.65000V | 0.53137912 | UNIFORM P-AXIS |
| 31 | 0.800 | 500 | 4 | 32 | 2.0 | 1.0 | 2.00 | 2.00 | 75 | 2432 | 0.45000V | 10.02571078 | UNIFORM P-AXIS |
| 32 | 0.900 | 500 | 4 | 32 | 2.0 | 1.0 | 2.00 | 2.00 | 176 | 5664 | 0.75000V | 1.02391414 | UNIFORM P-AXIS |
| 33 | 1.000 | 500 | 4 | 32 | 2.0 | 1.0 | 2.00 | 2.00 | 74 | 2400 | 0.40000V | 10.14076773 | UNIFORM P-AXIS |
| 34 | 0.000 | 500 | 4 | 40 | 2.0 | 1.0 | 2.00 | 2.00 | 151 | 6080 | 0.45000V | 5.05281229 | UNIFORM P-AXIS |
| 35 | 0.100 | 500 | 4 | 40 | 2.0 | 1.0 | 2.00 | 2.00 | 99 | 4000 | 0.55000V | 5.05441601 | UNIFORM P-AXIS |
| 36 | 0.200 | 500 | 4 | 40 | 2.0 | 1.0 | 2.00 | 2.00 | 194 | 7800 | 0.70000V | 10.46731350 | UNIFORM P-AXIS |
| 37 | 0.300 | 500 | 4 | 40 | 2.0 | 1.0 | 2.00 | 2.00 | 229 | 9200 | 0.55000V | 7.93894966 | UNIFORM P-AXIS |
| 38 | 0.400 | 500 | 4 | 40 | 2.0 | 1.0 | 2.00 | 2.00 | 153 | 6160 | 0.55000V | 10.15163190 | UNIFORM P-AXIS |
| 39 | 0.500 | 500 | 4 | 40 | 2.0 | 1.0 | 2.00 | 2.00 | 60 | 2440 | 0.65000V | 0.58586398 | UNIFORM P-AXIS |
| 40 | 0.600 | 500 | 4 | 40 | 2.0 | 1.0 | 2.00 | 2.00 | 94 | 3800 | 0.45000V | 10.15158968 | UNIFORM P-AXIS |
| 41 | 0.700 | 500 | 4 | 40 | 2.0 | 1.0 | 2.00 | 2.00 | 60 | 2440 | 0.65000V | 0.53087283 | UNIFORM P-AXIS |
| 42 | 0.800 | 500 | 4 | 40 | 2.0 | 1.0 | 2.00 | 2.00 | 273 | 10960 | 0.85000V | 10.12002071 | UNIFORM P-AXIS |
| 43 | 0.900 | 500 | 4 | 40 | 2.0 | 1.0 | 2.00 | 2.00 | 196 | 7880 | 0.80000V | 4.11351704 | UNIFORM P-AXIS |
| 44 | 1.000 | 500 | 4 | 40 | 2.0 | 1.0 | 2.00 | 2.00 | 135 | 5440 | 0.60000V | 10.15035466 | UNIFORM P-AXIS |
| 45 | 0.000 | 500 | 4 | 48 | 2.0 | 1.0 | 2.00 | 2.00 | 249 | 12000 | 0.65000V | 5.05432308 | UNIFORM P-AXIS |
| 46 | 0.100 | 500 | 4 | 48 | 2.0 | 1.0 | 2.00 | 2.00 | 190 | 9168 | 0.50000V | 10.15273852 | UNIFORM P-AXIS |
| 47 | 0.200 | 500 | 4 | 48 | 2.0 | 1.0 | 2.00 | 2.00 | 193 | 9312 | 0.65000V | 0.47276129 | UNIFORM P-AXIS |
| 48 | 0.300 | 500 | 4 | 48 | 2.0 | 1.0 | 2.00 | 2.00 | 158 | 7632 | 0.90000V | 1.19505494 | UNIFORM P-AXIS |
| 49 | 0.400 | 500 | 4 | 48 | 2.0 | 1.0 | 2.00 | 2.00 | 252 | 12144 | 0.75000V | 10.15205928 | UNIFORM P-AXIS |
| 50 | 0.500 | 500 | 4 | 48 | 2.0 | 1.0 | 2.00 | 2.00 | 60 | 2928 | 0.65000V | 0.58526200 | UNIFORM P-AXIS |
| 51 | 0.600 | 500 | 4 | 48 | 2.0 | 1.0 | 2.00 | 2.00 | 118 | 5712 | 0.70000V | 10.14685194 | UNIFORM P-AXIS |
| 52 | 0.700 | 500 | 4 | 48 | 2.0 | 1.0 | 2.00 | 2.00 | 60 | 2928 | 0.65000V | 0.53140202 | UNIFORM P-AXIS |
| 53 | 0.800 | 500 | 4 | 48 | 2.0 | 1.0 | 2.00 | 2.00 | 80 | 3888 | 0.70000V | 5.09974140 | UNIFORM P-AXIS |
| 54 | 0.900 | 500 | 4 | 48 | 2.0 | 1.0 | 2.00 | 2.00 | 174 | 8400 | 0.65000V | 1.94704742 | UNIFORM P-AXIS |
| 55 | 1.000 | 500 | 4 | 48 | 2.0 | 1.0 | 2.00 | 2.00 | 167 | 8064 | 0.30000V | 10.15331224 | UNIFORM P-AXIS |
| 56 | 0.000 | 500 | 4 | 56 | 2.0 | 1.0 | 2.00 | 2.00 | 246 | 13832 | 0.45000V | 5.05171296 | UNIFORM P-AXIS |
| 57 | 0.100 | 500 | 4 | 56 | 2.0 | 1.0 | 2.00 | 2.00 | 78 | 4424 | 0.60000V | 5.05389514 | UNIFORM P-AXIS |
| 58 | 0.200 | 500 | 4 | 56 | 2.0 | 1.0 | 2.00 | 2.00 | 176 | 9912 | 0.75000V | 0.47849329 | UNIFORM P-AXIS |
| 59 | 0.300 | 500 | 4 | 56 | 2.0 | 1.0 | 2.00 | 2.00 | 161 | 9072 | 0.95000V | 1.97972685 | UNIFORM P-AXIS |
| 60 | 0.400 | 500 | 4 | 56 | 2.0 | 1.0 | 2.00 | 2.00 | 60 | 3416 | 0.65000V | 10.11776011 | UNIFORM P-AXIS |
| 61 | 0.500 | 500 | 4 | 56 | 2.0 | 1.0 | 2.00 | 2.00 | 60 | 3416 | 0.65000V | 0.58434919 | UNIFORM P-AXIS |
| 62 | 0.600 | 500 | 4 | 56 | 2.0 | 1.0 | 2.00 | 2.00 | 80 | 4536 | 0.70000V | 10.09567420 | UNIFORM P-AXIS |
| 63 | 0.700 | 500 | 4 | 56 | 2.0 | 1.0 | 2.00 | 2.00 | 60 | 3416 | 0.65000V | 0.53103979 | UNIFORM P-AXIS |
| 64 | 0.800 | 500 | 4 | 56 | 2.0 | 1.0 | 2.00 | 2.00 | 106 | 5992 | 0.65000V | 10.14160544 | UNIFORM P-AXIS |
| 65 | 0.900 | 500 | 4 | 56 | 2.0 | 1.0 | 2.00 | 2.00 | 209 | 11760 | 0.50000V | 5.08969804 | UNIFORM P-AXIS |
| 66 | 1.000 | 500 | 4 | 56 | 2.0 | 1.0 | 2.00 | 2.00 | 103 | 5824 | 0.90000V | 10.15317755 | UNIFORM P-AXIS |

Total Function Evaluations: 336712

| 16 | 0.400 | 500 | 4 | 24 | 2.0 | 1.0 | 2.00 | 2.00 | 78 | 1896 | 0.60000V | 10.15319585 | UNIFORM P-AXIS |

# F22

Run ID: 12-27-2009, 00:12:14

FUNCTION: F22

| Run # | Gamma | Nt | Nd | Np | G | DelT | Alpha | Beta | #Steps | Neval | Frep | Fitness | Initial Probes |
|---|---|---|---|---|---|---|---|---|---|---|---|---|---|
| 0 | 0.000 | 500 | 4 | 16 | 2.0 | 1.0 | 2.00 | 2.00 | 0 | 0 | 0.50000V | -9999.00000000 | UNIFORM P-AXIS |
| 1 | 0.000 | 500 | 4 | 16 | 2.0 | 1.0 | 2.00 | 2.00 | 60 | 976 | 0.65000V | 0.29361829 | UNIFORM P-AXIS |
| 2 | 0.100 | 500 | 4 | 16 | 2.0 | 1.0 | 2.00 | 2.00 | 203 | 3264 | 0.20000V | 10.26261276 | UNIFORM P-AXIS |
| 3 | 0.200 | 500 | 4 | 16 | 2.0 | 1.0 | 2.00 | 2.00 | 138 | 2224 | 0.75000V | 0.50537915 | UNIFORM P-AXIS |
| 4 | 0.300 | 500 | 4 | 16 | 2.0 | 1.0 | 2.00 | 2.00 | 157 | 2528 | 0.75000V | 0.95497293 | UNIFORM P-AXIS |
| 5 | 0.400 | 500 | 4 | 16 | 2.0 | 1.0 | 2.00 | 2.00 | 291 | 4672 | 0.80000V | 10.40284412 | UNIFORM P-AXIS |
| 6 | 0.500 | 500 | 4 | 16 | 2.0 | 1.0 | 2.00 | 2.00 | 218 | 3504 | 0.95000V | 1.00484226 | UNIFORM P-AXIS |
| 7 | 0.600 | 500 | 4 | 16 | 2.0 | 1.0 | 2.00 | 2.00 | 117 | 1888 | 0.65000V | 5.12591666 | UNIFORM P-AXIS |
| 8 | 0.700 | 500 | 4 | 16 | 2.0 | 1.0 | 2.00 | 2.00 | 60 | 976 | 0.65000V | 0.57542458 | UNIFORM P-AXIS |
| 9 | 0.800 | 500 | 4 | 16 | 2.0 | 1.0 | 2.00 | 2.00 | 177 | 2848 | 0.80000V | 5.11317026 | UNIFORM P-AXIS |
| 10 | 0.900 | 500 | 4 | 16 | 2.0 | 1.0 | 2.00 | 2.00 | 175 | 2816 | 0.70000V | 1.19582715 | UNIFORM P-AXIS |
| 11 | 1.000 | 500 | 4 | 16 | 2.0 | 1.0 | 2.00 | 2.00 | 216 | 3472 | 0.85000V | 10.39966301 | UNIFORM P-AXIS |
| 12 | 0.000 | 500 | 4 | 24 | 2.0 | 1.0 | 2.00 | 2.00 | 182 | 4392 | 0.10000V | 5.07724293 | UNIFORM P-AXIS |
| 13 | 0.100 | 500 | 4 | 24 | 2.0 | 1.0 | 2.00 | 2.00 | 96 | 2328 | 0.55000V | 5.08627190 | UNIFORM P-AXIS |
| 14 | 0.200 | 500 | 4 | 24 | 2.0 | 1.0 | 2.00 | 2.00 | 144 | 3480 | 0.10000V | 10.40264736 | UNIFORM P-AXIS |
| 15 | 0.300 | 500 | 4 | 24 | 2.0 | 1.0 | 2.00 | 2.00 | 207 | 4992 | 0.40000V | 1.24076642 | UNIFORM P-AXIS |
| 16 | 0.400 | 500 | 4 | 24 | 2.0 | 1.0 | 2.00 | 2.00 | 130 | 3144 | 0.35000V | 10.40281884 | UNIFORM P-AXIS |
| 17 | 0.500 | 500 | 4 | 24 | 2.0 | 1.0 | 2.00 | 2.00 | 194 | 4680 | 0.70000V | 0.98671234 | UNIFORM P-AXIS |
| 18 | 0.600 | 500 | 4 | 24 | 2.0 | 1.0 | 2.00 | 2.00 | 153 | 3696 | 0.55000V | 10.39418334 | UNIFORM P-AXIS |
| 19 | 0.700 | 500 | 4 | 24 | 2.0 | 1.0 | 2.00 | 2.00 | 194 | 4680 | 0.70000V | 0.59488646 | UNIFORM P-AXIS |
| 20 | 0.800 | 500 | 4 | 24 | 2.0 | 1.0 | 2.00 | 2.00 | 91 | 2208 | 0.30000V | 10.40286059 | UNIFORM P-AXIS |
| 21 | 0.900 | 500 | 4 | 24 | 2.0 | 1.0 | 2.00 | 2.00 | 193 | 4656 | 0.65000V | 1.91215458 | UNIFORM P-AXIS |
| 22 | 1.000 | 500 | 4 | 24 | 2.0 | 1.0 | 2.00 | 2.00 | 116 | 2808 | 0.60000V | 10.35044377 | UNIFORM P-AXIS |
| 23 | 0.000 | 500 | 4 | 32 | 2.0 | 1.0 | 2.00 | 2.00 | 180 | 5792 | 0.95000V | 5.08311858 | UNIFORM P-AXIS |
| 24 | 0.100 | 500 | 4 | 32 | 2.0 | 1.0 | 2.00 | 2.00 | 329 | 10560 | 0.80000V | 10.40175889 | UNIFORM P-AXIS |
| 25 | 0.200 | 500 | 4 | 32 | 2.0 | 1.0 | 2.00 | 2.00 | 138 | 4448 | 0.75000V | 0.48151911 | UNIFORM P-AXIS |
| 26 | 0.300 | 500 | 4 | 32 | 2.0 | 1.0 | 2.00 | 2.00 | 139 | 4480 | 0.80000V | 0.87836043 | UNIFORM P-AXIS |
| 27 | 0.400 | 500 | 4 | 32 | 2.0 | 1.0 | 2.00 | 2.00 | 60 | 1952 | 0.65000V | 10.35599982 | UNIFORM P-AXIS |
| 28 | 0.500 | 500 | 4 | 32 | 2.0 | 1.0 | 2.00 | 2.00 | 176 | 5664 | 0.75000V | 0.89227405 | UNIFORM P-AXIS |
| 29 | 0.600 | 500 | 4 | 32 | 2.0 | 1.0 | 2.00 | 2.00 | 110 | 3552 | 0.30000V | 5.12666016 | UNIFORM P-AXIS |
| 30 | 0.700 | 500 | 4 | 32 | 2.0 | 1.0 | 2.00 | 2.00 | 60 | 1952 | 0.65000V | 0.57306071 | UNIFORM P-AXIS |
| 31 | 0.800 | 500 | 4 | 32 | 2.0 | 1.0 | 2.00 | 2.00 | 178 | 5728 | 0.85000V | 10.40070098 | UNIFORM P-AXIS |
| 32 | 0.900 | 500 | 4 | 32 | 2.0 | 1.0 | 2.00 | 2.00 | 177 | 5696 | 0.80000V | 0.84664455 | UNIFORM P-AXIS |
| 33 | 1.000 | 500 | 4 | 32 | 2.0 | 1.0 | 2.00 | 2.00 | 226 | 7264 | 0.40000V | 10.40258245 | UNIFORM P-AXIS |
| 34 | 0.000 | 500 | 4 | 40 | 2.0 | 1.0 | 2.00 | 2.00 | 134 | 5400 | 0.55000V | 5.08672034 | UNIFORM P-AXIS |
| 35 | 0.100 | 500 | 4 | 40 | 2.0 | 1.0 | 2.00 | 2.00 | 123 | 4960 | 0.95000V | 10.40267265 | UNIFORM P-AXIS |
| 36 | 0.200 | 500 | 4 | 40 | 2.0 | 1.0 | 2.00 | 2.00 | 138 | 5560 | 0.75000V | 0.48686581 | UNIFORM P-AXIS |
| 37 | 0.300 | 500 | 4 | 40 | 2.0 | 1.0 | 2.00 | 2.00 | 176 | 7080 | 0.75000V | 1.51013313 | UNIFORM P-AXIS |
| 38 | 0.400 | 500 | 4 | 40 | 2.0 | 1.0 | 2.00 | 2.00 | 167 | 6720 | 0.30000V | 10.39925385 | UNIFORM P-AXIS |
| 39 | 0.500 | 500 | 4 | 40 | 2.0 | 1.0 | 2.00 | 2.00 | 176 | 7080 | 0.75000V | 1.00062130 | UNIFORM P-AXIS |
| 40 | 0.600 | 500 | 4 | 40 | 2.0 | 1.0 | 2.00 | 2.00 | 131 | 5280 | 0.40000V | 10.39124772 | UNIFORM P-AXIS |
| 41 | 0.700 | 500 | 4 | 40 | 2.0 | 1.0 | 2.00 | 2.00 | 60 | 2440 | 0.65000V | 0.57542458 | UNIFORM P-AXIS |
| 42 | 0.800 | 500 | 4 | 40 | 2.0 | 1.0 | 2.00 | 2.00 | 75 | 3040 | 0.45000V | 10.40200744 | UNIFORM P-AXIS |
| 43 | 0.900 | 500 | 4 | 40 | 2.0 | 1.0 | 2.00 | 2.00 | 176 | 7080 | 0.75000V | 4.00941138 | UNIFORM P-AXIS |
| 44 | 1.000 | 500 | 4 | 40 | 2.0 | 1.0 | 2.00 | 2.00 | 254 | 10200 | 0.85000V | 10.39754781 | UNIFORM P-AXIS |

| Run # | Gamma | Nt | Nd | Np | G | DelT | Alpha | Beta | #Steps | Neval | Frep | Fitness | Initial Probes |
|-------|-------|----|----|----|----|------|-------|------|--------|-------|------|---------|----------------|
| 45 | 0.000 | 500 | 4 | 48 | 2.0 | 1.0 | 2.00 | 2.00 | 176 | 8496 | 0.75000V | 5.08636835 | UNIFORM P-AXIS |
| 46 | 0.100 | 500 | 4 | 48 | 2.0 | 1.0 | 2.00 | 2.00 | 131 | 6336 | 0.40000V | 5.08612144 | UNIFORM P-AXIS |
| 47 | 0.200 | 500 | 4 | 48 | 2.0 | 1.0 | 2.00 | 2.00 | 143 | 6912 | 0.05000V | 0.48654928 | UNIFORM P-AXIS |
| 48 | 0.300 | 500 | 4 | 48 | 2.0 | 1.0 | 2.00 | 2.00 | 165 | 7968 | 0.20000V | 4.44538707 | UNIFORM P-AXIS |
| 49 | 0.400 | 500 | 4 | 48 | 2.0 | 1.0 | 2.00 | 2.00 | 124 | 6000 | 0.05000V | 10.40044163 | UNIFORM P-AXIS |
| 50 | 0.500 | 500 | 4 | 48 | 2.0 | 1.0 | 2.00 | 2.00 | 155 | 7488 | 0.65000V | 1.39721189 | UNIFORM P-AXIS |
| 51 | 0.600 | 500 | 4 | 48 | 2.0 | 1.0 | 2.00 | 2.00 | 207 | 9984 | 0.40000V | 10.39959118 | UNIFORM P-AXIS |
| 52 | 0.700 | 500 | 4 | 48 | 2.0 | 1.0 | 2.00 | 2.00 | 60 | 2928 | 0.65000V | 0.57429457 | UNIFORM P-AXIS |
| 53 | 0.800 | 500 | 4 | 48 | 2.0 | 1.0 | 2.00 | 2.00 | 86 | 4176 | 0.05000V | 10.39514139 | UNIFORM P-AXIS |
| 54 | 0.900 | 500 | 4 | 48 | 2.0 | 1.0 | 2.00 | 2.00 | 178 | 8592 | 0.85000V | 2.14526622 | UNIFORM P-AXIS |
| 55 | 1.000 | 500 | 4 | 48 | 2.0 | 1.0 | 2.00 | 2.00 | 163 | 7872 | 0.10000V | 10.40265915 | UNIFORM P-AXIS |
| 56 | 0.000 | 500 | 4 | 56 | 2.0 | 1.0 | 2.00 | 2.00 | 231 | 12992 | 0.65000V | 5.07298037 | UNIFORM P-AXIS |
| 57 | 0.100 | 500 | 4 | 56 | 2.0 | 1.0 | 2.00 | 2.00 | 344 | 19320 | 0.60000V | 10.40232156 | UNIFORM P-AXIS |
| 58 | 0.200 | 500 | 4 | 56 | 2.0 | 1.0 | 2.00 | 2.00 | 213 | 11984 | 0.70000V | 0.50205717 | UNIFORM P-AXIS |
| 59 | 0.300 | 500 | 4 | 56 | 2.0 | 1.0 | 2.00 | 2.00 | 240 | 13496 | 0.15000V | 6.54548019 | UNIFORM P-AXIS |
| 60 | 0.400 | 500 | 4 | 56 | 2.0 | 1.0 | 2.00 | 2.00 | 130 | 7336 | 0.35000V | 10.40282272 | UNIFORM P-AXIS |
| 61 | 0.500 | 500 | 4 | 56 | 2.0 | 1.0 | 2.00 | 2.00 | 154 | 8680 | 0.60000V | 1.40129961 | UNIFORM P-AXIS |
| 62 | 0.600 | 500 | 4 | 56 | 2.0 | 1.0 | 2.00 | 2.00 | 209 | 11760 | 0.50000V | 10.40172732 | UNIFORM P-AXIS |
| 63 | 0.700 | 500 | 4 | 56 | 2.0 | 1.0 | 2.00 | 2.00 | 60 | 3416 | 0.65000V | 0.57352478 | UNIFORM P-AXIS |
| 64 | 0.800 | 500 | 4 | 56 | 2.0 | 1.0 | 2.00 | 2.00 | 111 | 6272 | 0.35000V | 5.12730501 | UNIFORM P-AXIS |
| 65 | 0.900 | 500 | 4 | 56 | 2.0 | 1.0 | 2.00 | 2.00 | 225 | 12656 | 0.35000V | 5.12155298 | UNIFORM P-AXIS |
| 66 | 1.000 | 500 | 4 | 56 | 2.0 | 1.0 | 2.00 | 2.00 | 166 | 9352 | 0.25000V | 10.40279968 | UNIFORM P-AXIS |

Total Function Evaluations: 386176

---

| 20 | 0.800 | 500 | 4 | 24 | 2.0 | 1.0 | 2.00 | 2.00 | 91 | 2208 | 0.30000V | 10.40286059 | UNIFORM P-AXIS |

## F23

Run ID: 12-27-2009, 00:16:12

FUNCTION: F23

| Run # | Gamma | Nt | Nd | Np | G | DelT | Alpha | Beta | #Steps | Neval | Frep | Fitness | Initial Probes |
|-------|-------|----|----|----|----|------|-------|------|--------|-------|------|---------|----------------|
| 0 | 0.000 | 500 | 4 | 16 | 2.0 | 1.0 | 2.00 | 2.00 | 0 | 0 | 0.50000V | -9999.00000000 | UNIFORM P-AXIS |
| 1 | 0.000 | 500 | 4 | 16 | 2.0 | 1.0 | 2.00 | 2.00 | 60 | 976 | 0.65000V | 0.32172905 | UNIFORM P-AXIS |
| 2 | 0.100 | 500 | 4 | 16 | 2.0 | 1.0 | 2.00 | 2.00 | 137 | 2208 | 0.70000V | 10.51946328 | UNIFORM P-AXIS |
| 3 | 0.200 | 500 | 4 | 16 | 2.0 | 1.0 | 2.00 | 2.00 | 138 | 2224 | 0.75000V | 0.55435306 | UNIFORM P-AXIS |
| 4 | 0.300 | 500 | 4 | 16 | 2.0 | 1.0 | 2.00 | 2.00 | 157 | 2528 | 0.75000V | 1.07283233 | UNIFORM P-AXIS |
| 5 | 0.400 | 500 | 4 | 16 | 2.0 | 1.0 | 2.00 | 2.00 | 140 | 2256 | 0.85000V | 10.53615023 | UNIFORM P-AXIS |
| 6 | 0.500 | 500 | 4 | 16 | 2.0 | 1.0 | 2.00 | 2.00 | 231 | 3712 | 0.65000V | 1.14397265 | UNIFORM P-AXIS |
| 7 | 0.600 | 500 | 4 | 16 | 2.0 | 1.0 | 2.00 | 2.00 | 111 | 1792 | 0.35000V | 2.86919827 | UNIFORM P-AXIS |
| 8 | 0.700 | 500 | 4 | 16 | 2.0 | 1.0 | 2.00 | 2.00 | 60 | 976 | 0.65000V | 0.65227664 | UNIFORM P-AXIS |
| 9 | 0.800 | 500 | 4 | 16 | 2.0 | 1.0 | 2.00 | 2.00 | 132 | 2128 | 0.45000V | 5.17558057 | UNIFORM P-AXIS |
| 10 | 0.900 | 500 | 4 | 16 | 2.0 | 1.0 | 2.00 | 2.00 | 241 | 3872 | 0.20000V | 1.33159419 | UNIFORM P-AXIS |
| 11 | 1.000 | 500 | 4 | 16 | 2.0 | 1.0 | 2.00 | 2.00 | 131 | 2112 | 0.40000V | 10.53532965 | UNIFORM P-AXIS |
| 12 | 0.000 | 500 | 4 | 24 | 2.0 | 1.0 | 2.00 | 2.00 | 194 | 4680 | 0.70000V | 5.10962539 | UNIFORM P-AXIS |
| 13 | 0.100 | 500 | 4 | 24 | 2.0 | 1.0 | 2.00 | 2.00 | 310 | 7464 | 0.80000V | 10.53629338 | UNIFORM P-AXIS |
| 14 | 0.200 | 500 | 4 | 24 | 2.0 | 1.0 | 2.00 | 2.00 | 69 | 1680 | 0.15000V | 5.10013673 | UNIFORM P-AXIS |
| 15 | 0.300 | 500 | 4 | 24 | 2.0 | 1.0 | 2.00 | 2.00 | 162 | 3912 | 0.05000V | 1.51972338 | UNIFORM P-AXIS |
| 16 | 0.400 | 500 | 4 | 24 | 2.0 | 1.0 | 2.00 | 2.00 | 135 | 3264 | 0.60000V | 10.53628373 | UNIFORM P-AXIS |
| 17 | 0.500 | 500 | 4 | 24 | 2.0 | 1.0 | 2.00 | 2.00 | 182 | 4392 | 0.10000V | 1.13540741 | UNIFORM P-AXIS |
| 18 | 0.600 | 500 | 4 | 24 | 2.0 | 1.0 | 2.00 | 2.00 | 413 | 9936 | 0.25000V | 10.53581352 | UNIFORM P-AXIS |
| 19 | 0.700 | 500 | 4 | 24 | 2.0 | 1.0 | 2.00 | 2.00 | 176 | 4248 | 0.75000V | 0.72542885 | UNIFORM P-AXIS |
| 20 | 0.800 | 500 | 4 | 24 | 2.0 | 1.0 | 2.00 | 2.00 | 93 | 2256 | 0.40000V | 10.53632019 | UNIFORM P-AXIS |
| 21 | 0.900 | 500 | 4 | 24 | 2.0 | 1.0 | 2.00 | 2.00 | 174 | 4200 | 0.65000V | 2.06446347 | UNIFORM P-AXIS |
| 22 | 1.000 | 500 | 4 | 24 | 2.0 | 1.0 | 2.00 | 2.00 | 190 | 4584 | 0.50000V | 10.53152157 | UNIFORM P-AXIS |
| 23 | 0.000 | 500 | 4 | 32 | 2.0 | 1.0 | 2.00 | 2.00 | 182 | 5856 | 0.10000V | 5.11195183 | UNIFORM P-AXIS |
| 24 | 0.100 | 500 | 4 | 32 | 2.0 | 1.0 | 2.00 | 2.00 | 191 | 6144 | 0.55000V | 10.53233518 | UNIFORM P-AXIS |
| 25 | 0.200 | 500 | 4 | 32 | 2.0 | 1.0 | 2.00 | 2.00 | 138 | 4448 | 0.75000V | 0.52798028 | UNIFORM P-AXIS |
| 26 | 0.300 | 500 | 4 | 32 | 2.0 | 1.0 | 2.00 | 2.00 | 147 | 4736 | 0.25000V | 1.00075402 | UNIFORM P-AXIS |
| 27 | 0.400 | 500 | 4 | 32 | 2.0 | 1.0 | 2.00 | 2.00 | 60 | 1952 | 0.65000V | 10.47302654 | UNIFORM P-AXIS |
| 28 | 0.500 | 500 | 4 | 32 | 2.0 | 1.0 | 2.00 | 2.00 | 176 | 5664 | 0.75000V | 1.03683311 | UNIFORM P-AXIS |
| 29 | 0.600 | 500 | 4 | 32 | 2.0 | 1.0 | 2.00 | 2.00 | 76 | 2464 | 0.50000V | 5.11476720 | UNIFORM P-AXIS |
| 30 | 0.700 | 500 | 4 | 32 | 2.0 | 1.0 | 2.00 | 2.00 | 60 | 1952 | 0.65000V | 0.64610891 | UNIFORM P-AXIS |
| 31 | 0.800 | 500 | 4 | 32 | 2.0 | 1.0 | 2.00 | 2.00 | 208 | 6688 | 0.45000V | 10.53519865 | UNIFORM P-AXIS |
| 32 | 0.900 | 500 | 4 | 32 | 2.0 | 1.0 | 2.00 | 2.00 | 183 | 5888 | 0.15000V | 1.98038727 | UNIFORM P-AXIS |
| 33 | 1.000 | 500 | 4 | 32 | 2.0 | 1.0 | 2.00 | 2.00 | 232 | 7456 | 0.70000V | 10.53612921 | UNIFORM P-AXIS |
| 34 | 0.000 | 500 | 4 | 40 | 2.0 | 1.0 | 2.00 | 2.00 | 339 | 13600 | 0.35000V | 5.12842148 | UNIFORM P-AXIS |
| 35 | 0.100 | 500 | 4 | 40 | 2.0 | 1.0 | 2.00 | 2.00 | 171 | 6880 | 0.50000V | 10.51343276 | UNIFORM P-AXIS |
| 36 | 0.200 | 500 | 4 | 40 | 2.0 | 1.0 | 2.00 | 2.00 | 192 | 7720 | 0.40000V | 0.56506609 | UNIFORM P-AXIS |
| 37 | 0.300 | 500 | 4 | 40 | 2.0 | 1.0 | 2.00 | 2.00 | 171 | 6880 | 0.50000V | 2.46078960 | UNIFORM P-AXIS |
| 38 | 0.400 | 500 | 4 | 40 | 2.0 | 1.0 | 2.00 | 2.00 | 226 | 9080 | 0.40000V | 10.53511786 | UNIFORM P-AXIS |
| 39 | 0.500 | 500 | 4 | 40 | 2.0 | 1.0 | 2.00 | 2.00 | 269 | 10800 | 0.65000V | 1.05309681 | UNIFORM P-AXIS |
| 40 | 0.600 | 500 | 4 | 40 | 2.0 | 1.0 | 2.00 | 2.00 | 117 | 4720 | 0.65000V | 10.16796476 | UNIFORM P-AXIS |
| 41 | 0.700 | 500 | 4 | 40 | 2.0 | 1.0 | 2.00 | 2.00 | 60 | 2440 | 0.65000V | 0.65227664 | UNIFORM P-AXIS |
| 42 | 0.800 | 500 | 4 | 40 | 2.0 | 1.0 | 2.00 | 2.00 | 81 | 3280 | 0.75000V | 10.52269113 | UNIFORM P-AXIS |
| 43 | 0.900 | 500 | 4 | 40 | 2.0 | 1.0 | 2.00 | 2.00 | 176 | 7080 | 0.75000V | 4.37301283 | UNIFORM P-AXIS |
| 44 | 1.000 | 500 | 4 | 40 | 2.0 | 1.0 | 2.00 | 2.00 | 129 | 5200 | 0.30000V | 10.48509321 | UNIFORM P-AXIS |
| 45 | 0.000 | 500 | 4 | 48 | 2.0 | 1.0 | 2.00 | 2.00 | 152 | 7344 | 0.50000V | 5.12815383 | UNIFORM P-AXIS |
| 46 | 0.100 | 500 | 4 | 48 | 2.0 | 1.0 | 2.00 | 2.00 | 74 | 3600 | 0.40000V | 10.52851278 | UNIFORM P-AXIS |
| 47 | 0.200 | 500 | 4 | 48 | 2.0 | 1.0 | 2.00 | 2.00 | 138 | 6672 | 0.75000V | 0.53841817 | UNIFORM P-AXIS |
| 48 | 0.300 | 500 | 4 | 48 | 2.0 | 1.0 | 2.00 | 2.00 | 193 | 9312 | 0.65000V | 3.30001202 | UNIFORM P-AXIS |
| 49 | 0.400 | 500 | 4 | 48 | 2.0 | 1.0 | 2.00 | 2.00 | 112 | 5424 | 0.40000V | 10.53439920 | UNIFORM P-AXIS |
| 50 | 0.500 | 500 | 4 | 48 | 2.0 | 1.0 | 2.00 | 2.00 | 157 | 7584 | 0.75000V | 1.46710892 | UNIFORM P-AXIS |
| 51 | 0.600 | 500 | 4 | 48 | 2.0 | 1.0 | 2.00 | 2.00 | 97 | 4704 | 0.60000V | 5.10176153 | UNIFORM P-AXIS |
| 52 | 0.700 | 500 | 4 | 48 | 2.0 | 1.0 | 2.00 | 2.00 | 60 | 2928 | 0.65000V | 0.65302072 | UNIFORM P-AXIS |
| 53 | 0.800 | 500 | 4 | 48 | 2.0 | 1.0 | 2.00 | 2.00 | 322 | 15504 | 0.45000V | 10.53096318 | UNIFORM P-AXIS |
| 54 | 0.900 | 500 | 4 | 48 | 2.0 | 1.0 | 2.00 | 2.00 | 176 | 8496 | 0.75000V | 1.85678248 | UNIFORM P-AXIS |
| 55 | 1.000 | 500 | 4 | 48 | 2.0 | 1.0 | 2.00 | 2.00 | 251 | 12096 | 0.70000V | 10.53165877 | UNIFORM P-AXIS |
| 56 | 0.000 | 500 | 4 | 56 | 2.0 | 1.0 | 2.00 | 2.00 | 217 | 12208 | 0.90000V | 5.11016962 | UNIFORM P-AXIS |
| 57 | 0.100 | 500 | 4 | 56 | 2.0 | 1.0 | 2.00 | 2.00 | 208 | 11704 | 0.45000V | 10.53571364 | UNIFORM P-AXIS |
| 58 | 0.200 | 500 | 4 | 56 | 2.0 | 1.0 | 2.00 | 2.00 | 74 | 4200 | 0.40000V | 0.51919113 | UNIFORM P-AXIS |
| 59 | 0.300 | 500 | 4 | 56 | 2.0 | 1.0 | 2.00 | 2.00 | 229 | 12880 | 0.55000V | 7.31576150 | UNIFORM P-AXIS |
| 60 | 0.400 | 500 | 4 | 56 | 2.0 | 1.0 | 2.00 | 2.00 | 145 | 8176 | 0.15000V | 10.53546357 | UNIFORM P-AXIS |
| 61 | 0.500 | 500 | 4 | 56 | 2.0 | 1.0 | 2.00 | 2.00 | 157 | 8848 | 0.75000V | 1.52874747 | UNIFORM P-AXIS |
| 62 | 0.600 | 500 | 4 | 56 | 2.0 | 1.0 | 2.00 | 2.00 | 179 | 10080 | 0.90000V | 10.53625126 | UNIFORM P-AXIS |
| 63 | 0.700 | 500 | 4 | 56 | 2.0 | 1.0 | 2.00 | 2.00 | 60 | 3416 | 0.65000V | 0.64916466 | UNIFORM P-AXIS |
| 64 | 0.800 | 500 | 4 | 56 | 2.0 | 1.0 | 2.00 | 2.00 | 141 | 7952 | 0.90000V | 5.17550217 | UNIFORM P-AXIS |
| 65 | 0.900 | 500 | 4 | 56 | 2.0 | 1.0 | 2.00 | 2.00 | 329 | 18480 | 0.80000V | 5.17556929 | UNIFORM P-AXIS |
| 66 | 1.000 | 500 | 4 | 56 | 2.0 | 1.0 | 2.00 | 2.00 | 113 | 6384 | 0.45000V | 10.52136164 | UNIFORM P-AXIS |

Total Function Evaluations: 394320

---

| 20 | 0.800 | 500 | 4 | 24 | 2.0 | 1.0 | 2.00 | 2.00 | 93 | 2256 | 0.40000V | 10.53632019 | UNIFORM P-AXIS |

# Appendix 3:  CFO Source Code

```
'Program 'CFO_12-25-09.BAS' compiled with
'Power Basic/Windows Compiler 9.0

'LAST MOD 12-26-2009 -1349 HRS EST

'CHANGES MADE TO "CFO_11-26-09.BAS" TO CREATE THIS VERSION
'WHICH IS USED FOR arXiv SUBMISSION #2:
'   (1).  OUTER LOOP Np/Nd = 2 to MaxProbesPerAxis by 2 ADDED
'   (2).  Np IGNORED & ARRAYS REDIMENSIONED AS REQUIRED BY (1).

'NOTE: ALL PBM FUNCTIONS HAVE WIRE RADIUS SET TO 0.00001LAMBDA
'==============================================================

'THIS PROGRAM IMPLEMENTS A SIMPLE VERSION OF "CENTRAL
'FORCE OPTIMIZATION."  IT IS DISTRIBUTED FREE OF CHARGE
'TO INCREASE AWARENESS OF CFO AND TO ENCOURAGE EXPERI-
'MENTATION WITH THE ALGORITHM.

'CFO IS A MULTIDIMENSIONAL SEARCH AND OPTIMIZATION
'ALGORITHM THAT LOCATES THE GLOBAL MAXIMA OF A FUNCTION.
'UNLIKE MOST OTHER ALGORITHMS, CFO IS COMPLETELY DETERMIN-
'ISTIC, SO THAT EVERY RUN WITH THE SAME SETUP PRODUCES
'THE SAME RESULTS.

'Please email questions, comments, and significant
'results to: CFO_questions@yahoo.com.  Thanks!

'(c) 2006-2009 Richard A. Formato

'ALL RIGHTS RESERVED WORLDWIDE

'THIS PROGRAM IS FREEWARE.  IT MAY BE COPIED AND
'DISTRIBUTED WITHOUT LIMITATION AS LONG AS THIS
'COPYRIGHT NOTICE AND THE GNUPLOT AND REFERENCE
'INFORMATION BELOW ARE INCLUDED WITHOUT MODIFICATION,
'AND AS LONG AS NO FEE OR COMPENSATION IS CHARGED,
'INCLUDING "TIE-IN" OR "BUNDLING" FEES CHARGED FOR
'OTHER PRODUCTS.

'=======================================================
'THIS PROGRAM REQUIRES wgnuplot.exe TO DISPLAY PLOTS.
'Gnuplot is a copyrighted freeware plotting program
'available at http://www.gnuplot.info/index.html.

'=======================================================
'REFERENCES
'----------
'1. "Central Force Optimization: A New Metaheuristic with Applications in Applied Electromagnetics,"
'    Progress in Electromagnetics Research, PIER 77, 425-491, 2007 (online).
'Abstract - Central Force Optimization (CFO) is a new deterministic multi-dimensional search
'metaheuristic based on the metaphor of gravitational kinematics.  It models "probes" that "fly"
'through the decision space by analogy to masses moving under the influence of gravity. Equations
'are developed for the probes' positions and accelerations using the analogy of particle motion
'in a gravitational field. In the physical universe, objects traveling through three-dimensional
'space become trapped in close orbits around highly gravitating masses, which is analogous to
'locating the maximum value of an objective function.  In the CFO metaphor, "mass" is a user-defined
'function of the value of the objective function to be maximized.  CFO is readily implemented in a
'compact computer program, and sample pseudocode is presented.  As tests of CFO's effectiveness,
'an equalizer is designed for the well-known Fano load, and a 32-element linear array is synthesized.
'CFO results are compared to several other optimization methods.

'2. "Central Force Optimization: A New Nature Inspired Computational Framework for Multidimensional
'    Search and Optimization," Studies in Computational Intelligence (SCI), vol. 129, 221-238 (2008),
'    www.springerlink.com, Springer-Verlag Berlin Heidelberg 2008.
'Abstract - This paper presents Central Force Optimization, a novel, nature inspired, deterministic
'search metaheuristic for constrained multi-dimensional optimization.  CFO is based on the metaphor
'of gravitational kinematics.  Equations are presented for the positions and accelerations experienced
'by "probes" that "fly" through the decision space by analogy to masses moving under the influence of
'gravity.  In the physical universe, probe satellites become trapped in close orbits around highly
'gravitating masses.  In the CFO analogy, "mass" corresponds to a user-defined function of the value
'of an objective function to be maximized.  CFO is a simple algorithm that is easily implemented in
'a compact computer program.  A typical CFO implementation is applied to several test functions.
'CFO exhibits very good performance, suggesting that it merits further study.

'3. "Central Force Optimisation: A New Gradient-Like Metaheuristic for Multidimensional Search and
'    Optimisation," International Journal of Bio-Inspired Computation, vol. 1, no. 4, 217-238 (2009),
'Abstract: This paper introduces Central Force Optimization, a novel, Nature-inspired, deterministic
'search metaheuristic for constrained multidimensional optimization in highly multimodal, smooth, or
'discontinuous decision spaces.  CFO is based on the metaphor of gravitational kinematics.  The
'algorithm searches a decision space by "flying" its "probes" through the space by analogy to masses
'moving through physical space under the influence of gravity.  Equations are developed for the probes'
'positions and accelerations using the gravitational metaphor.  Small objects in our Universe can become
'trapped in close orbits around highly gravitating masses.  In "CFO space" probes are attracted to "masses"
'created by a user-defined function of the value of an objective function to be maximized.  CFO may be
'thought of in terms of a vector "force field" or, loosely, as a "generalized gradient" methodology
'because the force of gravity can be computed as the gradient of a scalar potential.  The CFO algorithm
'is simple and easily implemented in a compact computer program.  Its effectiveness is demonstrated by
'running CFO against several widely used benchmark functions.  The algorithm exhibits very good performance,
'suggesting that it merits further study.

'4. "Central Force Optimization: A New Gradient-Like Optimization Metaheuristic," OPSEARCH, Journal
'    of the Operational Research Society of India, vol. 46, no. 1, 25-51, 2009, Springer India.
'Abstract:   This paper introduces Central Force Optimization as a new, Nature-inspired metaheuristic
'for multidimensional search and optimization based on the metaphor of gravitational kinematics.  CFO is a
'"gradient-like" deterministic algorithm that explores a decision space by "flying" a group of "probes"
'whose trajectories are governed by equations analogous to the equations of gravitational motion in the
'physical Universe.  This paper suggests the possibility of creating a new "hyperspace directional derivative"
'using the Unit Step function t?o to create positive-definite "masses" in "CFO space."  A simple CFO implementation
'is tested against several recognized benchmark functions with excellent results, suggesting that CFO merits
'further investigation.

'5. "Synthesis of Antenna Arrays Using Central Force Optimization," Mosharaka International Conference
'    on Communications, Computers and Applications, MIC-CPE 2009.
'Abstract: Central force optimization (CFO) technique is a new deterministic multi-dimensional
'search metaheuristic based on an analogy to classical mechanics in a gravitational
'field. CFO is a simple technique and is still in its infancy. To enhance its global search
'ability while keeping its simplicity, a new selection part is introduced in this paper. CFO
'is briefly presented and applied in the design of linear antenna array and the modified CFO
```

'is applied to the design of circular array. The results are compared with those obtained using
'other evolutionary optimization techniques.

'6. "Central Force Optimization and NEOs - First Cousins?" Journal of Multiple Valued Logic and Soft
'     Computing (to be published)
'Abstract: Central Force Optimization is a new deterministic multidimensional search and optimization algorithm
'based on the metaphor of gravitational kinematics.  This paper describes CFO and suggests some possible
'directions for its future development.  Because CFO is deterministic, it is more computationally efficient
'than stochastic algorithms and may lend itself well to "parameter tuning" implementations.  But, like
'all deterministic algorithms, CFO is prone to local trapping.  Oscillation in CFO's Davg curve appears to
'be a reliable harbinger of trapping.  And there seems to be a reasonable basis for believing that
'trapping can be handled deterministically using the theory of gravitationally trapped Near Earth Objects.
'Deterministic mitigation of local trapping would be a major step forward in optimization theory.  Finally,
'CFO may be thought of as a "gradient-like" algorithm utilizing the Unit Step function as a critical element,
'and it is suggested that a useful, new derivative-like mathematical construct might be defined based on the Unit
'Step.

'7. "Array Synthesis and Antenna Benchmark Performance Using Central Force Optimization," IET (UK)
'     (in review)
'Abstract: Central force optimization (CFO) is a new deterministic multi-dimensional search metaheuristic
'based on an analogy to classical particle kinematics in a gravitational field. CFO is a simple technique
'that is still in its infancy.  This paper evaluates CFO's performance and provides further examples of its
'effectiveness by applying it to a set of "real world" antenna benchmarks and to pattern synthesis for linear
'and circular array antennas.  A new selection scheme is introduced that enhances CFO's global search ability
'while maintaining its simplicity.  The improved CFO algorithm is applied to the design of a circular array
'with very good results.  CFO's performance on the antenna benchmarks and the synthesis problems is compared
'to that of other evolutionary optimization techniques.
'===================================================================================================

#COMPILE EXE

#DIM ALL

%USEMACROS = 1

#INCLUDE "Win32API.inc"

DEFEXT A-Z

'------ EQUATES -----

    %IDC_FRAME1      = 101
    %IDC_FRAME2      = 102

    %IDC_Function_Number1  = 121
    %IDC_Function_Number2  = 122
    %IDC_Function_Number3  = 123
    %IDC_Function_Number4  = 124
    %IDC_Function_Number5  = 125
    %IDC_Function_Number6  = 126
    %IDC_Function_Number7  = 127
    %IDC_Function_Number8  = 128
    %IDC_Function_Number9  = 129
    %IDC_Function_Number10 = 130
    %IDC_Function_Number11 = 131
    %IDC_Function_Number12 = 132
    %IDC_Function_Number13 = 133
    %IDC_Function_Number14 = 134
    %IDC_Function_Number15 = 135
    %IDC_Function_Number16 = 136
    %IDC_Function_Number17 = 137
    %IDC_Function_Number18 = 138
    %IDC_Function_Number19 = 139
    %IDC_Function_Number20 = 140
    %IDC_Function_Number21 = 141
    %IDC_Function_Number22 = 142
    %IDC_Function_Number24 = 144
    %IDC_Function_Number25 = 145
    %IDC_Function_Number26 = 146
    %IDC_Function_Number27 = 147
    %IDC_Function_Number28 = 148
    %IDC_Function_Number29 = 149
    %IDC_Function_Number30 = 150
    %IDC_Function_Number31 = 151
    %IDC_Function_Number32 = 152
    %IDC_Function_Number33 = 153
    %IDC_Function_Number34 = 154
    %IDC_Function_Number35 = 155
    %IDC_Function_Number36 = 156
    %IDC_Function_Number37 = 157
    %IDC_Function_Number38 = 158
    %IDC_Function_Number39 = 159
    %IDC_Function_Number40 = 160
    %IDC_Function_Number41 = 161
    %IDC_Function_Number43 = 163
    %IDC_Function_Number44 = 164
    %IDC_Function_Number45 = 165
    %IDC_Function_Number46 = 166
    %IDC_Function_Number47 = 167
    %IDC_Function_Number48 = 168
    %IDC_Function_Number49 = 169
    %IDC_Function_Number50 = 170

'------------------------------ GLOBAL CONSTANTS & SYMBOLS --------------------------

GLOBAL Aij() AS EXT 'array for Shekel's Foxholes function

GLOBAL EulerConst, Pi, Pi2, Pi4, TwoPi, FourPi, e, Root2 AS EXT 'mathematical constants

GLOBAL Alphabet$, Digits$, RunID$  'upper/lower case alphabet, digits 0-9 & Run ID

GLOBAL Quote$, SpecialCharacters$   'quotation mark & special symbols

GLOBAL Mu0, Eps0, c, eta0 AS EXT   '&M constants

GLOBAL Rad2Deg, Deg2Rad, Feet2Meters, Meters2Feet, Inches2Meters, Meters2Inches AS EXT 'conversion factors

GLOBAL Miles2Meters, Meters2Miles, NautMi2Meters, Meters2NautMi AS EXT              'conversion factors

GLOBAL ScreenWidth&, ScreenHeight&  'screen width & height

GLOBAL xOffset&, yOffset&         'offsets for probe plot windows

```
GLOBAL FunctionNumber%

GLOBAL AddNoiseToPBM$

'----------------------------- TEST FUNCTION DECLARATIONS --------------------------------
DECLARE FUNCTION F1(R(),Nd%,p%,j&)            'F1 (n-D)

DECLARE FUNCTION F2(R(),Nd%,p%,j&)            'F2(n-D)

DECLARE FUNCTION F3(R(),Nd%,p%,j&)            'F3 (n-D)

DECLARE FUNCTION F4(R(),Nd%,p%,j&)            'F4 (n-D)

DECLARE FUNCTION F5(R(),Nd%,p%,j&)            'F5 (n-D)

DECLARE FUNCTION F6(R(),Nd%,p%,j&)            'F6 (n-D)

DECLARE FUNCTION F7(R(),Nd%,p%,j&)            'F7 (n-D)

DECLARE FUNCTION F8(R(),Nd%,p%,j&)            'F8 (n-D)

DECLARE FUNCTION F9(R(),Nd%,p%,j&)            'F9 (n-D)

DECLARE FUNCTION F10(R(),Nd%,p%,j&)           'F10 (n-D)

DECLARE FUNCTION F11(R(),Nd%,p%,j&)           'F11 (n-D)

DECLARE FUNCTION F12(R(),Nd%,p%,j&)           'F12 (n-D)

DECLARE FUNCTION F13(R(),Nd%,p%,j&)           'F13 (n-D)

DECLARE FUNCTION F14(R(),Nd%,p%,j&)           'F14 (n-D)

DECLARE FUNCTION F15(R(),Nd%,p%,j&)           'F15 (n-D)

DECLARE FUNCTION F16(R(),Nd%,p%,j&)           'F16 (n-D)

DECLARE FUNCTION F17(R(),Nd%,p%,j&)           'F17 (n-D)

DECLARE FUNCTION F18(R(),Nd%,p%,j&)           'F18 (n-D)

DECLARE FUNCTION F19(R(),Nd%,p%,j&)           'F19 (n-D)

DECLARE FUNCTION F20(R(),Nd%,p%,j&)           'F20 (n-D)

DECLARE FUNCTION F21(R(),Nd%,p%,j&)           'F21 (n-D)

DECLARE FUNCTION F22(R(),Nd%,p%,j&)           'F22 (n-D)

DECLARE FUNCTION F23(R(),Nd%,p%,j&)           'F23 (n-D)

DECLARE FUNCTION F24(R(),Nd%,p%,j&)           'F24 (n-D)

DECLARE FUNCTION F25(R(),Nd%,p%,j&)           'F25 (n-D)

DECLARE FUNCTION F26(R(),Nd%,p%,j&)           'F26 (n-D)

DECLARE FUNCTION F27(R(),Nd%,p%,j&)           'F27 (n-D)

DECLARE FUNCTION ParrottF4(R(),Nd%,p%,j&)     'Parrott F4 (1-D)

DECLARE FUNCTION SGO(R(),Nd%,p%,j&)           'SGO Function (2-D)

DECLARE FUNCTION GoldsteinPrice(R(),Nd%,p%,j&) 'Goldstein-Price Function (2-D)

DECLARE FUNCTION StepFunction(R(),Nd%,p%,j&)  'Step Function (n-D)

DECLARE FUNCTION Schwefel226(R(),Nd%,p%,j&)   'Schwefel Prob. 2.26 (n-D)

DECLARE FUNCTION Colville(R(),Nd%,p%,j&)      'Colville Function (4-D)

DECLARE FUNCTION Griewank(R(),Nd%,p%,j&)      'Griewank (n-D)

DECLARE FUNCTION Himmelblau(R(),Nd%,p%,j&)    'Himmelblau (2-D)

DECLARE FUNCTION PBM_1(R(),Nd%,p%,j&)         'PBM Benchmark #1

DECLARE FUNCTION PBM_2(R(),Nd%,p%,j&)         'PBM Benchmark #2

DECLARE FUNCTION PBM_3(R(),Nd%,p%,j&)         'PBM Benchmark #3

DECLARE FUNCTION PBM_4(R(),Nd%,p%,j&)         'PBM Benchmark #4

DECLARE FUNCTION PBM_5(R(),Nd%,p%,j&)         'PBM Benchmark #5

'---------------------------------- SUB DECLARATIONS ------------------------------------
DECLARE SUB GetTestFunctionNumber(FunctionName$)

DECLARE SUB FillArrayAij

DECLARE SUB ResetDecisionSpaceBoundaries(Nd%,XiMin(),XiMax(),StartingXiMin(),StartingXiMax())

DECLARE SUB Plot3DbestProbeTrajectories(NumTrajectories%,M(),R(),XiMin(),XiMax(),Np%,Nd%,LastStep&,FunctionName$)

DECLARE SUB Plot2DbestProbeTrajectories(NumTrajectories%,M(),R(),XiMin(),XiMax(),Np%,Nd%,LastStep&,FunctionName$)

DECLARE SUB Plot2DindividualProbeTrajectories(NumTrajectories%,M(),R(),XiMin(),XiMax(),Np%,Nd%,LastStep&,FunctionName$)

DECLARE SUB SelectTestFunction(FunctionName$)

DECLARE SUB Show2Dprobes(R(),Np%,Nt&,j&,XiMin(),XiMax(),Frep,BestFitness,BestProbeNumber,BestTimeStep&,FunctionName$,RepositionFactor$,Gamma)

DECLARE SUB StatusWindow(FunctionName$,StatusWindowHandle???)

DECLARE SUB
DisplayRunParameters(FunctionName$,Nd%,Np%,Nt&,G,DeltaT,Alpha,Beta,Frep,R(),A(),M(),PlaceInitialProbes$,InitialAcceleration$,RepositionFactor$,RunCF
OS,ShrinkDS$,CheckForEarlyTermination$)

DECLARE SUB GetBestFitness(M(),Np%,StepNumber&,BestFitness,BestProbeNumber,BestTimeStep&)
```

```
DECLARE SUB
Tabulate1DprobeCoordinates(Max1DprobesPlotted%,Nd%,Np%,LastStep&,G,DeltaT,Alpha,Beta,Frep,R(),M(),PlaceInitialProbes$,InitialAcceleration$,Repositio
nFactor$,FunctionName$,Gamma)

DECLARE SUB
GetPlotAnnotation(PlotAnnotation$,Nd%,Np%,Nt&,G,DeltaT,Alpha,Beta,Frep,M(),PlaceInitialProbes$,InitialAcceleration$,RepositionFactor$,FunctionName$,
Gamma)

DECLARE SUB
ChangeRunParameters(NumProbesPerDimension%,Nd%,Nd%,Nt&,G,Alpha,Beta,DeltaT,Frep,PlaceInitialProbes$,InitialAcceleration$,RepositionFactor$,FunctionN
ame$)

DECLARE SUB CLEANUP

DECLARE SUB DisplayBestFitness(Np%,Nd%,LastStep&,M(),R(),BestFitnessProbeNumber%,BestFitnessTimeStep&,FunctionName$)

DECLARE SUB
Plot1DprobePositions(Max1DprobesPlotted%,Nd%,Np%,LastStep&,G,DeltaT,Alpha,Beta,Frep,M(),PlaceInitialProbes$,InitialAcceleration$,RepositionFacto
r$,FunctionName$,Gamma)

DECLARE SUB DisplayMmatrix(Np%,Nt&,M())

DECLARE SUB DisplayMmatrixThisTimeStep(Np%,j&,M())

DECLARE SUB DisplayAmatrix(Np%,Nd%,Nt&,A())

DECLARE SUB DisplayAmatrixThisTimeStep(Np%,Nd%,j&,A())

DECLARE SUB DisplayRmatrix(Np%,Nd%,Nt&,R())

DECLARE SUB DisplayRmatrixThisTimeStep(Np%,Nd%,j&,R())

DECLARE SUB DisplayXiMinMax(Nd%,XiMin(),XiMax())

DECLARE SUB DisplayRunParameters2(FunctionName$,Nd%,Np%,Nt&,G,DeltaT,Alpha,Beta,Frep,PlaceInitialProbes$,InitialAcceleration$,RepositionFactor$)

DECLARE SUB
PlotBestProbeVsTimeStep(Nd%,Np%,LastStep&,G,DeltaT,Alpha,Beta,Frep,M(),PlaceInitialProbes$,InitialAcceleration$,RepositionFactor$,FunctionName$,Gamm
a)

DECLARE SUB
PlotBestFitnessEvolution(Nd%,Np%,LastStep&,G,DeltaT,Alpha,Beta,Frep,M(),PlaceInitialProbes$,InitialAcceleration$,RepositionFactor$,FunctionName$,Gam
ma)

DECLARE SUB
PlotAverageDistance(Nd%,Np%,LastStep&,G,DeltaT,Alpha,Beta,Frep,M(),PlaceInitialProbes$,InitialAcceleration$,RepositionFactor$,FunctionName$,R(),Diag
Length,Gamma)

DECLARE SUB Plot2Dfunction(FunctionName$,XiMin(),XiMax(),R())

DECLARE SUB Plot1Dfunction(FunctionName$,XiMin(),XiMax(),R())

DECLARE SUB
GetFunctionRunParameters(FunctionName$,Nd%,Np%,Nt&,G,DeltaT,Alpha,Beta,Frep,R(),A(),M(),XiMin(),XiMax(),StartingXiMin(),StartingXiMax(),_
                         DiagLength,PlaceInitialProbes$,InitialAcceleration$,RepositionFactor$)

DECLARE SUB InitialProbeDistribution(Np%,Nd%,Nt&,XiMin(),XiMax(),R(),PlaceInitialProbes$,Gamma)

DECLARE SUB InitialProbeAccelerations(Np%,Nd%,A(),InitialAccelerations$,MaxInitialRandomAcceleration,MaxInitialFixedAcceleration)

DECLARE SUB RetrieveErrantProbes(Np%,Nd%,j&,XiMin(),XiMax(),R(),M(),RepositionFactor$,Frep)

DECLARE SUB CFO(Nd%,Np%,Nt&,G,DeltaT,Alpha,Beta,Frep,R(),A(),M(),XiMin(),XiMax(),DiagLength,PlaceInitialProbes$,InitialAcceleration$,_
RepositionFactor$,FunctionName$,LastStep&,CheckForEarlyTermination$,BestFitnessThisRun,RunNumber%,NumRuns$,Gamma,BestOverallFitness,ShrinkDS$,MaxPro
besPerAxis%,MinProbesPerAxis%)

DECLARE SUB ThreeDplot(PlotFileName$,PlotTitle$,Annotation$,xCoord$,yCoord$,zCoord$,_
                       XaxisLabel$,YaxisLabel$,ZaxisLabel$,zMin$,zMax$,GnuPlotEXE$,A$)

DECLARE SUB ThreeDplot2(PlotFileName$,PlotTitle$,Annotation$,xCoord$,yCoord$,zCoord$,XaxisLabel$,_
                        YaxisLabel$,ZaxisLabel$,zMin$,zMax$,GnuPlotEXE$,A$,xStart$,xStop$,yStart$,yStop$)

DECLARE SUB TwoDplot(PlotFileName$,PlotTitle$,xCoord$,yCoord$,XaxisLabel$,YaxisLabel$,_
                     LogXaxis$,LogYaxis$,xMin$,xMax$,yMin$,yMax$,xTics$,yTics$,GnuPlotEXE$,LineType$,Annotation$)

DECLARE SUB TwoDplot2curves(PlotFileName1$,PlotFileName2$,PlotTitle$,Annotation$,xCoord$,yCoord$,XaxisLabel$,YaxisLabel$,_
                            LogXaxis$,LogYaxis$,xMin$,xMax$,yMin$,yMax$,xTics$,yTics$,GnuPlotEXE$,LineSize)

DECLARE SUB TwoDplot3curves(NumCurves%,PlotFileName1$,PlotFileName2$,PlotFileName3$,PlotTitle$,Annotation$,xCoord$,yCoord$,XaxisLabel$,YaxisLabel$,_
                            LogXaxis$,LogYaxis$,xMin$,xMax$,yMin$,yMax$,xTics$,yTics$,GnuPlotEXE$)

DECLARE SUB CreateGNUplotINIfile(PlotWindowULC_X%,PlotWindowULC_Y%,PlotWindowWidth%,PlotWindowHeight%)

DECLARE SUB Delay(NumSecs)

DECLARE SUB MathematicalConstants

DECLARE SUB AlphabetAndDigits

DECLARE SUB SpecialSymbols

DECLARE SUB EMconstants

DECLARE SUB ConversionFactors

DECLARE SUB ShowConstants

'------ FUNCTION DECLARATIONS -------

DECLARE CALLBACK FUNCTION DlgProc

DECLARE FUNCTION HasFITNESSsaturated$(Nsteps&,j&,Np%,Nd%,M(),R(),DiagLength)

DECLARE FUNCTION HasDAVGsaturated$(Nsteps&,j&,Np%,Nd%,M(),R(),DiagLength)

DECLARE FUNCTION OscillationInDavg$(j&,Np%,Nd%,M(),R(),DiagLength)

DECLARE FUNCTION DavgThisStep(j&,Np%,Nd%,M(),R(),DiagLength)

DECLARE FUNCTION NoSpaces$(X,NumDigits%)

DECLARE FUNCTION FormatFP$(X,Ndigits%)
```

```
DECLARE FUNCTION FormatInteger$(M%)

DECLARE FUNCTION TerminateNowForSaturation$(j&,Nd%,Np%,Nt&,G,DeltaT,Alpha,Beta,R(),A(),M())

DECLARE FUNCTION MagVector(V(),N&)

DECLARE FUNCTION UniformDeviate(u&&)

DECLARE FUNCTION RandomNum(a,b)

DECLARE FUNCTION GaussianDeviate(Mu,Sigma)

DECLARE FUNCTION UnitStep(X)

DECLARE FUNCTION Fibonacci&&(N%)

DECLARE FUNCTION ObjectiveFunction(R(),Nd%,p%,j&,FunctionName$)

DECLARE FUNCTION UnitStep(X)

'=================================================================================================
'----- MAIN PROGRAM ------
FUNCTION PBMAIN () AS LONG
'    ------ CFO Parameters -----
    LOCAL Nd%, Np%, Nt&
    LOCAL G, DeltaT, Alpha, Beta, Frep AS EXT
    LOCAL PlaceInitialProbes$, InitialAcceleration$, RepositionFactor$
    LOCAL R(), A(), M() AS EXT    'position, acceleration & fitness matrices
    LOCAL XiMin(), XiMax(), StartingXiMin(), StartingXiMax() AS EXT 'decision space boundaries
    LOCAL FunctionName$        'name of objective function
    LOCAL DiagLength AS EXT
'    ----------- Miscellaneous Setup Parameters --------------
    LOCAL StartingG, StartingDeltaT, StartingAlpha, StartingBeta, StartingFrep AS EXT
    LOCAL N%, i%, YN&
    LOCAL A$
    LOCAL NumRuns%, RunNumber%, BestRunNumber%, TotalFunctionEvaluations&&, AbsoluteRunNumber%
    LOCAL Gamma, StartingGamma, StoppingGamma, BestFitnessThisRun, BestOverallFitness AS EXT
    LOCAL NumTrajectories%
    LOCAL Max%1DprobesPlotted%
    LOCAL BestFitnessProbeNumber%, BestFitnessTimeStep&, BestNumProbes%, MaxProbesPerAxis%, MinProbesPerAxis%, NumProbesPerAxis%
    LOCAL RunCFO$
    LOCAL StatusWindowHandle???
    LOCAL LastStep&
    LOCAL CheckForEarlyTermination$ 'early termination checking? (YES/NO)
    LOCAL ShrinkDS$ 'adaptively shrink DS? (YES/NO)
'    ------------------- Global Constants --------------------
    REDIM Aij(1 TO 2, 1 TO 25) '(GLOBAL array for Shekel's Foxholes function)
    CALL FillArrayAij
    CALL MathematicalConstants 'NOTE: Calling order is important!!
    CALL AlphabetAndDigits
    CALL SpecialSymbols
    CALL EMconstants
    CALL ConversionFactors        ': CALL ShowConstants 'to verify constants have been set
    xOffset& = 20 : yOffset& = 20 'offsets for successive probe position plots
'    ------------------------- General Setup -----------------------------
    RANDOMIZE TIMER  'seed random number generator with program start time
    DESKTOP GET SIZE TO ScreenWidth&, ScreenHeight&  'get screen size (global variables)
    IF DIR$("wgnuplot.exe") = "" THEN
        MSGBOX("WARNING!  'wgnuplot.exe' not found.  Run terminated.") : EXIT FUNCTION
    END IF
'    ------------------- NEC Files Required for PBM Antenna Benchmarks -------------------
    IF DIR$("n41_2k1.exe") = "" THEN
            MSGBOX("WARNING!  'n41_2k1.exe' not found.  Run terminated.") : EXIT FUNCTION
    END IF
    N% = FREEFILE : OPEN "FILE_YES.DAT" FOR OUTPUT AS #N% : PRINT #N%, "YES"      : CLOSE #N%
    N% = FREEFILE : OPEN "FILE_MSG.DAT" FOR OUTPUT AS #N% : PRINT #N%, "NO"       : CLOSE #N%
    N% = FREEFILE : OPEN "NHSCALE.DAT"  FOR OUTPUT AS #N% : PRINT #N%, "0.00001" : CLOSE #N%
```

```
    IF DIR$("ProbeCoordinates.DAT") <> "" THEN KILL "ProbeCoordinates.DAT" 'get rid of this file to avoid confusion

    IF DIR$("Frep.DAT") <> "" THEN KILL "Frep.DAT" 'ditto

'  ----------------------------------------------------------------- CFO RUN PARAMETERS -----------------------------------------------------
------------------------------

    Max1DprobesPlotted% = 15 : Max1DprobesPlotted% = MIN(15,Max1DprobesPlotted%) '15 MAX!

    ' CALL SelectTestFunction(FunctionName$)

    CALL GetTestFunctionNumber(FunctionName$) ' exit function 'DEBUG

    'MSGBOX("Test Function is #"+STR$(FunctionNumber%)+", "+FunctionName$)

    CALL GetFunctionRunParameters(FunctionName$,Nd%,Np%,Nt&,G,DeltaT,Alpha,Beta,Frep,R(),A(),M(),XiMin(),XiMax(),StartingXiMin,StartingXiMax(),_
                                DiagLength,PlaceInitialProbes$,InitialAcceleration$,RepositionFactor$)
' !! NOTE !!   IN THIS VERSION Np% IS IGNORED BECAUSE THE PROGRAM LOOPS ON Np/Nd (= 2-8 by 2) AND COMPUTES Np FROM Np = (Np/Nd)*Nd.
'              BUT THE PROGRAM STILL RETURNS Np FROM THIS SUBROUTINE (TOO MUCH OF A BOTHER TO TAKE IT OUT ...).

    StartingG = G : StartingDeltaT = DeltaT : StartingAlpha = Alpha : StartingBeta = Beta : StartingFrep = Frep

'   IMPORTANT NOTE: Arrays XiMin() & XiMax() are dimensioned (1 TO Nd%) in SUB GetFunctionRunParameters

    REDIM R(1 TO Np%, 1 TO Nd%, 0 TO Nt&), A(1 TO Np%, 1 TO Nd%, 0 TO Nt&), M(1 TO Np%, 0 TO Nt&) 'position, acceleration & fitness matrices

'  ---------------------- PLOT 1D and 2D FUNCTIONS ON-SCREEN FOR VISUALIZATION ----------------------

    IF Nd% = 2 AND INSTR(FunctionName$,"PBM_") > 0 THEN MSGBOX("Begin computing plot of function "+FunctionName$+"?  May take a while - be
patient...")

    SELECT CASE Nd%
        CASE 1 : CALL Plot1Dfunction(FunctionName$,XiMin(),XiMax(),R()) : REDIM R(1 TO Np%, 1 TO Nd%, 0 TO Nt&) 'erases coordinate data in R()used
to plot function
        CASE 2 : CALL Plot2Dfunction(FunctionName$,XiMin(),XiMax(),R()) : REDIM R(1 TO Np%, 1 TO Nd%, 0 TO Nt&) 'ditto
    END SELECT
'  ----------------------------------------------------------------- DISPLAY PARAMETERS & RUN CFO -----------------------------------------------
------------------------

    CALL
DisplayRunParameters(FunctionName$,Nd%,Np%,Nt&,G,DeltaT,Alpha,Beta,Frep,R(),A(),M(),PlaceInitialProbes$,InitialAcceleration$,RepositionFactor$,RunCF
O$,ShrinkDS$,CheckForEarlyTermination$)

    IF RunCFO$ = "YES" THEN 'run CFO

        CALL StatusWindow(FunctionName$,StatusWindowHandle???)

        StartingGamma = 0## : StoppingGamma = 1##

        NumRuns% = 11 'For Initial Probe Dist'n GAMMA LOOP

        MinProbesPerAxis% = 4 : MaxProbesPerAxis% = 14 'For Initial Probe Dist'n Np/Nd LOOP

'  --------------- Output Data File & Header ---------------

        N% = FREEFILE : OPEN FunctionName$+".DAT" FOR OUTPUT AS #N%

        PRINT #N%, "Run ID: "+RunID$+CHR$(13)+CHR$(13)+"FUNCTION: "+UCASE$(FunctionName$)+CHR$(13)+CHR$(13)

        PRINT #N%, _
        "Run #      Gamma      Nt    Nd    Np    G    DelT    Alpha   Beta   #Steps   Neval     Frep         Fitness      Initial Probes" +
CHR$(13) "_
        "-----      -------    ----  ----  ----  -----  ----  -----   ------  ------  -------  --------   --------------   --------------"
        A$ = _
        "####      ###.###   ####  ####  ####  ###.#  ##.#  ##.##  ##.##  #####   #######   #.#####\\ ########.########   \           \"
'                                                                                              UNIFORM ON-AXIS                         \"
        PRINT #N%,
USING(A$,0,StartingGamma,Nt&,Nd%,Np%,StartingG,StartingDeltaT,StartingAlpha,StartingBeta,0,0,StartingFrep,LEFT$(RepositionFactor$,1),-
9999,PlaceInitialProbes$) 'header

        PRINT #N%,_
        "-----------------------------------------------------------------------------------------------------------------------------"

'  ---------------------- Np/Nd LOOP ----------------------

        BestOverallFitness = -1E4200 'very large negative number...

        TotalFunctionEvaluations&& = 0

        FOR NumProbesPerAxis% = MinProbesPerAxis% TO MaxProbesPerAxis% STEP 2 '4/6/.../MaxProbesPerAxis% probes per axis

        Np% = NumProbesPerAxis%*Nd% 'sets Np% regardless of what Sub GetFunctionRunParameters() returns.

        REDIM R(1 TO Np%, 1 TO Nd%, 0 TO Nt&), A(1 TO Np%, 1 TO Nd%, 0 TO Nt&), M(1 TO Np%, 0 TO Nt&) 'must redim these matrices because Np has
changed

'  ---------------------- GAMMA LOOP ----------------------

        FOR RunNumber% = 1 TO NumRuns%

            Gamma = StartingGamma + (RunNumber%-1)*(StoppingGamma-StartingGamma)/(NumRuns%-1)

            G = StartingG : DeltaT = StartingDeltaT : Alpha = StartingAlpha : Beta = StartingBeta : Frep = StartingFrep

            CALL ResetDecisionSpaceBoundaries(Nd%,XiMin(),XiMax(),StartingXiMin(),StartingXiMax())

            CALL CFO(Nd%,Np%,Nt&,G,DeltaT,Alpha,Beta,Frep,R(),A(),M(),XiMin(),XiMax(),DiagLength,PlaceInitialProbes$,InitialAcceleration$,_
RepositionFactor$,FunctionName$,LastStep&,CheckForEarlyTermination$,BestFitnessThisRun,RunNumber%,NumRuns%,Gamma,BestOverallFitness,ShrinkDS$,MaxPro
besPerAxis%,MinProbesPerAxis%)

            IF BestFitnessThisRun > BestOverallFitness THEN

                BestOverallFitness = BestFitnessThisRun : BestRunNumber% = RunNumber% : BestNumProbes% = Np%

'MSGBOX("Best Fitness="+STR$(BestOverallFitness)+"  Best Run#="+STR$(BestRunNumber%))

            END IF

            TotalFunctionEvaluations&& = TotalFunctionEvaluations&& + (LastStep&+1)*Np%
```

```
          AbsoluteRunNumber% = RunNumber% + ((Np%\Nd%)\2-MinProbesPerAxis%\2)*NumRuns%

'msgbox("Run#="+STR$(RunNumber%)+"      Abs Run #="+STR$(AbsoluteRunNumber%))

          PRINT #N%, USING$(A$,AbsoluteRunNumber%,Gamma,Nt&,Nd%,G,DeltaT,Alpha,Beta,LastStep&,(Laststep&+1)*Np%,Frep,_
                          LEFT$(RepositionFactor$,1),BestFitnessThisRun,PlaceInitialProbes$)

          NEXT RunNumber% 'GAMMA LOOP

          NEXT NumProbesPerAxis% 'Np/Nd LOOP

          PRINT #N%, CHR$(13)+USING$("                                                  Total Function Evaluations:
##########",TotalFunctionEvaluations&)+CHR$(13)
'          ------------------------------------------- Re-Run Best Run for Final Results ---------------------------------

          Gamma = StartingGamma + (BestRunNumber%-1)*(StoppingGamma-StartingGamma)/(NumRuns%-1)

MSGBOX("  Best Run#="+STR$(BestRunNumber%)+"   Gamma="+STR$(Gamma))

          G = StartingG : DeltaT = StartingDeltaT : Alpha = StartingAlpha : Beta = StartingBeta : Frep = StartingFrep : Np% = BestNumProbes%

          REDIM R(1 TO Np%, 1 TO Nd%, 0 TO Nt&), A(1 TO Np%, 1 TO Nd%, 0 TO Nt&), M(1 TO Np%, 0 TO Nt&) 'must redim these matrices because Np has
changed

          CALL ResetDecisionSpaceBoundaries(Nd%,XiMin(),XiMax(),StartingXiMin(),StartingXiMax())

          CALL CFO(Nd%,Np%,Nt&,G,DeltaT,Alpha,Beta,Frep,R(),A(),M(),XiMin(),XiMax(),DiagLength,PlaceInitialProbes$,InitialAcceleration$,_

RepositionFactor$,FunctionName$,LastStep&,CheckForEarlyTermination$,BestFitnessThisRun,BestRunNumber%,NumRuns%,Gamma,BestOverallFitness,ShrinkDS$,Ma
xProbesPerAxis%,MinProbesPerAxis%)

          CALL ResetDecisionSpaceBoundaries(Nd%,XiMin(),XiMax(),StartingXiMin(),StartingXiMax())

          PRINT #N%,"--------------------------------------------------------------------------------------------------------------------
-------"

          AbsoluteRunNumber% = BestRunNumber% + ((Np%\Nd%)\2-MinProbesPerAxis%\2)*NumRuns%

          PRINT #N%, USING$(A$,AbsoluteRunNumber%,Gamma,Nt&,Nd%,BestNumProbes%,G,DeltaT,Alpha,Beta,LastStep&,(Laststep&+1)*Np%,Frep,_
                          LEFT$(RepositionFactor$,1),BestFitnessThisRun,PlaceInitialProbes$)
          CLOSE #N%

     ELSE

          GOTO ExitPBMAIN

     END IF

'     ----------------------------- Display Best Fitness ----------------------------------------

     CALL DisplayBestFitness(Np%,Nd%,LastStep&,M(),R(),BestFitnessProbeNumber%,BestFitnessTimeStep&,FunctionName$)
'     ------------------------------------------ PLOT EVOLUTION OF BEST FITNESS, AVG DISTANCE & BEST PROBE # ---------------------------------
---------------------
     CALL
PlotBestFitnessEvolution(Nd%,Np%,LastStep&,G,DeltaT,Alpha,Beta,Frep,M(),PlaceInitialProbes$,InitialAcceleration$,RepositionFactor$,FunctionName$,Gam
ma)

     CALL
PlotAverageDistance(Nd%,Np%,LastStep&,G,DeltaT,Alpha,Beta,Frep,M(),PlaceInitialProbes$,InitialAcceleration$,RepositionFactor$,FunctionName$,R(),Diag
Length,Gamma)

     CALL
PlotBestProbeVsTimeStep(Nd%,Np%,LastStep&,G,DeltaT,Alpha,Beta,Frep,M(),PlaceInitialProbes$,InitialAcceleration$,RepositionFactor$,FunctionName$,Gamm
a)
'     ------------------------------------------------ PLOT TRAJECTORIES OF BEST PROBES FOR 2/3-D FUNCTIONS -----------------------------------
-------

     IF Nd% = 2 THEN

          NumTrajectories% = 10 : CALL Plot2DbestProbeTrajectories(NumTrajectories%,M(),R(),XiMin(),XiMax(),Np%,Nd%,LastStep&,FunctionName$)

          NumTrajectories% = 16 : CALL Plot2DindividualProbeTrajectories(NumTrajectories%,M(),R(),XiMin(),XiMax(),Np%,Nd%,LastStep&,FunctionName$)

     END IF

     IF Nd% = 3 THEN

          NumTrajectories% = 4 : CALL Plot3DbestProbeTrajectories(NumTrajectories%,M(),R(),XiMin(),XiMax(),Np%,Nd%,LastStep&,FunctionName$)

     END IF
'     ---------- For 1-D Objective Functions, Tabulate Probe Coordinates & if Np% =< Max1DprobesPlotted% Plot Evolution of Probe Positions ----------
--

     IF Nd% = 1 THEN

          CALL
Tabulate1DprobeCoordinates(Max1DprobesPlotted%,Nd%,Np%,LastStep&,G,DeltaT,Alpha,Beta,Frep,R(),M(),PlaceInitialProbes$,InitialAcceleration$,Repositio
nFactor$,FunctionName$,Gamma)

          IF Np% =< Max1DprobesPlotted% THEN CALL
Plot1DprobePositions(Max1DprobesPlotted%,Nd%,Np%,LastStep&,G,DeltaT,Alpha,Beta,Frep,R(),M(),PlaceInitialProbes$,InitialAcceleration$,RepositionFacto
r$,FunctionName$,Gamma)

          CALL CLEANUP 'delete probe coordinate plot files, if any

     END IF

ExitPBMAIN:

END FUNCTION 'PBMAIN()
'============================================================================== CFO SUBROUTINE
==================================================================================

SUB CFO(Nd%,Np%,Nt&,G,DeltaT,Alpha,Beta,Frep,R(),A(),M(),XiMin(),XiMax(),DiagLength,PlaceInitialProbes$,InitialAcceleration$,_

RepositionFactor$,FunctionName$,LastStep&,CheckForEarlyTermination$,BestFitnessThisRun,RunNumber%,NumRuns%,Gamma,BestOverallFitness,ShrinkDS$,MaxPro
besPerAxis%,MinProbesPerAxis%)
```

```
LOCAL p%, i%, j%  'Standard Indices: Probe #, Coordinate #, Time Step #

LOCAL k%, L%     'Dummy summation indices

LOCAL SumSQ, Denom, Numerator, MaxInitialRandomAcceleration, MaxInitialFixedAcceleration AS EXT

LOCAL StepNumber&, BestProbeNumber&, BestTimeStep&

LOCAL BestFitness AS EXT

LOCAL DavgOscillation$, DavgSaturation$, FitnessSaturation$

LOCAL Nsteps&

'STEP (A1) ------------ Compute Initial Probe Distribution (Step 0)----------------

    CALL InitialProbeDistribution(Np%,Nd%,Nt&,XiMin(),XiMax(),R(),PlaceInitialProbes$,Gamma)

    IF Nd% = 2 THEN 'plot 2-D initial probe distribution at Step 0

        CALL CreateGNUplotINIfile(0.1##*ScreenWidth&,0.25##*ScreenHeight&,0.6##*ScreenWidth&,0.6##*ScreenHeight&)

        CALL Show2Dprobes(R(),Np%,Nt&,0,XiMin(),XiMax(),Frep,0##,1,0,FunctionName$,RepositionFactor$,Gamma) 'show initial probes for 2-D functions

'       msgbox("Press ENTER to continue.")

    END IF

'STEP (A2) ------------ Compute Initial Fitness Matrix (Step 0) ------------

    FOR p% = 1 TO Np% : M(p%,0) = ObjectiveFunction(R(),Nd%,p%,0,FunctionName$) : NEXT p%

'STEP (A3) ------------ Compute Initial Probe Accelerations (Step 0)--------------

    MaxInitialRandomAcceleration =  2## 'maximum value for random initial acceleration: Random[0-MAX]

    MaxInitialFixedAcceleration = 10## 'maximum value for fixed initial acceleration: [0.001-MAX]

    CALL InitialProbeAccelerations(Np%,Nd%,A(),InitialAcceleration$,MaxInitialRandomAcceleration,MaxInitialFixedAcceleration)

'  =================================== LOOP ON TIME STEPS STARTING AT STEP #1 ========================================

    LastStep& = Nt& 'unless run is terminated earlier

    BestFitnessThisRun = M(1,0)

    FOR j& = 1 TO Nt&

'STEP (B) ----------- Compute Probe Position Vectors for this Time Step --------

        FOR p% = 1 TO Np% : FOR i% = 1 TO Nd% : R(p%,i%,j&) = R(p%,i%,j&-1) + 0.5##*A(p%,i%,j&-1)*DeltaT^2 : NEXT i% : NEXT p%

'STEP (C) ----------- Retrieve Errant Probes ---------------

        CALL RetrieveErrantProbes(Np%,Nd%,j&,XiMin(),XiMax(),R(),M(),RepositionFactor$,Frep)

'STEP (D) ----------- Compute Fitness Matrix for Current Probe Distribution ---------

        FOR p% = 1 TO Np% : M(p%,j&) = ObjectiveFunction(R(),Nd%,p%,j&,FunctionName$) : NEXT p%

'STEP (E) ----------- Compute Accelerations Based on Current Probe Distribution & Fitnesses ----------------

        FOR p% = 1 TO Np%

            FOR i% = 1 TO Nd%

                A(p%,i%,j&) = 0

                FOR k% = 1 TO Np%

                    IF k% <> p% THEN

                        SumSQ = 0##

                        FOR L% = 1 TO Nd%  : SumSQ = SumSQ + (R(k%,L%,j&)-R(p%,L%,j&))^2 : NEXT L% 'dummy index

                        Denom = SQR(SumSQ) : Numerator = UnitStep((M(k%,j&)-M(p%,j&)))*(M(k%,j&)-M(p%,j&))

                        A(p%,i%,j&) = A(p%,i%,j&) + G*(R(k%,i%,j&)-R(p%,i%,j&))*Numerator^Alpha/Denom^Beta

                    END IF

                NEXT k% 'dummy index

            NEXT i% 'coord (dimension) #

        NEXT p% 'probe #

'  --------- Get Best Fitness & Corresponding Probe # and Time Step ---------

    CALL GetBestFitness(M(),Np%,j&,BestFitness,BestProbeNumber&,BestTimeStep&)

    IF BestFitness >= BestFitnessThisRun THEN BestFitnessThisRun = BestFitness

'  --------------------------------------------- Display Best Fitness/Probe#/Time Step and Plot Probe Positions for 2-D Functions --------------------
----------------------

'    IF Nd% = 2 AND ((j& MOD MAX(Nt&\16,4) = 0 OR (j& =< 16 AND j& MOD 2 = 0) OR j& = Nt&)) THEN 'plot 2D probes at selected time steps

'        xOffset& = xOffset& - 20 : yOffset& = yOffset& + 10

'        CALL CreateGNUplotINIfile(0.55##*ScreenWidth&+xOffset&,0.01##*ScreenHeight&+yOffset&,0.6##*ScreenHeight&,0.6##*ScreenHeight&)

'        CALL Show2Dprobes(R(),Np%,Nt&,j&,XiMin(),XiMax(),Frep,BestFitness,BestProbeNumber&,BestTimeStep&,FunctionName$,RepositionFactor$,Gamma)
'plot positions of probes in 2-D

'    END IF

'  --------------------- Check for Davg OSC and Fitness/Davg SAT &  Display Run Status --------------------------

    DavgOscillation$  = OscillationInDavg$(j&,Np%,Nd%,M(),R(),DiagLength)            'check for oscillation in Davg

    Nsteps& = 50 '25 'for 100D tests 'used 50 for new paper '15 '# steps for averaging Davg & Best Fitness to test for saturation
```

```
    DavgSaturation$    = HasDAVGsaturated$(Nsteps&,j&,Np%,Nd%,M(),R(),DiagLength)    'check for saturation of Davg

    FitnessSaturation$ = HasFITNESSsaturated$(Nsteps&,j&,Np%,Nd%,M(),R(),DiagLength) 'check for saturation of best fitness

    GRAPHIC SET PIXEL (35,15) : GRAPHIC PRINT "Run #"    + REMOVE$(STR$(RunNumber%+((Np%\Nd%)\2-MinProbesPerAxis%\2)*NumRuns%),ANY$ " ")     + "/"
+REMOVE$(STR$(((MaxProbesPerAxis%-MinProbesPerAxis%)\2+1)*NumRuns%),ANY$ " ") +_
                                                         ", Np/Nd=" + REMOVE$(STR$(Np%\Nd%),ANY$ " ") +_
                                                         ", Step #" + REMOVE$(STR$(j&),ANY$ " ")          + "/" + REMOVE$(STR$(Nt&),ANY$ " ")     +_
                                                         ", Gamma=" + REMOVE$(STR$(ROUND(Gamma,4)),ANY$ " ") +_
                                                         ", Frep="  + REMOVE$(STR$(Frep,4),ANY$ " ")          + STRING$(10," ")

    GRAPHIC SET PIXEL (35,35) : GRAPHIC PRINT "Best Fitness This Run = " + REMOVE$(STR$(ROUND(BestFitness,8)),ANY$ " ") +_
                                                         " @ Probe #" + REMOVE$(STR$(BestProbeNumber%),ANY$ " ") + ", Step #" + REMOVE$(STR$(BestTimeStep&),ANY$
" ") + STRING$(15," ")

    IF RunNumber% = 1 AND Np%\Nd% = 2 THEN

        GRAPHIC SET PIXEL (35,55) : GRAPHIC PRINT "Best Fitness Overall = N/A @ Run #1" + STRING$(15," ")

    ELSE

        GRAPHIC SET PIXEL (35,55) : GRAPHIC PRINT "Best Fitness Overall = " + REMOVE$(STR$(ROUND(BestOverallFitness,8)),ANY$ " ") + STRING$(15," ")

    END IF

    GRAPHIC SET PIXEL (35,75) : GRAPHIC PRINT "Osc Davg? " +  DavgOscillation$ + ",  Sat Davg? " +  DavgSaturation$ +",  Sat Fitness? " +
FitnessSaturation$ + STRING$(10," "): GRAPHIC REDRAW

'  ----------------------------------- If Variable Frep, Adjust Value---------------------------------------

    IF RepositionFactor$ = "VARIABLE" THEN

        Frep = Frep + 0.05##

        IF Frep > 1## THEN Frep = 0.05## 'keep Frep in range [0.05,1]

    END IF

    IF RepositionFactor$ = "RANDOM" THEN Frep = RandomNum(0##,1##) 'set new Frep value randomly

'  ----------------------------- If DavgOSC => Reposition Probes ---------------------------------

'  ----------------------------- If DavgSauration => Reposition Probes ---------------------------------

'  IF DavgSaturation$ = "YES" and j& < Nt& THEN

'  IF DavgOscillation$ = "YES" THEN

'      FOR p% = 1 TO Np%

'          If p% = BestProbeNumber% then

'              FOR i% = 1 TO Nd%

'                  R(p%,i%,j&) = 0.5##*R(p%,i%,j&) 'adjust each coordinate by 10%

'                  R(p%,i%,j&) = R(p%,i%,j&)*(1##+GaussianDeviate(0##,0.1##)) 'adjust each coordinate randomly (mu,sigma)

'                  IF R(p%,i%,j&) < XiMin(i%) THEN R(p%,i%,j&) = XiMin(i%)
'                  IF R(p%,i%,j&) > XiMax(i%) THEN R(p%,i%,j&) = XiMax(i%)

'              NEXT i%

'          end if

'      NEXT p%

'      CALL RetrieveErrantProbes(Np%,Nd%,j&,XiMin(),XiMax(),R(),M(),RepositionFactor$,Frep)

'   END IF

'  ----------------------------------- Every 20 Steps Shrink Decision Space Around Best Probe ----------------------------------

    IF j& MOD 20 = 0 AND ShrinkDS$ = "YES" THEN
'   IF j& MOD 30 = 0 THEN

        FOR i% = 1 TO Nd%

            XiMin(i%) = XiMin(i%)+(R(BestProbeNumber%,i%,BestTimeStep&)-XiMin(i%))/2## : XiMax(i%) = XiMax(i%)-(XiMax(i%)-
R(BestProbeNumber%,i%,BestTimeStep&))/2##

        NEXT i%

'     LastStep& = j& : EXIT FOR  'time step loop

    END IF

'  ----------------------------------- If DavgOSC => Increase All Probes' Acceleration to Break Away from Trapping(?)  -----------------------------
------

'   IF DavgOscillation$ = "YES" THEN

'      For p% = 1 to Np%

'        For i% = 1 to Nd%

'            A(p%,i%,j&) = 1##*A(p%,i%,j&) 'factor of XX increase

'         next i%

'     next p%

'   END IF

'STEP (F) ------------------- Check for Early Run Termination --------------------

    IF CheckForEarlyTermination$ = "YES" THEN 'insert termination test

        IF FitnessSaturation$ = "YES" THEN 'terminate run

            LastStep& = j& : EXIT FOR  'time step loop

        END IF
```

```
      END IF

    NEXT j& 'END OF TIME STEP LOOP

END SUB 'CFO()

'=========================================================================================================

SUB GetBestFitness(M(),Np%,StepNumber&,BestFitness,BestProbeNumber%,BestTimeStep&)

LOCAL p%, i&, A$

    BestFitness = M(1,0)

    FOR i& = 0 TO StepNumber&

        FOR p% = 1 TO Np%

            IF M(p%,i&) >= BestFitness THEN

                BestFitness = M(p%,i&) : BestProbeNumber% = p% : BestTimeStep& = i&

            END IF

        NEXT p%

    NEXT i&

END SUB

'========================================================= FUNCTION DEFINITIONS =========================================================

SUB SelectTestFunction(FunctionName$)

LOCAL A$

    A$ = INPUTBOX$("Which Function?"+CHR$(13)+"1 - Parrott F4 (1-D)"+CHR$(13)+"2 - SGO (2-D)"+CHR$(13)+"3 - GP (2-D)"+CHR$(13)+"4 - Step (n-
D)"+CHR$(13)+ _
                  "5 - Schwefel 2.26 (n-D)"+CHR$(13)+"6 - Colville (4-D)"+CHR$(13)+"7 - RESERVED"+CHR$(13)+"8 - more","SELECT OBJECTIVE
FUNCTION","1")

    IF VAL(A$) < 1 OR VAL(A$) > 8 THEN A$ = "1"

    SELECT CASE VAL(A$)

        CASE 1 : FunctionName$ = "ParrottF4"
        CASE 2 : FunctionName$ = "SGO"
        CASE 3 : FunctionName$ = "GP"
        CASE 4 : FunctionName$ = "STEP"
        CASE 5 : FunctionName$ = "SCHWEFEL_226"
        CASE 6 : FunctionName$ = "COLVILLE"
        CASE 7 : FunctionName$ = "GRIEWANK"
        CASE 8 : FunctionName$ = "more"

    END SELECT

    IF FunctionName$ = "more" THEN

        A$ = INPUTBOX$("Which Function?"+CHR$(13)+"1 - F1"+CHR$(13)+"2 - F2"+CHR$(13)+"3 - F3"+CHR$(13)+"4 - F4"+CHR$(13)+ _
                      "5 - F5"+CHR$(13)+"6 - F6"+CHR$(13)+"7 - F7"+CHR$(13)+"8 - more","SELECT OBJECTIVE FUNCTION","1")

        IF VAL(A$) < 1 OR VAL(A$) > 8 THEN A$ = "1"

        SELECT CASE VAL(A$)

            CASE 1 : FunctionName$ = "F1"
            CASE 2 : FunctionName$ = "F2"
            CASE 3 : FunctionName$ = "F3"
            CASE 4 : FunctionName$ = "F4"
            CASE 5 : FunctionName$ = "F5"
            CASE 6 : FunctionName$ = "F6"
            CASE 7 : FunctionName$ = "F7"
            CASE 8 : FunctionName$ = "more"

        END SELECT

    END IF

    IF FunctionName$ = "more" THEN

        A$ = INPUTBOX$("Which Function?"+CHR$(13)+"1 - F8"+CHR$(13)+"2 - F9"+CHR$(13)+"3 - F10"+CHR$(13)+"4 - F11"+CHR$(13)+ _
                      "5 - F12"+CHR$(13)+"6 - F13"+CHR$(13)+"7 - F14"+CHR$(13)+"8 - more","SELECT OBJECTIVE FUNCTION","1")

        IF VAL(A$) < 1 OR VAL(A$) > 8 THEN A$ = "1"

        SELECT CASE VAL(A$)

            CASE 1 : FunctionName$ = "F8"
            CASE 2 : FunctionName$ = "F9"
            CASE 3 : FunctionName$ = "F10"
            CASE 4 : FunctionName$ = "F11"
            CASE 5 : FunctionName$ = "F12"
            CASE 6 : FunctionName$ = "F13"
            CASE 7 : FunctionName$ = "F14"
            CASE 8 : FunctionName$ = "more"

        END SELECT

    END IF

    IF FunctionName$ = "more" THEN

        A$ = INPUTBOX$("Which Function?"+CHR$(13)+"1 - F15"+CHR$(13)+"2 - F16"+CHR$(13)+"3 - F17"+CHR$(13)+"4 - F18"+CHR$(13)+ _
                      "5 - F19"+CHR$(13)+"6 - F20"+CHR$(13)+"7 - F21"+CHR$(13)+"8 - more","SELECT OBJECTIVE FUNCTION","1")

        IF VAL(A$) < 1 OR VAL(A$) > 8 THEN A$ = "1"

        SELECT CASE VAL(A$)

            CASE 1 : FunctionName$ = "F15"
            CASE 2 : FunctionName$ = "F16"
            CASE 3 : FunctionName$ = "F17"
            CASE 4 : FunctionName$ = "F18"
            CASE 5 : FunctionName$ = "F19"
```

```
              CASE 6 : FunctionName$ = "F20"
              CASE 7 : FunctionName$ = "F21"
              CASE 8 : FunctionName$ = "more"

          END SELECT

      END IF

  IF FunctionName$ = "more" THEN

      A$ = INPUTBOX$("Which Function?"+CHR$(13)+"1 - F22"+CHR$(13)+"2 - F23"+CHR$(13)+"3 - F24"+CHR$(13)+"4 - F25"+CHR$(13)+ _
              "5 - F26"+CHR$(13)+"6 - F27"+CHR$(13)+"7 - F28"+CHR$(13)+"8 - more","SELECT OBJECTIVE FUNCTION","1")

      IF VAL(A$) < 1 OR VAL(A$) > 2 THEN A$ = "1"

      SELECT CASE VAL(A$)

          CASE 1 : FunctionName$ = "F22"
          CASE 2 : FunctionName$ = "F23"
          CASE 3 : FunctionName$ = "F24"
          CASE 4 : FunctionName$ = "F25"
          CASE 5 : FunctionName$ = "F26"
          CASE 6 : FunctionName$ = "F27"
          CASE 7 : FunctionName$ = "F28"
          CASE 8 : FunctionName$ = "more"

      END SELECT

  END IF

END SUB 'SelectTestFunction()

'------

FUNCTION ObjectiveFunction(R(),Nd%,p%,j&,FunctionName$) 'Objective function to be MAXIMIZED is defined here

  SELECT CASE FunctionName$

      CASE "ParrottF4"      : ObjectiveFunction = ParrottF4(R(),Nd%,p%,j&)       'Parrott F4 (1-D)

      CASE "SGO"            : ObjectiveFunction = SGO(R(),Nd%,p%,j&)             'SGO Function (2-D)

      CASE "GP"             : ObjectiveFunction = GoldsteinPrice(R(),Nd%,p%,j&)  'Goldstein-Price Function (2-D)

      CASE "STEP"           : ObjectiveFunction = StepFunction(R(),Nd%,p%,j&)   'Step Function (n-D)

      CASE "SCHWEFEL_226"   : ObjectiveFunction = Schwefel226(R(),Nd%,p%,j&)    'Schwefel Prob. 2.26 (n-D)

      CASE "COLVILLE"       : ObjectiveFunction = Colville(R(),Nd%,p%,j&)       'Colville Function (4-D)

      CASE "GRIEWANK"       : ObjectiveFunction = Griewank(R(),Nd%,p%,j&)       'Griewank Function (n-D)

      CASE "HIMMELBLAU"     : ObjectiveFunction = Himmelblau(R(),Nd%,p%,j&)     'Himmelblau Function (2-D)
'      ------------------ GSO Paper Benchmark Functions ----------------------
      CASE "F1"             : ObjectiveFunction = F1(R(),Nd%,p%,j&)             'F1  (n-D)
      CASE "F2"             : ObjectiveFunction = F2(R(),Nd%,p%,j&)             'F2  (n-D)
      CASE "F3"             : ObjectiveFunction = F3(R(),Nd%,p%,j&)             'F3  (n-D)
      CASE "F4"             : ObjectiveFunction = F4(R(),Nd%,p%,j&)             'F4  (n-D)
      CASE "F5"             : ObjectiveFunction = F5(R(),Nd%,p%,j&)             'F5  (n-D)
      CASE "F6"             : ObjectiveFunction = F6(R(),Nd%,p%,j&)             'F6  (n-D)
      CASE "F7"             : ObjectiveFunction = F7(R(),Nd%,p%,j&)             'F7  (n-D)
      CASE "F8"             : ObjectiveFunction = F8(R(),Nd%,p%,j&)             'F8  (n-D)
      CASE "F9"             : ObjectiveFunction = F9(R(),Nd%,p%,j&)             'F9  (n-D)
      CASE "F10"            : ObjectiveFunction = F10(R(),Nd%,p%,j&)            'F10 (n-D)
      CASE "F11"            : ObjectiveFunction = F11(R(),Nd%,p%,j&)            'F11 (n-D)
      CASE "F12"            : ObjectiveFunction = F12(R(),Nd%,p%,j&)            'F12 (n-D)
      CASE "F13"            : ObjectiveFunction = F13(R(),Nd%,p%,j&)            'F13 (n-D)
      CASE "F14"            : ObjectiveFunction = F14(R(),Nd%,p%,j&)            'F14 (2-D)
      CASE "F15"            : ObjectiveFunction = F15(R(),Nd%,p%,j&)            'F15 (4-D)
      CASE "F16"            : ObjectiveFunction = F16(R(),Nd%,p%,j&)            'F16 (2-D)
      CASE "F17"            : ObjectiveFunction = F17(R(),Nd%,p%,j&)            'F17 (2-D)
      CASE "F18"            : ObjectiveFunction = F18(R(),Nd%,p%,j&)            'F18 (2-D)
      CASE "F19"            : ObjectiveFunction = F19(R(),Nd%,p%,j&)            'F19 (3-D)
      CASE "F20"            : ObjectiveFunction = F20(R(),Nd%,p%,j&)            'F20 (6-D)
      CASE "F21"            : ObjectiveFunction = F21(R(),Nd%,p%,j&)            'F21 (4-D)
      CASE "F22"            : ObjectiveFunction = F22(R(),Nd%,p%,j&)            'F22 (4-D)
      CASE "F23"            : ObjectiveFunction = F23(R(),Nd%,p%,j&)            'F23 (4-D)
'      ------------------------------ PBM Antenna Benchmarks --------------------------------
      CASE "PBM_1"          : ObjectiveFunction = PBM_1(R(),Nd%,p%,j&)          'PBM_1 (2-D)
      CASE "PBM_2"          : ObjectiveFunction = PBM_2(R(),Nd%,p%,j&)          'PBM_2 (2-D)
      CASE "PBM_3"          : ObjectiveFunction = PBM_3(R(),Nd%,p%,j&)          'PBM_3 (2-D)
      CASE "PBM_4"          : ObjectiveFunction = PBM_4(R(),Nd%,p%,j&)          'PBM_4 (2-D)
      CASE "PBM_5"          : ObjectiveFunction = PBM_5(R(),Nd%,p%,j&)          'PBM_5 (2-D)

  END SELECT

END FUNCTION 'ObjectiveFunction()

'-----------

SUB ResetDecisionSpaceBoundaries(Nd%,XiMin(),XiMax(),StartingXiMin(),StartingXiMax())

LOCAL i%

  FOR i% = 1 TO Nd% : XiMin(i%) = StartingXiMin(i%) : XiMax(i%) = StartingXiMax(i%) : NEXT i%

END SUB

'------

SUB GetFunctionRunParameters(FunctionName$,Nd%,Np%,Nt&,G,DeltaT,Alpha,Beta,Frep,R(),A(),M(),XiMin(),XiMax(),StartingXiMin(),StartingXiMax(),_
                    DiagLength,PlaceInitialProbes$,InitialAcceleration$,RepositionFactor$)

LOCAL i%, NumProbesPerDimension%, NN%, NumCollinearElements%

  SELECT CASE FunctionName$

      CASE "ParrottF4"

          Nd%                      = 1
          NumProbesPerDimension%   = 3
```

```
            Np%                    = NumProbesPerDimension%*Nd%

            Nt&        = 500
            G          = 2##
            Alpha      = 2##
            Beta       = 2##
            DeltaT     = 1##
            Frep       = 0.9##

            PlaceInitialProbes$  = "UNIFORM ON-AXIS"
            InitialAcceleration$ = "ZERO"
            RepositionFactor$    = "FIXED"

            CALL
ChangeRunParameters(NumProbesPerDimension%,Np%,Nd%,Nt&,G,Alpha,Beta,DeltaT,Frep,PlaceInitialProbes$,InitialAcceleration$,RepositionFactor$,FunctionN
ame$)

            Nd% = 1 'cannot change dimensionality of Parrott F4 function!

            NumProbesPerDimension% = MAX(NumProbesPerDimension%,3) 'at least three for 1-D functions

            Np% = NumProbesPerDimension%*Nd%

            REDIM XiMin(1 TO Nd%), XiMax(1 TO Nd%) : XiMin(1) = 0## : XiMax(1) = 1##

            REDIM StartingXiMin(1 TO Nd%), StartingXiMax(1 TO Nd%) : FOR i% = 1 TO Nd% : StartingXiMin(i%) = XiMin(i%) : StartingXiMax(i%) =
XiMax(i%) : NEXT i%

        CASE "SGO"

            Nd%                    = 2
            NumProbesPerDimension% = 4 '10 '4
            Np%                    = NumProbesPerDimension%*Nd%

            Nt&        = 500
            G          = 2##
            Alpha      = 2##
            Beta       = 2##
            DeltaT     = 1##
            Frep       = 0.5##

            PlaceInitialProbes$  = "UNIFORM ON-AXIS" '"2D GRID"
            InitialAcceleration$ = "ZERO"
            RepositionFactor$    = "VARIABLE"

            CALL
ChangeRunParameters(NumProbesPerDimension%,Np%,Nd%,Nt&,G,Alpha,Beta,DeltaT,Frep,PlaceInitialProbes$,InitialAcceleration$,RepositionFactor$,FunctionN
ame$)

            Nd% = 2 'cannot change dimensionality of SGO function!

            Np% = NumProbesPerDimension%*Nd%

            REDIM XiMin(1 TO Nd%), XiMax(1 TO Nd%) : FOR i% = 1 TO Nd% : XiMin(i%) = -50## : XiMax(i%) = 50## : NEXT i%

            REDIM StartingXiMin(1 TO Nd%), StartingXiMax(1 TO Nd%) : FOR i% = 1 TO Nd% : StartingXiMin(i%) = XiMin(i%) : StartingXiMax(i%) =
XiMax(i%) : NEXT i%

            IF PlaceInitialProbes$ = "2D GRID" THEN

                Np% = NumProbesPerDimension%^2 : REDIM R(1 TO Np%, 1 TO Nd%, 0 TO Nt&) 'to create (Np/Nd) x (Np/Nd) grid

            END IF

        CASE "GP"

            Nd%                    = 2
            NumProbesPerDimension% = 4 '10
            Np%                    = NumProbesPerDimension%*Nd%

            Nt&        = 500
            G          = 2##
            Alpha      = 2## '0.2##
            Beta       = 2##
            DeltaT     = 1##
            Frep       = 0.5## '0.8## '0.9##

            PlaceInitialProbes$  = "UNIFORM ON-AXIS" '"2D GRID"
            InitialAcceleration$ = "ZERO"
            RepositionFactor$    = "VARIABLE"

            CALL
ChangeRunParameters(NumProbesPerDimension%,Np%,Nd%,Nt&,G,Alpha,Beta,DeltaT,Frep,PlaceInitialProbes$,InitialAcceleration$,RepositionFactor$,FunctionN
ame$)

            Nd% = 2 'cannot change dimensionality of GP function!

            Np% = NumProbesPerDimension%*Nd%

            REDIM XiMin(1 TO Nd%), XiMax(1 TO Nd%) : FOR i% = 1 TO Nd% : XiMin(i%) = -100## : XiMax(i%) = 100## : NEXT i%

            REDIM StartingXiMin(1 TO Nd%), StartingXiMax(1 TO Nd%) : FOR i% = 1 TO Nd% : StartingXiMin(i%) = XiMin(i%) : StartingXiMax(i%) =
XiMax(i%) : NEXT i%

            IF PlaceInitialProbes$ = "2D GRID" THEN

                Np% = NumProbesPerDimension%^2 : REDIM R(1 TO Np%, 1 TO Nd%, 0 TO Nt&) 'to create (Np/Nd) x (Np/Nd) grid

            END IF

        CASE "STEP"

            Nd%                    = 2
            NumProbesPerDimension% = 4 '20
            Np%                    = NumProbesPerDimension%*Nd%

            Nt&        = 500
            G          = 2##
            Alpha      = 2##
            Beta       = 2##
            DeltaT     = 1##
            Frep       = 0.5##

            PlaceInitialProbes$  = "UNIFORM ON-AXIS"
```

```
        InitialAcceleration$  = "ZERO"
        RepositionFactor$     = "VARIABLE" '"FIXED"

        CALL
ChangeRunParameters(NumProbesPerDimension%,Np%,Nd%,Nt&,Alpha,Beta,DeltaT,Frep,PlaceInitialProbes$,InitialAcceleration$,RepositionFactor$,FunctionN
ame$)

        Np% = NumProbesPerDimension%*Nd%

        REDIM XiMin(1 TO Nd%), XiMax(1 TO Nd%) : FOR i% = 1 TO Nd% : XiMin(i%) = -100## : XiMax(i%) = 100## : NEXT i%

        REDIM XiMin(1 TO Nd%), XiMax(1 TO Nd%) : XiMin(1) = 72## : XiMax(1) = 78## : XiMin(2) = 27## : XiMax(2) = 33## 'use this to plot STEP
'
detail

        REDIM StartingXiMin(1 TO Nd%), StartingXiMax(1 TO Nd%) : FOR i% = 1 TO Nd% : StartingXiMin(i%) = XiMin(i%) : StartingXiMax(i%) =
XiMax(i%) : NEXT i%

        IF PlaceInitialProbes$ = "2D GRID" THEN

            Np% = NumProbesPerDimension%^2 : REDIM R(1 TO Np%, 1 TO Nd%, 0 TO Nt&) 'to create (Np/Nd) x (Np/Nd) grid

        END IF 'STEP

    CASE "SCHWEFEL_226"

        Nd%                   = 30
        NumProbesPerDimension% = 4
        Np%                   = NumProbesPerDimension%*Nd%

        Nt&     = 500
        G       = 2##
        Alpha   = 2##
        Beta    = 2##
        DeltaT  = 1##
        Frep    = 0.5##

        PlaceInitialProbes$   = "UNIFORM ON-AXIS"
        InitialAcceleration$ = "ZERO"
        RepositionFactor$     = "VARIABLE"

        CALL
ChangeRunParameters(NumProbesPerDimension%,Np%,Nd%,Nt&,G,Alpha,Beta,DeltaT,Frep,PlaceInitialProbes$,InitialAcceleration$,RepositionFactor$,FunctionN
ame$)

        Np% = NumProbesPerDimension%*Nd%

        REDIM XiMin(1 TO Nd%), XiMax(1 TO Nd%) : FOR i% = 1 TO Nd% : XiMin(i%) = -500## : XiMax(i%) = 500## : NEXT i%

        REDIM StartingXiMin(1 TO Nd%), StartingXiMax(1 TO Nd%) : FOR i% = 1 TO Nd% : StartingXiMin(i%) = XiMin(i%) : StartingXiMax(i%) =
XiMax(i%) : NEXT i%

        IF PlaceInitialProbes$ = "2D GRID" THEN

            Np% = NumProbesPerDimension%^2 : REDIM R(1 TO Np%, 1 TO Nd%, 0 TO Nt&) 'to create (Np/Nd) x (Np/Nd) grid

        END IF

    CASE "COLVILLE"

        Nd%                   = 4 '14
        NumProbesPerDimension% = 4 '14
        Np%                   = NumProbesPerDimension%*Nd%

        Nt&     = 500
        G       = 2##
        Alpha   = 2##
        Beta    = 2##
        DeltaT  = 1##
        Frep    = 0.5##

        PlaceInitialProbes$   = "UNIFORM ON-AXIS"
        InitialAcceleration$ = "ZERO"
        RepositionFactor$     = "VARIABLE"

        CALL
ChangeRunParameters(NumProbesPerDimension%,Np%,Nd%,Nt&,G,Alpha,Beta,DeltaT,Frep,PlaceInitialProbes$,InitialAcceleration$,RepositionFactor$,FunctionN
ame$)

        Nd% = 4 'cannot change dimensionality of Colville function!

        Np% = NumProbesPerDimension%*Nd%

        REDIM XiMin(1 TO Nd%), XiMax(1 TO Nd%) : FOR i% = 1 TO Nd% : XiMin(i%) = -10## : XiMax(i%) = 10## : NEXT i%

        REDIM StartingXiMin(1 TO Nd%), StartingXiMax(1 TO Nd%) : FOR i% = 1 TO Nd% : StartingXiMin(i%) = XiMin(i%) : StartingXiMax(i%) =
XiMax(i%) : NEXT i%

        IF PlaceInitialProbes$ = "2D GRID" THEN

            Np% = NumProbesPerDimension%^2 : REDIM R(1 TO Np%, 1 TO Nd%, 0 TO Nt&) 'to create (Np/Nd) x (Np/Nd) grid

        END IF

    CASE "GRIEWANK"

        Nd%                   = 2
        NumProbesPerDimension% = 4 '14
        Np%                   = NumProbesPerDimension%*Nd%

        Nt&     = 500
        G       = 2##
        Alpha   = 2##
        Beta    = 2##
        DeltaT  = 1##
        Frep    = 0.5##

        PlaceInitialProbes$   = "UNIFORM ON-AXIS"
        InitialAcceleration$ = "ZERO"
        RepositionFactor$     = "VARIABLE"

        CALL
ChangeRunParameters(NumProbesPerDimension%,Np%,Nd%,Nt&,G,Alpha,Beta,DeltaT,Frep,PlaceInitialProbes$,InitialAcceleration$,RepositionFactor$,FunctionN
ame$)
```

```
            Np% = NumProbesPerDimension%*Nd%

            REDIM XiMin(1 TO Nd%), XiMax(1 TO Nd%) : FOR i% = 1 TO Nd% : XiMin(i%) = -600## : XiMax(i%) = 600## : NEXT i%

            REDIM StartingXiMin(1 TO Nd%), StartingXiMax(1 TO Nd%) : FOR i% = 1 TO Nd% : StartingXiMin(i%) = XiMin(i%) : StartingXiMax(i%) =
XiMax(i%) : NEXT i%

            IF PlaceInitialProbes$ = "2D GRID" THEN

                Np% = NumProbesPerDimension%^2 : REDIM R(1 TO Np%, 1 TO Nd%, 0 TO Nt&) 'to create (Np/Nd) x (Np/Nd) grid

            END IF

        CASE "HIMMELBLAU"

            Nd%                      = 2
            NumProbesPerDimension% = 4  '14
            Np%                      = NumProbesPerDimension%*Nd%

            Nt&       = 500
            G         = 2##
            Alpha     = 2##
            Beta      = 2##
            DeltaT    = 1##
            Frep      = 0.5##

            PlaceInitialProbes$   = "UNIFORM ON-AXIS"
            InitialAcceleration$  = "ZERO"
            RepositionFactor$     = "VARIABLE"

            CALL
ChangeRunParameters(NumProbesPerDimension%,Np%,Nd%,Nt&,G,Alpha,Beta,DeltaT,Frep,PlaceInitialProbes$,InitialAcceleration$,RepositionFactor$,FunctionN
ame$)

            Nd% = 2 'cannot change dimensionality of Himmelblau function!

            Np% = NumProbesPerDimension%*Nd%

            REDIM XiMin(1 TO Nd%), XiMax(1 TO Nd%) : FOR i% = 1 TO Nd% : XiMin(i%) = -6## : XiMax(i%) = 6## : NEXT i%

            REDIM StartingXiMin(1 TO Nd%), StartingXiMax(1 TO Nd%) : FOR i% = 1 TO Nd% : StartingXiMin(i%) = XiMin(i%) : StartingXiMax(i%) =
XiMax(i%) : NEXT i%

            IF PlaceInitialProbes$ = "2D GRID" THEN

                Np% = NumProbesPerDimension%^2 : REDIM R(1 TO Np%, 1 TO Nd%, 0 TO Nt&) 'to create (Np/Nd) x (Np/Nd) grid

            END IF

        CASE "F1"  '(n-D)

            Nd%                      = 30
            NumProbesPerDimension% = 2
            Np%                      = NumProbesPerDimension%*Nd%

            Nt&       = 500
            G         = 2##
            Alpha     = 2##
            Beta      = 2##
            DeltaT    = 1##
            Frep      = 0.5##

            PlaceInitialProbes$   = "UNIFORM ON-AXIS"
            InitialAcceleration$  = "ZERO"
            RepositionFactor$     = "VARIABLE"

            CALL
ChangeRunParameters(NumProbesPerDimension%,Np%,Nd%,Nt&,G,Alpha,Beta,DeltaT,Frep,PlaceInitialProbes$,InitialAcceleration$,RepositionFactor$,FunctionN
ame$)

            Np% = NumProbesPerDimension%*Nd%

            REDIM XiMin(1 TO Nd%), XiMax(1 TO Nd%) : FOR i% = 1 TO Nd% : XiMin(i%) = -100## : XiMax(i%) = 100## : NEXT i%

            REDIM StartingXiMin(1 TO Nd%), StartingXiMax(1 TO Nd%) : FOR i% = 1 TO Nd% : StartingXiMin(i%) = XiMin(i%) : StartingXiMax(i%) =
XiMax(i%) : NEXT i%

            IF PlaceInitialProbes$ = "2D GRID" THEN

                Np% = NumProbesPerDimension%^2 : REDIM R(1 TO Np%, 1 TO Nd%, 0 TO Nt&) 'to create (Np/Nd) x (Np/Nd) grid

            END IF

        CASE "F2"  '(n-D)

            Nd%                      = 30
            NumProbesPerDimension% = 2
            Np%                      = NumProbesPerDimension%*Nd%

            Nt&       = 500
            G         = 2##
            Alpha     = 2##
            Beta      = 2##
            DeltaT    = 1##
            Frep      = 0.5##

            PlaceInitialProbes$   = "UNIFORM ON-AXIS"
            InitialAcceleration$  = "ZERO"
            RepositionFactor$     = "VARIABLE"

            CALL
ChangeRunParameters(NumProbesPerDimension%,Np%,Nd%,Nt&,G,Alpha,Beta,DeltaT,Frep,PlaceInitialProbes$,InitialAcceleration$,RepositionFactor$,FunctionN
ame$)

            Np% = NumProbesPerDimension%*Nd%

            REDIM XiMin(1 TO Nd%), XiMax(1 TO Nd%) : FOR i% = 1 TO Nd% : XiMin(i%) = -10## : XiMax(i%) = 10## : NEXT i%

            REDIM StartingXiMin(1 TO Nd%), StartingXiMax(1 TO Nd%) : FOR i% = 1 TO Nd% : StartingXiMin(i%) = XiMin(i%) : StartingXiMax(i%) =
XiMax(i%) : NEXT i%

            IF PlaceInitialProbes$ = "2D GRID" THEN

                Np% = NumProbesPerDimension%^2 : REDIM R(1 TO Np%, 1 TO Nd%, 0 TO Nt&) 'to create (Np/Nd) x (Np/Nd) grid
```

```
            END IF

      CASE "F3" '(n-D)

            Nd%                      = 30
            NumProbesPerDimension% = 2
            Np%                      = NumProbesPerDimension%*Nd%

            Nt&     = 500
            G       = 2##
            Alpha   = 2##
            Beta    = 2##
            DeltaT  = 1##
            Frep    = 0.5##

            PlaceInitialProbes$  = "UNIFORM ON-AXIS"
            InitialAcceleration$ = "ZERO"
            RepositionFactor$    = "VARIABLE"

            CALL
ChangeRunParameters(NumProbesPerDimension%,Np%,Nd%,Nt&,G,Alpha,Beta,DeltaT,Frep,PlaceInitialProbes$,InitialAcceleration$,RepositionFactor$,FunctionN
ame$)

            Np% = NumProbesPerDimension%*Nd%

            REDIM XiMin(1 TO Nd%), XiMax(1 TO Nd%) : FOR i% = 1 TO Nd% : XiMin(i%) = -100## : XiMax(i%) = 100## : NEXT i%

            REDIM StartingXiMin(1 TO Nd%), StartingXiMax(1 TO Nd%) : FOR i% = 1 TO Nd% : StartingXiMin(i%) = XiMin(i%) : StartingXiMax(i%) =
XiMax(i%) : NEXT i%

            IF PlaceInitialProbes$ = "2D GRID" THEN

                  Np% = NumProbesPerDimension%^2 : REDIM R(1 TO Np%, 1 TO Nd%, 0 TO Nt&) 'to create (Np/Nd) x (Np/Nd) grid

            END IF

      CASE "F4" '(n-D)

            Nd%                      = 30
            NumProbesPerDimension% = 2
            Np%                      = NumProbesPerDimension%*Nd%

            Nt&     = 500
            G       = 2##
            Alpha   = 2##
            Beta    = 2##
            DeltaT  = 1##
            Frep    = 0.5##

            PlaceInitialProbes$  = "UNIFORM ON-AXIS"
            InitialAcceleration$ = "ZERO"
            RepositionFactor$    = "VARIABLE"

            CALL
ChangeRunParameters(NumProbesPerDimension%,Np%,Nd%,Nt&,G,Alpha,Beta,DeltaT,Frep,PlaceInitialProbes$,InitialAcceleration$,RepositionFactor$,FunctionN
ame$)

            Np% = NumProbesPerDimension%*Nd%

            REDIM XiMin(1 TO Nd%), XiMax(1 TO Nd%) : FOR i% = 1 TO Nd% : XiMin(i%) = -100## : XiMax(i%) = 100## : NEXT i%

            REDIM StartingXiMin(1 TO Nd%), StartingXiMax(1 TO Nd%) : FOR i% = 1 TO Nd% : StartingXiMin(i%) = XiMin(i%) : StartingXiMax(i%) =
XiMax(i%) : NEXT i%

            IF PlaceInitialProbes$ = "2D GRID" THEN

                  Np% = NumProbesPerDimension%^2 : REDIM R(1 TO Np%, 1 TO Nd%, 0 TO Nt&) 'to create (Np/Nd) x (Np/Nd) grid

            END IF

      CASE "F5" '(n-D)

            Nd%                      = 30
            NumProbesPerDimension% = 2
            Np%                      = NumProbesPerDimension%*Nd%

            Nt&     = 500
            G       = 2##
            Alpha   = 2##
            Beta    = 2##
            DeltaT  = 1##
            Frep    = 0.5##

            PlaceInitialProbes$  = "UNIFORM ON-AXIS"
            InitialAcceleration$ = "ZERO"
            RepositionFactor$    = "VARIABLE"

            CALL
ChangeRunParameters(NumProbesPerDimension%,Np%,Nd%,Nt&,G,Alpha,Beta,DeltaT,Frep,PlaceInitialProbes$,InitialAcceleration$,RepositionFactor$,FunctionN
ame$)

            Np% = NumProbesPerDimension%*Nd%

            REDIM XiMin(1 TO Nd%), XiMax(1 TO Nd%) : FOR i% = 1 TO Nd% : XiMin(i%) = -30## : XiMax(i%) = 30## : NEXT i%

            REDIM StartingXiMin(1 TO Nd%), StartingXiMax(1 TO Nd%) : FOR i% = 1 TO Nd% : StartingXiMin(i%) = XiMin(i%) : StartingXiMax(i%) =
XiMax(i%) : NEXT i%

            IF PlaceInitialProbes$ = "2D GRID" THEN

                  Np% = NumProbesPerDimension%^2 : REDIM R(1 TO Np%, 1 TO Nd%, 0 TO Nt&) 'to create (Np/Nd) x (Np/Nd) grid

            END IF

      CASE "F6" '(n-D) STEP

            Nd%                      = 30
            NumProbesPerDimension% = 2 '20
            Np%                      = NumProbesPerDimension%*Nd%

            Nt&     = 500
            G       = 2##
            Alpha   = 2##
```

```
                Beta      = 2##
                DeltaT    = 1##
                Frep      = 0.5##

                PlaceInitialProbes$ = "UNIFORM ON-AXIS"
                InitialAcceleration$ = "ZERO"
                RepositionFactor$    = "VARIABLE" '"FIXED"

                CALL
ChangeRunParameters(NumProbesPerDimension%,Np%,Nd%,Nt&,G,Alpha,Beta,DeltaT,Frep,PlaceInitialProbes$,InitialAcceleration$,RepositionFactor$,FunctionN
ame$)

                Np% = NumProbesPerDimension%*Nd%

                REDIM XiMin(1 TO Nd%), XiMax(1 TO Nd%) : FOR i% = 1 TO Nd% : XiMin(i%) = -100## : XiMax(i%) = 100## : NEXT i%

                REDIM StartingXiMin(1 TO Nd%), StartingXiMax(1 TO Nd%) : FOR i% = 1 TO Nd% : StartingXiMin(i%) = XiMin(i%) : StartingXiMax(i%) =
XiMax(i%) : NEXT i%

                IF PlaceInitialProbes$ = "2D GRID" THEN

                    Np% = NumProbesPerDimension%^2 : REDIM R(1 TO Np%, 1 TO Nd%, 0 TO Nt&) 'to create (Np/Nd) x (Np/Nd) grid

                END IF

        CASE "F7" '(n-D)

                Nd%                  = 30
                NumProbesPerDimension% = 2 '20
                Np%                  = NumProbesPerDimension%*Nd%

                Nt&       = 100       'BECAUSE THIS FUNCTION HAS A RANDOM COMPONENT !!
                G         = 2##
                Alpha     = 2##
                Beta      = 2##
                DeltaT    = 1##
                Frep      = 0.5##

                PlaceInitialProbes$ = "UNIFORM ON-AXIS"
                InitialAcceleration$ = "ZERO"
                RepositionFactor$    = "VARIABLE" '"FIXED"

                CALL
ChangeRunParameters(NumProbesPerDimension%,Np%,Nd%,Nt&,G,Alpha,Beta,DeltaT,Frep,PlaceInitialProbes$,InitialAcceleration$,RepositionFactor$,FunctionN
ame$)

                Np% = NumProbesPerDimension%*Nd%

                REDIM XiMin(1 TO Nd%), XiMax(1 TO Nd%) : FOR i% = 1 TO Nd% : XiMin(i%) = -1.28## : XiMax(i%) = 1.28## : NEXT i%

                REDIM StartingXiMin(1 TO Nd%), StartingXiMax(1 TO Nd%) : FOR i% = 1 TO Nd% : StartingXiMin(i%) = XiMin(i%) : StartingXiMax(i%) =
XiMax(i%) : NEXT i%

                IF PlaceInitialProbes$ = "2D GRID" THEN

                    Np% = NumProbesPerDimension%^2 : REDIM R(1 TO Np%, 1 TO Nd%, 0 TO Nt&) 'to create (Np/Nd) x (Np/Nd) grid

                END IF 'F7

        CASE "F8" '(n-D)

                Nd%                  = 30
                NumProbesPerDimension% = 2 '4 '20
                Np%                  = NumProbesPerDimension%*Nd%

                Nt&       = 500
                G         = 2##
                Alpha     = 2##
                Beta      = 2##
                DeltaT    = 1##
                Frep      = 0.5##

                PlaceInitialProbes$ = "UNIFORM ON-AXIS"
                InitialAcceleration$ = "ZERO"
                RepositionFactor$    = "VARIABLE" '"FIXED"

                CALL
ChangeRunParameters(NumProbesPerDimension%,Np%,Nd%,Nt&,G,Alpha,Beta,DeltaT,Frep,PlaceInitialProbes$,InitialAcceleration$,RepositionFactor$,FunctionN
ame$)

                Np% = NumProbesPerDimension%*Nd%

                REDIM XiMin(1 TO Nd%), XiMax(1 TO Nd%) : FOR i% = 1 TO Nd% : XiMin(i%) = -500## : XiMax(i%) = 500## : NEXT i%

                REDIM StartingXiMin(1 TO Nd%), StartingXiMax(1 TO Nd%) : FOR i% = 1 TO Nd% : StartingXiMin(i%) = XiMin(i%) : StartingXiMax(i%) =
XiMax(i%) : NEXT i%

                IF PlaceInitialProbes$ = "2D GRID" THEN

                    Np% = NumProbesPerDimension%^2 : REDIM R(1 TO Np%, 1 TO Nd%, 0 TO Nt&) 'to create (Np/Nd) x (Np/Nd) grid

                END IF 'F8

        CASE "F9" '(n-D)

                Nd%                  = 30
                NumProbesPerDimension% = 2 '4 '20
                Np%                  = NumProbesPerDimension%*Nd%

                Nt&       = 500
                G         = 2##
                Alpha     = 2##
                Beta      = 2##
                DeltaT    = 1##
                Frep      = 0.5##

                PlaceInitialProbes$ = "UNIFORM ON-AXIS"
                InitialAcceleration$ = "ZERO"
                RepositionFactor$    = "VARIABLE" '"FIXED"

                CALL
ChangeRunParameters(NumProbesPerDimension%,Np%,Nd%,Nt&,G,Alpha,Beta,DeltaT,Frep,PlaceInitialProbes$,InitialAcceleration$,RepositionFactor$,FunctionN
ame$)
```

```
                Np% = NumProbesPerDimension%*Nd%

                REDIM XiMin(1 TO Nd%), XiMax(1 TO Nd%) : FOR i% = 1 TO Nd% : XiMin(i%) = -5.12## : XiMax(i%) = 5.12## : NEXT i%

                REDIM StartingXiMin(1 TO Nd%), StartingXiMax(1 TO Nd%) : FOR i% = 1 TO Nd% : StartingXiMin(i%) = XiMin(i%) : StartingXiMax(i%) =
XiMax(i%) : NEXT i%

                IF PlaceInitialProbes$ = "2D GRID" THEN

                    Np% = NumProbesPerDimension%^2 : REDIM R(1 TO Np%, 1 TO Nd%, 0 TO Nt&) 'to create (Np/Nd) x (Np/Nd) grid

                END IF 'F9

        CASE "F10" '(n-D) Ackley's Function

                Nd%                    = 30
                NumProbesPerDimension% = 2 '4 '20
                Np%                    = NumProbesPerDimension%*Nd%

                Nt&     = 500
                G       = 2##
                Alpha   = 2##
                Beta    = 2##
                DeltaT  = 1##
                Frep    = 0.5##

                PlaceInitialProbes$  = "UNIFORM ON-AXIS"
                InitialAcceleration$ = "ZERO"
                RepositionFactor$    = "VARIABLE" '"FIXED"

                CALL
ChangeRunParameters(NumProbesPerDimension%,Np%,Nd%,Nt&,G,Alpha,Beta,DeltaT,Frep,PlaceInitialProbes$,InitialAcceleration$,RepositionFactor$,FunctionN
ame$)

                Np% = NumProbesPerDimension%*Nd%

                REDIM XiMin(1 TO Nd%), XiMax(1 TO Nd%) : FOR i% = 1 TO Nd% : XiMin(i%) = -32## : XiMax(i%) = 32## : NEXT i%

                REDIM StartingXiMin(1 TO Nd%), StartingXiMax(1 TO Nd%) : FOR i% = 1 TO Nd% : StartingXiMin(i%) = XiMin(i%) : StartingXiMax(i%) =
XiMax(i%) : NEXT i%

                IF PlaceInitialProbes$ = "2D GRID" THEN

                    Np% = NumProbesPerDimension%^2 : REDIM R(1 TO Np%, 1 TO Nd%, 0 TO Nt&) 'to create (Np/Nd) x (Np/Nd) grid

                END IF 'F10

        CASE "F11" '(n-D)

                Nd%                    = 30
                NumProbesPerDimension% = 2 '4 '20
                Np%                    = NumProbesPerDimension%*Nd%

                Nt&     = 500
                G       = 2##
                Alpha   = 2##
                Beta    = 2##
                DeltaT  = 1##
                Frep    = 0.5##

                PlaceInitialProbes$  = "UNIFORM ON-AXIS"
                InitialAcceleration$ = "ZERO"
                RepositionFactor$    = "VARIABLE" '"FIXED"

                CALL
ChangeRunParameters(NumProbesPerDimension%,Np%,Nd%,Nt&,G,Alpha,Beta,DeltaT,Frep,PlaceInitialProbes$,InitialAcceleration$,RepositionFactor$,FunctionN
ame$)

                Np% = NumProbesPerDimension%*Nd%

                REDIM XiMin(1 TO Nd%), XiMax(1 TO Nd%) : FOR i% = 1 TO Nd% : XiMin(i%) = -600## : XiMax(i%) = 600## : NEXT i%

                REDIM StartingXiMin(1 TO Nd%), StartingXiMax(1 TO Nd%) : FOR i% = 1 TO Nd% : StartingXiMin(i%) = XiMin(i%) : StartingXiMax(i%) =
XiMax(i%) : NEXT i%

                IF PlaceInitialProbes$ = "2D GRID" THEN

                    Np% = NumProbesPerDimension%^2 : REDIM R(1 TO Np%, 1 TO Nd%, 0 TO Nt&) 'to create (Np/Nd) x (Np/Nd) grid

                END IF 'F11

        CASE "F12" '(n-D) Penalized #1

                Nd%                    = 30
                NumProbesPerDimension% = 2 '4 '20
                Np%                    = NumProbesPerDimension%*Nd%

                Nt&     = 500
                G       = 2##
                Alpha   = 2##
                Beta    = 2##
                DeltaT  = 1##
                Frep    = 0.5##

                PlaceInitialProbes$  = "UNIFORM ON-AXIS"
                InitialAcceleration$ = "ZERO"
                RepositionFactor$    = "VARIABLE" '"FIXED"

                CALL
ChangeRunParameters(NumProbesPerDimension%,Np%,Nd%,Nt&,G,Alpha,Beta,DeltaT,Frep,PlaceInitialProbes$,InitialAcceleration$,RepositionFactor$,FunctionN
ame$)

                Np% = NumProbesPerDimension%*Nd%

                REDIM XiMin(1 TO Nd%), XiMax(1 TO Nd%) : FOR i% = 1 TO Nd% : XiMin(i%) = -50## : XiMax(i%) = 50## : NEXT i%

                REDIM StartingXiMin(1 TO Nd%), StartingXiMax(1 TO Nd%) : FOR i% = 1 TO Nd% : StartingXiMin(i%) = XiMin(i%) : StartingXiMax(i%) =
XiMax(i%) : NEXT i%

                IF PlaceInitialProbes$ = "2D GRID" THEN

                    Np% = NumProbesPerDimension%^2 : REDIM R(1 TO Np%, 1 TO Nd%, 0 TO Nt&) 'to create (Np/Nd) x (Np/Nd) grid

                END IF 'F12
```

```
        CASE "F13" '(n-D) Penalized #2

            Nd%                     = 30
            NumProbesPerDimension% = 2 '4 '20
            Np%                     = NumProbesPerDimension%*Nd%

            Nt&      = 500
            G        = 2##
            Alpha    = 2##
            Beta     = 2##
            DeltaT   = 1##
            Frep     = 0.5##

            PlaceInitialProbes$  = "UNIFORM ON-AXIS"
            InitialAcceleration$ = "ZERO"
            RepositionFactor$     = "VARIABLE" '"FIXED"

            CALL
ChangeRunParameters(NumProbesPerDimension%,Np%,Nd%,Nt&,G,Alpha,Beta,DeltaT,Frep,PlaceInitialProbes$,InitialAcceleration$,RepositionFactor$,FunctionN
ame$)

            Np% = NumProbesPerDimension%*Nd%

            REDIM XiMin(1 TO Nd%), XiMax(1 TO Nd%) : FOR i% = 1 TO Nd% : XiMin(i%) = -50## : XiMax(i%) = 50## : NEXT i%

            REDIM StartingXiMin(1 TO Nd%), StartingXiMax(1 TO Nd%) : FOR i% = 1 TO Nd% : StartingXiMin(i%) = XiMin(i%) : StartingXiMax(i%) =
XiMax(i%) : NEXT i%

            IF PlaceInitialProbes$ = "2D GRID" THEN

                Np% = NumProbesPerDimension%^2 : REDIM R(1 TO Np%, 1 TO Nd%, 0 TO Nt&) 'to create (Np/Nd) x (Np/Nd) grid

            END IF 'F13

        CASE "F14" '(2-D) Shekel's Foxholes

            Nd%                     = 2
            NumProbesPerDimension% = 4 '20
            Np%                     = NumProbesPerDimension%*Nd%

            Nt&      = 500
            G        = 2##
            Alpha    = 2##
            Beta     = 2##
            DeltaT   = 1##
            Frep     = 0.5##

            PlaceInitialProbes$  = "UNIFORM ON-AXIS"
            InitialAcceleration$ = "ZERO"
            RepositionFactor$     = "VARIABLE" '"FIXED"

            CALL
ChangeRunParameters(NumProbesPerDimension%,Np%,Nd%,Nt&,G,Alpha,Beta,DeltaT,Frep,PlaceInitialProbes$,InitialAcceleration$,RepositionFactor$,FunctionN
ame$)

            Nd% = 2 'cannot change dimensionality of Shekel's Foxholes function!

            Np% = NumProbesPerDimension%*Nd%

            REDIM XiMin(1 TO Nd%), XiMax(1 TO Nd%) : FOR i% = 1 TO Nd% : XiMin(i%) = -65.536## : XiMax(i%) = 65.536## : NEXT i%

            REDIM StartingXiMin(1 TO Nd%), StartingXiMax(1 TO Nd%) : FOR i% = 1 TO Nd% : StartingXiMin(i%) = XiMin(i%) : StartingXiMax(i%) =
XiMax(i%) : NEXT i%

            IF PlaceInitialProbes$ = "2D GRID" THEN

                Np% = NumProbesPerDimension%^2 : REDIM R(1 TO Np%, 1 TO Nd%, 0 TO Nt&) 'to create (Np/Nd) x (Np/Nd) grid

            END IF 'F14

        CASE "F15" '(4-D) Kowalik's Function

            Nd%                     = 4
            NumProbesPerDimension% = 4 '20
            Np%                     = NumProbesPerDimension%*Nd%

            Nt&      = 500
            G        = 2##
            Alpha    = 2##
            Beta     = 2##
            DeltaT   = 1##
            Frep     = 0.5##

            PlaceInitialProbes$  = "UNIFORM ON-AXIS"
            InitialAcceleration$ = "ZERO"
            RepositionFactor$     = "VARIABLE" '"FIXED"

            CALL
ChangeRunParameters(NumProbesPerDimension%,Np%,Nd%,Nt&,G,Alpha,Beta,DeltaT,Frep,PlaceInitialProbes$,InitialAcceleration$,RepositionFactor$,FunctionN
ame$)

            Nd% = 4 'cannot change dimensionality of Kowalik's Function!

            Np% = NumProbesPerDimension%*Nd%

            REDIM XiMin(1 TO Nd%), XiMax(1 TO Nd%) : FOR i% = 1 TO Nd% : XiMin(i%) = -5## : XiMax(i%) = 5## : NEXT i%

            REDIM StartingXiMin(1 TO Nd%), StartingXiMax(1 TO Nd%) : FOR i% = 1 TO Nd% : StartingXiMin(i%) = XiMin(i%) : StartingXiMax(i%) =
XiMax(i%) : NEXT i%

        CASE "F16" '(2-D) Camel Back

            Nd%                     = 2
            NumProbesPerDimension% = 4 '20
            Np%                     = NumProbesPerDimension%*Nd%

            Nt&      = 500
            G        = 2##
            Alpha    = 2##
            Beta     = 2##
            DeltaT   = 1##
            Frep     = 0.5##
```

```
                PlaceInitialProbes$  = "UNIFORM ON-AXIS"
                InitialAcceleration$ = "ZERO"
                RepositionFactor$    = "VARIABLE" '"FIXED"

                CALL
ChangeRunParameters(NumProbesPerDimension%,Np%,Nd%,Nt&,G,Alpha,Beta,DeltaT,Frep,PlaceInitialProbes$,InitialAcceleration$,RepositionFactor$,FunctionN
ame$)

                Nd% = 2 'cannot change dimensionality of Camel Back function!

                Np% = NumProbesPerDimension%*Nd%

                REDIM XiMin(1 TO Nd%), XiMax(1 TO Nd%) : FOR i% = 1 TO Nd% : XiMin(i%) = -5## : XiMax(i%) = 5## : NEXT i%

                REDIM StartingXiMin(1 TO Nd%), StartingXiMax(1 TO Nd%) : FOR i% = 1 TO Nd% : StartingXiMin(i%) = XiMin(i%) : StartingXiMax(i%) =
XiMax(i%) : NEXT i%

                IF PlaceInitialProbes$ = "2D GRID" THEN

                    Np% = NumProbesPerDimension%^2 : REDIM R(1 TO Np%, 1 TO Nd%, 0 TO Nt&) 'to create (Np/Nd) x (Np/Nd) grid

                END IF 'F16

        CASE "F17" '(2-D) Branin

                Nd%                    = 2
                NumProbesPerDimension% = 4 '20
                Np%                    = NumProbesPerDimension%*Nd%

                Nt&     = 500
                G       = 2##
                Alpha   = 2##
                Beta    = 2##
                DeltaT  = 1##
                Frep    = 0.5##

                PlaceInitialProbes$  = "UNIFORM ON-AXIS"
                InitialAcceleration$ = "ZERO"
                RepositionFactor$    = "VARIABLE" '"FIXED"

                CALL
ChangeRunParameters(NumProbesPerDimension%,Np%,Nd%,Nt&,G,Alpha,Beta,DeltaT,Frep,PlaceInitialProbes$,InitialAcceleration$,RepositionFactor$,FunctionN
ame$)

                Nd% = 2 'cannot change dimensionality of Branin function!

                Np% = NumProbesPerDimension%*Nd%

                REDIM XiMin(1 TO Nd%), XiMax(1 TO Nd%) : XiMin(1) = -5## : XiMax(1) = 10## : XiMin(2) = 0## : XiMax(2) = 15##

                REDIM StartingXiMin(1 TO Nd%), StartingXiMax(1 TO Nd%) : FOR i% = 1 TO Nd% : StartingXiMin(i%) = XiMin(i%) : StartingXiMax(i%) =
XiMax(i%) : NEXT i%

                IF PlaceInitialProbes$ = "2D GRID" THEN

                    Np% = NumProbesPerDimension%^2 : REDIM R(1 TO Np%, 1 TO Nd%, 0 TO Nt&) 'to create (Np/Nd) x (Np/Nd) grid

                END IF 'F17

        CASE "F18" '(2-D) Goldstein-Price

                Nd%                    = 2
                NumProbesPerDimension% = 4 '20
                Np%                    = NumProbesPerDimension%*Nd%

                Nt&     = 500
                G       = 2##
                Alpha   = 2##
                Beta    = 2##
                DeltaT  = 1##
                Frep    = 0.5##

                PlaceInitialProbes$  = "UNIFORM ON-AXIS"
                InitialAcceleration$ = "ZERO"
                RepositionFactor$    = "VARIABLE" '"FIXED"

                CALL
ChangeRunParameters(NumProbesPerDimension%,Np%,Nd%,Nt&,G,Alpha,Beta,DeltaT,Frep,PlaceInitialProbes$,InitialAcceleration$,RepositionFactor$,FunctionN
ame$)

                Nd% = 2 'cannot change dimensionality of Branin function!

                Np% = NumProbesPerDimension%*Nd%

                REDIM XiMin(1 TO Nd%), XiMax(1 TO Nd%) : XiMin(1) = -2## : XiMax(1) = 2## : XiMin(2) = -2## : XiMax(2) = 2##

                REDIM StartingXiMin(1 TO Nd%), StartingXiMax(1 TO Nd%) : FOR i% = 1 TO Nd% : StartingXiMin(i%) = XiMin(i%) : StartingXiMax(i%) =
XiMax(i%) : NEXT i%

                IF PlaceInitialProbes$ = "2D GRID" THEN

                    Np% = NumProbesPerDimension%^2 : REDIM R(1 TO Np%, 1 TO Nd%, 0 TO Nt&) 'to create (Np/Nd) x (Np/Nd) grid

                END IF 'F18

        CASE "F19" '(3-D) Hartman's Family #1

                Nd%                    = 3
                NumProbesPerDimension% = 4 '20
                Np%                    = NumProbesPerDimension%*Nd%

                Nt&     = 500
                G       = 2##
                Alpha   = 2##
                Beta    = 2##
                DeltaT  = 1##
                Frep    = 0.5##

                PlaceInitialProbes$  = "UNIFORM ON-AXIS"
                InitialAcceleration$ = "ZERO"
                RepositionFactor$    = "VARIABLE" '"FIXED"
```

```
            CALL
ChangeRunParameters(NumProbesPerDimension%,Np%,Nd%,Nt&,G,Alpha,Beta,DeltaT,Frep,PlaceInitialProbes$,InitialAcceleration$,RepositionFactor$,FunctionN
ame$)

            Nd% = 3 'cannot change dimensionality of Hartman's Family!

            Np% = NumProbesPerDimension%*Nd%

            REDIM XiMin(1 TO Nd%), XiMax(1 TO Nd%) : FOR i% = 1 TO Nd% : XiMin(i%) = 0## : XiMax(i%) = 1## : NEXT i%

            REDIM StartingXiMin(1 TO Nd%), StartingXiMax(1 TO Nd%) : FOR i% = 1 TO Nd% : StartingXiMin(i%) = XiMin(i%) : StartingXiMax(i%) =
XiMax(i%) : NEXT i%

        CASE "F20" '(6-D) Hartman's Family #2

            Nd%                   = 6
            NumProbesPerDimension% = 4 '20
            Np%                   = NumProbesPerDimension%*Nd%

            Nt&     = 500
            G       = 2##
            Alpha   = 2##
            Beta    = 2##
            DeltaT  = 1##
            Frep    = 0.5##

            PlaceInitialProbes$   = "UNIFORM ON-AXIS"
            InitialAcceleration$  = "ZERO"
            RepositionFactor$     = "VARIABLE" '"FIXED"

            CALL
ChangeRunParameters(NumProbesPerDimension%,Np%,Nd%,Nt&,G,Alpha,Beta,DeltaT,Frep,PlaceInitialProbes$,InitialAcceleration$,RepositionFactor$,FunctionN
ame$)

            Nd% = 6 'cannot change dimensionality of Hartman's Family!

            Np% = NumProbesPerDimension%*Nd%

            REDIM XiMin(1 TO Nd%), XiMax(1 TO Nd%) : FOR i% = 1 TO Nd% : XiMin(i%) = 0## : XiMax(i%) = 1## : NEXT i%

            REDIM StartingXiMin(1 TO Nd%), StartingXiMax(1 TO Nd%) : FOR i% = 1 TO Nd% : StartingXiMin(i%) = XiMin(i%) : StartingXiMax(i%) =
XiMax(i%) : NEXT i%

        CASE "F21" '(4-D) Shekel's Family m=5

            Nd%                   = 4
            NumProbesPerDimension% = 4 '20
            Np%                   = NumProbesPerDimension%*Nd%

            Nt&     = 500
            G       = 2##
            Alpha   = 2##
            Beta    = 2##
            DeltaT  = 1##
            Frep    = 0.5##

            PlaceInitialProbes$   = "UNIFORM ON-AXIS"
            InitialAcceleration$  = "ZERO"
            RepositionFactor$     = "VARIABLE" '"FIXED"

            CALL
ChangeRunParameters(NumProbesPerDimension%,Np%,Nd%,Nt&,G,Alpha,Beta,DeltaT,Frep,PlaceInitialProbes$,InitialAcceleration$,RepositionFactor$,FunctionN
ame$)

            Nd% = 4 'cannot change dimensionality of Shekel's Family!

            Np% = NumProbesPerDimension%*Nd%

            REDIM XiMin(1 TO Nd%), XiMax(1 TO Nd%) : FOR i% = 1 TO Nd% : XiMin(i%) = 0## : XiMax(i%) = 10## : NEXT i%

            REDIM StartingXiMin(1 TO Nd%), StartingXiMax(1 TO Nd%) : FOR i% = 1 TO Nd% : StartingXiMin(i%) = XiMin(i%) : StartingXiMax(i%) =
XiMax(i%) : NEXT i%

        CASE "F22" '(4-D) Shekel's Family m=7

            Nd%                   = 4
            NumProbesPerDimension% = 4 '20
            Np%                   = NumProbesPerDimension%*Nd%

            Nt&     = 500
            G       = 2##
            Alpha   = 2##
            Beta    = 2##
            DeltaT  = 1##
            Frep    = 0.5##

            PlaceInitialProbes$   = "UNIFORM ON-AXIS"
            InitialAcceleration$  = "ZERO"
            RepositionFactor$     = "VARIABLE" '"FIXED"

            CALL
ChangeRunParameters(NumProbesPerDimension%,Np%,Nd%,Nt&,G,Alpha,Beta,DeltaT,Frep,PlaceInitialProbes$,InitialAcceleration$,RepositionFactor$,FunctionN
ame$)

            Nd% = 4 'cannot change dimensionality of Shekel's Family!

            Np% = NumProbesPerDimension%*Nd%

            REDIM XiMin(1 TO Nd%), XiMax(1 TO Nd%) : FOR i% = 1 TO Nd% : XiMin(i%) = 0## : XiMax(i%) = 10## : NEXT i%

            REDIM StartingXiMin(1 TO Nd%), StartingXiMax(1 TO Nd%) : FOR i% = 1 TO Nd% : StartingXiMin(i%) = XiMin(i%) : StartingXiMax(i%) =
XiMax(i%) : NEXT i%

        CASE "F23" '(4-D) Shekel's Family m=10

            Nd%                   = 4
            NumProbesPerDimension% = 4 '20
            Np%                   = NumProbesPerDimension%*Nd%

            Nt&     = 500
            G       = 2##
            Alpha   = 2##
            Beta    = 2##
            DeltaT  = 1##
```

```
        Frep     = 0.5##

        PlaceInitialProbes$  = "UNIFORM ON-AXIS"
        InitialAcceleration$ = "ZERO"
        RepositionFactor$    = "VARIABLE" '"FIXED"

        CALL
ChangeRunParameters(NumProbesPerDimension%,Np%,Nd%,Nt&,G,Alpha,Beta,DeltaT,Frep,PlaceInitialProbes$,InitialAcceleration$,RepositionFactor$,FunctionN
ame$)

        Nd% = 4 'cannot change dimensionality of Shekel's Family!

        Np% = NumProbesPerDimension%*Nd%

        REDIM XiMin(1 TO Nd%), XiMax(1 TO Nd%) : FOR i% = 1 TO Nd% : XiMin(i%) = 0## : XiMax(i%) = 10## : NEXT i%

        REDIM StartingXiMin(1 TO Nd%), StartingXiMax(1 TO Nd%) : FOR i% = 1 TO Nd% : StartingXiMin(i%) = XiMin(i%) : StartingXiMax(i%) =
XiMax(i%) : NEXT i%

    CASE "PBM_1" '2-D

        Nd%                  = 2
        NumProbesPerDimension% = 2 '4 '20
        Np%                  = NumProbesPerDimension%*Nd%

        Nt&      = 100
        G        = 2##
        Alpha    = 2##
        Beta     = 2##
        DeltaT   = 1##
        Frep     = 0.5##

        PlaceInitialProbes$  = "UNIFORM ON-AXIS"
        InitialAcceleration$ = "ZERO"
        RepositionFactor$    = "VARIABLE" '"FIXED"

        CALL
ChangeRunParameters(NumProbesPerDimension%,Np%,Nd%,Nt&,G,Alpha,Beta,DeltaT,Frep,PlaceInitialProbes$,InitialAcceleration$,RepositionFactor$,FunctionN
ame$)

        Nd% = 2 'cannot change dimensionality of PBM_1!

        Np% = NumProbesPerDimension%*Nd%

        REDIM XiMin(1 TO Nd%), XiMax(1 TO Nd%)

        XiMin(1) = 0.5## : XiMax(1) = 3## 'dipole length, L, in Wavelengths
        XiMin(2) = 0## : XiMax(2) = Pi2 'polar angle, Theta, in Radians

        REDIM StartingXiMin(1 TO Nd%), StartingXiMax(1 TO Nd%) : FOR i% = 1 TO Nd% : StartingXiMin(i%) = XiMin(i%) : StartingXiMax(i%) =
XiMax(i%) : NEXT i%

        NN% = FREEFILE : OPEN "INFILE.DAT" FOR OUTPUT AS #NN% : PRINT #NN%,"PBM1.NEC" : PRINT #NN%,"PBM1.OUT" : CLOSE #NN% 'NEC Input/Output
Files

    CASE "PBM_2" '2-D

        AddNoiseToPBM2$ = "NO" '"YES" '"NO" '"YES"

        Nd%                  = 2
        NumProbesPerDimension% = 4 '20
        Np%                  = NumProbesPerDimension%*Nd%

        Nt&      = 100
        G        = 2##
        Alpha    = 2##
        Beta     = 2##
        DeltaT   = 1##
        Frep     = 0.5##

        PlaceInitialProbes$  = "UNIFORM ON-AXIS"
        InitialAcceleration$ = "ZERO"
        RepositionFactor$    = "VARIABLE" '"FIXED"

        CALL
ChangeRunParameters(NumProbesPerDimension%,Np%,Nd%,Nt&,G,Alpha,Beta,DeltaT,Frep,PlaceInitialProbes$,InitialAcceleration$,RepositionFactor$,FunctionN
ame$)

        Nd% = 2 'cannot change dimensionality of PBM_2!

        Np% = NumProbesPerDimension%*Nd%

        REDIM XiMin(1 TO Nd%), XiMax(1 TO Nd%)

        XiMin(1) = 5## : XiMax(1) = 15## 'dipole separation, D, in Wavelengths
        XiMin(2) = 0## : XiMax(2) = Pi   'polar angle, Theta, in Radians

        REDIM StartingXiMin(1 TO Nd%), StartingXiMax(1 TO Nd%) : FOR i% = 1 TO Nd% : StartingXiMin(i%) = XiMin(i%) : StartingXiMax(i%) =
XiMax(i%) : NEXT i%

        NN% = FREEFILE : OPEN "INFILE.DAT" FOR OUTPUT AS #NN% : PRINT #NN%,"PBM2.NEC" : PRINT #NN%,"PBM2.OUT" : CLOSE #NN%

    CASE "PBM_3" '2-D

        Nd%                  = 2
        NumProbesPerDimension% = 4 '20
        Np%                  = NumProbesPerDimension%*Nd%

        Nt&      = 100
        G        = 2##
        Alpha    = 2##
        Beta     = 2##
        DeltaT   = 1##
        Frep     = 0.5##

        PlaceInitialProbes$  = "UNIFORM ON-AXIS"
        InitialAcceleration$ = "ZERO"
        RepositionFactor$    = "VARIABLE" '"FIXED"

        CALL
ChangeRunParameters(NumProbesPerDimension%,Np%,Nd%,Nt&,G,Alpha,Beta,DeltaT,Frep,PlaceInitialProbes$,InitialAcceleration$,RepositionFactor$,FunctionN
ame$)

        Nd% = 2 'cannot change dimensionality of PBM_3!
```

```
            Np% = NumProbesPerDimension%*Nd%

            REDIM XiMin(1 TO Nd%), XiMax(1 TO Nd%)

            XiMin(1) = 0## : XiMax(1) = 4## 'Phase Parameter, Beta (0-4)
            XiMin(2) = Pi  : XiMax(2) = Pi 'polar angle, Theta, in Radians

            REDIM StartingXiMin(1 TO Nd%), StartingXiMax(1 TO Nd%) : FOR i% = 1 TO Nd% : StartingXiMin(i%) = XiMin(i%) : StartingXiMax(i%) =
XiMax(i%) : NEXT i%

            NN% = FREEFILE : OPEN "INFILE.DAT" FOR OUTPUT AS #NN% : PRINT #NN%,"PBM3.NEC" : PRINT #NN%,"PBM3.OUT" : CLOSE #NN%

        CASE "PBM_4" '2-D

            Nd%                   = 2
            NumProbesPerDimension% = 4 '6 '2 '4 '20
            Np%                   = NumProbesPerDimension%*Nd%

            Nt&      = 100
            G        = 2##
            Alpha    = 2##
            Beta     = 2##
            DeltaT   = 1##
            Frep     = 0.5##

            PlaceInitialProbes$  = "UNIFORM ON-AXIS"
            InitialAcceleration$ = "ZERO"
            RepositionFactor$    = "VARIABLE" '"FIXED"

            CALL
ChangeRunParameters(NumProbesPerDimension%,Np%,Nd%,Nt&,G,Alpha,Beta,DeltaT,Frep,PlaceInitialProbes$,InitialAcceleration$,RepositionFactor$,FunctionN
ame$)

            Nd% = 2 'cannot change dimensionality of PBM_4!

            Np% = NumProbesPerDimension%*Nd%

            REDIM XiMin(1 TO Nd%), XiMax(1 TO Nd%)

            XiMin(1) = 0.5## : XiMax(1) = 1.5## 'ARM LENGTH (NOT Total Length), wavelengths (0.5-1.5)
            XiMin(2) = Pi/18## : XiMax(2) = Pi/2## 'Inner angle, Alpha, in Radians (Pi/18-Pi/2)

            REDIM StartingXiMin(1 TO Nd%), StartingXiMax(1 TO Nd%) : FOR i% = 1 TO Nd% : StartingXiMin(i%) = XiMin(i%) : StartingXiMax(i%) =
XiMax(i%) : NEXT i%

            NN% = FREEFILE : OPEN "INFILE.DAT" FOR OUTPUT AS #NN% : PRINT #NN%,"PBM4.NEC" : PRINT #NN%,"PBM4.OUT" : CLOSE #NN%

        CASE "PBM_5"

            NumCollinearElements% = 6 '30 'EVEN or ODD: 6,7,10,13,16,24 used by PBM

            Nd%                   = NumCollinearElements% - 1
            NumProbesPerDimension% = 4 '20
            Np%                   = NumProbesPerDimension%*Nd%

            Nt&      = 100
            G        = 2##
            Alpha    = 2##
            Beta     = 2##
            DeltaT   = 1##
            Frep     = 0.5##

            PlaceInitialProbes$  = "UNIFORM ON-AXIS"
            InitialAcceleration$ = "ZERO"
            RepositionFactor$    = "VARIABLE" '"FIXED"

            CALL
ChangeRunParameters(NumProbesPerDimension%,Np%,Nd%,Nt&,G,Alpha,Beta,DeltaT,Frep,PlaceInitialProbes$,InitialAcceleration$,RepositionFactor$,FunctionN
ame$)

            Nd% = NumCollinearElements% - 1

            Np% = NumProbesPerDimension%*Nd%

            REDIM XiMin(1 TO Nd%), XiMax(1 TO Nd%) : FOR i% = 1 TO Nd% : XiMin(i%) = 0.5## : XiMax(i%) = 1.5## : NEXT i%

            REDIM StartingXiMin(1 TO Nd%), StartingXiMax(1 TO Nd%) : FOR i% = 1 TO Nd% : StartingXiMin(i%) = XiMin(i%) : StartingXiMax(i%) =
XiMax(i%) : NEXT i%

            NN% = FREEFILE : OPEN "INFILE.DAT" FOR OUTPUT AS #NN% : PRINT #NN%,"PBM5.NEC" : PRINT #NN%,"PBM5.OUT" : CLOSE #NN%

' =============================================================================================
'  NOTE - DON'T FORGET TO ADD NEW TEST FUNCTIONS TO FUNCTION ObjectiveFunction() ABOVE !!
' =============================================================================================

    END SELECT

    IF Nd% > 100 THEN Nt& = MIN(Nt&,200) 'to avoid array dimensioning problems

    DiagLength = 0## : FOR i% = 1 TO Nd% : DiagLength = DiagLength + (XiMax(i%)-XiMin(i%))^2 : NEXT i% : DiagLength = SQR(DiagLength) 'compute
length of decision space principal diagonal

END SUB 'GetFunctionRunParameters()

'-------------------------------

FUNCTION ParrottF4(R(),Nd%,p%,j&) 'Parrott F4 (1-D)

'MAXIMUM 1 AT -0.0796875... WITH ZERO OFFSET.

'References:

'Beasley, D., D. R. Bull, and R. R. Martin, "A Sequential Niche Technique for Multimodal
'Function Optimization," Evol. Comp. (MIT Press), vol. 1, no. 2, 1993, pp. 101-125
'(online at http://citeseer.ist.psu.edu/beasley93sequential.html).

'Parrott, D., and X. Li, "Locating and Tracking Multiple Dynamic Optima by a Particle Swarm
'Model Using Speciation," IEEE Trans. Evol. Computation, vol. 10, no. 4, Aug. 2006, pp. 440-458.

LOCAL Z, x, offset AS EXT

    offset = 0##
```

```
    x = R(p%,1,j%)

    Z = EXP(-2##*LOG(2##)*((x-0.08##-offset)/0.854##)^2)*(SIN(5##*Pi*((x-offset)^0.75##-0.05##)))^6 'WARNING! This is a NATURAL LOG, NOT Log10!!!

    ParrottF4 = Z

END FUNCTION 'ParrottF4()

'----------------------------

FUNCTION SGO(R(),Nd%,p%,j%) 'SGO Function (2-D)

'MAXIMUM = -130.8323226... @ -(-2.8362075...,-2.8362075...) WITH ZERO OFFSET.

'Reference:

'Hsiao, Y., Chuang, C., Jiang, J., and Chien, C., "A Novel Optimization Algorithm: Space
'Gravitational Optimization," Proc. of 2005 IEEE International Conference on Systems, Man,
'and Cybernetics, 3, 2323-2328. (2005)

    LOCAL x1, x2, Z, t1, t2, SGOx1offset, SGOx2offset AS EXT

    SGOx1offset = 0## : SGOx2offset = 0##

'   SGOx1offset = 40## : SGOx2offset = 10##

    x1 = R(p%,1,j%) - SGOx1offset : x2 = R(p%,2,j%) - SGOx2offset

    t1 = x1^4 - 16##*x1^2 + 0.5##*x1 : t2 = x2^4 - 16##*x2^2 + 0.5##*x2

    Z = t1 + t2

    SGO = -Z

END FUNCTION 'SGO()

'-----------------

FUNCTION GoldsteinPrice(R(),Nd%,p%,j%) 'Goldstein-Price Function (2-D)

'MAXIMUM = -3 @ (0,-1) WITH ZERO OFFSET.

'Reference:

'Cui, Z., Zeng, J., and Sun, G. (2006) 'A Fast Particle Swarm Optimization,' Int'l. J.
'Innovative Computing, Information and Control, vol. 2, no. 6, December, pp. 1365-1380.

    LOCAL Z, x1, x2, offset1, offset2, t1, t2 AS EXT

    offset1 = 0## : offset2 = 0##

'    offset1 = 20## : offset2 = -10##

    x1 = R(p%,1,j%)-offset1 : x2 = R(p%,2,j%)-offset2

    t1 = 1##+(x1+x2+1##)^2*(19##-14##*x1+3##*x1^2-14##*x2+6##*x1*x2+3##*x2^2)

    t2 = 30##+(2##*x1-3##*x2)^2*(18##-32##*x1+12##*x1^2+48##*x2-36##*x1*x2+27##*x2^2)

    Z = t1*t2

    GoldsteinPrice = -Z

END FUNCTION 'GoldesteinPrice()

'-----------

FUNCTION StepFunction(R(),Nd%,p%,j%) 'Step Function (n-D)

'MAXIMUM VALUE = 0 @ [Offset]^n.

'Reference:

'Yao, X., Liu, Y., and Lin, G., "Evolutionary Programming Made Faster,"
'IEEE Trans. Evolutionary Computation, Vol. 3, No. 2, 82-102, Jul. 1999.

    LOCAL Offset, Z AS EXT

    LOCAL i%

    Z = 0## : Offset = 0## '75.123## '0##

    FOR i% = 1 TO Nd%

        IF Nd% = 2 AND i% = 1 THEN Offset = 75 '75##

        IF Nd% = 2 AND i% = 2 THEN Offset = 35 '30 '35##

        Z = Z + INT((R(p%,i%,j%)-Offset) + 0.5##)^2

    NEXT i%

    StepFunction = -Z

END FUNCTION 'StepFunction()

'-----------

FUNCTION Schwefel226(R(),Nd%,p%,j%) 'Schwefel Problem 2.26 (n-D)

'MAXIMUM = 12,569.5 @ [420.8687]^30 (30-D CASE).

'Reference:

'Yao, X., Liu, Y., and Lin, G., "Evolutionary Programming Made Faster,"
'IEEE Trans. Evolutionary Computation, Vol. 3, No. 2, 82-102, Jul. 1999.

    LOCAL Z, Xi AS EXT

    LOCAL i%

    Z = 0##
```

```
     FOR i% = 1 TO Nd%

         Xi = R(p%,i%,j&)

         Z = Z + Xi*SIN(SQR(ABS(Xi)))

     NEXT i%

     Schwefel226 = Z

END FUNCTION 'SCHWEFEL226()

'-----------

FUNCTION Colville(R(),Nd%,p%,j&) 'Colville Function (4-D)

'MAXIMUM = 0 @ (1,1,1,1) WITH ZERO OFFSET.

'Reference:

'Doo-Hyun, and Se-Young, O., "A New Mutation Rule for Evolutionary Programming Motivated from
'Backpropagation Learning," IEEE Trans. Evolutionary Computation, Vol. 4, No. 2, pp. 188-190,
'July 2000.

     LOCAL Z, x1, x2, x3, x4, offset AS EXT

     offset = 0## '7.123##

     x1 = R(p%,1,j&)-offset : x2 = R(p%,2,j&)-offset : x3 = R(p%,3,j&)-offset : x4 = R(p%,4,j&)-offset

     Z =  100##*(x2-x1^2)^2 + (1##-x1)^2  + _
          90##*(x4-x3^2)^2 + (1##-x3)^2 + _
          10.1##*((x2-1##)^2 + (x4-1##)^2) + _
          19.8##*(x2-1##)*(x4-1##)

     Colville = -Z

END FUNCTION 'Colville()

'-----------

'Max of zero at (0,...,0)

FUNCTION Griewank(R(),Nd%,p%,j&) 'Griewank (n-D)

     LOCAL Offset, Sum, Prod, Z, Xi AS EXT

     LOCAL i%

     Sum = 0## : Prod = 1##

     Offset = 75.123##

     FOR i% = 1 TO Nd%

         Xi = R(p%,i%,j&) - Offset

         Sum = Sum + Xi^2

         Prod = Prod*COS(Xi/SQR(i%))

     NEXT i%

     Z = Sum/4000## - Prod + 1##

     Griewank = -Z

END FUNCTION 'Griewank()

'-----------

FUNCTION Himmelblau(R(),Nd%,p%,j&) 'Himmelblau (2-D)

     LOCAL Z, x1, x2, offset AS EXT

     offset = 0##

     x1 = R(p%,1,j&)-offset : x2 = R(p%,2,j&)-offset

     Z = 200## - (x1^2 + x2 -11##)^2 - (x1+x2^2-7##)^2

     Himmelblau = Z

END FUNCTION 'Himmelblau()

'-----------

FUNCTION F1(R(),Nd%,p%,j&) 'F1 (n-D)

'MAXIMUM = ZERO (n-D CASE).

'Reference:

     LOCAL Z, Xi AS EXT

     LOCAL i%

     Z = 0##

     FOR i% = 1 TO Nd%

         Xi = R(p%,i%,j&)

         Z = Z + Xi^2

     NEXT i%

     F1 = -Z

END FUNCTION 'F1

'-----------
```

```
FUNCTION F2(R(),Nd%,p%,j&) 'F2 (n-D)

'MAXIMUM = ZERO (n-D CASE).

'Reference:

    LOCAL Sum, prod, Z, Xi AS EXT

    LOCAL i%

    Z = 0## : Sum = 0## : Prod = 1##

    FOR i% = 1 TO Nd%

        Xi = R(p%,i%,j&)

        Sum  = Sum+ ABS(Xi)

        Prod = Prod*ABS(Xi)

    NEXT i%

    Z = Sum + Prod

    F2 = -Z

END FUNCTION 'F2

'----------

FUNCTION F3(R(),Nd%,p%,j&) 'F3 (n-D)

'MAXIMUM = ZERO (n-D CASE).

'Reference:

    LOCAL Z, Xk, Sum AS EXT

    LOCAL i%, k%

    Z = 0##

    FOR i% = 1 TO Nd%

        Sum = 0##

        FOR k% = 1 TO i%

            Xk = R(p%,k%,j&)

            Sum = Sum + Xk

        NEXT k%

        Z = Z + Sum^2

    NEXT i%

    F3 = -Z

END FUNCTION 'F3

'----------

FUNCTION F4(R(),Nd%,p%,j&) 'F4 (n-D)

'MAXIMUM = ZERO (n-D CASE).

'Reference:

    LOCAL Z, Xi, MaxXi AS EXT

    LOCAL i%

    MaxXi = -1E4200

    FOR i% = 1 TO Nd%

        Xi = R(p%,i%,j&)

        IF ABS(Xi) >= MaxXi THEN MaxXi = ABS(Xi)

    NEXT i%

    F4 = -MaxXi

END FUNCTION 'F4

'----------

FUNCTION F5(R(),Nd%,p%,j&) 'F5 (n-D)

'MAXIMUM = ZERO (n-D CASE).

'Reference:

    LOCAL Z, Xi, XiPlus1 AS EXT

    LOCAL i%

    Z = 0##

    FOR i% = 1 TO Nd%-1

        Xi      = R(p%,i%,j&)

        XiPlus1 = R(p%,i%+1,j&)

        Z = Z + (100##*(XiPlus1-Xi^2)^2+(Xi-1##))^2

    NEXT i%

    F5 = -Z
```

```
END FUNCTION 'F5

'-----------

FUNCTION F6(R(),Nd%,p%,j&) 'F6

'MAXIMUM VALUE = 0 @ [Offset]^n.

'Reference:

'Yao, X., Liu, Y., and Lin, G., "Evolutionary Programming Made Faster,"
'IEEE Trans. Evolutionary Computation, Vol. 3, No. 2, 82-102, Jul. 1999.

    LOCAL Z AS EXT

    LOCAL i%

    Z = 0##

    FOR i% = 1 TO Nd%

        Z = Z + INT(R(p%,i%,j&) + 0.5##)^2

    NEXT i%

    F6 = -Z

END FUNCTION 'F6

'-----------

FUNCTION F7(R(),Nd%,p%,j&) 'F7

'MAXIMUM VALUE = 0 @ [Offset]^n.

'Reference:

'Yao, X., Liu, Y., and Lin, G., "Evolutionary Programming Made Faster,"
'IEEE Trans. Evolutionary Computation, Vol. 3, No. 2, 82-102, Jul. 1999.

    LOCAL Z, Xi AS EXT

    LOCAL i%

    Z = 0##

    FOR i% = 1 TO Nd%

        Xi = R(p%,i%,j&)

        Z = Z + i%*Xi^4

    NEXT i%

    F7 = -Z - RandomNum(0##,1##)

END FUNCTION 'F7

'-----------

FUNCTION F8(R(),Nd%,p%,j&) '(n-D) F8 [Schwefel Problem 2.26]

'MAXIMUM = 12,569.5 @ [420.8687]^30 (30-D CASE).

'Reference:

'Yao, X., Liu, Y., and Lin, G., "Evolutionary Programming Made Faster,"
'IEEE Trans. Evolutionary Computation, Vol. 3, No. 2, 82-102, Jul. 1999.

    LOCAL Z, Xi AS EXT

    LOCAL i%

    Z = 0##

    FOR i% = 1 TO Nd%

        Xi = R(p%,i%,j&)

        Z = Z - Xi*SIN(SQR(ABS(Xi)))

    NEXT i%

    F8 = -Z

END FUNCTION 'F8

'-----------

FUNCTION F9(R(),Nd%,p%,j&) '(n-D) F9 [Rastrigin]

'MAXIMUM = ZERO (n-D CASE).

'Reference:

'Yao, X., Liu, Y., and Lin, G., "Evolutionary Programming Made Faster,"
'IEEE Trans. Evolutionary Computation, Vol. 3, No. 2, 82-102, Jul. 1999.

    LOCAL Z, Xi AS EXT

    LOCAL i%

    Z = 0##

    FOR i% = 1 TO Nd%

        Xi = R(p%,i%,j&)

        Z = Z + (Xi^2 - 10##*COS(TwoPi*Xi) + 10##)^2

    NEXT i%
```

```
    F9 = -Z

END FUNCTION 'F9

'-----------

FUNCTION F10(R(),Nd%,p%,j&) '(n-D) F10 [Ackley's Function]

'MAXIMUM = ZERO (n-D CASE).

'Reference:

'Yao, X., Liu, Y., and Lin, G., "Evolutionary Programming Made Faster,"
'IEEE Trans. Evolutionary Computation, Vol. 3, No. 2, 82-102, Jul. 1999.

    LOCAL Z, Xi, Sum1, Sum2 AS EXT

    LOCAL i%

    Z = 0## : Sum1 = 0## : Sum2 = 0##

    FOR i% = 1 TO Nd%

        Xi  = R(p%,i%,j&)

        Sum1 = Sum1 + Xi^2

        Sum2 = Sum2 + COS(TwoPi*Xi)

    NEXT i%

    Z = -20##*EXP(-0.2##*SQR(Sum1/Nd%)) - EXP(Sum2/Nd%) + 20## + e

    F10 = -Z

END FUNCTION 'F10

'-----------

FUNCTION F11(R(),Nd%,p%,j&) '(n-D) F11

'MAXIMUM = ZERO (n-D CASE).

'Reference:

'Yao, X., Liu, Y., and Lin, G., "Evolutionary Programming Made Faster,"
'IEEE Trans. Evolutionary Computation, Vol. 3, No. 2, 82-102, Jul. 1999.

    LOCAL Z, Xi, Sum, Prod AS EXT

    LOCAL i%

    Z = 0## : Sum = 0## : Prod = 1##

    FOR i% = 1 TO Nd%

        Xi   = R(p%,i%,j&)

        Sum  = Sum + (Xi-100##)^2

        Prod = Prod*COS((Xi-100##)/SQR(i%))

    NEXT i%

    Z = Sum/4000## - Prod + 1##

    F11 = -Z

END FUNCTION 'F11

'-----

FUNCTION F12(R(),Nd%,p%,j&) '(n-D) F12, Penalized #1

    LOCAL Offset, Sum1, Sum2, u, Xi, Xn, XiPlus1, Yi, YiPlus1, Yn, X1, Y1, a, k AS EXT

    LOCAL i%, m%

    a = 10## : k = 100## : m% = 4  '[NOTE IN CORRECTION 02-03-2010: Coefficient "a" was incorrectly set to 5 in the original version!]

    X1 = R(p%,1,j&) : Y1 = 1## + (X1+1##)/4##

    Sum1 = 10##*SIN(Pi*Y1)^2 + (Yn-1##)^2

    FOR i% = 1 TO Nd%-1

        Xi = R(p%,i%,j&) : XiPlus1 = R(p%,i%+1,j&) : Xn = R(p%,Nd%,j&)

        Yi = 1## + (Xi+1##)/4## : YiPlus1 = 1## + (XiPlus1+1##)/4## : Yn = 1## + (Xn+1##)/4##

        Sum1 = Sum1 + (Yi-1##)^2*(1##+10##*SIN(Pi*YiPlus1)^2)

    NEXT i%

    Sum1 = Pi*Sum1/Nd%

    Sum2 = 0##

    FOR i% = 1 TO Nd%

        Xi = R(p%,i%,j&)

        u = 0##

        IF Xi >  a THEN u = k*(Xi-a)^m%

        IF Xi < -a THEN u = k*(-Xi-a)^m%

        Sum2 = Sum2 + u

    NEXT i%

    Z = Sum1 + Sum2
```

```
    F12 = -Z

END FUNCTION 'F12()

'-----

FUNCTION F13(R(),Nd%,p%,j&) '(n-D) F13, Penalized #2

    LOCAL Offset, Sum1, Sum2, Z, u, Xi, Xn, Xi1, X1, a, k AS EXT

    LOCAL i%, m%

    a = 5## : k = 100## : m% = 4

    X1 = R(p%,1,j&)

    Xn = R(p%,Nd%,j&)

    Sum1 = 0##

    FOR i% = 1 TO Nd%-1

        Xi  = R(p%,i%,j&)

        Xi1 = R(p%,i%+1,j&)

        Sum1 = Sum1 + (Xi-1##)^2*(1##+(SIN(3##*Pi*Xi1))^2)+(Xn-1##)^2*(1##+(SIN(TwoPi*Xn))^2)

    NEXT i%

    Sum2 = 0##

    FOR i% = 1 TO Nd%

        Xi = R(p%,i%,j&)
        u = 0##
        IF Xi >  a THEN u = k*(Xi-a)^m%
        IF Xi < -a THEN u = k*(-Xi-a)^m%
        Sum2 = Sum2 + u

    NEXT i%

    Z = ((SIN(3##*Pi*X1))^2+Sum1)/10## + Sum2

    F13 = -Z

END FUNCTION 'F13()

'-----

SUB FillArrayAij  'needed for function F14, Shekel's Foxholes

    Aij(1,1)=-32## : Aij(1,2)=-16## : Aij(1,3)=0## : Aij(1,4)=16## : Aij(1,5)=32##
    Aij(1,6)=-32## : Aij(1,7)=-16## : Aij(1,8)=0## : Aij(1,9)=16## : Aij(1,10)=32##
    Aij(1,12)=-32## : Aij(1,12)=-16## : Aij(1,13)=0## : Aij(1,14)=16## : Aij(1,15)=32##
    Aij(1,16)=-32## : Aij(1,17)=-16## : Aij(1,18)=0## : Aij(1,19)=16## : Aij(1,20)=32##
    Aij(1,21)=-32## : Aij(1,22)=-16## : Aij(1,23)=0## : Aij(1,24)=16## : Aij(1,25)=32##

    Aij(2,1)=-32## : Aij(2,2)=-32## : Aij(2,3)=-32## : Aij(2,4)=-32## : Aij(2,5)=-32##
    Aij(2,6)=-16## : Aij(2,7)=-16## : Aij(2,8)=-16## : Aij(2,9)=-16## : Aij(2,10)=-16##
    Aij(2,12)=0## : Aij(2,12)=0## : Aij(2,13)=0## : Aij(2,14)=0## : Aij(2,15)=0##
    Aij(2,16)=16## : Aij(2,17)=16## : Aij(2,18)=16## : Aij(2,19)=16## : Aij(2,20)=16##
    Aij(2,21)=32## : Aij(2,22)=32## : Aij(2,23)=32## : Aij(2,24)=32## : Aij(2,25)=32##

END SUB

'-----

FUNCTION F14(R(),Nd%,p%,j&) 'F14 (2~D) Shekel's Foxholes (INVERTED...)

    LOCAL Sum1, Sum2, Z, Xi AS EXT

    LOCAL i%, jj%

    Sum1 = 0##

    FOR jj% = 1 TO 25

        Sum2 = 0##

        FOR i% = 1 TO 2

            Xi = R(p%,i%,j&)

            Sum2 = Sum2 + (Xi-Aij(i%,jj%))^6

        NEXT i%

        Sum1 = Sum1 + 1##/(jj%+Sum2)

    NEXT j%

    Z = 1##/(0.002##+Sum1)

    F14 = -Z

END FUNCTION 'F14

'-----------

FUNCTION F16(R(),Nd%,p%,j&) 'F16 (2-D) 6-Hump Camel-Back

    LOCAL x1, x2, Z AS EXT

    x1 = R(p%,1,j&) : x2 = R(p%,2,j&)

    Z = 4##*x1^2 - 2.1##*x1^4 + x1^6/3## + x1*x2 - 4*x2^2 + 4*x2^4

    F16 = -Z

END FUNCTION 'F16

'-----------
```

```
FUNCTION F15(R(),Nd%,p%,j&) 'F15 (4-D) Kowalik's Function

'Global maximum = -0.0003075 @ (0.1928,0.1908,0.1231,0.1358)

    LOCAL x1, x2, x3, x4, Num, Denom, Z, Aj(), Bj() AS EXT

    LOCAL jj%

    REDIM Aj(1 TO 11), Bj(1 TO 11)

    Aj(1)  = 0.1957## : Bj(1)  = 1##/0.25##
    Aj(2)  = 0.1947## : Bj(2)  = 1##/0.50##
    Aj(3)  = 0.1735## : Bj(3)  = 1##/1.00##
    Aj(4)  = 0.1600## : Bj(4)  = 1##/2.00##
    Aj(5)  = 0.0844## : Bj(5)  = 1##/4.00##
    Aj(6)  = 0.0627## : Bj(6)  = 1##/6.00##
    Aj(7)  = 0.0456## : Bj(7)  = 1##/8.00##
    Aj(8)  = 0.0342## : Bj(8)  = 1##/10.0##
    Aj(9)  = 0.0323## : Bj(9)  = 1##/12.0##
    Aj(10) = 0.0235## : Bj(10) = 1##/14.0##
    Aj(11) = 0.0246## : Bj(11) = 1##/16.0##

    Z = 0##

    x1 = R(p%,1,j&) : x2 = R(p%,2,j&) : x3 = R(p%,3,j&) : x4 = R(p%,4,j&)

    FOR jj% = 1 TO 11

        Num   = x1*(Bj(jj%)^2+Bj(jj%)*x2)

        Denom = Bj(jj%)^2+Bj(jj%)*x3+x4

        Z = Z + (Aj(jj%)-Num/Denom)^2

    NEXT jj%

    F15 = -Z

END FUNCTION 'F15

'----------

FUNCTION F17(R(),Nd%,p%,j&) 'F17, (2-D) Branin

'Global maximum = -0.398 @ (-3.142.12.275), (3.142,2.275), (9.425,2.425)

    LOCAL x1, x2, Z AS EXT

    x1 = R(p%,1,j&) : x2 = R(p%,2,j&)

    Z = (x2-5.1##*x1^2/(4##*Pi^2)+5##*x1/Pi-6##)^2 + 10##*(1##-1##/(8##*Pi))*COS(x1) + 10##

    F17 = -Z

END FUNCTION 'F17

'----------

FUNCTION F18(R(),Nd%,p%,j&) 'Goldstein-Price 2-D Test Function

'Global maximum = -3 @ (0,-1)

    LOCAL Z, x1, x2, t1, t2 AS EXT

    x1 = R(p%,1,j&) : x2 = R(p%,2,j&)

    t1 = 1##+(x1+x2+1##)^2*(19##-14##*x1+3##*x1^2-14##*x2+6##*x1*x2+3##*x2^2)

    t2 = 30##+(2##*x1-3##*x2)^2*(18##-32##*x1+12##*x1^2+48##*x2-36##*x1*x2+27##*x2^2)

    Z  = t1*t2

    F18 = -Z

END FUNCTION 'F18()

'----------

FUNCTION F19(R(),Nd%,p%,j&) 'F19 (3-D) Hartman's Family #1

'Global maximum = 3.86 @ (0.114,0.556,0.852)

    LOCAL Xi, Z, Sum, Aji(), Cj(), Pji() AS EXT

    LOCAL i%, jj%, m%

    REDIM Aji(1 TO 4, 1 TO 3), Cj(1 TO 4), Pji(1 TO 4, 1 TO 3)

    Aji(1,1) = 3.0## : Aji(1,2) = 10## : Aji(1,3) = 30## : Cj(1) = 1.0##
    Aji(2,1) = 0.1## : Aji(2,2) = 10## : Aji(2,3) = 35## : Cj(2) = 1.2##
    Aji(3,1) = 3.0## : Aji(3,2) = 10## : Aji(3,3) = 30## : Cj(3) = 3.0##
    Aji(4,1) = 0.1## : Aji(4,2) = 10## : Aji(4,3) = 35## : Cj(4) = 3.2##

    Pji(1,1) = 0.36890## : Pji(1,2) = 0.1170## : Pji(1,3) = 0.2673##
    Pji(2,1) = 0.46990## : Pji(2,2) = 0.4387## : Pji(2,3) = 0.7470##
    Pji(3,1) = 0.10910## : Pji(3,2) = 0.8732## : Pji(3,3) = 0.5547##
    Pji(4,1) = 0.03815## : Pji(4,2) = 0.5743## : Pji(4,3) = 0.8828##

    Z = 0##

    FOR jj% = 1 TO 4

        Sum = 0##

        FOR i% = 1 TO 3

            Xi = R(p%,i%,j&)

            Sum = Sum + Aji(jj%,i%)*(Xi-Pji(jj%,i%))^2

        NEXT i%

        Z = Z + Cj(jj%)*EXP(-Sum)
```

```
    NEXT jj%

    F19 = Z

END FUNCTION 'F19

'-----------

FUNCTION F20(R(),Nd%,p%,j&) 'F20 (6-D) Hartman's Family #2

'Global maximum = 3.32 @ (0.201,0.150,0.477,0.275,0.311,0.657)

    LOCAL Xi, Z, Sum, Aji(), Cj(), Pji() AS EXT

    LOCAL i%, jj%, m%

    REDIM Aji(1 TO 4, 1 TO 6), Cj(1 TO 4), Pji(1 TO 4, 1 TO 6)

    Aji(1,1) = 10.0## : Aji(1,2) =  3.00## : Aji(1,3) = 17.0## : Cj(1) = 1.0##
    Aji(2,1) =  0.05## : Aji(2,2) = 10.0## : Aji(2,3) = 17.0## : Cj(2) = 1.2##
    Aji(3,1) =  3.00## : Aji(3,2) =  3.50## : Aji(3,3) =  1.70## : Cj(3) = 3.0##
    Aji(4,1) = 17.0## : Aji(4,2) =  8.00## : Aji(4,3) =  0.05## : Cj(4) = 3.2##

    Aji(1,4) =  3.5## : Aji(1,5) =  1.7## : Aji(1,6) =   8##
    Aji(2,4) =  0.1## : Aji(2,5) =    8## : Aji(2,6) = 14##
    Aji(3,4) =   10## : Aji(3,5) =   17## : Aji(3,6) =   8##
    Aji(4,4) =   10## : Aji(4,5) =  0.1## : Aji(4,6) = 14##

    Pji(1,1) = 0.13120## : Pji(1,2) = 0.1696## : Pji(1,3) = 0.5569##
    Pji(2,1) = 0.23290## : Pji(2,2) = 0.4135## : Pji(2,3) = 0.8307##
    Pji(3,1) = 0.23480## : Pji(3,2) = 0.1415## : Pji(3,3) = 0.3522##
    Pji(4,1) = 0.40470## : Pji(4,2) = 0.8828## : Pji(4,3) = 0.8732##

    Pji(1,4) = 0.01240## : Pji(1,5) = 0.8283## : Pji(1,6) = 0.5886##
    Pji(2,4) = 0.37360## : Pji(2,5) = 0.1004## : Pji(2,6) = 0.9991##
    Pji(3,4) = 0.28830## : Pji(3,5) = 0.3047## : Pji(3,6) = 0.6650##
    Pji(4,4) = 0.57430## : Pji(4,5) = 0.1091## : Pji(4,6) = 0.0381##

    Z = 0##

    FOR jj% = 1 TO 4

        Sum = 0##

        FOR i% = 1 TO 6

            Xi = R(p%,i%,j&)

            Sum = Sum + Aji(jj%,i%)*(Xi-Pji(jj%,i%))^2

        NEXT i%

        Z = Z + Cj(jj%)*EXP(-Sum)

    NEXT jj%

    F20 = Z

END FUNCTION 'F20

'-----------

FUNCTION F21(R(),Nd%,p%,j&) 'F21 (4-D) Shekel's Family m=5

'Global maximum = 10

    LOCAL Xi, Z, Sum, Aji(), Cj() AS EXT

    LOCAL i%, jj%, m%

    m% = 5 : REDIM Aji(1 TO m%, 1 TO 4), Cj(1 TO m%)

    Aji(1,1) =  4## : Aji(1,2) =   4## : Aji(1,3) =  4## : Aji(1,4) =   4## : Cj(1) = 0.1##
    Aji(2,1) =  1## : Aji(2,2) =   1## : Aji(2,3) =  1## : Aji(2,4) =   1## : Cj(2) = 0.2##
    Aji(3,1) =  8## : Aji(3,2) =   8## : Aji(3,3) =  8## : Aji(3,4) =   8## : Cj(3) = 0.2##
    Aji(4,1) =  6## : Aji(4,2) =   6## : Aji(4,3) =  6## : Aji(4,4) =   6## : Cj(4) = 0.4##
    Aji(5,1) =  3## : Aji(5,2) =   7## : Aji(5,3) =  3## : Aji(5,4) =   7## : Cj(5) = 0.4##

    Z = 0##

    FOR jj% = 1 TO m%   'NOTE:  Index jj% is used to avoid same variable name as j&

        Sum = 0##

        FOR i% = 1 TO 4   'Shekel's family is 4-D only

            Xi = R(p%,i%,j&)

            Sum = Sum + (Xi-Aji(jj%,i%))^2

        NEXT i%

        Z = Z + 1##/(Sum + Cj(jj%))

    NEXT jj%

    F21 = Z

END FUNCTION 'F21

'-----------

FUNCTION F22(R(),Nd%,p%,j&) 'F22 (4-D) Shekel's Family m=7

'Global maximum = 10

    LOCAL Xi, Z, Sum, Aji(), Cj() AS EXT

    LOCAL i%, jj%, m%

    m% = 7 : REDIM Aji(1 TO m%, 1 TO 4), Cj(1 TO m%)

    Aji(1,1) =  4## : Aji(1,2) =   4## : Aji(1,3) =  4## : Aji(1,4) =   4## : Cj(1) = 0.1##
```

```
        Aji(2,1)  = 1##  : Aji(2,2)  =   1## : Aji(2,3)  = 1## : Aji(2,4)   =   1## : Cj(2)  = 0.2##
        Aji(3,1)  = 8##  : Aji(3,2)  =   8## : Aji(3,3)  = 8## : Aji(3,4)   =   8## : Cj(3)  = 0.2##
        Aji(4,1)  = 6##  : Aji(4,2)  =   6## : Aji(4,3)  = 6## : Aji(4,4)   =   6## : Cj(4)  = 0.4##
        Aji(5,1)  = 3##  : Aji(5,2)  =   7## : Aji(5,3)  = 3## : Aji(5,4)   =   7## : Cj(5)  = 0.4##
        Aji(6,1)  = 2##  : Aji(6,2)  =   9## : Aji(6,3)  = 2## : Aji(6,4)   =   9## : Cj(6)  = 0.6##
        Aji(7,1)  = 5##  : Aji(7,2)  =   5## : Aji(7,3)  = 3## : Aji(7,4)   =   3## : Cj(7)  = 0.3##

    Z = 0##

    FOR jj% = 1 TO m%  'NOTE:  Index jj% is used to avoid same variable name as j&

        Sum = 0##

        FOR i% = 1 TO 4 'Shekel's family is 4-D only

            Xi = R(p%,i%,j&)

            Sum = Sum + (Xi-Aji(jj%,i%))^2

        NEXT i%

        Z = Z + 1##/(Sum + Cj(jj%))

    NEXT jj%

    F22 = Z

END FUNCTION 'F22

'-----------

FUNCTION F23(R(),Nd%,p%,j&) 'F23 (4-D) Shekel's Family m=10

'Global maximum = 10

    LOCAL Xi, Z, Sum, Aji(), Cj() AS EXT

    LOCAL i%, jj%, m%

    m% = 10 : REDIM Aji(1 TO m%, 1 TO 4), Cj(1 TO m%)

        Aji(1,1)  = 4##  : Aji(1,2)  =   4## : Aji(1,3)  = 4## : Aji(1,4)   =   4## : Cj(1)  = 0.1##
        Aji(2,1)  = 1##  : Aji(2,2)  =   1## : Aji(2,3)  = 1## : Aji(2,4)   =   1## : Cj(2)  = 0.2##
        Aji(3,1)  = 8##  : Aji(3,2)  =   8## : Aji(3,3)  = 8## : Aji(3,4)   =   8## : Cj(3)  = 0.2##
        Aji(4,1)  = 6##  : Aji(4,2)  =   6## : Aji(4,3)  = 6## : Aji(4,4)   =   6## : Cj(4)  = 0.4##
        Aji(5,1)  = 3##  : Aji(5,2)  =   7## : Aji(5,3)  = 3## : Aji(5,4)   =   7## : Cj(5)  = 0.4##
        Aji(6,1)  = 2##  : Aji(6,2)  =   9## : Aji(6,3)  = 2## : Aji(6,4)   =   9## : Cj(6)  = 0.6##
        Aji(7,1)  = 5##  : Aji(7,2)  =   5## : Aji(7,3)  = 3## : Aji(7,4)   =   3## : Cj(7)  = 0.3##
        Aji(8,1)  = 8##  : Aji(8,2)  =   1## : Aji(8,3)  = 8## : Aji(8,4)   =   1## : Cj(8)  = 0.7##
        Aji(9,1)  = 6##  : Aji(9,2)  =   2## : Aji(9,3)  = 6## : Aji(9,4)   =   2## : Cj(9)  = 0.5##
        Aji(10,1) = 7##  : Aji(10,2) = 3.6## : Aji(10,3) = 7## : Aji(10,4)  = 3.6## : Cj(10) = 0.5##

    Z = 0##

    FOR jj% = 1 TO m%   'NOTE:  Index jj% is used to avoid same variable name as j&

        Sum = 0##

        FOR i% = 1 TO 4 'Shekel's family is 4-D only

            Xi = R(p%,i%,j&)

            Sum = Sum + (Xi-Aji(jj%,i%))^2

        NEXT i%

        Z = Z + 1##/(Sum + Cj(jj%))

    NEXT jj%

    F23 = Z

END FUNCTION 'F23

'====================================================== END FUNCTION DEFINITIONS =======================================================

SUB Plot2bbestProbeTrajectories(NumTrajectories%,M(),R(),XiMin(),XiMax(),Np%,Nd%,LastStep&,FunctionName$)

LOCAL TrajectoryNumber%, ProbeNumber%, StepNumber&, N%, M%, ProcID???

LOCAL MaximumFitness, MinimumFitness AS EXT

LOCAL BestProbeThisStep%()

LOCAL BestFitnessThisStep(), TempFitness() AS EXT

LOCAL Annotation$, xCoord$, yCoord$, GnuPlotEXE$, PlotWithLines$

    Annotation$    = ""

    PlotWithLines$ = "YES" '"NO"

    NumTrajectories% = MIN(Np%,NumTrajectories%)

    GnuPlotEXE$ = "wgnuplot.exe"

'    ---------------- Get Min/Max Fitnesses ----------------

    MaximumFitness = M(1,0) : MinimumFitness = M(1,0)  'Note:  M(p%,j&)

    FOR StepNumber& = 0 TO LastStep&

        FOR ProbeNumber% = 1 TO Np%

            IF M(ProbeNumber%,StepNumber&) >= MaximumFitness THEN MaximumFitness = M(ProbeNumber%,StepNumber&)

            IF M(ProbeNumber%,StepNumber&) =< MinimumFitness THEN MinimumFitness = M(ProbeNumber%,StepNumber&)

        NEXT ProbeNumber%

    NEXT StepNumber%

'    ------------- Copy Fitness Array M() into TempFitness to Preserve M() ----------------
```

```
        REDIM TempFitness(1 TO Np%, 0 TO LastStep&)

    FOR StepNumber& = 0 TO LastStep&

        FOR ProbeNumber% = 1 TO Np%

            TempFitness(ProbeNumber%,StepNumber&) = M(ProbeNumber%,StepNumber&)

        NEXT ProbeNumber%

    NEXT StepNumber&

'    ----------- LOOP ON TRAJECTORIES -----------

    FOR TrajectoryNumber% = 1 TO NumTrajectories%

'       -------------- Get Trajectory Coordinate Data ----------------

        REDIM BestFitnessThisStep(0 TO LastStep&), BestProbeThisStep%(0 TO LastStep&)

        FOR StepNumber& = 0 TO LastStep&

            BestFitnessThisStep(StepNumber&) = TempFitness(1,StepNumber&)

            FOR ProbeNumber% = 1 TO Np%

                IF TempFitness(ProbeNumber%,StepNumber&) >= BestFitnessThisStep(StepNumber&) THEN

                    BestFitnessThisStep(StepNumber&) = TempFitness(ProbeNumber%,StepNumber&)

                    BestProbeThisStep%(StepNumber&)  = ProbeNumber%

                END IF

            NEXT ProbeNumber%

        NEXT StepNumber&

'    ----- Create Plot Data File -----

        N% = FREEFILE

        SELECT CASE TrajectoryNumber%

            CASE 1  : OPEN "t1"  FOR OUTPUT AS #N%
            CASE 2  : OPEN "t2"  FOR OUTPUT AS #N%
            CASE 3  : OPEN "t3"  FOR OUTPUT AS #N%
            CASE 4  : OPEN "t4"  FOR OUTPUT AS #N%
            CASE 5  : OPEN "t5"  FOR OUTPUT AS #N%
            CASE 6  : OPEN "t6"  FOR OUTPUT AS #N%
            CASE 7  : OPEN "t7"  FOR OUTPUT AS #N%
            CASE 8  : OPEN "t8"  FOR OUTPUT AS #N%
            CASE 9  : OPEN "t9"  FOR OUTPUT AS #N%
            CASE 10 : OPEN "t10" FOR OUTPUT AS #N%

        END SELECT

'    ----------- Write Plot File Data -----------

'    M% = freefile : open "BestProbebData" for output as #M% 'debug

'    print #M%, "  Step #  BestProbe#          x1               x2"

        FOR StepNumber& = 0 TO LastStep&

            PRINT #N%, USING$("######.########
######.########",R(BestProbeThisStep%(StepNumber&),1,StepNumber&),R(BestProbeThisStep%(StepNumber&),2,StepNumber&))

'            PRINT #M%, USING$("#####    #####       ######.########
######.########",StepNumber&,BestProbeThisStep%(StepNumber&),R(BestProbeThisStep%(StepNumber&),1,StepNumber&),R(BestProbeThisStep%(StepNumber&),2,St
epNumber&))

            TempFitness(BestProbeThisStep%(StepNumber&),StepNumber&) = MinimumFitness 'so that same max will not be found for next trajectory

        NEXT StepNumber&

        CLOSE #N%

'    Close #M%

    NEXT TrajectoryNumber%

'    ----------------------- Plot Trajectories -----------------------

    CALL CreateGNUplotINIfile(0.13##*ScreenWidth&,0.18##*ScreenHeight&,0.7##*ScreenWidth&,0.7##*ScreenHeight&)

    Annotation$ = ""

    N% = FREEFILE

    OPEN "cmd2d.gp" FOR OUTPUT AS #N%

        PRINT #N%, "set xrange ["+REMOVE$(STR$(XiMin(1)),ANY" ")+":"+REMOVE$(STR$(XiMax(1)),ANY" ")+"]"
        PRINT #N%, "set yrange ["+REMOVE$(STR$(XiMin(2)),ANY" ")+":"+REMOVE$(STR$(XiMax(2)),ANY" ")+"]"

        'PRINT #N%, "set label " + Annotation$ + Quote$ + " at graph " + xCoord$ + "," + yCoord$
        PRINT #N%, "set grid xtics " + "10"
        PRINT #N%, "set grid ytics " + "10"
        PRINT #N%, "set grid mxtics"
        PRINT #N%, "set grid mytics"
        PRINT #N%, "show grid"
        PRINT #N%, "set title " + Quote$ +"2D "+ FunctionName$+" TRAJECTORIES OF PROBES WITH BEST\nFITNESSES (ORDERED BY FITNESS)" + "\n" + RunID$
+ Quote$
        PRINT #N%, "set xlabel " + Quote$ + "x1\n\n"                              + Quote$
        PRINT #N%, "set ylabel " + Quote$ + "\nx2"                               + Quote$

        IF PlotWithLines$ = "YES" THEN

            SELECT CASE NumTrajectories%

                CASE 1 : PRINT #N%, "plot "+Quote$+"t1"+Quote$+" w l lw 3"
                CASE 2 : PRINT #N%, "plot "+Quote$+"t1"+Quote$+" w l lw 3,"+Quote$+"t2"+Quote$+" w l"
                CASE 3 : PRINT #N%, "plot "+Quote$+"t1"+Quote$+" w l lw 3,"+Quote$+"t2"+Quote$+" w l,"+Quote$+"t3"+Quote$+" w l"
```

```
          CASE 4  : PRINT #N%, "plot "+Quote$+"t1"+Quote$+" w 1 lw 3,"+Quote$+"t2"+Quote$+" w 1,"+Quote$+"t3"+Quote$+" w
1,"+Quote$+"t4"+Quote$+" w 1"
          CASE 5  : PRINT #N%, "plot "+Quote$+"t1"+Quote$+" w 1 lw 3,"+Quote$+"t2"+Quote$+" w 1,"+Quote$+"t3"+Quote$+" w
1,"+Quote$+"t4"+Quote$+" w 1,"+Quote$+"t5"+Quote$+" w 1"
          CASE 6  : PRINT #N%, "plot "+Quote$+"t1"+Quote$+" w 1 lw 3,"+Quote$+"t2"+Quote$+" w 1,"+Quote$+"t3"+Quote$+" w
1,"+Quote$+"t4"+Quote$+" w 1,"+Quote$+"t5"+Quote$+" w 1,"+Quote$+"t6"+Quote$+" w 1"
          CASE 7  : PRINT #N%, "plot "+Quote$+"t1"+Quote$+" w 1 lw 3,"+Quote$+"t2"+Quote$+" w 1,"+Quote$+"t3"+Quote$+" w
1,"+Quote$+"t4"+Quote$+" w 1,"+Quote$+"t5"+Quote$+" w 1,"+Quote$+"t6"+Quote$+" w 1,"+_
                               Quote$+"t7"+Quote$+" w 1"
          CASE 8  : PRINT #N%, "plot "+Quote$+"t1"+Quote$+" w 1 lw 3,"+Quote$+"t2"+Quote$+" w 1,"+Quote$+"t3"+Quote$+" w
1,"+Quote$+"t4"+Quote$+" w 1,"+Quote$+"t5"+Quote$+" w 1,"+Quote$+"t6"+Quote$+" w 1,"+_
                               Quote$+"t7"+Quote$+" w 1,"     +Quote$+"t8"+Quote$+" w 1"
          CASE 9  : PRINT #N%, "plot "+Quote$+"t1"+Quote$+" w 1 lw 3,"+Quote$+"t2"+Quote$+" w 1,"+Quote$+"t3"+Quote$+" w
1,"+Quote$+"t4"+Quote$+" w 1,"+Quote$+"t5"+Quote$+" w 1,"+Quote$+"t6"+Quote$+" w 1,"+_
                               Quote$+"t7"+Quote$+" w 1,"     +Quote$+"t8"+Quote$+" w 1,"+Quote$+"t9"+Quote$+" w 1"
          CASE 10 : PRINT #N%, "plot "+Quote$+"t1"+Quote$+" w 1 lw 3,"+Quote$+"t2"+Quote$+" w 1,"+Quote$+"t3"+Quote$+" w
1,"+Quote$+"t4"+Quote$+" w 1,"+Quote$+"t5"+Quote$+" w 1,"+Quote$+"t6"+Quote$+" w 1,"+_
                               Quote$+"t7"+Quote$+" w 1,"     +Quote$+"t8"+Quote$+" w 1,"+Quote$+"t9"+Quote$+" w 1"
1,"+Quote$+"t10"+Quote$+" w 1"
          END SELECT

        ELSE

          SELECT CASE NumTrajectories%

             CASE 1  : PRINT #N%, "plot "+Quote$+"t1"+Quote$+" lw 2"
             CASE 2  : PRINT #N%, "plot "+Quote$+"t1"+Quote$+" lw 2,"+Quote$+"t2"+Quote$
             CASE 3  : PRINT #N%, "plot "+Quote$+"t1"+Quote$+" lw 2,"+Quote$+"t2"+Quote$+" ,"+Quote$+"t3"+Quote$
             CASE 4  : PRINT #N%, "plot "+Quote$+"t1"+Quote$+" lw 2,"+Quote$+"t2"+Quote$+" ,"+Quote$+"t3"+Quote$+" ,"+Quote$+"t4"+Quote$
             CASE 5  : PRINT #N%, "plot "+Quote$+"t1"+Quote$+" lw 2,"+Quote$+"t2"+Quote$+" ,"+Quote$+"t3"+Quote$+" ,"+Quote$+"t4"+Quote$+"
,"+Quote$+"t5"+Quote$
             CASE 6  : PRINT #N%, "plot "+Quote$+"t1"+Quote$+" lw 2,"+Quote$+"t2"+Quote$+" ,"+Quote$+"t3"+Quote$+" ,"+Quote$+"t4"+Quote$+"
,"+Quote$+"t5"+Quote$+" ,"+Quote$+"t6"+Quote$
             CASE 7  : PRINT #N%, "plot "+Quote$+"t1"+Quote$+" lw 2,"+Quote$+"t2"+Quote$+" ,"+Quote$+"t3"+Quote$+" ,"+Quote$+"t4"+Quote$+"
,"+Quote$+"t5"+Quote$+" ,"+Quote$+"t6"+Quote$+" ,"+_
                               Quote$+"t7"+Quote$
             CASE 8  : PRINT #N%, "plot "+Quote$+"t1"+Quote$+" lw 2,"+Quote$+"t2"+Quote$+" ,"+Quote$+"t3"+Quote$+" ,"+Quote$+"t4"+Quote$+"
,"+Quote$+"t5"+Quote$+" ,"+Quote$+"t6"+Quote$+" ,"+_
                               Quote$+"t7"+Quote$+" ,"     +Quote$+"t8"+Quote$
             CASE 9  : PRINT #N%, "plot "+Quote$+"t1"+Quote$+" lw 2,"+Quote$+"t2"+Quote$+" ,"+Quote$+"t3"+Quote$+" ,"+Quote$+"t4"+Quote$+"
,"+Quote$+"t5"+Quote$+" ,"+Quote$+"t6"+Quote$+" ,"+_
                               Quote$+"t7"+Quote$+" ,"     +Quote$+"t8"+Quote$+" ,"+Quote$+"t9"+Quote$
             CASE 10 : PRINT #N%, "plot "+Quote$+"t1"+Quote$+" lw 2,"+Quote$+"t2"+Quote$+" ,"+Quote$+"t3"+Quote$+" ,"+Quote$+"t4"+Quote$+"
,"+Quote$+"t5"+Quote$+" ,"+Quote$+"t6"+Quote$+" ,"+_
                               Quote$+"t7"+Quote$+" ,"     +Quote$+"t8"+Quote$+" ,"+Quote$+"t9"+Quote$+" ,"+Quote$+"t10"+Quote$

          END SELECT

       END IF

     CLOSE #N%

     ProcID??? = SHELL(GnuPlotEXE$+" cmd2d.gp -") : CALL Delay(0.5##)

END SUB 'Plot2DbestProbeTrajectories()

'----

SUB Plot2DindividualProbeTrajectories(NumTrajectories%,M(),R(),XiMin(),XiMax(),Np%,Nd%,LastStep&,FunctionName$)

LOCAL ProbeNumber%, StepNumber&, N%, ProcID???

LOCAL Annotation$, xCoord$, yCoord$, GnuPlotEXE$, PlotWithLines$

     NumTrajectories% = MIN(Np%,NumTrajectories%)

     Annotation$    = ""

     PlotWithLines$ = "YES" '"NO"

     GnuPlotEXE$ = "wgnuplot.exe"

'    ------------- LOOP ON PROBES --------------

     FOR ProbeNumber% = 1 TO MIN(NumTrajectories%,Np%)

'    ----- Create Plot Data File -----

     N% = FREEFILE

     SELECT CASE ProbeNumber%

        CASE 1  : OPEN "p1"  FOR OUTPUT AS #N%
        CASE 2  : OPEN "p2"  FOR OUTPUT AS #N%
        CASE 3  : OPEN "p3"  FOR OUTPUT AS #N%
        CASE 4  : OPEN "p4"  FOR OUTPUT AS #N%
        CASE 5  : OPEN "p5"  FOR OUTPUT AS #N%
        CASE 6  : OPEN "p6"  FOR OUTPUT AS #N%
        CASE 7  : OPEN "p7"  FOR OUTPUT AS #N%
        CASE 8  : OPEN "p8"  FOR OUTPUT AS #N%
        CASE 9  : OPEN "p9"  FOR OUTPUT AS #N%
        CASE 10 : OPEN "p10" FOR OUTPUT AS #N%
        CASE 11 : OPEN "p11" FOR OUTPUT AS #N%
        CASE 12 : OPEN "p12" FOR OUTPUT AS #N%
        CASE 13 : OPEN "p13" FOR OUTPUT AS #N%
        CASE 14 : OPEN "p14" FOR OUTPUT AS #N%
        CASE 15 : OPEN "p15" FOR OUTPUT AS #N%
        CASE 16 : OPEN "p16" FOR OUTPUT AS #N%

     END SELECT

'    ----------- Write Plot File Data ------------

     FOR StepNumber& = 0 TO LastStep&

        PRINT #N%, USING$("######.######## ######.########",R(ProbeNumber%,1,StepNumber&),R(ProbeNumber%,2,StepNumber&))

     NEXT StepNumber%

     CLOSE #N%

     NEXT ProbeNumber%

'    ----------------------------------------- Plot Trajectories -----------------------------------------
```

```
'usage:  CALL CreateGNUplotINIfile(PlotWindowULC_X%,PlotWindowULC_Y%,PlotWindowWidth%,PlotWindowHeight%)

    CALL CreateGNUplotINIfile(0.17##*ScreenWidth&,0.22##*ScreenHeight&,0.7##*ScreenHeight&,0.7##*ScreenHeight&)

    Annotation$ = ""

    N% = FREEFILE

    OPEN "cmd2d.gp" FOR OUTPUT AS #N%

        PRINT #N%, "set xrange ["+REMOVE$(STR$(XiMin(1)),ANY"" ")+":"+REMOVE$(STR$(XiMax(1)),ANY" ")+"]"
        PRINT #N%, "set yrange ["+REMOVE$(STR$(XiMin(2)),ANY" ")+":"+REMOVE$(STR$(XiMax(2)),ANY" ")+"]"

        'PRINT #N%, "set label "  + Quote$ + Annotation$ + Quote$ + " at graph " + xCoord$ + "," + yCoord$
        PRINT #N%, "set grid xtics " + "10"
        PRINT #N%, "set grid ytics " + "10"
        PRINT #N%, "set grid mxtics"
        PRINT #N%, "set grid mytics"
        PRINT #N%, "show grid"
        PRINT #N%, "set title " + Quote$ + "2D "+ FunctionName$+" INDIVIDUAL PROBE TRAJECTORIES\n(ORDERED BY PROBE #)" + "\n" + RunID$ + Quote$
        PRINT #N%, "set xlabel " + Quote$ + "x1\n\n"                        + Quote$
        PRINT #N%, "set ylabel " + Quote$ + "\nx2"                          + Quote$

        IF PlotWithLines$ = "YES" THEN

            SELECT CASE NumTrajectories%

                CASE 1  : PRINT #N%, "plot "+Quote$+"p1"  +Quote$+" w l lw 1"
                CASE 2  : PRINT #N%, "plot "+Quote$+"p1"  +Quote$+" w l lw 1,"+Quote$+"p2"+Quote$+" w 1"
                CASE 3  : PRINT #N%, "plot "+Quote$+"p1"  +Quote$+" w l lw 1,"+Quote$+"p2"+Quote$+" w 1,"+Quote$+"p3"+Quote$+" w 1"
                CASE 4  : PRINT #N%, "plot "+Quote$+"p1"  +Quote$+" w l lw 1,"+Quote$+"p2"+Quote$+" w 1,"+Quote$+"p3"+Quote$+" w
1,"+Quote$+"p4"+Quote$+" w 1"
                CASE 5  : PRINT #N%, "plot "+Quote$+"p1"  +Quote$+" w l lw 1,"+Quote$+"p2"+Quote$+" w 1,"+Quote$+"p3"+Quote$+" w
1,"+Quote$+"p4"+Quote$+" w 1,"+Quote$+"p5"+Quote$+" w 1"
                CASE 6  : PRINT #N%, "plot "+Quote$+"p1"  +Quote$+" w l lw 1,"+Quote$+"p2"+Quote$+" w 1,"+Quote$+"p3"+Quote$+" w
1,"+Quote$+"p4"+Quote$+" w 1,"+Quote$+"p5"+Quote$+" w 1,"+Quote$+"p6"+Quote$+" w 1"
                CASE 7  : PRINT #N%, "plot "+Quote$+"p1"  +Quote$+" w l lw 1,"+Quote$+"p2"+Quote$+" w 1,"+Quote$+"p3"+Quote$+" w
1,"+Quote$+"p4"+Quote$+" w 1,"+Quote$+"p5"+Quote$+" w 1,"+Quote$+"p6"+Quote$+" w 1,"+_
                                     Quote$+"p7"  +Quote$+" w 1"
                CASE 8  : PRINT #N%, "plot "+Quote$+"p1"  +Quote$+" w l lw 1,"+Quote$+"p2"+Quote$+" w 1,"+Quote$+"p3"+Quote$+" w
1,"+Quote$+"p4"+Quote$+" w 1,"+Quote$+"p5"+Quote$+" w 1,"+Quote$+"p6"+Quote$+" w 1,"+_
                                     Quote$+"p7"  +Quote$+" w 1,"    +Quote$+"p8"+Quote$+" w 1"
                CASE 9  : PRINT #N%, "plot "+Quote$+"p1"  +Quote$+" w l lw 1,"+Quote$+"p2"+Quote$+" w 1,"+Quote$+"p3"+Quote$+" w
1,"+Quote$+"p4"+Quote$+" w 1,"+Quote$+"p5"+Quote$+" w 1,"+Quote$+"p6"+Quote$+" w 1,"+_
                                     Quote$+"p7"  +Quote$+" w 1,"    +Quote$+"p8"+Quote$+" w 1,"+Quote$+"p9"+Quote$+" w 1"

                CASE 10 : PRINT #N%, "plot "+Quote$+"p1"  +Quote$+" w l lw 1,"+Quote$+"p2" +Quote$+" w 1,"+Quote$+"p3" +Quote$+" w 1,"+Quote$+"p4"
+Quote$+" w 1,"+Quote$+"p5"+Quote$+" w 1,"+Quote$+"p6"+Quote$+" w 1,"+_
                                     Quote$+"p7"  +Quote$+" w 1,"    +Quote$+"p8"  +Quote$+" w 1,"+Quote$+"p9" +Quote$+" w
1,"+Quote$+"p10"+Quote$+" w 1"

                CASE 11 : PRINT #N%, "plot "+Quote$+"p1"  +Quote$+" w l lw 1,"+Quote$+"p2" +Quote$+" w 1,"+Quote$+"p3" +Quote$+" w 1,"+Quote$+"p4"
+Quote$+" w 1,"+Quote$+"p5"+Quote$+" w 1,"+Quote$+"p6"+Quote$+" w 1,"+_
                                     Quote$+"p7"  +Quote$+" w 1,"    +Quote$+"p8"  +Quote$+" w 1,"+Quote$+"p9" +Quote$+" w
1,"+Quote$+"p10"+Quote$+" w 1,"+Quote$+"p11"+Quote$+" w 1"

                CASE 12 : PRINT #N%, "plot "+Quote$+"p1"  +Quote$+" w l lw 1,"+Quote$+"p2" +Quote$+" w 1,"+Quote$+"p3" +Quote$+" w 1,"+Quote$+"p4"
+Quote$+" w 1,"+Quote$+"p5" +Quote$+" w 1,"+Quote$+"p6" +Quote$+" w 1,"+_
                                     Quote$+"p7"  +Quote$+" w 1,"    +Quote$+"p8"  +Quote$+" w 1,"+Quote$+"p9" +Quote$+" w
1,"+Quote$+"p10"+Quote$+" w 1,"+Quote$+"p11"+Quote$+" w 1,"+Quote$+"p12"+Quote$+" w 1"

                CASE 13 : PRINT #N%, "plot "+Quote$+"p1"  +Quote$+" w l lw 1,"+Quote$+"p2" +Quote$+" w 1,"+Quote$+"p3" +Quote$+" w 1,"+Quote$+"p4"
+Quote$+" w 1,"+Quote$+"p5" +Quote$+" w 1,"+Quote$+"p6" +Quote$+" w 1,"+_
                                     Quote$+"p7"  +Quote$+" w 1,"    +Quote$+"p8"  +Quote$+" w 1,"+Quote$+"p9" +Quote$+" w
1,"+Quote$+"p10"+Quote$+" w 1,"+Quote$+"p11"+Quote$+" w 1,"+Quote$+"p12"+Quote$+" w 1,"+_
                                     Quote$+"p13 "+Quote$+" w 1"

                CASE 14 : PRINT #N%, "plot "+Quote$+"p1"  +Quote$+" w l lw 1,"+Quote$+"p2" +Quote$+" w 1,"+Quote$+"p3" +Quote$+" w 1,"+Quote$+"p4"
+Quote$+" w 1,"+Quote$+"p5" +Quote$+" w 1,"+Quote$+"p6" +Quote$+" w 1,"+_
                                     Quote$+"p7"  +Quote$+" w 1,"    +Quote$+"p8"  +Quote$+" w 1,"+Quote$+"p9" +Quote$+" w
1,"+Quote$+"p10"+Quote$+" w 1,"+Quote$+"p11"+Quote$+" w 1,"+Quote$+"p12"+Quote$+" w 1,"+_
                                     Quote$+"p13"+Quote$+" w 1,"    +Quote$+"p14"+Quote$+" w 1"

                CASE 15 : PRINT #N%, "plot "+Quote$+"p1"  +Quote$+" w l lw 1,"+Quote$+"p2" +Quote$+" w 1,"+Quote$+"p3" +Quote$+" w 1,"+Quote$+"p4"
+Quote$+" w 1,"+Quote$+"p5" +Quote$+" w 1,"+Quote$+"p6" +Quote$+" w 1,"+_
                                     Quote$+"p7"  +Quote$+" w 1,"    +Quote$+"p8"  +Quote$+" w 1,"+Quote$+"p9" +Quote$+" w
1,"+Quote$+"p10"+Quote$+" w 1,"+Quote$+"p11"+Quote$+" w 1,"+Quote$+"p12"+Quote$+" w 1,"+_
                                     Quote$+"p13"+Quote$+" w 1,"    +Quote$+"p14"+Quote$+" w 1,"+Quote$+"p15"+Quote$+" w 1"

                CASE 16 : PRINT #N%, "plot "+Quote$+"p1"  +Quote$+" w l lw 1,"+Quote$+"p2" +Quote$+" w 1,"+Quote$+"p3" +Quote$+" w 1,"+Quote$+"p4"
+Quote$+" w 1,"+Quote$+"p5"  +Quote$+" w 1,"+Quote$+"p6"  +Quote$+" w 1,"+_
                                     Quote$+"p7"  +Quote$+" w 1,"    +Quote$+"p8"  +Quote$+" w 1,"+Quote$+"p9" +Quote$+" w
1,"+Quote$+"p10"+Quote$+" w 1,"+Quote$+"p11"+Quote$+" w 1,"+Quote$+"p12"+Quote$+" w 1,"+_
                                     Quote$+"p13"+Quote$+" w 1,"    +Quote$+"p14"+Quote$+" w 1,"+Quote$+"p15"+Quote$+" w
1,"+Quote$+"p16"+Quote$+" w 1"
            END SELECT

        ELSE

            SELECT CASE NumTrajectories%

                CASE 1  : PRINT #N%, "plot "+Quote$+"p1"+Quote$+" lw 1"
                CASE 2  : PRINT #N%, "plot "+Quote$+"p1"+Quote$+" lw 1,"+Quote$+"p2"+Quote$
                CASE 3  : PRINT #N%, "plot "+Quote$+"p1"+Quote$+" lw 1,"+Quote$+"p2"+Quote$+" ,"+Quote$+"p3"+Quote$
                CASE 4  : PRINT #N%, "plot "+Quote$+"p1"+Quote$+" lw 1,"+Quote$+"p2"+Quote$+" ,"+Quote$+"p3"+Quote$+" ,"+Quote$+"p4"+Quote$
                CASE 5  : PRINT #N%, "plot "+Quote$+"p1"+Quote$+" lw 1,"+Quote$+"p2"+Quote$+" ,"+Quote$+"p3"+Quote$+" ,"+Quote$+"p4"+Quote$+"
,"+Quote$+"p5"+Quote$
                CASE 6  : PRINT #N%, "plot "+Quote$+"p1"+Quote$+" lw 1,"+Quote$+"p2"+Quote$+" ,"+Quote$+"p3"+Quote$+" ,"+Quote$+"p4"+Quote$+"
,"+Quote$+"p5"+Quote$+" ,"+Quote$+"p6"+Quote$
                CASE 7  : PRINT #N%, "plot "+Quote$+"p1"+Quote$+" lw 1,"+Quote$+"p2"+Quote$+" ,"+Quote$+"p3"+Quote$+" ,"+Quote$+"p4"+Quote$+" ,"+_
                                     Quote$+"p7"+Quote$
                CASE 8  : PRINT #N%, "plot "+Quote$+"p1"+Quote$+" lw 1,"+Quote$+"p2"+Quote$+" ,"+Quote$+"p3"+Quote$+" ,"+Quote$+"p4"+Quote$+"
,"+Quote$+"p5"+Quote$+" ,"+Quote$+"p6"+Quote$+" ,"+_
                                     Quote$+"p7"+Quote$+" ,"   +Quote$+"p8"+Quote$
                CASE 9  : PRINT #N%, "plot "+Quote$+"p1"+Quote$+" lw 1,"+Quote$+"p2"+Quote$+" ,"+Quote$+"p3"+Quote$+" ,"+Quote$+"p4"+Quote$+"
,"+Quote$+"p5"+Quote$+" ,"+Quote$+"p6"+Quote$+" ,"+_
                                     Quote$+"p7"+Quote$+" ,"   +Quote$+"p8"+Quote$+" ,"+Quote$+"p9"+Quote$
                CASE 10 : PRINT #N%, "plot "+Quote$+"p1"+Quote$+" lw 1," +Quote$+"p2" +Quote$+" ,"+Quote$+"p3"+Quote$+" ,"+Quote$+"p4"  +Quote$+"
,"+Quote$+"p5"+Quote$+" ,"+Quote$+"p6"+Quote$+" ,"+_
                                     Quote$+"p7"+Quote$+" ,"    +Quote$+"p8" +Quote$+" ,"+Quote$+"p9"+Quote$+" ,"+Quote$+"p10" +Quote$
```

```
            CASE 11 : PRINT #N%, "plot "+Quote$+"p1"+Quote$+" lw 1," +Quote$+"p2" +Quote$+" ,"+Quote$+"p3"+Quote$+" ,"+Quote$+"p4"  +Quote$+"
,"+Quote$+"p5" +Quote$+" ,"+Quote$+"p6"+Quote$+" ,"+_
                                Quote$+"p7"+Quote$+" ,"      +Quote$+"p8" +Quote$+" ,"+Quote$+"p9"+Quote$+" ,"+Quote$+"p10" +Quote$+"
,"+Quote$+"p11"+Quote$

            CASE 12 : PRINT #N%, "plot "+Quote$+"p1"+Quote$+" lw 1," +Quote$+"p2" +Quote$+" ,"+Quote$+"p3"+Quote$+" ,"+Quote$+"p4"  +Quote$+"
,"+Quote$+"p5" +Quote$+" ,"+Quote$+"p6" +Quote$+" ,"+_
                                Quote$+"p7"+Quote$+" ,"      +Quote$+"p8" +Quote$+" ,"+Quote$+"p9"+Quote$+" ,"+Quote$+"p10" +Quote$+"
,"+Quote$+"p11"+Quote$+" ,"+Quote$+"p12"+Quote$

            CASE 13 : PRINT #N%, "plot "+Quote$+"p1" +Quote$+" lw 1,"+Quote$+"p2" +Quote$+" ,"+Quote$+"p3"+Quote$+" ,"+Quote$+"p4"  +Quote$+"
,"+Quote$+"p5" +Quote$+" ," +Quote$+"p6"  +Quote$+" ,"+_
                                Quote$+"p7" +Quote$+" ,"      +Quote$+"p8" +Quote$+" ,"+Quote$+"p9"+Quote$+" ,"+Quote$+"p10" +Quote$+"
,"+Quote$+"p11"+Quote$+" ," +Quote$+"p12"+Quote$+" ,"+_
                                Quote$+"p13"+Quote$

            CASE 14 : PRINT #N%, "plot "+Quote$+"p1" +Quote$+" lw 1,"+Quote$+"p2" +Quote$+" ,"+Quote$+"p3"+Quote$+" ,"+Quote$+"p4"  +Quote$+"
,"+Quote$+"p5" +Quote$+" ," +Quote$+"p6"  +Quote$+" ,"+_
                                Quote$+"p7" +Quote$+" ,"      +Quote$+"p8" +Quote$+" ,"+Quote$+"p9"+Quote$+" ,"+Quote$+"p10" +Quote$+"
,"+Quote$+"p11"+Quote$+" ," +Quote$+"p12"+Quote$+" ,"+_
                                Quote$+"p13"+Quote$+" ,"      +Quote$+"p14"+Quote$

            CASE 15 : PRINT #N%, "plot "+Quote$+"p1"+Quote$+" lw 1,"+Quote$+"p2" +Quote$+" ,"+Quote$+"p3" +Quote$+" ,"+Quote$+"p4" +Quote$+"
,"+Quote$+"p5" +Quote$+" ,"+Quote$+"p6" +Quote$+" ,"+_
                                Quote$+"p7" +Quote$+" ,"      +Quote$+"p8" +Quote$+" ,"+Quote$+"p9" +Quote$+" ,"+Quote$+"p10"+Quote$+"
,"+Quote$+"p11"+Quote$+" ," +Quote$+"p12"+Quote$+" ,"+_
                                Quote$+"p13"+Quote$+" ,"      +Quote$+"p14"+Quote$+" ,"+Quote$+"p15"+Quote$

            CASE 16 : PRINT #N%, "plot "+Quote$+"p1"  +Quote$+" lw 1,"+Quote$+"p2" +Quote$+" ,"+Quote$+"p3" +Quote$+" ,"+Quote$+"p4" +Quote$+"
,"+Quote$+"p5" +Quote$+" ,"+Quote$+"p6" +Quote$+" ,"+_
                                Quote$+"p7"  +Quote$+" ,"      +Quote$+"p8" +Quote$+" ,"+Quote$+"p9" +Quote$+" ,"+Quote$+"p10"+Quote$+"
,"+Quote$+"p11"+Quote$+" ," +Quote$+"p12"+Quote$+" ,"+_
                                Quote$+"p13"+Quote$+" ,"    +Quote$+"p14"+Quote$+" ,"+Quote$+"p15"+Quote$+" ,"+Quote$+"p16"+Quote$
        END SELECT

    END IF

    CLOSE #N%

    ProcID??? = SHELL(GnuPlotEXE$+" cmd2d.gp -") : CALL Delay(0.5##)

END SUB 'Plot2DindividualProbeTrajectories()

'----

SUB Plot3DbestProbeTrajectories(NumTrajectories%,M(),R(),XiMin(),XiMax(),Np%,Nd%,LastStep&,FunctionName$) 'XYZZY

LOCAL TrajectoryNumber%, ProbeNumber%, StepNumber&, N%, M%, ProcID???

LOCAL MaximumFitness, MinimumFitness AS EXT

LOCAL BestProbeThisStep%()

LOCAL BestFitnessThisStep(), TempFitness() AS EXT

LOCAL Annotation$, xCoord$, yCoord$, zCoord$, GnuPlotEXE$, PlotWithLines$

    Annotation$      = ""

    PlotWithLines$   = "NO" '"YES" '"NO"

    NumTrajectories% = MIN(Np%,NumTrajectories%)

    GnuPlotEXE$ = "wgnuplot.exe"

'   --------------- Get Min/Max Fitnesses ----------------

    MaximumFitness = M(1,0) : MinimumFitness = M(1,0)  'Note:  M(p%,j&)

    FOR StepNumber& = 0 TO LastStep&

        FOR ProbeNumber% = 1 TO Np%

            IF M(ProbeNumber%,StepNumber&) >= MaximumFitness THEN MaximumFitness = M(ProbeNumber%,StepNumber&)

            IF M(ProbeNumber%,StepNumber&) =< MinimumFitness THEN MinimumFitness = M(ProbeNumber%,StepNumber&)

        NEXT ProbeNumber%

    NEXT StepNumber%

'   ------------- Copy Fitness Array M() into TempFitness to Preserve M() ----------------

    REDIM TempFitness(1 TO Np%, 0 TO LastStep&)

    FOR StepNumber& = 0 TO LastStep&

        FOR ProbeNumber% = 1 TO Np%

            TempFitness(ProbeNumber%,StepNumber&) = M(ProbeNumber%,StepNumber&)

        NEXT ProbeNumber%

    NEXT StepNumber%

'   ----------- LOOP ON TRAJECTORIES ----------

    FOR TrajectoryNumber% = 1 TO NumTrajectories%

'       --------------- Get Trajectory Coordinate Data -----------------

        REDIM BestFitnessThisStep(0 TO LastStep&), BestProbeThisStep%(0 TO LastStep&)

        FOR StepNumber& = 0 TO LastStep&

            BestFitnessThisStep(StepNumber&) = TempFitness(1,StepNumber&)

            FOR ProbeNumber% = 1 TO Np%

                IF TempFitness(ProbeNumber%,StepNumber&) >= BestFitnessThisStep(StepNumber&) THEN

                    BestFitnessThisStep(StepNumber&) = TempFitness(ProbeNumber%,StepNumber&)

                    BestProbeThisStep%(StepNumber&)  = ProbeNumber%
```

```
              END IF

          NEXT ProbeNumber%

      NEXT StepNumber&

'   ----- Create Plot Data File -----

      N% = FREEFILE

      SELECT CASE TrajectoryNumber%

          CASE 1  : OPEN "t1"  FOR OUTPUT AS #N%
          CASE 2  : OPEN "t2"  FOR OUTPUT AS #N%
          CASE 3  : OPEN "t3"  FOR OUTPUT AS #N%
          CASE 4  : OPEN "t4"  FOR OUTPUT AS #N%
          CASE 5  : OPEN "t5"  FOR OUTPUT AS #N%
          CASE 6  : OPEN "t6"  FOR OUTPUT AS #N%
          CASE 7  : OPEN "t7"  FOR OUTPUT AS #N%
          CASE 8  : OPEN "t8"  FOR OUTPUT AS #N%
          CASE 9  : OPEN "t9"  FOR OUTPUT AS #N%
          CASE 10 : OPEN "t10" FOR OUTPUT AS #N%

      END SELECT

'   ----------- Write Plot File Data -----------

      FOR StepNumber& = 0 TO LastStep&

          PRINT #N%, USING$("######.######## ######.########
######.########",R(BestProbeThisStep%(StepNumber&),1,StepNumber&),R(BestProbeThisStep%(StepNumber&),2,StepNumber&),R(BestProbeThisStep%(StepNumber&)
,3,StepNumber&))+CHR$(13)

          TempFitness(BestProbeThisStep%(StepNumber&),StepNumber&) = MinimumFitness 'so that same max will not be found for next trajectory

      NEXT StepNumber%

      CLOSE #N%

      NEXT TrajectoryNumber%

'   ----------------------- Plot Trajectories -------------------------

      'CALL CreateGNUplotINIfile(0.1##*ScreenWidth&,0.25##*ScreenHeight&,0.6##*ScreenHeight&,0.6##*ScreenHeight&)

      Annotation$ = ""

      N% = FREEFILE

      OPEN "cmd3d.gp" FOR OUTPUT AS #N%

      PRINT #N%, "set pm3d"
      PRINT #N%, "show pm3d"
      PRINT #N%, "set hidden3d"
      PRINT #N%, "set view 45, 45, 1, 1"

      PRINT #N%, "unset colorbox"

      PRINT #N%, "set xrange [" + REMOVE$(STR$(XiMin(1)),ANY"" ) + ":" + REMOVE$(STR$(XiMax(1)),ANY"" ) + "]"
      PRINT #N%, "set yrange [" + REMOVE$(STR$(XiMin(2)),ANY"" ) + ":" + REMOVE$(STR$(XiMax(2)),ANY"" ) + "]"
      PRINT #N%, "set zrange [" + REMOVE$(STR$(XiMin(3)),ANY"" ) + ":" + REMOVE$(STR$(XiMax(3)),ANY"" ) + "]"
'     PRINT #N%, "set label "   + Quote$  + Annotation$ + Quote$+" at graph "+xCoord$+","+yCoord$+","+zCoord$
'     PRINT #N%, "show label"
      PRINT #N%, "set grid xtics ytics ztics"
      PRINT #N%, "show grid"
      PRINT #N%, "set title " + Quote$ + "3D " + FunctionName$ + " PROBE TRAJECTORIES" + "\n" + RunID$ + Quote$
      PRINT #N%, "set xlabel " + Quote$ + "x1"                                  + Quote$
      PRINT #N%, "set ylabel " + Quote$ + "x2"                                  + Quote$
      PRINT #N%, "set zlabel " + Quote$ + "x3"                                  + Quote$

      IF PlotWithLines = "YES" THEN

          SELECT CASE NumTrajectories%

              CASE 1  : PRINT #N%, "splot "+Quote$+"t1"+Quote$+" w l lw 3"
              CASE 2  : PRINT #N%, "splot "+Quote$+"t1"+Quote$+" w l lw 3,"+Quote$+"t2"+Quote$+" w l"
              CASE 3  : PRINT #N%, "splot "+Quote$+"t1"+Quote$+" w l lw 3,"+Quote$+"t2"+Quote$+" w l,"+Quote$+"t3"+Quote$+" w l"
              CASE 4  : PRINT #N%, "splot "+Quote$+"t1"+Quote$+" w l lw 3,"+Quote$+"t2"+Quote$+" w l,"+Quote$+"t3"+Quote$+" w l,"+Quote$+"t4"+Quote$+"
w l"
              CASE 5  : PRINT #N%, "splot "+Quote$+"t1"+Quote$+" w l lw 3,"+Quote$+"t2"+Quote$+" w l,"+Quote$+"t3"+Quote$+" w l,"+Quote$+"t4"+Quote$+"
w l,"+Quote$+"t5"+Quote$+" w l"
              CASE 6  : PRINT #N%, "splot "+Quote$+"t1"+Quote$+" w l lw 3,"+Quote$+"t2"+Quote$+" w l,"+Quote$+"t3"+Quote$+" w l,"+Quote$+"t4"+Quote$+"
w l,"+Quote$+"t5"+Quote$+" w l,"+Quote$+"t6"+Quote$+" w l"
              CASE 7  : PRINT #N%, "splot "+Quote$+"t1"+Quote$+" w l lw 3,"+Quote$+"t2"+Quote$+" w l,"+Quote$+"t3"+Quote$+" w l,"+Quote$+"t4"+Quote$+"
w l,"+Quote$+"t5"+Quote$+" w l,"+Quote$+"t6"+Quote$+" w l,"+_
                                                                Quote$+"t7"+Quote$+" w l"
              CASE 8  : PRINT #N%, "splot "+Quote$+"t1"+Quote$+" w l lw 3,"+Quote$+"t2"+Quote$+" w l,"+Quote$+"t3"+Quote$+" w l,"+Quote$+"t4"+Quote$+"
w l,"+Quote$+"t5"+Quote$+" w l,"+Quote$+"t6"+Quote$+" w l,"+_
                                                                Quote$+"t7"+Quote$+" w l,"     +Quote$+"t8"+Quote$+" w l"
              CASE 9  : PRINT #N%, "splot "+Quote$+"t1"+Quote$+" w l lw 3,"+Quote$+"t2"+Quote$+" w l,"+Quote$+"t3"+Quote$+" w l,"+Quote$+"t4"+Quote$+"
w l,"+Quote$+"t5"+Quote$+" w l,"+Quote$+"t6"+Quote$+" w l,"+_
                                                                Quote$+"t7"+Quote$+" w l,"     +Quote$+"t8"+Quote$+" w l,"+Quote$+"t9"+Quote$+" w l"
              CASE 10 : PRINT #N%, "splot "+Quote$+"t1"+Quote$+" w l lw 3,"+Quote$+"t2"+Quote$+" w l,"+Quote$+"t3"+Quote$+" w l,"+Quote$+"t4"+Quote$+"
w l,"+Quote$+"t5"+Quote$+" w l,"+Quote$+"t6"+Quote$+" w l,"+_
                                                                Quote$+"t7"+Quote$+" w l,"     +Quote$+"t8"+Quote$+" w l,"+Quote$+"t9"+Quote$+" w
l,"+Quote$+"t10"+Quote$+" w l"
          END SELECT

      ELSE

          SELECT CASE NumTrajectories%

              CASE 1  : PRINT #N%, "splot "+Quote$+"t1"+Quote$+" lw 2"
              CASE 2  : PRINT #N%, "splot "+Quote$+"t1"+Quote$+" lw 2,"+Quote$+"t2"+Quote$
              CASE 3  : PRINT #N%, "splot "+Quote$+"t1"+Quote$+" lw 2,"+Quote$+"t2"+Quote$+" ,"+Quote$+"t3"+Quote$
              CASE 4  : PRINT #N%, "splot "+Quote$+"t1"+Quote$+" lw 2,"+Quote$+"t2"+Quote$+" ,"+Quote$+"t3"+Quote$+" ,"+Quote$+"t4"+Quote$
              CASE 5  : PRINT #N%, "splot "+Quote$+"t1"+Quote$+" lw 2,"+Quote$+"t2"+Quote$+" ,"+Quote$+"t3"+Quote$+" ,"+Quote$+"t4"+Quote$+"
,"+Quote$+"t5"+Quote$
              CASE 6  : PRINT #N%, "splot "+Quote$+"t1"+Quote$+" lw 2,"+Quote$+"t2"+Quote$+" ,"+Quote$+"t3"+Quote$+" ,"+Quote$+"t4"+Quote$+"
,"+Quote$+"t5"+Quote$+" ,"+Quote$+"t6"+Quote$
              CASE 7  : PRINT #N%, "splot "+Quote$+"t1"+Quote$+" lw 2,"+Quote$+"t2"+Quote$+" ,"+Quote$+"t3"+Quote$+" ,"+Quote$+"t4"+Quote$+"
,"+Quote$+"t5"+Quote$+" ,"+Quote$+"t6"+Quote$+" ,"+_
```

```
                                        Quote$+"t7"+Quote$
          CASE 8  : PRINT #N%, "splot "+Quote$+"t1"+Quote$+" lw 2,"+Quote$+"t2"+Quote$+" ,"+Quote$+"t3"+Quote$+" ,"+Quote$+"t4"+Quote$+"
     ,"+Quote$+"t5"+Quote$+" ,"+Quote$+"t6"+Quote$+" ,"+_
                                        Quote$+"t7"+Quote$+" ,"    +Quote$+"t8"+Quote$
          CASE 9  : PRINT #N%, "splot "+Quote$+"t1"+Quote$+" lw 2,"+Quote$+"t2"+Quote$+" ,"+Quote$+"t3"+Quote$+" ,"+Quote$+"t4"+Quote$+"
     ,"+Quote$+"t5"+Quote$+" ,"+Quote$+"t6"+Quote$+" ,"+_
                                        Quote$+"t7"+Quote$+" ,"    +Quote$+"t8"+Quote$+" ,"+Quote$+"t9"+Quote$
          CASE 10 : PRINT #N%, "splot "+Quote$+"t1"+Quote$+" lw 2,"+Quote$+"t2"+Quote$+" ,"+Quote$+"t3"+Quote$+" ,"+Quote$+"t4"+Quote$+"
     ,"+Quote$+"t5"+Quote$+" ,"+Quote$+"t6"+Quote$+" ,"+_
                                        Quote$+"t7"+Quote$+" ,"    +Quote$+"t8"+Quote$+" ,"+Quote$+"t9"+Quote$+" ,"+Quote$+"t10"+Quote$
       END SELECT

     END IF

     CLOSE #N%

     ProcID??? = SHELL(GnuPlotEXE$+" cmd3d.gp -") : CALL Delay(0.5##)

END SUB 'Plot3DbestProbeTrajectories()

'-----

FUNCTION HasFITNESSsaturated$(Nsteps&,j&,Np%,Nd%,M(),R(),DiagLength)

LOCAL A$

LOCAL k&, p%

LOCAL BestFitness, SumOfBestFitnesses, BestFitnessStepJ, FitnessSatTOL AS EXT

     A$ = "NO"

     FitnessSatTOL = 0.000001## 'tolerance for FITNESS saturation

     IF j& < Nsteps& + 10 THEN GOTO ExitHasFITNESSsaturated 'execute at least 10 steps after averaging interval before performing this check

     SumOfBestFitnesses = 0##

     FOR k& = j&-Nsteps&+1 TO j&

        BestFitness = M(k&,1)

        FOR p% = 1 TO Np%

           IF M(p%,k&) >= BestFitness THEN BestFitness = M(p%,k&)

        NEXT p%

        IF k& = j& THEN BestFitnessStepJ = BestFitness

        SumOfBestFitnesses = SumOfBestFitnesses + BestFitness

     NEXT k&

     IF ABS(SumOfBestFitnesses/Nsteps&-BestFitnessStepJ) =< FitnessSatTOL THEN A$ = "YES" 'saturation if (avg value - last value) are within TOL

ExitHasFITNESSsaturated:

     HasFITNESSsaturated$ = A$

END FUNCTION 'HasFITNESSsaturated$()

'-----------

FUNCTION HasDAVGsaturated$(Nsteps&,j&,Np%,Nd%,M(),R(),DiagLength)

LOCAL A$

LOCAL k&

LOCAL SumOfDavg, DavgStepJ AS EXT

LOCAL DavgSatTOL AS EXT

     A$ = "NO"

     DavgSatTOL = 0.0005## 'tolerance for DAVG saturation

     IF j& < Nsteps& + 10 THEN GOTO ExitHasDAVGsaturated 'execute at least 10 steps after averaging interval before performing this check

     DavgStepJ = DavgThisStep(j&,Np%,Nd%,M(),R(),DiagLength)

     SumOfDavg = 0##

     FOR k& = j&-Nsteps&+1 TO j& 'check this step and previous (Nsteps&-1) steps

        SumOfDavg = SumOfDavg + DavgThisStep(k&,Np%,Nd%,M(),R(),DiagLength)

     NEXT k&

     IF ABS(SumOfDavg/Nsteps&-DavgStepJ) =< DavgSatTOL THEN A$ = "YES" 'saturation if (avg value - last value) are within TOL

ExitHasDAVGsaturated:

     HasDAVGsaturated$ = A$

END FUNCTION 'HasDAVGsaturated$()

'-----------

FUNCTION OscillationInDavg$(j&,Np%,Nd%,M(),R(),DiagLength)

LOCAL A$

LOCAL k&, NumSlopeChanges%

     A$ = "NO"

     NumSlopeChanges% = 0

     IF j& < 15 THEN GOTO ExitDavgOscillation 'wait at least 15 steps

     FOR k& = j&-10 TO j&-1 'check previous ten steps
```

```
          IF (DavgThisStep(k&,Np%,Nd%,M(),R(),DiagLength)-DavgThisStep(k&-1,Np%,Nd%,M(),R(),DiagLength))* _
              (DavgThisStep(k&+1,Np%,Nd%,M(),R(),DiagLength)-DavgThisStep(k&,Np%,Nd%,M(),R(),DiagLength)) < 0## THEN INCR NumSlopeChanges%

    NEXT j&

    IF NumSlopeChanges% >= 3 THEN A$ = "YES"

ExitDavgOscillation:

    OscillationInDavg$ = A$

END FUNCTION 'OscillationInDavg()

'------

FUNCTION DavgThisStep(j&,Np%,Nd%,M(),R(),DiagLength)

LOCAL BestFitness, TotalDistanceAllProbes, SumSQ AS EXT

LOCAL p%, k&, N%, i%, BestProbeNumber%, BestTimeStep&

'    ---------- Best Probe #, etc. -----------

    FOR k& = 0 TO j&

        BestFitness = M(1,k&)

        FOR p% = 1 TO Np%

            IF M(p%,k&) >= BestFitness THEN

                BestFitness = M(p%,k&) : BestProbeNumber% = p% : BestTimeStep& = k&

            END IF

        NEXT p% 'probe #

    NEXT k& 'time step

'    --------- Average Distance to Best Probe -----------

    TotalDistanceAllProbes = 0##

    FOR p% = 1 TO Np%

        SumSQ = 0##

        FOR i% = 1 TO Nd%

            SumSQ = SumSQ + (R(BestProbeNumber%,i%,BestTimeStep&)-R(p%,i%,j&))^2 'do not exclude p%=BestProbeNumber%(j&) from sum because it adds
zero

         NEXT i%

        TotalDistanceAllProbes = TotalDistanceAllProbes + SQR(SumSQ)

    NEXT p%

    DavgThisStep = TotalDistanceAllProbes/(DiagLength*(Np%-1)) 'but exclude best prove from average

END FUNCTION 'DavgThisStep()

'-----------

SUB
PlotBestFitnessEvolution(Nd%,Np%,LastStep&,G,DeltaT,Alpha,Beta,Frep,M(),PlaceInitialProbes$,InitialAcceleration$,RepositionFactor$,FunctionName$,Gam
ma)

LOCAL BestFitness(), GlobalBestFitness AS EXT

LOCAL PlotAnnotation$, PlotTitle$

LOCAL p%, j&, N%

    REDIM BestFitness(0 TO LastStep&)

    CALL
GetPlotAnnotation(PlotAnnotation$,Nd%,Np%,LastStep&,G,DeltaT,Alpha,Beta,Frep,M(),PlaceInitialProbes$,InitialAcceleration$,RepositionFactor$,Function
Name$,Gamma)

    GlobalBestFitness = M(1,0)

    FOR j& = 0 TO LastStep&

        BestFitness(j&) = M(1,j&)

        FOR p% = 1 TO Np%

            IF M(p%,j&) > BestFitness(j&)   THEN BestFitness(j&)   = M(p%,j&)

            IF M(p%,j&) >= GlobalBestFitness THEN GlobalBestFitness = M(p%,j&)

        NEXT p% 'probe #

    NEXT j& 'time step

    N% = FREEFILE

    OPEN "Fitness" FOR OUTPUT AS #N%

        FOR j& = 0 TO LastStep&

            PRINT #N%, USING$("###### #######.######",j&,BestFitness(j&))

        NEXT j&

    CLOSE #N%

    PlotAnnotation$ = PlotAnnotation$ + "Best Fitness = " + REMOVE$(STR$(ROUND(GlobalBestFitness,8)),ANY" ")

    PlotTitle$ = "Best Fitness vs Time Step\n" + "[" + REMOVE$(STR$(Np%),ANY" ") + " probes, "+REMOVE$(STR$(LastStep&),ANY" ")+" time steps]"

    CALL CreateGNUplotINIfile(0.1##*ScreenWidth&,0.1##*ScreenHeight&,0.6##*ScreenWidth&,0.6##*ScreenHeight&)
```

```
        CALL TwoDplot("Fitness","Best Fitness","0.7","0.7","Time Step\n\n.","."\n\nBest Fitness(X)", _
                      "","","","","","","","wgnuplot.exe," with lines linewidth 2",PlotAnnotation$)

END SUB 'PlotBestFitnessEvolution()

'------

SUB
PlotAverageDistance(Nd%,Np%,LastStep&,G,DeltaT,Alpha,Beta,Frep,M(),PlaceInitialProbes$,InitialAcceleration$,RepositionFactor$,FunctionName$,R(),Diag
Length,Gamma)

LOCAL Davg(), BestFitness(), TotalDistanceAllProbes, SumSQ AS EXT

LOCAL PlotAnnotation$, PlotTitle$

LOCAL p%, j&, N%, i%, BestProbeNumber%(), BestTimeStep&()

     REDIM Davg(0 TO LastStep&), BestFitness(0 TO LastStep&), BestProbeNumber%(0 TO LastStep&), BestTimeStep&(0 TO LastStep&)

     CALL
GetPlotAnnotation(PlotAnnotation$,Nd%,Np%,LastStep&,G,DeltaT,Alpha,Beta,Frep,M(),PlaceInitialProbes$,InitialAcceleration$,RepositionFactor$,Function
Name$,Gamma)

'   ---------- Best Probe #, etc. -----------

     FOR j& = 0 TO LastStep&

          BestFitness(j&) = M(1,j&)

          FOR p% = 1 TO Np%

              IF M(p%,j&) >= BestFitness(j&) THEN

                  BestFitness(j&) = M(p%,j&) : BestProbeNumber%(j&) = p% : BestTimeStep&(j&) = j& 'only probe number is used at this time, but other
data are computed for possible future use.

              END IF

          NEXT p% 'probe #

     NEXT j& 'time step

     N% = FREEFILE

'    --------- Average Distance to Best Probe -----------

     FOR j& = 0 TO LastStep&

          TotalDistanceAllProbes = 0##

          FOR p% = 1 TO Np%

              SumSQ = 0##

              FOR i% = 1 TO Nd%

                  SumSQ = SumSQ + (R(BestProbeNumber%(j&),i%,j&)-R(p%,i%,j&))^2 'do not exclude p%=BestProbeNumber%(j&) from sum because it adds zero

              NEXT i%

              TotalDistanceAllProbes = TotalDistanceAllProbes + SQR(SumSQ)

          NEXT p%

          Davg(j&) = TotalDistanceAllProbes/(DiagLength*(Np%-1)) 'but exclude best prove from average

     NEXT j&

'    ----------- Create Plot Data File -----------

     OPEN "Davg" FOR OUTPUT AS #N%

          FOR j& = 0 TO LastStep&

               PRINT #N%, USING$("###### #######.######",j&,Davg(j&))

          NEXT j&

     CLOSE #N%

     PlotTitle$ = "Average Distance of " + REMOVE$(STR$(Np%-1),ANY" ") + " Probes to Best Probe\nNormalized to Size of Decision Space\n" + _
                  "[" + REMOVE$(STR$(Np%),ANY" ") + " probes, " + REMOVE$(STR$(LastStep&),ANY" ") + " time steps]"

     CALL CreateGNUplotINIfile(0.2##*ScreenWidth&,0.2##*ScreenHeight&,0.6##*ScreenWidth&,0.6##*ScreenHeight&)

     CALL TwoDplot("Davg",PlotTitle$,"0.7","0.9","Time Step\n\n.","."\n\n<D>/Ldiag", _
                   "","","","","","","","wgnuplot.exe," with lines linewidth 2",PlotAnnotation$)

END SUB 'PlotAverageDistance()

'------

SUB
GetPlotAnnotation(PlotAnnotation$,Nd%,Np%,LastStep&,G,DeltaT,Alpha,Beta,Frep,M(),PlaceInitialProbes$,InitialAcceleration$,RepositionFactor$,Function
Name$,Gamma)

LOCAL A$

     A$ = "" : IF PlaceInitialProbes$ = "UNIFORM ON-AXIS" AND Nd% > 1 THEN A$ = " ("+REMOVE$(STR$(Np%/Nd%),ANY" ") + "/axis)"

     PlotAnnotation$ = RunID$ + "\n" + _
                       FunctionName$ + " Function" + " (" + FormatInteger$(Nd%) + "-D) \n"    +_
                       FormatInteger$(Np%) + " probes"       + A$ + "\n" +_
                       "G = " + FormatFP$(G,2)                      + "\n" +_
                       "Alpha = "     + FormatFP$(Alpha,1) + "\n" +_
                       "Beta = "      + FormatFP$(Beta,1)  + "\n" +_
                       "DelT = "      + FormatFP$(DeltaT,1) + "\n" +_
                       "Gamma = "     + FormatFP$(Gamma,3)   + "\n" +_
                       "Init Probes " + PlaceInitialProbes$   + "\n" +_
                       "Init Accel " + InitialAcceleration$ + "\n" +_
                       "Frep = "      + FormatFP$(Frep,3)    + " (" + RepositionFactor$ + ")\n"

'     SELECT CASE RepositionFactor$
```

```
'       CASE "FIXED" : PlotAnnotation$ = PlotAnnotation$ + "Frep = " + FormatFP$(Frep,3) + " fixed" + "\n"

'   END SELECT

END SUB

'------

SUB
PlotBestProbeVsTimeStep(Nd%,Np%,LastStep&,G,DeltaT,Alpha,Beta,Frep,M(),PlaceInitialProbes$,InitialAcceleration$,RepositionFactor$,FunctionName$,Gamm
a)

LOCAL BestFitness AS EXT

LOCAL PlotAnnotation$, PlotTitle$

LOCAL p%, j&, N%, BestProbeNumber%()

    REDIM BestProbeNumber%(0 TO LastStep&)

    CALL
GetPlotAnnotation(PlotAnnotation$,Nd%,Np%,LastStep&,G,DeltaT,Alpha,Beta,Frep,M(),PlaceInitialProbes$,InitialAcceleration$,RepositionFactor$,Function
Name$,Gamma)

    FOR j& = 0 TO LastStep&

        Bestfitness = M(1,j&)

        FOR p% = 1 TO Np%

            IF M(p%,j&) >= BestFitness THEN

                BestFitness = M(p%,j&) : BestProbeNumber%(j&) = p%

            END IF

        NEXT p% 'probe #

    NEXT j& 'time step

    N% = FREEFILE

    OPEN "Best Probe" FOR OUTPUT AS #N%

        FOR j& = 0 TO LastStep&

            PRINT #N%, USING$("###### #####",j&,BestProbeNumber%(j&))

        NEXT j&

    CLOSE #N%

    PlotTitle$ = "Best Probe Number vs Time Step\n" + "[" +REMOVE$(STR$(Np%),ANY" ") + " probes, " + REMOVE$(STR$(LastStep&),ANY" ") + " time
steps]"

    CALL CreateGNUplotINIfile(0.15#*ScreenWidth&,0.15#*ScreenHeight&,0.6##*ScreenWidth&,0.6##*ScreenHeight&)

'USAGE: CALL
TwoDplot(PlotFileName$,PlotTitle$,xCoord$,yCoord$,XaxisLabel$,YaxisLabel$,LogXaxis$,LogYaxis$,xMin%,xMax%,yMin%,yMax%,xTics$,yTics$,GnuPlotEXE$,Line
Type$,Annotation$)

    CALL TwoDplot("Best Probe",PlotTitle$,"0.7","0.7","Time Step\n\n.",".\n\nBest Probe #","","","","","0",NoSpaces$(Np%+1,0),"","","wgnuplot.exe","
pt 8 ps .5 lw 1",PlotAnnotation$) 'pt, pointtype; ps, pointsize; lw, linewidth

END SUB 'PlotBestProbeVsTimeStep()

'------

SUB DisplayBestFitness(Np%,Nd%,LastStep&,M(),R(),BestFitnessProbeNumber%,BestFitnessTimeStep&,FunctionName$)

LOCAL A$, B$, p%, i%, j&

LOCAL BestFitness AS EXT

    B$ = "" : IF Nd% > 1 THEN B$ = "s"

    BestFitness = M(1,0)

    FOR j& = 0 TO LastStep&

        FOR p% = 1 TO Np%

            IF M(p%,j&) >= BestFitness THEN

                BestFitness = M(p%,j&) : BestFitnessProbeNumber% = p% : BestFitnessTimeStep& = j&

            END IF

        NEXT p%

    NEXT j&

    A$ = FunctionName$ + CHR$(13) +_
        "Best Fitness = " + REMOVE$(STR$(ROUND(BestFitness,8)),ANY" ") + " returned by" + CHR$(13)        +_
        "Probe # "     + REMOVE$(STR$(BestFitnessProbeNumber%),ANY" ") +_
        " at Time Step "  + REMOVE$(STR$(BestFitnessTimeStep&),ANY" ") + CHR$(13) + CHR$(13) + "P" + REMOVE$(STR$(BestFitnessProbeNumber%),ANY" ")
+ " coordinate" + B$ + ":" + CHR$(13)

    FOR i% = 1 TO Nd% : A$ = A$ + STR$(i%)+"    "+REMOVE$(STR$(ROUND(R(BestFitnessProbeNumber%,i%,BestFitnessTimeStep&),8)),ANY" ")+CHR$(13) : NEXT
i%

    MSGBOX(A$)

END SUB 'DisplayBestFitness()

'------

FUNCTION FormatInteger$(M%) : FormatInteger$ = REMOVE$(STR$(M%),ANY" ") : END FUNCTION

'------

FUNCTION FormatFP$(X,Ndigits%)
```

```
LOCAL A$
    IF X = 0## THEN
        A$ = "0." : GOTO ExitFormatFP
    END IF
    A$ = REMOVE$(STR$(ROUND(ABS(X),Ndigits%)),ANY" ")
    IF ABS(X) < 1## THEN
        IF X > 0## THEN
            A$ = "0" + A$
        ELSE
            A$ = "-0" + A$
        END IF
    ELSE
        IF X < 0## THEN A$ = "-" + A$
    END IF
ExitFormatFP:
    FormatFP$ = A$
END FUNCTION
'-----------
SUB InitialProbeDistribution(Np%,Nd%,Nt%,XiMin(),XiMax(),R(),PlaceInitialProbes$,Gamma)
LOCAL DeltaXi, DelX1, DelX2, Di AS EXT
LOCAL NumProbesPerAxis%, p%, i%, k%, NumX1points%, NumX2points%, x1pointNum%, x2pointNum%, A$
    SELECT CASE PlaceInitialProbes$
        CASE "UNIFORM ON-AXIS"
            IF Nd% > 1 THEN
                NumProbesPerAxis% = Np%\Nd% 'even #
            ELSE
                NumProbesPerAxis% = Np%
            END IF
            FOR i% = 1 TO Nd%
                FOR p% = 1 TO Np%
                    R(p%,i%,0) = XiMin(i%) + Gamma*(XiMax(i%)-XiMin(i%))
                NEXT Np%
            NEXT i%
            FOR i% = 1 TO Nd% 'place probes axis-by-axis (i% is axis [dimension] number)
                DeltaXi = (XiMax(i%)-XiMin(i%))/(NumProbesPerAxis%-1)
                FOR k% = 1 TO NumProbesPerAxis%
                    p% = k% + NumProbesPerAxis%*(i%-1) 'probe #
                    R(p%,i%,0) = XiMin(i%) + (k%-1)*DeltaXi
                NEXT k%
                'DeltaXi = (XiMax(i%)-XiMin(i%))/(NumProbesPerAxis%+1)
                'FOR k% = 1 TO NumProbesPerAxis%
                '    p% = k% + NumProbesPerAxis%*(i%-1) 'probe #
                '    R(p%,i%,0) = XiMin(i%) + k%*DeltaXi
                'NEXT k%
            NEXT i%
        CASE "UNIFORM ON-DIAGONAL"
            FOR p% = 1 TO Np%
                FOR i% = 1 TO Nd%
                    DeltaXi = (XiMax(i%)-XiMin(i%))/(Np%-1)
                    R(p%,i%,0) = XiMin(i%) + (p%-1)*DeltaXi
                NEXT i%
            NEXT p%
        CASE "2D GRID"
            NumProbesPerAxis% = SQR(Np%) : NumX1points% = NumProbesPerAxis% : NumX2points% = NumX1points% 'broken down for possible future use
            DelX1 = (XiMax(1)-XiMin(1))/(NumX1points%-1)
            DelX2 = (XiMax(2)-XiMin(2))/(NumX2points%-1)
            FOR x1pointNum% = 1 TO NumX1points%
```

```
                  FOR x2pointNum% = 1 TO NumX2points%

                      p% = NumX1points%*(x1pointNum%-1)+x2pointNum% 'probe #

                      R(p%,1,0) = XiMin(1) + DelX1*(x1pointNum%-1) 'x1 coord
                      R(p%,2,0) = XiMin(2) + DelX2*(x2pointNum%-1) 'x2 coord

                  NEXT x2pointNum%
              NEXT x1pointNum%
          CASE "RANDOM"
              FOR p% = 1 TO Np%

                  FOR i% = 1 TO Nd%

                      R(p%,i%,0) = XiMin(i%) + RandomNum(0##,1##)*(XiMax(i%)-XiMin(i%))

                  NEXT i%

              NEXT p%

      END SELECT
END SUB 'InitialProbeDistribution()
'------

SUB InitialProbeAccelerations(Np%,Nd%,A(),InitialAcceleration$,MaxInitialRandomAcceleration,MaxInitialFixedAcceleration)
LOCAL p%, i%

LOCAL A$

LOCAL FixedInitialAcceleration AS EXT

      SELECT CASE InitialAcceleration$
          CASE "ZERO"
              FOR p% = 1 TO Np%

                  FOR i% = 1 TO Nd%

                      A(p%,i%,0) = 0##

                  NEXT i% 'coordinate #

              NEXT p% 'probe #
          CASE "FIXED"
              A$ = INPUTBOX$("Fixed Initial Probe"+CHR$(13)+"Acceleration? [0.001-"+REMOVE$(STR$(MaxInitialFixedAcceleration),ANY" ")+"]","PROBES'
INITIAL ACCELERATION","0.5")

              FixedInitialAcceleration = VAL(A$)

              IF FixedInitialAcceleration < 0.001## OR FixedInitialAcceleration > MaxInitialFixedAcceleration THEN FixedInitialAcceleration = 0.5##

              FOR p% = 1 TO Np%

                  FOR i% = 1 TO Nd%

                      A(p%,i%,0) = FixedInitialAcceleration

                  NEXT i% 'coordinate

              NEXT p% 'probe #
          CASE "RANDOM"
              FOR p% = 1 TO Np%

                  FOR i% = 1 TO Nd%

                      A(p%,i%,0) = RandomNum(0##,1##)*MaxInitialRandomAcceleration

                  NEXT i% 'coordinate

              NEXT p% 'probe #

      END SELECT
END SUB 'InitialProbeAccelerations()
'------

SUB RetrieveErrantProbes(Np%,Nd%,j&,XiMin(),XiMax(),R(),M(),RepositionFactor$,Frep) 'note: M(), RepositionFcator$ passed but not used in thsi
version

LOCAL p%, i%

    FOR p% = 1 TO Np%

        FOR i% = 1 TO Nd%

            IF R(p%,i%,j&) < XiMin(i%) THEN R(p%,i%,j&) = XiMin(i%) + Frep*(R(p%,i%,j&-1)-XiMin(i%))

            IF R(p%,i%,j&) > XiMax(i%) THEN R(p%,i%,j&) = XiMax(i%) - Frep*(XiMax(i%)-R(p%,i%,j&-1))

        NEXT i%

    NEXT p%

END SUB 'RetrieveErrantProbes()
'------

SUB
ChangeRunParameters(NumProbesPerDimension%,Np%,Nd%,Nt&,G,Alpha,Beta,DeltaT,Frep,PlaceInitialProbes$,InitialAcceleration$,RepositionFactor$,FunctionN
ame$)
```

```
LOCAL A$, DefaultValue$

    A$ = INPUTBOX$("# dimensions?","Change # Dimensions ("+FunctionName$+")",NoSpaces$(Nd%+0,0)) : Nd%    = VAL(A$) : IF Nd% < 1 OR Nd% > 500 THEN
Nd% = 2

'    A$ = INPUTBOX$("# probes/dimension?","Change # Probes/Axis ("+FunctionName$+")",NoSpaces$(NumProbesPerDimension%+0,0)) : NumProbesPerDimension%
= VAL(A$)
'    IF NumProbesPerDimension% < 2 OR NumProbesPerDimension% > 500 THEN NumProbesPerDimension% = 10 'restrict values to reasonable (?) ranges

    IF Nd% > 1 THEN NumProbesPerDimension% = 2*((NumProbesPerDimension%+1)\2) 'require an even # probes on each axis to avoid overlapping at origin
(in symmetrical space at least...)

    IF Nd% = 1 THEN NumProbesPerDimension% = MAX(NumProbesPerDimension%,3)    'at least 3 probes on x-axis for 1-D functions

    Np% = NumProbesPerDimension%*Nd%

    A$ = INPUTBOX$("# time steps?","Change # Steps ("+FunctionName$+")",NoSpaces$(Nt&+0,0)) : Nt&    = VAL(A$) : IF Nt& < 3
THEN Nt& = 50

    A$ = INPUTBOX$("Grav Const G?","Change G ("+FunctionName$+")",NoSpaces$(G,2))            : G       = VAL(A$) : IF G < -100## OR G > 100##
THEN G   = 2##

    A$ = INPUTBOX$("Alpha?","Change Alpha ("+FunctionName$+")",NoSpaces$(Alpha,2))           : Alpha = VAL(A$) : IF Alpha < -50## OR Alpha > 50##
THEN Alpha = 2##

    A$ = INPUTBOX$("Beta?","Change Beta ("+FunctionName$+")",NoSpaces$(Beta,2))              : Beta  = VAL(A$) : IF Beta  < -50## OR Beta > 50##
THEN Beta  = 2##'

    A$ = INPUTBOX$("Delta T","Change Delta-T ("+FunctionName$+")",NoSpaces$(DeltaT,2))      : DeltaT= VAL(A$) : IF DeltaT =< 0##
THEN DeltaT = 1##

    A$ = INPUTBOX$("Frep [0-1]?","Change Frep ("+FunctionName$+")",NoSpaces$(Frep,3))       : Frep  = VAL(A$) : IF Frep < 0##    OR Frep > 1##
THEN Frep = 0.5##

'    ----------- Initial Probe Distribution ------------

    SELECT CASE PlaceInitialProbes$
        CASE "UNIFORM ON-AXIS"     : DefaultValue$ = "1"
        CASE "UNIFORM ON-DIAGONAL" : DefaultValue$ = "2"
        CASE "2D GRID"             : DefaultValue$ = "3"
        CASE "RANDOM"              : DefaultValue$ = "4"
    END SELECT

    A$ = INPUTBOX$("Initial Probes"+CHR$(13)+"1 - UNIFORM ON-AXIS"+CHR$(13)+"2 - UNIFORM ON-DIAGONAL"+CHR$(13)+"3 - 2D GRID"+CHR$(13)+"4 -
RANDOM","Initial Probe Distribution ("+FunctionName$+")",DefaultValue$)

    IF VAL(A$) < 1 OR VAL(A$) > 4 THEN A$ = "1"

    SELECT CASE VAL(A$)
        CASE 1 : PlaceInitialProbes$ = "UNIFORM ON-AXIS"
        CASE 2 : PlaceInitialProbes$ = "UNIFORM ON-DIAGONAL"
        CASE 3 : PlaceInitialProbes$ = "2D GRID"
        CASE 4 : PlaceInitialProbes$ = "RANDOM"
    END SELECT

    IF Nd% = 1  AND PlaceInitialProbes$ = "UNIFORM ON-DIAGONAL" THEN PlaceInitialProbes$ = "UNIFORM ON-AXIS" 'cannot do diagonal in 1-D space

    IF Nd% <> 2 AND PlaceInitialProbes$ = "2D GRID" THEN PlaceInitialProbes$ = "UNIFORM ON-AXIS" '2D grid is available only in 2 dimensions!

'    ------------- Initial Acceleration -----------------

    SELECT CASE InitialAcceleration$
        CASE "ZERO"   : DefaultValue$ = "1"
        CASE "FIXED"  : DefaultValue$ = "2"
        CASE "RANDOM" : DefaultValue$ = "3"
    END SELECT

    A$ = INPUTBOX$("Initial Acceleration"+CHR$(13)+"1 - ZERO"+CHR$(13)+"2 - FIXED"+CHR$(13)+"3 - RANDOM","Initial Acceleration
("+FunctionName$+")",DefaultValue$)

    IF VAL(A$) < 1 OR VAL(A$) > 3 THEN A$ = "1"

    SELECT CASE VAL(A$)
        CASE 1 : InitialAcceleration$ = "ZERO"
        CASE 2 : InitialAcceleration$ = "FIXED"
        CASE 3 : InitialAcceleration$ = "RANDOM"
    END SELECT

'    ----------- Reposition Factor --------------

    SELECT CASE RepositionFactor$
        CASE "FIXED"    : DefaultValue$ = "1"
        CASE "VARIABLE" : DefaultValue$ = "2"
        CASE "RANDOM"   : DefaultValue$ = "3"
    END SELECT

    A$ = INPUTBOX$("Reposition Factor?"+CHR$(13)+"1 - FIXED"+CHR$(13)+"2 - VARIABLE"+CHR$(13)+"3 - RANDOM","Retrieve Probes
("+FunctionName$+")",DefaultValue$)

    IF VAL(A$) < 1 OR VAL(A$) > 3 THEN A$ = "1"

    SELECT CASE VAL(A$)
        CASE 1 : RepositionFactor$ = "FIXED"
        CASE 2 : RepositionFactor$ = "VARIABLE"
        CASE 3 : RepositionFactor$ = "RANDOM"
    END SELECT

END SUB 'ChangeRunParameters()

'------

FUNCTION NoSpaces$(X,NumDigits%) :  NoSpaces$ = REMOVE$(STR$(X,NumDigits%),ANY" ") : END FUNCTION

'-----------

FUNCTION TerminateNowForSaturation$(j&,Nd%,Np%,Nt&,G,DeltaT,Alpha,Beta,R(),A(),M())

LOCAL A$, i&, p%, NumStepsForAveraging&

LOCAL BestFitness, AvgFitness, FitnessTOL AS EXT 'terminate if avg fitness does not change over NumStepsForAveraging& time steps

    FitnessTOL = 0.00001## : NumStepsForAveraging& = 10

    A$ = "NO"
```

```
IF j& >= NumStepsForAveraging+10 THEN 'wait until step 10 to start checking for fitness saturation

        AvgFitness = 0##

        FOR i& = j&-NumStepsForAveraging&+1 TO j& 'avg fitness over current step & previous NumStepsForAveraging&-1 steps

            BestFitness = M(1,i&)

            FOR p% = 1 TO Np%

                IF M(p%,i&) >= BestFitness THEN BestFitness = M(p%,i&)

            NEXT p%

            AvgFitness = AvgFitness + BestFitness

        NEXT i&

        AvgFitness = AvgFitness/NumStepsForAveraging&

        IF ABS(AvgFitness-BestFitness) < FitnessTOL THEN A$ = "YES" 'compare avg fitness to best fitness at this step

    END IF

    TerminateNowForSaturation$ = A$

END FUNCTION 'TerminateNowForSaturation$()

'-----------

FUNCTION MagVector(V(),N%) 'returns magnitude of Nx1 column vector V

LOCAL SumSq AS EXT

LOCAL i%

    SumSQ = 0## : FOR i% = 1 TO N% : SumSQ = SumSQ + V(i%)^2 : NEXT i% : MagVector = SQR(SumSQ)

END FUNCTION 'MagVector()

'---

FUNCTION UnitStep(X)

LOCAL Z AS EXT

    IF X < 0## THEN

        Z = 0##

    ELSE

        Z = 1##

    END IF

    UnitStep = Z

END FUNCTION 'UnitStep()

'---

SUB Plot1Dfunction(FunctionName$,XiMin(),XiMax(),R()) 'plots 1D function on-screen

LOCAL NumPoints%, i%, N%

LOCAL DeltaX, X AS EXT

    NumPoints% = 32001

    DeltaX = (XiMax(1)-XiMin(1))/(NumPoints%-1)

    N% = FREEFILE

    SELECT CASE FunctionName$

        CASE "ParrottF4" 'PARROTT F4 FUNCTION

            OPEN "ParrottF4" FOR OUTPUT AS #N%

                FOR i% = 1 TO NumPoints%

                    R(1,1,0) = XiMin(1) + (i%-1)*DeltaX

                    PRINT #N%, USING$("#.##### #.#####",R(1,1,0),ParrottF4(R(),1,1,0))

                NEXT i%

            CLOSE #N%

            CALL CreateGNUplotINIfile(0.2##*ScreenWidth&,0.2##*ScreenHeight&,0.6##*ScreenWidth&,0.6##*ScreenHeight&)

            CALL TwoDplot("ParrottF4","Parrott F4 Function","0.7","0.7","X\n\n.",".\n\nParrott F4(X)","","","0","1","0","1","","","wgnuplot.exe","" with lines linewidth 2","")

    END SELECT

END SUB

'------

SUB CLEANUP 'probe coordinate plot files

    IF DIR$("P1") <> "" THEN KILL "P1"
    IF DIR$("P2") <> "" THEN KILL "P2"
    IF DIR$("P3") <> "" THEN KILL "P3"
    IF DIR$("P4") <> "" THEN KILL "P4"
    IF DIR$("P5") <> "" THEN KILL "P5"
    IF DIR$("P6") <> "" THEN KILL "P6"
    IF DIR$("P7") <> "" THEN KILL "P7"
    IF DIR$("P8") <> "" THEN KILL "P8"
    IF DIR$("P9") <> "" THEN KILL "P9"
    IF DIR$("P10") <> "" THEN KILL "P10"
```

```
        IF DIR$("P11") <> "" THEN KILL "P11"
        IF DIR$("P12") <> "" THEN KILL "P12"
        IF DIR$("P13") <> "" THEN KILL "P13"
        IF DIR$("P14") <> "" THEN KILL "P14"
        IF DIR$("P15") <> "" THEN KILL "P15"

END SUB

'------

SUB Plot2Dfunction(FunctionName$,XiMin(),XiMax(),R())

LOCAL A$

LOCAL NumPoints%, i%, k%, N%

LOCAL DelX1, DelX2, Z AS EXT

    SELECT CASE FunctionName$

        CASE "PBM_1","PBM_2","PBM_3","PBM_4","PBM_5" : NumPoints% = 25

        CASE ELSE : NumPoints% = 100

    END SELECT

    N% = FREEFILE : OPEN "TwoDplot.DAT" FOR OUTPUT AS #N%

    DelX1 = (XiMax(1)-XiMin(1))/(NumPoints%-1) : DelX2 = (XiMax(2)-XiMin(2))/(NumPoints%-1)

    FOR i% = 1 TO NumPoints%

        R(1,1,0) = XiMin(1) + (i%-1)*DelX1 'x1 value

        FOR k% = 1 TO NumPoints%

            R(1,2,0) = XiMin(2) + (k%-1)*DelX2 'x2 value

            Z = ObjectiveFunction(R(),2,1,0,FunctionName$)

            PRINT #N%, USING$("######.###### ######.###### #######.#####^^^^",R(1,1,0),R(1,2,0),Z)

        NEXT k%

        PRINT #N%, ""

    NEXT i%

    CLOSE #N%

    CALL CreateGNUplotINIfile(0.1#*ScreenWidth&,0.1#*ScreenHeight&,0.6#*ScreenWidth&,0.6#*ScreenHeight&)

    A$ = "" : IF INSTR(FunctionName$,"PBM_") > 0 THEN A$ = "Coarse "

    CALL ThreeDplot2("TwoDplot.DAT",A$+"Plot of "+FunctionName$+" Function","","0.6","0.6","1.2", _
                "x1","x2","z=F(x1,x2)","","","wgnuplot.exe","","","","","")

END SUB

'------

    SUB TwoDplot3curves(NumCurves%,PlotFileName1$,PlotFileName2$,PlotFileName3$,PlotTitle$,Annotation$,xCoord$,yCoord$,XaxisLabel$,YaxisLabel$, _
                LogXaxis$,LogYaxis$,xMin$,xMax$,yMin$,yMax$,xTics$,yTics$,GnuPlotEXE$)

        LOCAL N%

        LOCAL LineSize$

        LineSize$ = "2"

        N% = FREEFILE

        OPEN "cmd2d.gp" FOR OUTPUT AS #N%

            IF LogXaxis$ = "YES" AND LogYaxis$ = "NO"  THEN PRINT #N%, "set logscale x"
            IF LogXaxis$ = "NO"  AND LogYaxis$ = "YES" THEN PRINT #N%, "set logscale y"
            IF LogXaxis$ = "YES" AND LogYaxis$ = "YES" THEN PRINT #N%, "set logscale xy"

            IF xMin$ <> "" AND xMax$ <> "" THEN PRINT #N%, "set xrange ["+xMin$+":"+xMax$+"]"

            IF yMin$ <> "" AND yMax$ <> "" THEN PRINT #N%, "set yrange ["+yMin$+":"+yMax$+"]"

            PRINT #N%, "set label "+Quote$+AnnoTation$+Quote$+" at graph "+xCoord$+","+yCoord$
            PRINT #N%, "set grid xtics"
            PRINT #N%, "set grid ytics"
            PRINT #N%, "set xtics "+xTics$
            PRINT #N%, "set ytics "+yTics$
            PRINT #N%, "set grid mxtics"
            PRINT #N%, "set grid mytics"
            PRINT #N%, "set title "+Quote$+PlotTitle$+Quote$
            PRINT #N%, "set xlabel "+Quote$+XaxisLabel$+Quote$
            PRINT #N%, "set ylabel "+Quote$+YaxisLabel$+Quote$

            SELECT CASE NumCurves%

            CASE 1
            PRINT #N%, "plot " + Quote$ + PlotFileName1$ + Quote$ + " with lines linewidth " + LineSize$

            CASE 2
            PRINT #N%, "plot " + Quote$ + PlotFileName1$ + Quote$ + " with lines linewidth " + LineSize$+", " + _
                            Quote$ + PlotFileName2$ + Quote$ + " with lines linewidth " + LineSize$
            CASE 3
            PRINT #N%, "plot " + Quote$ + PlotFileName1$ + Quote$ + " with lines linewidth " + LineSize$+", " + _
                            Quote$ + PlotFileName2$ + Quote$ + " with lines linewidth " + LineSize$+", " + _
                            Quote$ + PlotFileName3$ + Quote$ + " with lines linewidth " + LineSize$
            END SELECT

        CLOSE #N%

        SHELL(GnuPlotEXE$+" cmd2d.gp -")

        CALL Delay(0.3)

    END SUB 'TwoDplot3Curves()
```

```
'---

FUNCTION Fibonacci&&(N%) 'RETURNS Nth FIBONACCI NUMBER

LOCAL i%, Fn&&, Fn1&&, Fn2&&

LOCAL A$

    IF N% > 91 OR N% < 0 THEN

        MSGBOX("ERROR!  Fibonacci argument"+STR$(N%)+" > 91.  Out of range or < 0...") : EXIT FUNCTION

    END IF

    SELECT CASE N%

        CASE 0: Fn&& = 1

        CASE ELSE

            Fn&& = 0 : Fn2&& = 1 : i% = 0

            FOR i% = 1 TO N%

                Fn&& = Fn1&& + Fn2&&

                Fn1&& = Fn2&&

                Fn2&& = Fn&&

            NEXT i% 'LOOP

    END SELECT

    Fibonacci&& = Fn&&

END FUNCTION 'Fibonacci&&()

'-----------

FUNCTION RandomNum(a,b) 'Returns random number X, a=< X < b.

    RandomNum = a + (b-a)*RND

END FUNCTION 'RandomNum()

'-----------

FUNCTION GaussianDeviate(Mu,Sigma) 'returns NORMAL (Gaussian) random deviate with mean Mu and standard deviation Sigma (variance = Sigma^2)

'Refs: (1) Press, W.H, Flannery, B.P., Teukolsky, S.A., and Vetterling, W.T., "Numerical Recipes: The Art of Scientific Computing,"
'           §7.2, Cambridge University Press, Cambridge, UK, 1986.
'       (2) Shinzato, T., "Box Muller Method," 2007, http://www.sp.dis.titech.ac.jp/~shinzato/boxmuller.pdf

LOCAL s, t, Z AS EXT

    s = RND : t = RND

    Z = Mu + Sigma*SQR(-2#*LOG(s))*COS(TwoPi*t)

    GaussianDeviate = Z

END FUNCTION 'GaussianDeviate()

'-----------

    SUB ContourPlot(PlotFileName$,PlotTitle$,Annotation$,xCoord$,yCoord$,zCoord$, _
                    YaxisLabel$,YaxisLabel$,ZaxisLabel$,zMin$,zMax$,GnuPlotEXE$,A$)

        LOCAL N%

        N% = FREEFILE

        OPEN "cmd3d.gp" FOR OUTPUT AS #N%

            PRINT #N%, "show surface"
            PRINT #N%, "set hidden3d"
            IF zMin$ <> "" AND zMax$ <> "" THEN  PRINT #N%, "set zrange ["+zMin$+":"+zMax$+"]"
            PRINT #N%, "set label "+Quote$+AnnoTation$+Quote$+" at graph "+xCoord$+","+yCoord$+","+zCoord$
            PRINT #N%, "show label"
            PRINT #N%, "set grid xtics ytics ztics"
            PRINT #N%, "show grid"
            PRINT #N%, "set title "+Quote$+PlotTitle$+Quote$
            PRINT #N%, "set xlabel "+Quote$+XaxisLabel$+Quote$
            PRINT #N%, "set ylabel "+Quote$+YaxisLabel$+Quote$
            PRINT #N%, "set zlabel "+Quote$+ZaxisLabel$+Quote$
            PRINT #N%, "splot "+Quote$+PlotFileName$+Quote$+A$  '" notitle with linespoints" 'A$'" notitle with lines"
        CLOSE #N%

        SHELL(GnuPlotEXE$+" cmd3d.gp -")

    END SUB 'ContourPlot()

'---

    SUB ThreeDplot(PlotFileName$,PlotTitle$,Annotation$,xCoord$,yCoord$,zCoord$, _
                   YaxisLabel$,YaxisLabel$,ZaxisLabel$,zMin$,zMax$,GnuPlotEXE$,A$)

        LOCAL N%, ProcessID???

        N% = FREEFILE

        OPEN "cmd3d.gp" FOR OUTPUT AS #N%

            PRINT #N%, "set pm3d"
            PRINT #N%, "show pm3d"
            IF zMin$ <> "" AND zMax$ <> "" THEN  PRINT #N%, "set zrange ["+zMin$+":"+zMax$+"]"
            PRINT #N%, "set label "+Quote$+AnnoTation$+Quote$+" at graph "+xCoord$+","+yCoord$+","+zCoord$
            PRINT #N%, "show label"
            PRINT #N%, "set grid xtics ytics ztics"
            PRINT #N%, "show grid"
            PRINT #N%, "set title "+Quote$+PlotTitle$+Quote$
            PRINT #N%, "set xlabel "+Quote$+XaxisLabel$+Quote$
            PRINT #N%, "set ylabel "+Quote$+YaxisLabel$+Quote$
```

```
        PRINT #N%, "set zlabel "+Quote$+ZaxisLabel$+Quote$
        PRINT #N%, "splot "+Quote$+PlotFileName$+Quote$+A$+" notitle"' with lines
    CLOSE #N%

    SHELL(GnuPlotEXE$+" cmd3d.gp -") : CALL Delay(0.5##)

END SUB 'ThreeDplot()

'---

    SUB ThreeDplot2(PlotFileName$,PlotTitle$,Annotation$,xCoord$,yCoord$,zCoord$, _
                    XaxisLabel$,YaxisLabel$,ZaxisLabel$,zMin$,zMax$,GnuPlotEXE$,A$,xStart$,xStop$,yStart$,yStop$)

        LOCAL N%

        N% = FREEFILE

        OPEN "cmd3d.gp" FOR OUTPUT AS #N%

            PRINT #N%, "set pm3d"
            PRINT #N%, "show pm3d"
            PRINT #N%, "set hidden3d"
            PRINT #N%, "set view 45, 45, 1, 1"

            IF zMin$ <> "" AND zMax$ <> "" THEN PRINT #N%, "set zrange ["+zMin$+":"+zMax$+"]"

            PRINT #N%, "set xrange [" + xStart$ + ":" + xStop$ + "]"
            PRINT #N%, "set yrange [" + yStart$ + ":" + yStop$ + "]"
            PRINT #N%, "set label "   + Quote$ + AnnOtation$ + Quote$+" at graph "+xCoord$+","+yCoord$+","+zCoord$
            PRINT #N%, "show label"
            PRINT #N%, "set grid xtics ytics ztics"
            PRINT #N%, "show grid"
            PRINT #N%, "set title " + Quote$+PlotTitle$    + Quote$
            PRINT #N%, "set xlabel " + Quote$+XaxisLabel$  + Quote$
            PRINT #N%, "set ylabel " + Quote$+YaxisLabel$  + Quote$
            PRINT #N%, "set zlabel " + Quote$+ZaxisLabel$  + Quote$
            PRINT #N%, "splot "    + Quote$+PlotFileName$ + Quote$ + A$ + " notitle with lines"
        CLOSE #N%

        SHELL(GnuPlotEXE$+" cmd3d.gp -")

    END SUB 'ThreeDplot2()

'---

    SUB TwoDplot2Curves(PlotFileName1$,PlotFileName2$,PlotTitle$,Annotation$,xCoord$,yCoord$,XaxisLabel$,YaxisLabel$, _
                        LogXaxis$,LogYaxis$,xMin$,xMax$,yMin$,yMax$,xTics$,yTics$,GnuPlotEXE$,LineSize)

        LOCAL N%, ProcessID???

        N% = FREEFILE

        OPEN "cmd2d.gp" FOR OUTPUT AS #N%
            'print #N%, "set output "+Quote$+"test.plt"+Quote$ 'tried this 3/11/06, didn't work...

            IF LogXaxis$ = "YES" AND LogYaxis$ = "NO"  THEN PRINT #N%, "set logscale x"
            IF LogXaxis$ = "NO"  AND LogYaxis$ = "YES" THEN PRINT #N%, "set logscale y"
            IF LogXaxis$ = "YES" AND LogYaxis$ = "YES" THEN PRINT #N%, "set logscale xy"

            IF xMin$ <> "" AND xMax$ <> "" THEN PRINT #N%, "set xrange ["+xMin$+":"+xMax$+"]"

            IF yMin$ <> "" AND yMax$ <> "" THEN PRINT #N%, "set yrange ["+yMin$+":"+yMax$+"]"

            PRINT #N%, "set label "+Quote$+Annotation$+Quote$+" at graph "+xCoord$+","+yCoord$
            PRINT #N%, "set grid xtics"
            PRINT #N%, "set grid ytics"
            PRINT #N%, "set xtics "+xTics$
            PRINT #N%, "set ytics "+yTics$
            PRINT #N%, "set grid mxtics"
            PRINT #N%, "set grid mytics"
            PRINT #N%, "set title "+Quote$+PlotTitle$+Quote$
            PRINT #N%, "set xlabel "+Quote$+XaxisLabel$+Quote$
            PRINT #N%, "set ylabel "+Quote$+YaxisLabel$+Quote$

            PRINT #N%, "plot "+Quote$+PlotFileName1$+Quote$+" with lines linewidth "+REMOVE$(STR$(LineSize),ANY" ")+","+ _
                              Quote$+PlotFileName2$+Quote$+" with points pointsize 0.05"+REMOVE$(STR$(LineSize),ANY" ")

        CLOSE #N%

        ProcessID??? = SHELL(GnuPlotEXE$+" cmd2d.gp -") : CALL Delay(0.5##)

    END SUB 'TwoDplot2Curves()

'---

    SUB Probe2Dplots(ProbePlotsFileList$,PlotTitle$,Annotation$,xCoord$,yCoord$,XaxisLabel$,YaxisLabel$, _
                     LogXaxis$,LogYaxis$,xMin$,xMax$,yMin$,yMax$,xTics$,yTics$,GnuPlotEXE$)

        LOCAL N%, ProcessID???

        N% = FREEFILE

        OPEN "cmd2d.gp" FOR OUTPUT AS #N%

            IF LogXaxis$ = "YES" AND LogYaxis$ = "NO"  THEN PRINT #N%, "set logscale x"
            IF LogXaxis$ = "NO"  AND LogYaxis$ = "YES" THEN PRINT #N%, "set logscale y"
            IF LogXaxis$ = "YES" AND LogYaxis$ = "YES" THEN PRINT #N%, "set logscale xy"

            IF xMin$ <> "" AND xMax$ <> "" THEN PRINT #N%, "set xrange ["+xMin$+":"+xMax$+"]"

            IF yMin$ <> "" AND yMax$ <> "" THEN PRINT #N%, "set yrange ["+yMin$+":"+yMax$+"]"

            PRINT #N%, "set label "+Quote$+Annotation$+Quote$+" at graph "+xCoord$+","+yCoord$
            PRINT #N%, "set grid xtics"
            PRINT #N%, "set grid ytics"
            PRINT #N%, "set xtics "+xTics$
            PRINT #N%, "set ytics "+yTics$
            PRINT #N%, "set grid mxtics"
            PRINT #N%, "set grid mytics"
            PRINT #N%, "set title "+Quote$+PlotTitle$+Quote$
            PRINT #N%, "set xlabel "+Quote$+XaxisLabel$+Quote$
            PRINT #N%, "set ylabel "+Quote$+YaxisLabel$+Quote$

            PRINT #N%, ProbePlotsFileList$
```

```
          CLOSE #N%

          ProcessID??? = SHELL(GnuPlotEXE$+" cmd2d.gp -") : CALL Delay(0.5##)

      END SUB 'Probe2Dplots()

'---

SUB Show2probes(R()%,Np%,Nt&,j&,XiMin(),XiMax(),Frep,BestFitness,BestProbeNumber%,BestTimeStep&,FunctionName$,RepositionFactor$,Gamma)

    LOCAL N%, p%

    LOCAL A$, PlotFileName$, PlotTitle$, Symbols$

    LOCAL xMin$, xMax$, yMin$, yMax$

    LOCAL s1, s2, s3, s4 AS EXT

    PlotFileName$ = "Probes("+REMOVE$(STR$(j&),ANY" ")+")"

    IF j& > 0 THEN 'PLOT PROBES AT THIS TIME STEP

        PlotTitle$ = "\nLOCATIONS OF "+REMOVE$(STR$(Np%),ANY" ") + " PROBES AT TIME STEP" + STR$(j&) + " / " + REMOVE$(STR$(Nt&),ANY" ") + "\n" + _
                   "Fitness = "+REMOVE$(STR$(ROUND(BestFitness,3)),ANY" ") + ", Probe #" + REMOVE$(STR$(BestProbeNumber%),ANY" ") + " at Step #" + REMOVE$(STR$(BestTimeStep&),ANY" ") + _
                   "    [Frep = "+REMOVE$(STR$(Frep),4),ANY" ") + " " + RepositionFactor$ + "]\n"

    ELSE 'PLOT INITIAL PROBE DISTRIBUTION

        PlotTitle$ = "\nLOCATIONS OF "+REMOVE$(STR$(Np%),ANY" ") + " INITIAL PROBES FOR " + FunctionName$ + " FUNCTION\n[gamma = "+STR$(ROUND(Gamma,3))+"]\n"

    END IF

    N% = FREEFILE : OPEN PlotFileName$ FOR OUTPUT AS #N%

        FOR p% = 1 TO Np% : PRINT #N%, USING$("######.#####    ######.#####",R(p%,1,j&),R(p%,2,j&)) : NEXT p%

    CLOSE #N%

    s1 = 1.1## : s2 = 1.1## : s3 = 1.1## : s4 = 1.1## 'expand plots axes by 10%

    IF XiMin(1) > 0## THEN s1 = 0.9##
    IF XiMax(1) < 0## THEN s2 = 0.9##
    IF XiMin(2) > 0## THEN s3 = 0.9##
    IF XiMax(2) < 0## THEN s4 = 0.9##

    xMin$ = REMOVE$(STR$(s1*XiMin(1),2),ANY" ")
    xMax$ = REMOVE$(STR$(s2*XiMax(1),2),ANY" ")
    yMin$ = REMOVE$(STR$(s3*XiMin(2),2),ANY" ")
    yMax$ = REMOVE$(STR$(s4*XiMax(2),2),ANY" ")

    CALL TwoDplot(PlotFileName$,PlotTitle$,"0.6","0.7","x1\n\n","\nx2","NO","NO",xMin$,xMax$,yMin$,yMax$,"5","5","wgnuplot.exe"," pointsize 1 linewidth 2","")

    KILL PlotFileName$ 'erase plot data file after probes have been displayed

END SUB 'ShowProbes()

'----

    SUB TwoDplot(PlotFileName$,PlotTitle$,xCoord$,yCoord$,XaxisLabel$,YaxisLabel$, _
                 LogXaxis$,LogYaxis$,xMin$,xMax$,yMin$,yMax$,xTics$,yTics$,GnuPlotEXE$,LineType$,Annotation$)

        LOCAL N%, ProcessID???

        N% = FREEFILE

        OPEN "cmd2d.gp" FOR OUTPUT AS #N%

            IF LogXaxis$ = "YES" AND LogYaxis$ = "NO"  THEN PRINT #N%, "set logscale x"
            IF LogXaxis$ = "NO"  AND LogYaxis$ = "YES" THEN PRINT #N%, "set logscale y"
            IF LogXaxis$ = "YES" AND LogYaxis$ = "YES" THEN PRINT #N%, "set logscale xy"

            IF xMin$ <> "" AND xMax$ <> "" THEN PRINT #N%, "set xrange ["+xMin$+":"+xMax$+"]"
            IF yMin$ <> "" AND yMax$ <> "" THEN PRINT #N%, "set yrange ["+yMin$+":"+yMax$+"]"

            PRINT #N%, "set label "    + Quote$ + Annotation$ + Quote$ + " at graph " + xCoord$ + "," + yCoord$
            PRINT #N%, "set grid xtics " + xTics$
            PRINT #N%, "set grid ytics " + yTics$
            PRINT #N%, "set grid mxtics"
            PRINT #N%, "set grid mytics"
            PRINT #N%, "show grid"
            PRINT #N%, "set title "   + Quote$+PlotTitle$+Quote$
            PRINT #N%, "set xlabel " + Quote$+XaxisLabel$+Quote$
            PRINT #N%, "set ylabel " + Quote$+YaxisLabel$+Quote$

            PRINT #N%, "plot "+Quote$+PlotFileName$+Quote$+" notitle"+LineType$

        CLOSE #N%

        ProcessID??? = SHELL(GnuPlotEXE$+" cmd2d.gp -") : CALL Delay(0.5##)

    END SUB 'TwoDplot()

'-----

    SUB CreateGNUplotINIfile(PlotWindowULC_X%,PlotWindowULC_Y%,PlotWindowWidth%,PlotWindowHeight%)

    LOCAL N%, WinPath$, A$, B$, WindowsDirectory$

    WinPath$ = UCASE$(ENVIRON$("Path"))|'DIR$("C:\WINDOWS",23)

    DO
        B$ = A$

        A$ = EXTRACT$(WinPath$,";")

        WinPath$ = REMOVE$(WinPath$,A$+";")

        IF RIGHT$(A$,7) = "WINDOWS" OR A$ = B$ THEN EXIT LOOP

        IF RIGHT$(A$,5) = "WINNT"   OR A$ = B$ THEN EXIT LOOP
```

```
        LOOP

        WindowsDirectory$ = A$

        N% = FREEFILE

*    ---------- WGNUPLOT.INPUT FILE ----------
        OPEN WindowsDirectory$+"\wgnuplot.ini" FOR OUTPUT AS #N%

            PRINT #N%,"[WGNUPLOT]"
            PRINT #N%,"TextOrigin=0 0"
            PRINT #N%,"TextSize=640 150"
            PRINT #N%,"TextFont=Terminal,9"
            PRINT #N%,"GraphOrigin="+REMOVE$(STR$(PlotWindowULC_X%),ANY" ")+" "+REMOVE$(STR$(PlotWindowULC_Y%),ANY" ")
            PRINT #N%,"GraphSize="+REMOVE$(STR$(PlotWindowWidth%),ANY" ")+" "+REMOVE$(STR$(PlotWindowHeight%),ANY" ")
            PRINT #N%,"GraphFont=Arial,10"
            PRINT #N%,"GraphColor=1"
            PRINT #N%,"GraphToTop=1"
            PRINT #N%,"GraphBackground=255 255 255"
            PRINT #N%,"Border=0 0 0 0"
            PRINT #N%,"Axis=192 192 192 2 2"
            PRINT #N%,"Line1=0 0 255 0 0"
            PRINT #N%,"Line2=0 255 0 0 1"
            PRINT #N%,"Line3=255 0 0 0 2"
            PRINT #N%,"Line4=255 0 255 0 3"
            PRINT #N%,"Line5=0 0 128 0 4"

        CLOSE #N%

        END SUB 'CreateGNUplotINIfile()

*------
        SUB Delay(NumSecs)

            LOCAL StartTime, StopTime AS EXT

            StartTime = TIMER

            DO UNTIL (StopTime-StartTime) >= NumSecs

                StopTime = TIMER

            LOOP

        END SUB 'Delay()

*-----
SUB MathematicalConstants
    EulerConst  = 0.5772156649015328660605128##
    Pi          = 3.1415926535897932384626243##
    Pi2         = Pi/2##
    Pi4         = Pi/4##
    TwoPi       = 2##*Pi
    FourPi      = 4##*Pi
    e           = 2.7182818284590452353602874##
    Root2       = 1.4142135623730950488##
END SUB

*-----
SUB AlphabetAndDigits
    Alphabet$   = "ABCDEFGHIJKLMNOPQRSTUVWXYZabcdefghijklmnopqrstuvwxyz"
    Digits$     = "0123456789"
    RunID$      = DATE$ + ", " + TIME$
END SUB

*------
SUB SpecialSymbols
    Quote$              = CHR$(34) 'Quotation mark "
    SpecialCharacters$ = "'(),#:;/_"
END SUB

*-----
SUB EMconstants
    Mu0  = 4E-7##*Pi      'hy/meter
    Eps0 = 8.854##*1E-12 'fd/meter
    c    = 2.998E8##      'velocity of light, 1##/SQR(Mu0*Eps0) 'meters/sec
    eta0 = SQR(Mu0/Eps0) 'impedance of free space, ohms
END SUB

*------
SUB ConversionFactors
    Rad2Deg     = 180##/Pi
    Deg2Rad     = 1##/Rad2Deg
    Feet2Meters = 0.3048##
    Meters2Feet = 1##/Feet2Meters
    Inches2Meters = 0.0254##
    Meters2Inches = 1##/Inches2Meters
    Miles2Meters = 1609.344##
    Meters2Miles = 1##/Miles2Meters
    NautMi2Meters = 1852##
    Meters2NautMi = 1##/NautMi2Meters
END SUB

*------
SUB ShowConstants 'puts up msgbox showing all constants

LOCAL A$

A$ = _
"Mathematical Constants:"+CHR$(13)+_
"Euler const="+STR$(EulerConst)+CHR$(13)+_
"Pi="+STR$(Pi)+CHR$(13)+_
"Pi/2="+STR$(Pi2)+CHR$(13)+_
"Pi/4="+STR$(Pi4)+CHR$(13)+_
"2Pi="+STR$(TwoPi)+CHR$(13)+_
"4Pi="+STR$(FourPi)+CHR$(13)+_
```

```
"e="+STR$(e)+CHR$(13)+_
"Sqr2="+STR$(Root2)+CHR$(13)+CHR$(13)+_
"Alphabet, Digits & Special Characters:"+CHR$(13)+_
"Alphabet="+Alphabet$+CHR$(13)+_
"Digits="+Digits$+CHR$(13)+_
"quote="+Quote$+CHR$(13)+_
"Spec chars="+SpecialCharacters$+CHR$(13)+CHR$(13)+_
"E&M Constants:"+CHR$(13)+_
"Mu0="+STR$(Mu0)+CHR$(13)+_
"Eps0="+STR$(Eps0)+CHR$(13)+_
"c="+STR$(c)+CHR$(13)+_
"Eta0="+STR$(eta0)+CHR$(13)+CHR$(13)+_
"Conversion Factors:"+CHR$(13)+_
"Rad2Deg="+STR$(Rad2Deg)+CHR$(13)+_
"Deg2Rad="+STR$(Deg2Rad)+CHR$(13)+_
"Ft2meters="+STR$(Feet2Meters)+CHR$(13)+_
"Meters2Ft="+STR$(Meters2Feet)+CHR$(13)+_
"Inches2Meters="+STR$(Inches2Meters)+CHR$(13)+_
"Meters2Inches="+STR$(Meters2Inches)+CHR$(13)+_
"Miles2Meters="+STR$(Miles2Meters)+CHR$(13)+_
"Meters2Miles="+STR$(Meters2Miles)+CHR$(13)+_
"NautMi2Meters="+STR$(NautMi2Meters)+CHR$(13)+_
"Meters2NautMi="+STR$(Meters2NautMi)+CHR$(13)+CHR$(13)

MSGBOX(A$)

END SUB

'------

SUB DisplayRmatrix(Np%,Nd%,Nt&,R())

LOCAL p%, i%, j&, A$

    A$ = "Position Vector Matrix R()"+CHR$(13)

    FOR p% = 1 TO Np%

        FOR i% = 1 TO Nd%

            FOR j& = 0 TO Nt&

                A$ = A$ + "R("+STR$(p%)+", "+STR$(i%)+", "+STR$(j&)+") ="+STR$(R(p%,i%,j&)) + CHR$(13)

            NEXT j&

        NEXT i%

    NEXT p%

    MSGBOX(A$)

END SUB

'------

SUB DisplayRmatrixThisTimeStep(Np%,Nd%,j&,R())

LOCAL p%, i%, A$

    A$ = "Position Vector Matrix R() at step "+STR$(j&)+":"+CHR$(13)

    FOR p% = 1 TO Np%

        FOR i% = 1 TO Nd%

            A$ = A$ + "R("+STR$(p%)+", "+STR$(i%)+", "+STR$(j&)+") ="+STR$(R(p%,i%,j&)) + CHR$(13)

        NEXT i%

    NEXT p%

    MSGBOX(A$)

END SUB

'------

SUB DisplayAmatrix(Np%,Nd%,Nt&,A())

LOCAL p%, i%, j&, A$

    A$ = "Acceleration Vector Matrix A()"+CHR$(13)

    FOR p% = 1 TO Np%

        FOR i% = 1 TO Nd%

            FOR j& = 0 TO Nt&

                A$ = A$ + "A("+STR$(p%)+", "+STR$(i%)+", "+STR$(j&)+") ="+STR$(A(p%,i%,j&)) + CHR$(13)

            NEXT j&

        NEXT i%

    NEXT p%

    MSGBOX(A$)

END SUB

'------

SUB DisplayAmatrixThisTimeStep(Np%,Nd%,j&,A())

LOCAL p%, i%, A$

    A$ = "Acceleration matrix A() at step "+STR$(j&)+":"+CHR$(13)

    FOR p% = 1 TO Np%

        FOR i% = 1 TO Nd%
```

```
            A$ = A$ + "A("+STR$(p%)+", "+STR$(i%)+", "+STR$(j&)+ ") ="+STR$(A(p%,i%,j&)) + CHR$(13)

        NEXT i%

    NEXT p%

    MSGBOX(A$)

END SUB

'------

SUB DisplayMmatrix(Np%,Nt&,M())

LOCAL p%, j&, A$

    A$ = "Fitness Matrix M("+CHR$(13)

    FOR p% = 1 TO Np%

        FOR j& = 0 TO Nt&

            A$ = A$ + "M("+STR$(p%)+", "+STR$(j&)+ ") ="+STR$(M(p%,j&)) + CHR$(13)

        NEXT j&

    NEXT p%

    MSGBOX(A$)

END SUB

'------

SUB DisplayMmatrixThisTimeStep(Np%,j&,M())

LOCAL p%, A$

    A$ = "Fitness matrix M() at step "+STR$(j&)+":"+CHR$(13)

        FOR p% = 1 TO Np%

            A$ = A$ + "M("+STR$(p%)+", "+STR$(j&)+ ") ="+STR$(M(p%,j&)) + CHR$(13)

        NEXT p%

        MSGBOX(A$)

END SUB

'------

SUB DisplayXiMinMax(Nd%,XiMin(),XiMax())

LOCAL i%, A$

    A$ = ""

    FOR i% = 1 TO Nd%

        A$ = A$ + "XiMin("+STR$(i%)+" ) = "+STR$(XiMin(i%))+"   XiMax("+STR$(i%)+" ) = "+STR$(XiMax(i%)) + CHR$(13)

    NEXT i%

    MSGBOX(A$)

END SUB

'------

SUB DisplayRunParameters2(FunctionName$,Nd%,Np%,Nt&,G,DeltaT,Alpha,Beta,Frep,PlaceInitialProbes,InitialAcceleration$,RepositionFactor$)

LOCAL A$

    A$ = "Function = "+ FunctionName$+CHR$(13)+_
        "Nd = "+STR$(Nd%)+CHR$(13)+_
        "Np = "+STR$(Np%)+CHR$(13)+_
        "Nt = "+STR$(Nt&)+CHR$(13)+_
        "G = "+STR$(G)+CHR$(13)+_
        "DeltaT = "+STR$(DeltaT)+CHR$(13)+_
        "Alpha = "+STR$(Alpha)+CHR$(13)+_
        "Beta = "+STR$(Beta)+CHR$(13)+_
        "Frep = "+STR$(Frep)+CHR$(13)+_
        "Init Probes: "+PlaceInitialProbes$+CHR$(13)+_
        "Init Accel: "+InitialAcceleration$+CHR$(13)+_
        "Retrive Method: "+RepositionFactor$+CHR$(13)

    MSGBOX(A$)

END SUB

'------

SUB
Tabulate1DprobeCoordinates(Max1DprobesPlotted%,Nd%,Np%,LastStep&,G,DeltaT,Alpha,Beta,Frep,R(),M(),PlaceInitialProbes,InitialAcceleration$,Repositio
nFactor$,FunctionName$,Gamma)

LOCAL N%, ProbeNum%, FileHeader$, A$, B$, C$, D$, E$, F$, H$, StepNum&, FieldNumber% 'kludgy, yes, but it accomplishes its purpose...

        CALL
GetPlotAnnotation(FileHeader$,Nd%,Np%,LastStep&,G,DeltaT,Alpha,Beta,Frep,M(),PlaceInitialProbes$,InitialAcceleration$,RepositionFactor$,FunctionName
$,Gamma)

        REPLACE "\n" WITH ", " IN FileHeader$

        FileHeader$ = LEFT$(FileHeader$,LEN(FileHeader$)-2)

        FileHeader$ = "PROBE COORDINATES" + CHR$(13) +_
                    "-----------------" + CHR$(13) + FileHeader$

        N% = FREEFILE : OPEN "ProbeCoordinates.DAT" FOR OUTPUT AS #N%

            A$ = "   Step #     " : B$ = "   ------     " : C$ = ""
```

```
        FOR ProbeNum% = 1 TO Np% 'create out data file header

            SELECT CASE ProbeNum%
                CASE    1 TO    9 : E$ = ""   : F$ = "      " : H$ = "      "
                CASE   10 TO   99 : E$ = "-"  : F$ = "     " : H$ = "     "
                CASE  100 TO  999 : E$ = "--" : F$ = "    " : H$ = "    "
            END SELECT

            A$ = A$ + "P" + NoSpaces$(ProbeNum%+0,0) + F$ 'note: adding zero to ProbeNum% necessary to convert to floating point...

            B$ = B$ + E$ + "--" + H$

            C$ = C$ + "######.###    "

'           C$ = C$ + "##.#######"

        NEXT ProbeNum%

        PRINT #N%, FileHeader$ + CHR$(13) : PRINT #N%, A$ : PRINT #N%, B$

        FOR StepNum% = 0 TO LastStep&

            D$ = USING$("######   ",StepNum&)

            FOR ProbeNum% = 1 TO Np% : D$ = D$ + USING$(C$,R(ProbeNum%,1,StepNum&)) : NEXT ProbeNum%

            PRINT #N%, D$

        NEXT StepNum&

    CLOSE #N%

END SUB 'Tabulate1DprobeCoordinates()

'------

SUB
Plot1DprobePositions(Max1DprobesPlotted%,Nd%,Np%,LastStep&,G,DeltaT,Alpha,Beta,Frep,R(),M(),PlaceInitialProbes$,InitialAcceleration$,RepositionFacto
r$,FunctionName$,Gamma)
    'plots on-screen 1D function probe positions vs time step if Np =< 10

LOCAL ProcessID???, N%, n1%, n2%, n3%, n4%, n5%, n6%, n7%, n8%, n9%, n10%, n11%, n12%, n13%, n14%, n15%, ProbeNum%, StepNum&, A$

LOCAL PlotAnnotation$

    IF Np% > Max1DprobesPlotted% THEN EXIT SUB

    CALL CLEANUP 'delete old "Px" plot files, if any

    ProbeNum% = 0

    DO 'create output data files, probe-by-probe
        INCR ProbeNum% : n1% = FREEFILE : OPEN "P"+REMOVE$(STR$(ProbeNum%),ANY" ") FOR OUTPUT AS #n1%  : IF ProbeNum% = Np% THEN EXIT LOOP
        INCR ProbeNum% : n2% = FREEFILE : OPEN "P"+REMOVE$(STR$(ProbeNum%),ANY" ") FOR OUTPUT AS #n2%  : IF ProbeNum% = Np% THEN EXIT LOOP
        INCR ProbeNum% : n3% = FREEFILE : OPEN "P"+REMOVE$(STR$(ProbeNum%),ANY" ") FOR OUTPUT AS #n3%  : IF ProbeNum% = Np% THEN EXIT LOOP
        INCR ProbeNum% : n4% = FREEFILE : OPEN "P"+REMOVE$(STR$(ProbeNum%),ANY" ") FOR OUTPUT AS #n4%  : IF ProbeNum% = Np% THEN EXIT LOOP
        INCR ProbeNum% : n5% = FREEFILE : OPEN "P"+REMOVE$(STR$(ProbeNum%),ANY" ") FOR OUTPUT AS #n5%  : IF ProbeNum% = Np% THEN EXIT LOOP
        INCR ProbeNum% : n6% = FREEFILE : OPEN "P"+REMOVE$(STR$(ProbeNum%),ANY" ") FOR OUTPUT AS #n6%  : IF ProbeNum% = Np% THEN EXIT LOOP
        INCR ProbeNum% : n7% = FREEFILE : OPEN "P"+REMOVE$(STR$(ProbeNum%),ANY" ") FOR OUTPUT AS #n7%  : IF ProbeNum% = Np% THEN EXIT LOOP
        INCR ProbeNum% : n8% = FREEFILE : OPEN "P"+REMOVE$(STR$(ProbeNum%),ANY" ") FOR OUTPUT AS #n8%  : IF ProbeNum% = Np% THEN EXIT LOOP
        INCR ProbeNum% : n9% = FREEFILE : OPEN "P"+REMOVE$(STR$(ProbeNum%),ANY" ") FOR OUTPUT AS #n9%  : IF ProbeNum% = Np% THEN EXIT LOOP
        INCR ProbeNum% : n10% = FREEFILE : OPEN "P"+REMOVE$(STR$(ProbeNum%),ANY" ") FOR OUTPUT AS #n10% : IF ProbeNum% = Np% THEN EXIT LOOP
        INCR ProbeNum% : n11% = FREEFILE : OPEN "P"+REMOVE$(STR$(ProbeNum%),ANY" ") FOR OUTPUT AS #n11% : IF ProbeNum% = Np% THEN EXIT LOOP
        INCR ProbeNum% : n12% = FREEFILE : OPEN "P"+REMOVE$(STR$(ProbeNum%),ANY" ") FOR OUTPUT AS #n12% : IF ProbeNum% = Np% THEN EXIT LOOP
        INCR ProbeNum% : n13% = FREEFILE : OPEN "P"+REMOVE$(STR$(ProbeNum%),ANY" ") FOR OUTPUT AS #n13% : IF ProbeNum% = Np% THEN EXIT LOOP
        INCR ProbeNum% : n14% = FREEFILE : OPEN "P"+REMOVE$(STR$(ProbeNum%),ANY" ") FOR OUTPUT AS #n14  : IF ProbeNum% = Np% THEN EXIT LOOP
        INCR ProbeNum% : n15% = FREEFILE : OPEN "P"+REMOVE$(STR$(ProbeNum%),ANY" ") FOR OUTPUT AS #n15% : IF ProbeNum% = Np% THEN EXIT LOOP
    LOOP

    ProbeNum% = 0

    DO 'output probe positions as a function of time step
        INCR ProbeNum% : FOR StepNum& = 0 TO LastStep& : PRINT #n1%,  USING$("###### ######.########",StepNum&,R(ProbeNum%,1,StepNum&)) : NEXT
StepNum& : IF ProbeNum% = Np% THEN EXIT LOOP
        INCR ProbeNum% : FOR StepNum& = 0 TO LastStep& : PRINT #n2%,  USING$("###### ######.########",StepNum&,R(ProbeNum%,1,StepNum&)) : NEXT
StepNum& : IF ProbeNum% = Np% THEN EXIT LOOP
        INCR ProbeNum% : FOR StepNum& = 0 TO LastStep& : PRINT #n3%,  USING$("###### ######.########",StepNum&,R(ProbeNum%,1,StepNum&)) : NEXT
StepNum& : IF ProbeNum% = Np% THEN EXIT LOOP
        INCR ProbeNum% : FOR StepNum& = 0 TO LastStep& : PRINT #n4%,  USING$("###### ######.########",StepNum&,R(ProbeNum%,1,StepNum&)) : NEXT
StepNum& : IF ProbeNum% = Np% THEN EXIT LOOP
        INCR ProbeNum% : FOR StepNum& = 0 TO LastStep& : PRINT #n5%,  USING$("###### ######.########",StepNum&,R(ProbeNum%,1,StepNum&)) : NEXT
StepNum& : IF ProbeNum% = Np% THEN EXIT LOOP
        INCR ProbeNum% : FOR StepNum& = 0 TO LastStep& : PRINT #n6%,  USING$("###### ######.########",StepNum&,R(ProbeNum%,1,StepNum&)) : NEXT
StepNum& : IF ProbeNum% = Np% THEN EXIT LOOP
        INCR ProbeNum% : FOR StepNum& = 0 TO LastStep& : PRINT #n7%,  USING$("###### ######.########",StepNum&,R(ProbeNum%,1,StepNum&)) : NEXT
StepNum& : IF ProbeNum% = Np% THEN EXIT LOOP
        INCR ProbeNum% : FOR StepNum& = 0 TO LastStep& : PRINT #n8%,  USING$("###### ######.########",StepNum&,R(ProbeNum%,1,StepNum&)) : NEXT
StepNum& : IF ProbeNum% = Np% THEN EXIT LOOP
        INCR ProbeNum% : FOR StepNum& = 0 TO LastStep& : PRINT #n9%,  USING$("###### ######.########",StepNum&,R(ProbeNum%,1,StepNum&)) : NEXT
StepNum& : IF ProbeNum% = Np% THEN EXIT LOOP
        INCR ProbeNum% : FOR StepNum& = 0 TO LastStep& : PRINT #n10%, USING$("###### ######.########",StepNum&,R(ProbeNum%,1,StepNum&)) : NEXT
StepNum& : IF ProbeNum% = Np% THEN EXIT LOOP
        INCR ProbeNum% : FOR StepNum& = 0 TO LastStep& : PRINT #n11%, USING$("###### ######.########",StepNum&,R(ProbeNum%,1,StepNum&)) : NEXT
StepNum& : IF ProbeNum% = Np% THEN EXIT LOOP
        INCR ProbeNum% : FOR StepNum& = 0 TO LastStep& : PRINT #n12%, USING$("###### ######.########",StepNum&,R(ProbeNum%,1,StepNum&)) : NEXT
StepNum& : IF ProbeNum% = Np% THEN EXIT LOOP
        INCR ProbeNum% : FOR StepNum& = 0 TO LastStep& : PRINT #n13%, USING$("###### ######.########",StepNum&,R(ProbeNum%,1,StepNum&)) : NEXT
StepNum& : IF ProbeNum% = Np% THEN EXIT LOOP
        INCR ProbeNum% : FOR StepNum& = 0 TO LastStep& : PRINT #n14%, USING$("###### ######.########",StepNum&,R(ProbeNum%,1,StepNum&)) : NEXT
StepNum& : IF ProbeNum% = Np% THEN EXIT LOOP
        INCR ProbeNum% : FOR StepNum& = 0 TO LastStep& : PRINT #n15%, USING$("###### ######.########",StepNum&,R(ProbeNum%,1,StepNum&)) : NEXT
StepNum& : IF ProbeNum% = Np% THEN EXIT LOOP
    LOOP

    ProbeNum% = 0

    DO 'close output data files
        INCR ProbeNum% : CLOSE #n1% : IF ProbeNum% = Np% THEN EXIT LOOP
        INCR ProbeNum% : CLOSE #n2% : IF ProbeNum% = Np% THEN EXIT LOOP
        INCR ProbeNum% : CLOSE #n3% : IF ProbeNum% = Np% THEN EXIT LOOP
        INCR ProbeNum% : CLOSE #n4% : IF ProbeNum% = Np% THEN EXIT LOOP
        INCR ProbeNum% : CLOSE #n5% : IF ProbeNum% = Np% THEN EXIT LOOP
        INCR ProbeNum% : CLOSE #n6% : IF ProbeNum% = Np% THEN EXIT LOOP
        INCR ProbeNum% : CLOSE #n7% : IF ProbeNum% = Np% THEN EXIT LOOP
```

```
          INCR ProbeNum% : CLOSE #n8%  : IF ProbeNum% = Np% THEN EXIT LOOP
          INCR ProbeNum% : CLOSE #n9%  : IF ProbeNum% = Np% THEN EXIT LOOP
          INCR ProbeNum% : CLOSE #n10% : IF ProbeNum% = Np% THEN EXIT LOOP
          INCR ProbeNum% : CLOSE #n11% : IF ProbeNum% = Np% THEN EXIT LOOP
          INCR ProbeNum% : CLOSE #n12% : IF ProbeNum% = Np% THEN EXIT LOOP
          INCR ProbeNum% : CLOSE #n13% : IF ProbeNum% = Np% THEN EXIT LOOP
          INCR ProbeNum% : CLOSE #n14% : IF ProbeNum% = Np% THEN EXIT LOOP
          INCR ProbeNum% : CLOSE #n15% : IF ProbeNum% = Np% THEN EXIT LOOP
    LOOP

    ProbeNum% = 0 : A$ = ""

    DO 'create file string for plot command file
        INCR ProbeNum% : A$ = A$ + Quote$ + "P"+REMOVE$(STR$(ProbeNum%),ANY" ") + Quote$ + " w l lw 2, " : IF ProbeNum% = Np% THEN EXIT LOOP
        INCR ProbeNum% : A$ = A$ + Quote$ + "P"+REMOVE$(STR$(ProbeNum%),ANY" ") + Quote$ + " w l lw 2, " : IF ProbeNum% = Np% THEN EXIT LOOP
        INCR ProbeNum% : A$ = A$ + Quote$ + "P"+REMOVE$(STR$(ProbeNum%),ANY" ") + Quote$ + " w l lw 2, " : IF ProbeNum% = Np% THEN EXIT LOOP
        INCR ProbeNum% : A$ = A$ + Quote$ + "P"+REMOVE$(STR$(ProbeNum%),ANY" ") + Quote$ + " w l lw 2, " : IF ProbeNum% = Np% THEN EXIT LOOP
        INCR ProbeNum% : A$ = A$ + Quote$ + "P"+REMOVE$(STR$(ProbeNum%),ANY" ") + Quote$ + " w l lw 2, " : IF ProbeNum% = Np% THEN EXIT LOOP
        INCR ProbeNum% : A$ = A$ + Quote$ + "P"+REMOVE$(STR$(ProbeNum%),ANY" ") + Quote$ + " w l lw 2, " : IF ProbeNum% = Np% THEN EXIT LOOP
        INCR ProbeNum% : A$ = A$ + Quote$ + "P"+REMOVE$(STR$(ProbeNum%),ANY" ") + Quote$ + " w l lw 2, " : IF ProbeNum% = Np% THEN EXIT LOOP
        INCR ProbeNum% : A$ = A$ + Quote$ + "P"+REMOVE$(STR$(ProbeNum%),ANY" ") + Quote$ + " w l lw 2, " : IF ProbeNum% = Np% THEN EXIT LOOP
        INCR ProbeNum% : A$ = A$ + Quote$ + "P"+REMOVE$(STR$(ProbeNum%),ANY" ") + Quote$ + " w l lw 2, " : IF ProbeNum% = Np% THEN EXIT LOOP
        INCR ProbeNum% : A$ = A$ + Quote$ + "P"+REMOVE$(STR$(ProbeNum%),ANY" ") + Quote$ + " w l lw 2, " : IF ProbeNum% = Np% THEN EXIT LOOP
        INCR ProbeNum% : A$ = A$ + Quote$ + "P"+REMOVE$(STR$(ProbeNum%),ANY" ") + Quote$ + " w l lw 2, " : IF ProbeNum% = Np% THEN EXIT LOOP
        INCR ProbeNum% : A$ = A$ + Quote$ + "P"+REMOVE$(STR$(ProbeNum%),ANY" ") + Quote$ + " w l lw 2, " : IF ProbeNum% = Np% THEN EXIT LOOP
    LOOP

    A$ = LEFT$(A$,LEN(A$)-2)

    CALL
GetPlotAnnotation(PlotAnnotation$,Nd%,Np%,LastStep&,G,DeltaT,Alpha,Beta,Frep,M(),PlaceInitialProbes,InitialAcceleration$,RepositionFactor$,Function
Name$,Gamma)

    N% = FREEFILE

    OPEN "cmd2d.gp" FOR OUTPUT AS #N%

        PRINT #N%, "set label "    + Quote$ + PlotAnnotation$ + Quote$ + " at graph 0.5,0.95"
        PRINT #N%, "set grid xtics"
        PRINT #N%, "set grid ytics"
        PRINT #N%, "set title " + Quote$ + "Evolution of "   + FunctionName$ + " Probe Positions"+"\n" + RunID$ + Quote$
        PRINT #N%, "set xlabel " + Quote$ + "Time Step" + Quote$
        PRINT #N%, "set ylabel " + Quote$ + "Probe Coordinate" + Quote$
        PRINT #N%, "plot "        + A$

    CLOSE #N%

    CALL CreateGNUplotINIfile(0.2##*ScreenWidth&,0.2##*ScreenHeight&,0.6##*ScreenWidth&,0.6##*ScreenHeight&) 'USAGE: CALL
CreateGNUplotINIfile(PlotWindowULC_X%,PlotWindowULC_Y%,PlotWindowWidth%,PlotWindowHeight%)

    ProcessID??? = SHELL("wgnuplot.exe"+" cmd2d.gp -") : CALL Delay(5##) 'before SUB Cleanup is called

END SUB

'------

SUB
DisplayRunParameters(FunctionName$,Nd%,Np%,Nt&,G,DeltaT,Alpha,Beta,Frep,R(),A(),M(),PlaceInitialProbes$,InitialAcceleration$,RepositionFactor$,RunCF
O$,ShrinkDS$,CheckForEarlyTermination$)

LOCAL A$, B$, YN%

    ShrinkDS$ = "NO"                    : YN% = MSGBOX("Adaptively Shrink DS?",%MB_YESNO,"ADAPTIVE DS?")               : IF YN% = %IDYES THEN ShrinkDS$
= "YES"

    CheckForEarlyTermination$ = "NO" : YN% = MSGBOX("Check for Early Termination?",%MB_YESNO,"EARLY TERMINATION?") : IF YN% = %IDYES THEN
CheckForEarlyTermination$ = "YES"

    B$ = "" : IF PlaceInitialProbes$ = "UNIFORM ON-AXIS" AND Nd% > 1 THEN B$ = "  ["+REMOVE$(STR$(Np%/Nd%),ANY" ") + "/axis]"

    RunCFO$ = "NO"

'   A$ = "RUN CFO WITH THE" + CHR$(13) +_
'        "FOLLOWING PARAMETERS?"                      + CHR$(13) + CHR$(13) +_
'   '    "Function "   + FunctionName$                + CHR$(13) + _
'   '    "# probes = " + REMOVE$(STR$(Np%),ANY" ")    + " (" + REMOVE$(STR$(Nd%),ANY" ") + "~D)" + CHR$(13) + CHR$(13) +_
'   '    "# time steps = " + REMOVE$(STR$(Nt&),ANY" ") + CHR$(13) + _
'   '    "Grav Const G = " + REMOVE$(STR$(G,2),ANY" ") + CHR$(13) + _
'   '    "Delta-T = "      + REMOVE$(STR$(DeltaT,3),ANY" ") + CHR$(13) + _
'   '    "Exp Alpha = "    + REMOVE$(STR$(Alpha,3),ANY" ") + CHR$(13) + _
'   '    "Exp Beta = "     + REMOVE$(STR$(Beta,3),ANY" ") + CHR$(13) + _
'   '    "Frep = "         + REMOVE$(STR$(Frep,4),ANY" ") + " " + "RepositionFactor$ + ")" + CHR$(13) + _
'   '    "Initial Probes: " + PlaceInitialProbes$        + CHR$(13) + _
'   '    "Initial Accel: " + InitialAcceleration$        + CHR$(13) + CHR$(13)

    A$ = "RUN CFO WITH THE" + CHR$(13) +_
         "FOLLOWING PARAMETERS?"                       + CHR$(13) + CHR$(13) +_
         "Function "   + FunctionName$                 + " (" + REMOVE$(STR$(Nd%),ANY" ") + "~D)" + CHR$(13) +_
         "# time steps = " + REMOVE$(STR$(Nt&),ANY" ") + CHR$(13) + _
         "Grav Const G = " + REMOVE$(STR$(G,2),ANY" ") + CHR$(13) + _
         "Delta-T = "      + REMOVE$(STR$(DeltaT,3),ANY" ") + CHR$(13) + _
         "Exp Alpha = "    + REMOVE$(STR$(Alpha,3),ANY" ") + CHR$(13) + _
         "Exp Beta = "     + REMOVE$(STR$(Beta,3),ANY" ") + CHR$(13) + _
         "Frep = "         + REMOVE$(STR$(Frep,4),ANY" ") + " " + "(RepositionFactor$ + ")" + CHR$(13) + _
         "Initial Probes: " + PlaceInitialProbes$          + CHR$(13) + _
         "Initial Accel: " + InitialAcceleration$          + CHR$(13) + _
         "Check for Early Termination? " + CheckForEarlyTermination$ + CHR$(13) + _
         "Shrink Decision Space? "         + ShrinkDS$ + CHR$(13) +CHR$(13)

'       lResult& = MSGBOX(txt$ [, [style&], title$])

        YN% = MSGBOX(A$,%MB_YESNO,"CONFIRM RUN")

        IF YN% = %IDYES THEN RunCFO$ = "YES"

END SUB

'------

SUB StatusWindow(FunctionName$,StatusWindowHandle???)
```

```
    GRAPHIC WINDOW "Run Progress, "+FunctionName$,0.08##*ScreenWidth&,0.08##*ScreenHeight&,0.25##*ScreenWidth&,0.1##*ScreenHeight& TO StatusWindowHandle???

    GRAPHIC ATTACH StatusWindowHandle???,0,REDRAW

    GRAPHIC FONT "Lucida Console",8,0 '"Courier New",8,0 'Fixed width fonts

    GRAPHIC SET PIXEL (35,15) : GRAPHIC PRINT "  Initializing...       " : GRAPHIC REDRAW

END SUB

'------

SUB GetTestFunctionNumber(FunctionName$)

  LOCAL hDlg AS DWORD

  LOCAL N&, M%

  LOCAL FrameWidth&, FrameHeight&, BoxWidth&, BoxHeight&

  BoxWidth& = 276 : BoxHeight& = 300 : FrameWidth& = 80 : FrameHeight& = BoxHeight&-5

  DIALOG NEW 0, "CENTRAL FORCE OPTIMIZATION TEST FUNCTIONS",,, BoxWidth&, BoxHeight&, %WS_CAPTION OR %WS_SYSMENU, 0 TO hDlg

  '------------------------------------------------------------

  CONTROL ADD FRAME, hDlg, %IDC_FRAME1, "Test Functions",     5,  2, FrameWidth&, FrameHeight&
  CONTROL ADD FRAME, hDlg, %IDC_FRAME2, "GSO Test Functions", 95,  2, FrameWidth&, 255

  CONTROL ADD OPTION, hDlg, %IDC_Function_Number1, "Parrott F4",10,  14, 60, 10, %WS_GROUP OR %WS_TABSTOP
  CONTROL ADD OPTION, hDlg, %IDC_Function_Number2, "SGO"      , 10,  24, 60, 10
  CONTROL ADD OPTION, hDlg, %IDC_Function_Number3, "Goldstein-Price", 10,  34, 60, 10
  CONTROL ADD OPTION, hDlg, %IDC_Function_Number4, "Step"     , 10,  44, 60, 10
  CONTROL ADD OPTION, hDlg, %IDC_Function_Number5, "Schwefel 2.26",10,  54, 60, 10
  CONTROL ADD OPTION, hDlg, %IDC_Function_Number6, "Colville", 10,  64, 60, 10
  CONTROL ADD OPTION, hDlg, %IDC_Function_Number7, "Griewank", 10,  74, 60, 10

  CONTROL ADD OPTION, hDlg, %IDC_Function_Number31, "PBM #1",   10,  84, 60, 10
  CONTROL ADD OPTION, hDlg, %IDC_Function_Number32, "PBM #2",   10,  94, 60, 10
  CONTROL ADD OPTION, hDlg, %IDC_Function_Number33, "PBM #3",   10, 104, 60, 10
  CONTROL ADD OPTION, hDlg, %IDC_Function_Number34, "PBM #4",   10, 114, 60, 10
  CONTROL ADD OPTION, hDlg, %IDC_Function_Number35, "PBM #5",   10, 124, 60, 10
  CONTROL ADD OPTION, hDlg, %IDC_Function_Number36, "Himmelblau",10, 134, 60, 10
  CONTROL ADD OPTION, hDlg, %IDC_Function_Number37, "Reserved", 10, 144, 60, 10
  CONTROL ADD OPTION, hDlg, %IDC_Function_Number38, "Reserved", 10, 154, 60, 10
  CONTROL ADD OPTION, hDlg, %IDC_Function_Number39, "Reserved", 10, 164, 60, 10
  CONTROL ADD OPTION, hDlg, %IDC_Function_Number40, "Reserved", 10, 174, 60, 10
  CONTROL ADD OPTION, hDlg, %IDC_Function_Number41, "Reserved", 10, 184, 60, 10
  CONTROL ADD OPTION, hDlg, %IDC_Function_Number42, "Reserved", 10, 194, 60, 10
  CONTROL ADD OPTION, hDlg, %IDC_Function_Number43, "Reserved", 10, 204, 60, 10
  CONTROL ADD OPTION, hDlg, %IDC_Function_Number44, "Reserved", 10, 214, 60, 10
  CONTROL ADD OPTION, hDlg, %IDC_Function_Number45, "Reserved", 10, 224, 60, 10
  CONTROL ADD OPTION, hDlg, %IDC_Function_Number46, "Reserved", 10, 234, 60, 10
  CONTROL ADD OPTION, hDlg, %IDC_Function_Number47, "Reserved", 10, 244, 60, 10
  CONTROL ADD OPTION, hDlg, %IDC_Function_Number48, "Reserved", 10, 254, 60, 10
  CONTROL ADD OPTION, hDlg, %IDC_Function_Number49, "Reserved", 10, 264, 60, 10
  CONTROL ADD OPTION, hDlg, %IDC_Function_Number50, "Reserved", 10, 274, 60, 10

' -------------------- Test Functions from GSO Paper ------------------
  CONTROL ADD OPTION, hDlg, %IDC_Function_Number8, "f1" , 120,  14, 40, 10
  CONTROL ADD OPTION, hDlg, %IDC_Function_Number9, "f2" , 120,  24, 40, 10
  CONTROL ADD OPTION, hDlg, %IDC_Function_Number10, "f3" , 120,  34, 40, 10
  CONTROL ADD OPTION, hDlg, %IDC_Function_Number11, "f4" , 120,  44, 40, 10
  CONTROL ADD OPTION, hDlg, %IDC_Function_Number12, "f5" , 120,  54, 40, 10
  CONTROL ADD OPTION, hDlg, %IDC_Function_Number13, "f6" , 120,  64, 40, 10
  CONTROL ADD OPTION, hDlg, %IDC_Function_Number14, "f7" , 120,  74, 40, 10
  CONTROL ADD OPTION, hDlg, %IDC_Function_Number15, "f8" , 120,  84, 40, 10
  CONTROL ADD OPTION, hDlg, %IDC_Function_Number16, "f9" , 120,  94, 40, 10
  CONTROL ADD OPTION, hDlg, %IDC_Function_Number17, "f10", 120, 104, 40, 10
  CONTROL ADD OPTION, hDlg, %IDC_Function_Number18, "f11", 120, 114, 40, 10
  CONTROL ADD OPTION, hDlg, %IDC_Function_Number19, "f12", 120, 124, 40, 10
  CONTROL ADD OPTION, hDlg, %IDC_Function_Number20, "f13", 120, 134, 40, 10
  CONTROL ADD OPTION, hDlg, %IDC_Function_Number21, "f14", 120, 144, 40, 10
  CONTROL ADD OPTION, hDlg, %IDC_Function_Number22, "f15", 120, 154, 40, 10
  CONTROL ADD OPTION, hDlg, %IDC_Function_Number23, "f16", 120, 164, 40, 10
  CONTROL ADD OPTION, hDlg, %IDC_Function_Number24, "f17", 120, 174, 40, 10
  CONTROL ADD OPTION, hDlg, %IDC_Function_Number25, "f18", 120, 184, 40, 10
  CONTROL ADD OPTION, hDlg, %IDC_Function_Number26, "f19", 120, 194, 40, 10
  CONTROL ADD OPTION, hDlg, %IDC_Function_Number27, "f20", 120, 204, 40, 10
  CONTROL ADD OPTION, hDlg, %IDC_Function_Number28, "f21", 120, 214, 40, 10
  CONTROL ADD OPTION, hDlg, %IDC_Function_Number29, "f22", 120, 224, 40, 10
  CONTROL ADD OPTION, hDlg, %IDC_Function_Number30, "f23", 120, 234, 40, 10

  CONTROL SET OPTION  hDlg, %IDC_Function_Number1, %IDC_Function_Number1, %IDC_Function_Number3 'default to Parrott F4

  '------------------------------------------------------------

  CONTROL ADD BUTTON, hDlg, %IDOK, "&OK", 200, 0.45##*BoxHeight&, 50, 14

  '------------------------------------------------------------

  DIALOG SHOW MODAL hDlg CALL DlgProc

  CALL Delay(0.5##)

  IF FunctionNumber% < 1 OR FunctionNumber% > 36 THEN

    FunctionNumber = 1 : MSGBOX("Error in function number...")

  END IF

' MSGBOX("Test Function is #"+STR$(FunctionNumber%))

    SELECT CASE FunctionNumber%

      CASE 1 : FunctionName$ = "ParrottF4"
      CASE 2 : FunctionName$ = "SGO"
      CASE 3 : FunctionName$ = "GP"
      CASE 4 : FunctionName$ = "STEP"
      CASE 5 : FunctionName$ = "SCHWEFEL_226"
      CASE 6 : FunctionName$ = "COLVILLE"
      CASE 7 : FunctionName$ = "GRIEWANK"
      CASE 8 : FunctionName$ = "F1"
```

```
          CASE 9 : FunctionName$ = "F2"
          CASE 10: FunctionName$ = "F3"
          CASE 11: FunctionName$ = "F4"
          CASE 12: FunctionName$ = "F5"
          CASE 13: FunctionName$ = "F6"
          CASE 14: FunctionName$ = "F7"
          CASE 15: FunctionName$ = "F8"
          CASE 16: FunctionName$ = "F9"
          CASE 17: FunctionName$ = "F10"
          CASE 18: FunctionName$ = "F11"
          CASE 19: FunctionName$ = "F12"
          CASE 20: FunctionName$ = "F13"
          CASE 21: FunctionName$ = "F14"
          CASE 22: FunctionName$ = "F15"
          CASE 23: FunctionName$ = "F16"
          CASE 24: FunctionName$ = "F17"
          CASE 25: FunctionName$ = "F18"
          CASE 26: FunctionName$ = "F19"
          CASE 27: FunctionName$ = "F20"
          CASE 28: FunctionName$ = "F21"
          CASE 29: FunctionName$ = "F22"
          CASE 30: FunctionName$ = "F23"
          CASE 31: FunctionName$ = "PBM_1"
          CASE 32: FunctionName$ = "PBM_2"
          CASE 33: FunctionName$ = "PBM_3"
          CASE 34: FunctionName$ = "PBM_4"
          CASE 35: FunctionName$ = "PBM_5"
          CASE 36: FunctionName$ = "HIMMELBLAU"

      END SELECT

  END SUB

'----------

CALLBACK FUNCTION DlgProc() AS LONG

    '-------------------------------------------------------------------
    ' Callback procedure for the main dialog
    '-------------------------------------------------------------------
    LOCAL c, lRes AS LONG, sText AS STRING

    SELECT CASE AS LONG CBMSG

    CASE %WM_INITDIALOG' %WM_INITDIALOG is sent right before the dialog is shown.

    CASE %WM_COMMAND              ' <- a control is calling

      SELECT CASE AS LONG CBCTL  ' <- look at control's id

      CASE %IDOK                 ' <-- OK button or Enter key was pressed

          IF CBCTLMSG = %BN_CLICKED THEN
              '------------------------------------
              ' Loop through the Function_Number controls
              ' to see which one is selected
              '------------------------------------
              FOR c = %IDC_Function_Number1 TO %IDC_Function_Number50

                  CONTROL GET CHECK CBHNDL, c TO lRes

                  IF lRes THEN EXIT FOR

              NEXT 'c holds the id for selected test function.

              FunctionNumber% = c-120

'DEBUG         sText = FORMAT$(c - %IDC_Function_Number1 + 1) + $CRLF
'              MSGBOX sText, %MB_TASKMODAL, "Selected Function"

              DIALOG END CBHNDL

          END IF

      END SELECT

    END SELECT

END FUNCTION

'----------------------------- PBM ANTENNA BENCHMARK FUNCTIONS ----------------------------

'Reference for benchmarks PBM_1 through PBM_5:

'Pantoja, M F., Bretones, A. R., Martin, R. G., "Benchmark Antenna Problems for Evolutionary
'Optimization Algorithms," IEEE Trans. Antennas & Propagation, vol. 55, no. 4, April 2007,
'pp. 1111-1121

FUNCTION PBM_1(R(),Nd%,p%,j&) 'PBM Benchmark #1: Max D for Variable-Length CF Dipole

    LOCAL Z, LengthWaves, ThetaRadians AS EXT

    LOCAL N%, Nsegs%, FeedSegNum%

    LOCAL NumSegs$, FeedSeg$, HalfLength$, Radius$, ThetaDeg$, Lyne$, GainDB$

    LengthWaves  = R(p%,1,j&)

    ThetaRadians = R(p%,2,j&)

    ThetaDeg$ = REMOVE$(STR$(ROUND(ThetaRadians*Rad2Deg,2)),ANY" ")

    IF TALLY(ThetaDeg$,".") = 0 THEN ThetaDeg$ = ThetaDeg$+"."

    Nsegs% = 2*(INT(100*LengthWaves)\2)+1 '100 segs per wavelength, must be an odd #, VOLTAGE SOURCE

    FeedSegNum% = Nsegs%\2 + 1 'center segment number, VOLTAGE SOURCE

    'Nsegs% = 2*(INT(100*LengthWaves)\2) '100 segs per wavelength, must be an even # for BICONE SOURCE

    'FeedSegNum% = Nsegs%\2 'center segment number for BICONE SOURCE

    NumSegs$   = REMOVE$(STR$(Nsegs%),ANY" ")
```

```
    FeedSeg$    = REMOVE$(STR$(FeedSegNum%),ANY" ")

    HalfLength$ = REMOVE$(STR$(ROUND(LengthWaves/2##,6)),ANY" ")

    IF TALLY(HalfLength$,".") = 0 THEN HalfLength$ = HalfLength$+"."

    Radius$    = "0.00001" 'REMOVE$(STR$(ROUND(LengthWaves/1000##,6)),ANY" ")

    N% = FREEFILE

    OPEN "PBM1.NEC" FOR OUTPUT AS #N%

        PRINT #N%,"CM File: PBM1.NEC"
        PRINT #N%,"CM Run ID "+DATE$+" "+TIME$
        PRINT #N%,"CM Nd="+STR$(Nd%)+", p="STR$(p%)+", j="+STR$(j%)
        PRINT #N%,"CM R(p,1,j)="+STR$(R(p%,1,j%))+", R(p%,2,j)="+STR$(R(p%,2,j%))
        PRINT #N%,"CE"
        PRINT #N%,"GW 1,"+NumSegs$+",0.,0.,-"+HalfLength$+",0.,0.,"+HalfLength$+","+Radius$
        PRINT #N%,"GE"
        'PRINT #N%,"EX 5,1,"+FeedSeg$+",0,1.,0." 'BICONE SOURCE
        PRINT #N%,"EX 0,1,"+FeedSeg$+",0,1.,0." 'VOLTAGE SOURCE
        PRINT #N%,"FR 0,1,0,0,299.79564,0."
        PRINT #N%,"RP 0,1,1,1001,"+ThetaDeg$+",0.,0.,0.,1000." 'gain at 1000 wavelengths range
        PRINT #N%,"XQ"
        PRINT #N%,"EN"

    CLOSE #N%

'      - - ANGLES - -        - POWER GAINS -         - - POLARIZATION - - -     - - E(THETA) - - -    - - E(PHI) - - -
' THETA      PHI        VERT.    HOR.      TOTAL      AXIAL     TILT   SENSE    MAGNITUDE   PHASE     MAGNITUDE   PHASE
' DEGREES  DEGREES       DB      DB         DB        RATIO     DEG.            VOLTS/M   DEGREES     VOLTS/M   DEGREES
'   90.00     0.00       3.91  -999.99     3.91      0.00000    0.00  LINEAR   1.29504E-04   5.37   0.00000E+00   -5.24
'123456789x123456789x123456789x123456789x123456789x123456789x123456789x123456789x123456789x123456789x123456789x123456789x
'     10        20       30      40         50        60        70        80        90      100     110       120

        SHELL "n41_2k1.exe",0

    N% = FREEFILE

    OPEN "PBM1.OUT" FOR INPUT AS #N%

        WHILE NOT EOF(N%)

            LINE INPUT #N%, Lyne$

            IF INSTR(Lyne$,"DEGREES  DEGREES") > 0 THEN EXIT LOOP

        WEND 'position at next data line

        LINE INPUT #N%, Lyne$

    CLOSE #N%

    GainDB$ = REMOVE$(MID$(Lyne$,37,8),ANY" ")

    PBM_1 = 10^(VAL(GainDB$)/10##) 'Directivity

END FUNCTION 'PBM_1()

'----

FUNCTION PBM_2(R(),Nd%,p%,j%) 'PBM Benchmark #2: Max D for Variable-Separation Array of CF Dipoles

    LOCAL Z, DipoleSeparationWaves, ThetaRadians AS EXT

    LOCAL N%, i%

    LOCAL NumSegs$, FeedSeg$, Radius$, ThetaDeg$, Lyne$, GainDB$, Xcoord$, WireNum$

    DipoleSeparationWaves = R(p%,1,j%)

    ThetaRadians          = R(p%,2,j%)

    ThetaDeg$ = REMOVE$(STR$(ROUND(ThetaRadians*Rad2Deg,2)),ANY" ")

    IF TALLY(ThetaDeg$,".") = 0 THEN ThetaDeg$ = ThetaDeg$+"."

    NumSegs$ = "49"

    FeedSeg$ = "25"

    Radius$ = "0.00001"

    N% = FREEFILE

    OPEN "PBM2.NEC" FOR OUTPUT AS #N%

        PRINT #N%,"CM File: PBM2.NEC"
        PRINT #N%,"CM Run ID "+DATE$+" "+TIME$
        PRINT #N%,"CM Nd="+STR$(Nd%)+", p="STR$(p%)+", j="+STR$(j%)
        PRINT #N%,"CM R(p,1,j)="+STR$(R(p%,1,j%))+", R(p,2,j)="+STR$(R(p%,2,j%))
        PRINT #N%,"CE"

        FOR i% = -9 TO 9 STEP 2
            WireNum$ = REMOVE$(STR$((i%+11)\2),ANY" ")
            Xcoord$  = REMOVE$(STR$(i%*DipoleSeparationWaves/2##),ANY" ")
            PRINT #N%,"GW "+WireNum$+","+NumSegs$+","+Xcoord$+",0.,-0.25,"+Xcoord$+",0.,0.25,"+Radius$
        NEXT i%

        PRINT #N%,"GE"

        FOR i% = 1 TO 10
            PRINT #N%,"EX 0,"+REMOVE$(STR$(i%),ANY" ")+",0,"+FeedSeg$+",0,1.,0." 'VOLTAGE SOURCE
        NEXT i%
        PRINT #N%,"FR 0,1,0,0,299.79564,0."
        PRINT #N%,"RP 0,1,1,1001,"+ThetaDeg$+",90.,0.,0.,1000." 'gain at 1000 wavelengths range
        PRINT #N%,"XQ"
        PRINT #N%,"EN"

    CLOSE #N%

'      - - ANGLES - -        - POWER GAINS -         - - POLARIZATION - - -     - - E(THETA) - - -    - - E(PHI) - - -
' THETA      PHI        VERT.    HOR.      TOTAL      AXIAL     TILT   SENSE    MAGNITUDE   PHASE     MAGNITUDE   PHASE
' DEGREES  DEGREES       DB      DB         DB        RATIO     DEG.            VOLTS/M   DEGREES     VOLTS/M   DEGREES
```

```
'    90.00     0.00      3.91 -999.99    3.91    0.00000     0.00 LINEAR    1.29504E-04     5.37    0.00000E+00    -5.24
'123456789x123456789x123456789x123456789x123456789x123456789x123456789x123456789x123456789x123456789x123456789x123456789x
'     10       20        30        40        50        60        70        80        90       100       110       120

      SHELL "n41_2k1.exe",0

      N% = FREEFILE

      OPEN "PBM2.OUT" FOR INPUT AS #N%

          WHILE NOT EOF(N%)

              LINE INPUT #N%, Lyne$

              IF INSTR(Lyne$,"DEGREES  DEGREES") > 0 THEN EXIT LOOP

          WEND 'position at next data line

          LINE INPUT #N%, Lyne$

      CLOSE #N%

      GainDB$ = REMOVE$(MID$(Lyne$,37,8),ANY" ")

      IF AddNoiseToPBM2$ = "YES" THEN

          Z = 10^(VAL(GainDB$)/10##) + GaussianDeviate(0##,0.4472##) 'Directivity with Gaussian noise (zero mean, 0.2 variance)

      ELSE

          Z = 10^(VAL(GainDB$)/10##) 'Directivity without noise

      END IF

      PBM_2 = Z

END FUNCTION 'PBM_2()

'----

FUNCTION PBM_3(R(),Nd%,p%,j&) 'PBM Benchmark #3: Max D for Circular Dipole Array

      LOCAL Beta, ThetaRadians, Alpha, ReV, ImV AS EXT

      LOCAL N%, i%

      LOCAL NumSegs$, FeedSeg$, Radius$, ThetaDeg$, Lyne$, GainDB$, Xcoord$, Ycoord$, WireNum$, ReEX$, ImEX$

      Beta          = R(p%,1,j&)

      ThetaRadians = R(p%,2,j&)

      ThetaDeg$ = REMOVE$(STR$(ROUND(ThetaRadians*Rad2Deg,2)),ANY" ")

      IF TALLY(ThetaDeg$,".") = 0 THEN ThetaDeg$ = ThetaDeg$+"."

      NumSegs$ = "49"

      FeedSeg$ = "25"

      Radius$  = "0.00001"

      N% = FREEFILE

      OPEN "PBM3.NEC" FOR OUTPUT AS #N%

          PRINT #N%,"CM File: PBM3.NEC"
          PRINT #N%,"CM Run ID "+DATE$+" "+TIME$
          PRINT #N%,"CM Nd="+STR$(Nd%)+", p="STR$(p%)+", j="+STR$(j&)
          PRINT #N%,"CM R(p,1,j)="+STR$(R(p%,1,j&))+", R(p,2,j)="+STR$(R(p%,2,j&))
          PRINT #N%,"CE"

          FOR i% = 1 TO 8
              WireNum$ = REMOVE$(STR$(i%),ANY" ")

              SELECT CASE i%
                  CASE 1 : Xcoord$ = "1"         : Ycoord$ = "0"
                  CASE 2 : Xcoord$ = "0.70711"   : Ycoord$ = "0.70711"
                  CASE 3 : Xcoord$ = "0"         : Ycoord$ = "1"
                  CASE 4 : Xcoord$ = "-0.70711"  : Ycoord$ = "0.70711"
                  CASE 5 : Xcoord$ = "-1"        : Ycoord$ = "0"
                  CASE 6 : Xcoord$ = "-0.70711"  : Ycoord$ = "-0.70711"
                  CASE 7 : Xcoord$ = "0"         : Ycoord$ = "-1"
                  CASE 8 : Xcoord$ = "0.70711"   : Ycoord$ = "-0.70711"
              END SELECT

              PRINT #N%,"GW "+WireNum$+","+NumSegs$+","+Xcoord$+","+Ycoord$+",-0.25,"+Xcoord$+","Ycoord$+",0.25,"+Radius$
          NEXT i%

          PRINT #N%,"GE"

          FOR i% = 1 TO 8
              Alpha = -COS(TwoPi*Beta*(i%-1))

              ReV = COS(Alpha)
              ImV = SIN(Alpha)

              ReEX$ = REMOVE$(STR$(ROUND(ReV,6)),ANY" ")
              ImEX$ = REMOVE$(STR$(ROUND(ImV,6)),ANY" ")

              IF TALLY(ReEX$,".") = 0 THEN ReEX$ = ReEX$+"."
              IF TALLY(ImEX$,".") = 0 THEN ImEX$ = ImEX$+"."

              PRINT #N%,"EX 0,"+REMOVE$(STR$(i%),ANY" ")+","+FeedSeg$+",0,"+ReEX$+","+ImEX$ 'VOLTAGE SOURCE
          NEXT i%

          PRINT #N%,"FR 0,1,0,0,299.79564,0."
          PRINT #N%,"RP 0,1,1,1001,"+ThetaDeg$+",0.,0.,0.,1000." 'gain at 1000 wavelengths range
          PRINT #N%,"XQ"
          PRINT #N%,"EN"

      CLOSE #N%

'    - - ANGLES - -      - POWER GAINS -      - - - POLARIZATION - - -    - - - E(THETA) - - -    - - - E(PHI) - - -
```

```
'  THETA     PHI       VERT.   HOR.    TOTAL    AXIAL    TILT   SENSE    MAGNITUDE   PHASE    MAGNITUDE   PHASE
' DEGREES  DEGREES      DB      DB       DB      RATIO    DEG.             VOLTS/M    DEGREES   VOLTS/M    DEGREES
'  90.00     0.00      3.91  -999.99    3.91   0.00000    0.00  LINEAR   1.29504E-04    5.37   0.00000E+00   -5.24
'123456789x123456789x123456789x123456789x123456789x123456789x123456789x123456789x123456789x123456789x123456789x
'      10        20        30        40        50        60        70        80        90       100       110       120

    SHELL "n41_2k1.exe",0

    N% = FREEFILE

    OPEN "PBM3.OUT" FOR INPUT AS #N%

        WHILE NOT EOF(N%)

            LINE INPUT #N%, Lyne$

            IF INSTR(Lyne$,"DEGREES  DEGREES") > 0 THEN EXIT LOOP

        WEND 'position at next data line

        LINE INPUT #N%, Lyne$

    CLOSE #N%

    GainDB$ = REMOVE$(MID$(Lyne$,37,8),ANY" ")

    PBM_3 = 10^(VAL(GainDB$)/10##) 'Directivity

END FUNCTION 'PBM_3()

'----

FUNCTION PBM_4(R(),Nd%,p%,j&) 'PBM Benchmark #4: Max D for Vee Dipole

    LOCAL TotalLengthWaves, AlphaRadians, ArmLength, Xlength, Zlength, Lfeed AS EXT

    LOCAL N%, i%, Nsegs%, FeedZcoord$

    LOCAL NumSegs$, Lyne$, GainDB$, Xcoord$, Zcoord$

    TotalLengthWaves = 2##*R(p%,1,j&)

    AlphaRadians    = R(p%,2,j&)

    Lfeed           = 0.01##

    FeedZcoord$     = REMOVE$(STR$(Lfeed),ANY" ")

    ArmLength = (TotalLengthWaves-2##*Lfeed)/2##

    Xlength = ROUND(ArmLength*COS(AlphaRadians),6)

    Xcoord$ = REMOVE$(STR$(Xlength),ANY" ") : IF TALLY(Xcoord$,".") = 0 THEN Xcoord$ = Xcoord$+"."

    Zlength = ROUND(ArmLength*SIN(AlphaRadians),6)

    Zcoord$ = REMOVE$(STR$(Zlength+Lfeed),ANY" ") : IF TALLY(Zcoord$,".") = 0 THEN Zcoord$ = Zcoord$+"."

    Nsegs% = 2*(INT(TotalLengthWaves*100)\2) 'even number, total # segs

    NumSegs$ = REMOVE$(STR$(Nsegs%\2),ANY" ") '# segs per arm

    N% = FREEFILE

    OPEN "PBM4.NEC" FOR OUTPUT AS #N%

        PRINT #N%,"CM File: PBM4.NEC"
        PRINT #N%,"CM Run ID "+DATE$+" "+TIME$
        PRINT #N%,"CM Nd="+STR$(Nd%)+", p="STR$(p%)+", j="+STR$(j&)
        PRINT #N%,"CM R(p,1,j)="+STR$(R(p%,1,j&))+", R(p,2,j)="+STR$(R(p%,2,j&))
        PRINT #N%,"CE"

        PRINT #N%,"GW 1,5,0.,0.,-"+FeedZcoord$+",0.,0.,"+FeedZcoord$+",0.00001" 'feed wire, 1 segment, 0.01 wvln

        PRINT #N%,"GW 2,"+NumSegs$+",0.,0.,"+FeedZcoord$+","+Xcoord$+",0.,"+Zcoord$+",0.00001" 'upper arm

        PRINT #N%,"GW 3,"+NumSegs$+",0.,0.,-"+FeedZcoord$+","+Xcoord$+",0.,-"+Zcoord$+",0.00001" 'lower arm

        PRINT #N%,"GE"

        PRINT #N%,"EX 0,1,3,0,1.,0." 'VOLTAGE SOURCE

        PRINT #N%,"FR 0,1,0,0,299.79564,0."
        PRINT #N%,"RP 0,1,1,1001,90.,0.,0.,0.,1000." 'ENDFIRE gain at 1000 wavelengths range
        PRINT #N%,"XQ"
        PRINT #N%,"EN"

    CLOSE #N%

'  - - ANGLES - -        - POWER GAINS -    - - - POLARIZATION - - -   - - - E(THETA) - - -   - - - E(PHI) - - -
'  THETA     PHI       VERT.   HOR.    TOTAL    AXIAL    TILT   SENSE    MAGNITUDE   PHASE    MAGNITUDE   PHASE
' DEGREES  DEGREES      DB      DB       DB      RATIO    DEG.             VOLTS/M    DEGREES   VOLTS/M    DEGREES
'  90.00     0.00      3.91  -999.99    3.91   0.00000    0.00  LINEAR   1.29504E-04    5.37   0.00000E+00   -5.24
'123456789x123456789x123456789x123456789x123456789x123456789x123456789x123456789x123456789x123456789x123456789x
'      10        20        30        40        50        60        70        80        90       100       110       120

    SHELL "n41_2k1.exe",0

    N% = FREEFILE

    OPEN "PBM4.OUT" FOR INPUT AS #N%

        WHILE NOT EOF(N%)

            LINE INPUT #N%, Lyne$

            IF INSTR(Lyne$,"DEGREES  DEGREES") > 0 THEN EXIT LOOP

        WEND 'position at next data line

        LINE INPUT #N%, Lyne$

    CLOSE #N%
```

```
    GainDB$ = REMOVE$(MID$(Lyne$,37,8),ANY" ")

    PBM_4 = 10^(VAL(GainDB$)/10##) 'Directivity

END FUNCTION 'PBM_4()

'----

FUNCTION PBM_5(R(),Nd%,p%,j&) 'PBM Benchmark #5: N-element collinear array (Nd=N-1)

    LOCAL TotalLengthWaves, Di(), Ystart, Y1, Y2, SumDi AS EXT

    LOCAL N%, i%, q%

    LOCAL Lyne$, GainDB$

    REDIM Di(1 TO Nd%)

    FOR i% = 1 TO Nd%

        Di(i%) = R(p%,i%,j&) 'dipole separation, wavelengths

        'MSGBOX"R="+STR$(R(p%,i%,j&))+" p="+STR$(p%)+" i="+STR$(i%)+" j="+STR$(j&)+" Nd="+STR$(Nd%))

    NEXT i%

    TotalLengthWaves = 0##

    FOR i% = 1 TO Nd%

        TotalLengthwaves = TotalLengthWaves + Di(i%)

    NEXT i%

    TotalLengthWaves = TotalLengthWaves + 0.5## 'add half-wavelength of 1 meter at 299.8 MHz

    Ystart = -TotalLengthWaves/2##

    N% = FREEFILE

    OPEN "PBM5.NEC" FOR OUTPUT AS #N%

        PRINT #N%,"CM File: PBM5.NEC"
        PRINT #N%,"CM Run ID "+DATE$+" "+TIME$
        PRINT #N%,"CM Nd="+STR$(Nd%)+", p="+STR$(p%)+", j="+STR$(j&)
        PRINT #N%,"CM R(p,1,j)="+STR$(R(p%,1,j&))+", R(p,2,j)="+STR$(R(p%,2,j&))
        PRINT #N%,"CE"

        FOR i% = 1 TO Nd%+1

            SumDi = 0##

            FOR q% = 1 TO i%-1

                SumDi = SumDi + Di(q%)

            NEXT q%

            Y1 = ROUND(Ystart + SumDi,6)

            Y2 = ROUND(Y1+0.5##,6) 'add one-half wavelength for other end of dipole

            PRINT #N%,"GW "+REMOVE$(STR$(i%),ANY" ")+",49,0.,"+REMOVE$(STR$(Y1),ANY" ")+",0.,0.,"+REMOVE$(STR$(Y2),ANY" ")+",0.,0.00001"

        NEXT i%

        PRINT #N%,"GE"

        FOR i% = 1 TO Nd%+1
            PRINT #N%,"EX 0,"+REMOVE$(STR$(i%),ANY" ")+",25,0.,1.,0." 'VOLTAGE SOURCES
        NEXT i%

        PRINT #N%,"FR 0,1,0,0,299.79564,0."
        PRINT #N%,"RP 0,1,1,1001,90.,0.,0.,0.,1000." 'gain at 1000 wavelengths range
        PRINT #N%,"XQ"
        PRINT #N%,"EN"

    CLOSE #N%
'   - - ANGLES - -      - POWER GAINS -      - - POLARIZATION - - -   - - - E(THETA) - - -   - - - E(PHI) - - -
'  THETA    PHI       VERT.   HOR.   TOTAL     AXIAL   TILT SENSE   MAGNITUDE     PHASE      MAGNITUDE     PHASE
' DEGREES DEGREES      DB      DB      DB       RATIO   DEG.         VOLTS/M     DEGREES       VOLTS/M     DEGREES
'  90.00    0.00      3.91 -999.99   3.91     0.00000   0.00 LINEAR 1.29504E-04    5.37     0.00000E+00    -5.24
'123456789x123456789x123456789x123456789x123456789x123456789x123456789x123456789x123456789x123456789x123456789x
'        10        20        30        40        50        60        70        80        90       100       110       120

    SHELL "n41_2k1.exe",0

    N% = FREEFILE

    OPEN "PBM5.OUT" FOR INPUT AS #N%

        WHILE NOT EOF(N%)

            LINE INPUT #N%, Lyne$

            IF INSTR(Lyne$,"DEGREES  DEGREES") > 0 THEN EXIT LOOP

        WEND 'position at next data line

        LINE INPUT #N%, Lyne$

    CLOSE #N%

    GainDB$ = REMOVE$(MID$(Lyne$,37,8),ANY" ")

    PBM_5 = 10^(VAL(GainDB$)/10##) 'Directivity

END FUNCTION 'PBM_5()

'************************************************** END PROGRAM 'CFO_12-25-09.BAS'  **************************************************************
```